%% file: PBH.tex
\documentclass[%
aps,
onecolumn,
12pt,
tightenlines,
nofootinbib,
superscriptaddress,
floatfix,
prd,
]{revtex4-2}

\input{PBH-header}

\begin{document}

\title{\Large {\huge P}rimordial {\huge B}lack {\huge H}oles}

\author{Albert Escriv{\`a}}
\email{escriva.manas.albert.y0@a.mail.nagoya-u.ac.jp}
\affiliation{Department of Physics, 
    Nagoya University, 
    Furo-cho Chikusa-ku,
    Nagoya 464-8602,
    Japan
    }
\author{Florian K{\"u}hnel}
\email{fkuehnel@mpp.mpg.de}
\affiliation{
	Max-Planck-Institut f{\"u}r Physik,
	Boltzmannstr.~8,
	85748 Garching,
	Germany,}
\affiliation{
	Arnold Sommerfeld Center,
	Ludwig-Maximilians-Universit{\"a}t,
	Theresienstra{\ss}e 37,
	80333 M{\"u}nchen,
	Germany}
\author{Yuichiro Tada}
\email{tada.yuichiro.y8@f.mail.nagoya-u.ac.jp}
\affiliation{Institute for Advanced Research, 
    Nagoya University,
    Furo-cho Chikusa-ku, 
    Nagoya 464-8601, 
    Japan
    }
\affiliation{Department of Physics, 
    Nagoya University, 
    Furo-cho Chikusa-ku,
    Nagoya 464-8602,
    Japan
    }
\affiliation{Theory Center, 
    IPNS, KEK, 1-1 Oho, Tsukuba,
    Ibaraki 305-0801,
    Japan
    }

\date{\formatdate{\day}{\month}{\year}, \currenttime}

\begin{abstract}
\vs{4mm}
\begin{tcolorbox}
    Aspects of primordial black holes, \ie~black holes formed in the early Universe, are reviewed. Special emphasis is put on their formation, their r{\^o}le as dark matter candidates and their manifold signatures, particularly through gravitational waves.
\end{tcolorbox}
\end{abstract}

\begin{center}
    \phantom{\fontsize{50}{50}\selectfont I}\\
    {\fontsize{40}{10}\selectfont {\fontsize{50}{20}\selectfont P}rimordial\\[14mm]
    {\fontsize{50}{20}\selectfont B}lack\,
    {\fontsize{50}{20}\selectfont H}oles}\\[100mm]
    {\noindent\makebox[\linewidth]{\resizebox{0.3333\linewidth}{1pt}{$\bullet$}}\bigskip}\\[5mm]
    {\fontsize{18}{5}\selectfont Albert Escriv{\`a}}\\[8mm]
    {\fontsize{18}{5}\selectfont Florian K{\"u}hnel}\\[8mm]
    {\fontsize{18}{5}\selectfont Yuichiro Tada}
\end{center}
\newpage
\maketitle

\newpage
\tableofcontents
\newpage

\section{Introduction}
\label{sec:Introduction}
\vs{-3mm}
\lettrine[lines=3, slope=0.0em, findent=0em, nindent=0.2em, lhang=0.1, loversize=0.1]{B}{} lack holes formed in the early Universe through a non-stellar way are called {\it primordial black holes} (PBHs). After an initial negative and erroneous discussion in the late 1960s by Zel'dovich and Novikov~\cite{1967SvA....10..602Z}, the first solid and ground-breaking work on PBHs has been put forward by Hawking~\cite{Hawking:1971ei} and Carr \& Hawking~\cite{1974MNRAS.168..399C, 1975ApJ...201....1C} in the early 1970s. Soon afterwards, it was realised that PBHs could constitute (possibly all of the) dark matter~\cite{Chapline:1975ojl} (see References~\cite{2016PhRvD..94h3504C, 2017JPhCS.840a2032G, 2020ARNPS..70..355C, 2021RPPh...84k6902C, 2021JPhG...48d3001G, Carr:2021bzv} for reviews). This exciting possibility was further substantiated when it was understood that PBHs are a rather natural consequence of many inflationary scenarios~\cite{1993PhRvD..48..543C, 1994PhRvD..50.7173I, 1996PhRvD..54.6040G, 1996NuPhB.472..377R}. Further strong recent tailwind came from the milestone discovery of black hole mergers by the Laser Interferometer Gravitational-Wave Observatory (LIGO) and Virgo~\cite{2016PhRvX...6d1015A, 2023PhRvX..13d1039A}, which could conceivably have primordial origin~\cite{Murgia:2019duy}.

Despite its importance, the conundrum of the origin of the dark matter is by far not the only one which primordial black holes could naturally resolve. Particularly, they could explain
\vs{-0.5mm}
\begin{tcolorbox}
\begin{flushleft}
\begin{enumerate}

    \item
        microlensing events towards the Galactic bulge generated by planetary-mass objects with about $1\mspace{0.5mu}\%$ of the dark matter density~\cite{2019PhRvD..99h3503N}, well above expectations for free-floating planets;

    \item
        microlensing of quasars~\cite{2017ApJ...836L..18M}, including those that are so mis-\\
        aligned with the lensing galaxy that the probability of lensing\\
        by a star is very low;

    \item
        the unexpected high number of microlensing events towards\\
        the Galactic bulge by dark objects in the mass gap between approximately $2$ and $5\.\Msun$~\cite{2020A&A...636A..20W}, where stellar evolution models\\
        fail to form black holes~\cite{1999astro.ph..9270B}; 

    \item
        unexplained correlations in the source-subtracted X-ray and\\
        cosmic infrared-background fluctuations~\cite{2005Natur.438...45K};

    \item
        the non-observation of ultrafaint dwarf galaxies below some\\
        critical radius~\cite{2018PDU....22..137C}; 

    \item
        the recent detection of galaxies at high redshifts (above $z = 10$; possibly up to $z \approx 18$~\cite{2023Natur.616..266L}), being in increasing tension with standard particle dark matter cosmologies;

\end{enumerate}
\end{flushleft}
\end{tcolorbox}

\newpage

\begin{tcolorbox}
\begin{flushleft}
\begin{enumerate}\addtocounter{enumi}{+6}
    \item
        the recently-observed supernova population which does\\
        not trace the stellar density, but well follows the expected distribution of dark matter/white dwarf interactions~\cite{2022arXiv221100013S};

    \item
        the masses, spins and coalescence rates for the black holes\\
        found by LIGO/Virgo~\cite{2021arXiv211103634T}, including multiple events with\\
        black holes in the upper or lower mass gaps; 

    \item
        the unexplained relationship between the mass of a galaxy\\
        and that of its central black hole.

\end{enumerate}
\end{flushleft}
\vs{-2mm}
\end{tcolorbox}
\vs{2mm}

It is remarkable that all of the above conundra are fully explained through the thermal history of the Universe as has been pointed out in Reference~\cite{Carr:2019kxo}. Therein various events which change the number of relativistic degrees of freedom, such as the Quantum Chromo Dynamics (QCD) phase transition\./\.cross over, naturally induce peaks in the PBH mass function around planetary mass, a solar mass, a few ten solar masses and around a million solar masses. Indeed, PBHs could serve as probes for the physics at those epochs, helping for instance to understand the nature of cosmic phase transitions/crossovers~\cite{2021PhRvD.103f3506B}. Furthermore, they provide a link to the physics of inflation, thereby allowing us to study the earliest times.

Of course, there is a distinction between primordial black holes constituting the entirety of the dark matter or part of it. Dark matter could well be both {\it micro-} and {\it macroscopic}, with very rich interplay (\cf~Reference~\cite{2021MNRAS.506.3648C}). Trivially, whenever PBHs are not $100\mspace{0.5mu}\%$ of the dark matter, the latter necessarily consists of at least one additional ingredient. In the case of particles, these could conceivably be weakly-interacting massive particles (WIMPs) (see Reference~\cite{2018RPPh...81f6201R} for a review), sterile neutrinos~\cite{2005PhLB..620...17A, 2013PhRvD..87i3006C}, axions and axion-like particles~\cite{1978PhRvL..40..223w, 1978PhRvL..40..279W, 1977PhRvL..38.1440P} as well as ultralight bosons~\cite{2000PhRvL..85.1158H}. All of those have been subject to intense dark matter studies, allowing us to formulate stringent constraints on fundamental parameters (\cf~Chapter 27 of Reference~\cite{ParticleDataGroup:2020ssz} for a review). Soon after PBH formation, the particles would form halos around the holes, leading to amplified detection signatures, for instance through an enhanced annihilation rate in the case of WIMPs (\cf~Reference~\cite{2016AstL...42..347E}). In fact, the constraints are so strong for solar-mass PBHs that, in case these were responsible for the LIGO/Virgo mergers, all standard WIMP scenarios would be ruled out~\cite{2019PhRvD.100b3506A, 2021MNRAS.506.3648C}.\footnote{\setstretch{0.9}In fact, there is even a parameter window in which neither WIMPs nor PBHs could be a dominant dark matter candidate, hence pointing even to yet a third dark matter component~\cite{2021MNRAS.506.3648C}.}
\newpage

Many constraints on the primordial black holes abundance have been discussed throughout the past decades, being due to evaporation, gravitational lensing, disruption or are dynamical in nature (see Reference~\cite{2020ARNPS..70..355C} for a recent review). However, all constraints are subject to multiple assumptions, such that if relaxed, the associated constraints might be significantly weakened or even disappear entirely. More reliable than constraints are positive evidences which include those connected to the nine conundra listed above. Particularly, microlensing surveys~(see \eg~References~\cite{2019PhRvD..99h3503N, 2020A&A...633A.107H, 2022MNRAS.512.5706H}) as well as high-redshift observations~(see, for instance, References~\cite{2022ApJ...926..205C, 2022ApJ...937L..30L}) give outstanding support to the PBH dark matter hypothesis, such that it only appears to be a question of time when the first confirmed detection of a primordial black hole will be made. In fact, the LIGO/Virgo observations allow us to already identify several candidate mergers in each of which at least one of the progenitors has subsolar mass~\cite{2021arXiv210511449P, 2022arXiv221201477T}{\,---\,}an extraordinary hint for a primordial nature.\footnote{\setstretch{0.9}The recent Reference~\cite{Carr:2023tpt} extensively discusses the manifold strong positive evidences for PBHs.}

Undoubtedly, the discovery of primordial black holes will bring us much information about the early Universe and high-energy physics. For example, if primordial black holes are generated by large primordial perturbations during cosmic inflation, the specific nature of the inflationary dynamics can be revealed in an overwhelmingly clear way. To this end, one must precisely know how the physics of the early Universe leads to the formation of primordial black holes, to which we devote a large part of this review.

Let us finally remark that{\,---\,}contrary to some folklore which says that primordial black hole dark matter requires significant fine-tuning, particularly of the inflaton potential{\,---\,}they are in fact a likely outcome of many cosmological and particle-physics models. There are numerous scenarios (see Section~\ref{sec:Formation-1}) in which PBHs neither 
    ({\it i}$\mspace{1.5mu}$) 
        originate from density fluctuations of inflationary origin nor
    ({\it ii}$\mspace{1.5mu}$) 
        are exponentially sensitive to the choice of the model parameters (such as in the quark-confinement scenario~\cite{2021PhRvD.104l3507D}; see Section~\ref{sec:Quark-Confinement}). It is clearly not less natural to have PBHs than particles as dark matter. In fact, since black holes are maximal capacitors of information (see Reference~\cite{2021JHEP...03..126D}), which can be entirely read out over their lifetime~\cite{Dvali:2011aa}, they are by far {\it the} dark matter candidate which allows to learn most of their formation environment, at times inaccessible by any other means.

This review is devoted to an introduction and detailed discussion of the formation, signatures and observational hints to and prospects of the conceivably most natural dark matter candidate{\,---\,}{\it primordial black holes}.
\newpage
\vphantom{.}
\newpage

\section{Formation}
\label{sec:Formation-1}
\vs{-3mm}
\lettrine[lines=3, slope=0em, findent=0em, nindent=0.2em, lhang=0.1, loversize=0.1]{C}{}\,ollapse of superhorizon fluctuations is one of the main origins of primordial black holes. Since their advent by Hawking and Carr~\cite{Hawking:1971ei, 1974MNRAS.168..399C, 1975ApJ...201....1C}, the physics of overdensity collapse has been extensively studied and significantly developed. Here we put special focus on the development and application of a state-of-the-art PBH formation criterion as well as on the nature of large fluctuations generated by cosmic inflation. Several other primordial black hole formation mechanisms are also summarised.
\vs{2mm}

\subsection{Collapse of Inflationary Perturbations}
\label{sec:Collapse-of-Inflationary-Perturbations}
\vs{-1mm}
As already pointed out, primordial black holes could have been formed in the early Universe, for instance in the era of radiation domination due to the gravitational collapse of large curvature perturbations generated during inflation~\cite{1974MNRAS.168..399C, Hawking:1971ei}. 

This scenario assumes an enhancement of the primordial fluctuations at small scales with an amplitude significantly above the value required to match cosmic microwave background (CMB) observations at larger scales. From the standard peak theory of Gau{\ss}ian random fields~\cite{1986ApJ...304...15B}, those large (and rare) peaks of the primordial fluctuations are nearly spherical. It is therefore a good approximation to consider those cosmological fluctuations which gravitationally collapse to black holes as spherical.\footnote{\setstretch{0.9}In Section~\ref{sec:Threshold-Values} we will briefly discuss the effect of deviations from sphericity on primordial black hole formation.}

After being redshifted outside of the Hubble horizon during inflation, the fluctuations remain frozen (\ie~the gauge-invariant comoving curvature perturbations remain constant) until in a later era (for instance radiation domination) their scales become sub-Hubble again. If those perturbations exceed a (shape- and environment-dependent) threshold (see Section~\ref{sec:Threshold--Estimation-Scheme}), they will start to collapse and in turn, form a black hole; otherwise they will disperse because of pressure gradients which prevent the collapse. Therefore, those gradients crucially influence the collapse threshold for the perturbation. As we will see in Section~\ref{sec:Statistics}, the abundance of PBHs is exponentially sensitive to the threshold value, and hence to the environment within which the collapse takes place.
\newpage

\begin{figure}[t]
    \centering
    \includegraphics[width = 0.23675\hsize]{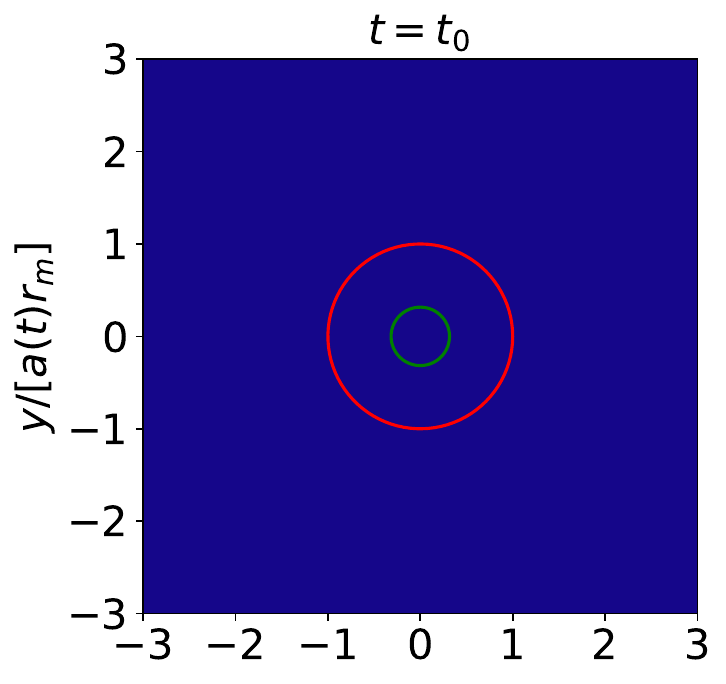}
    \includegraphics[width = 0.22\hsize]{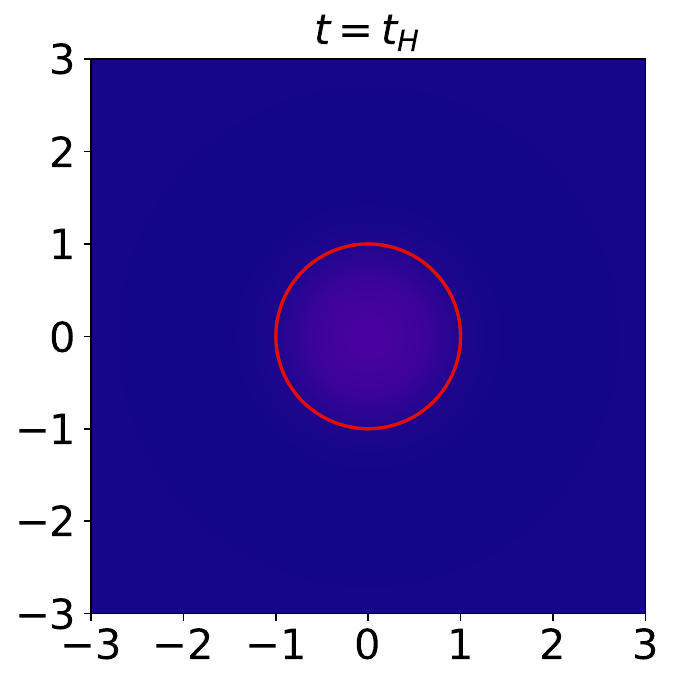}
    \includegraphics[width = 0.22\hsize]{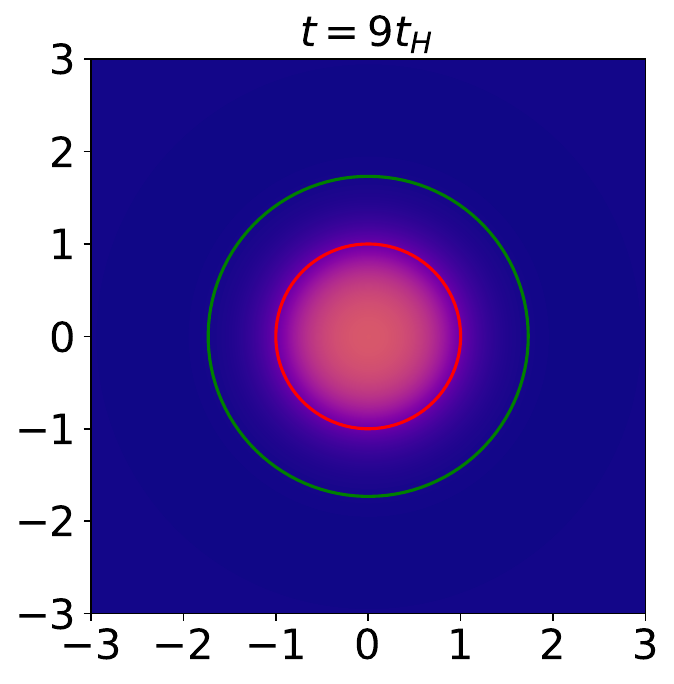}
    \includegraphics[width = 0.2675\hsize]{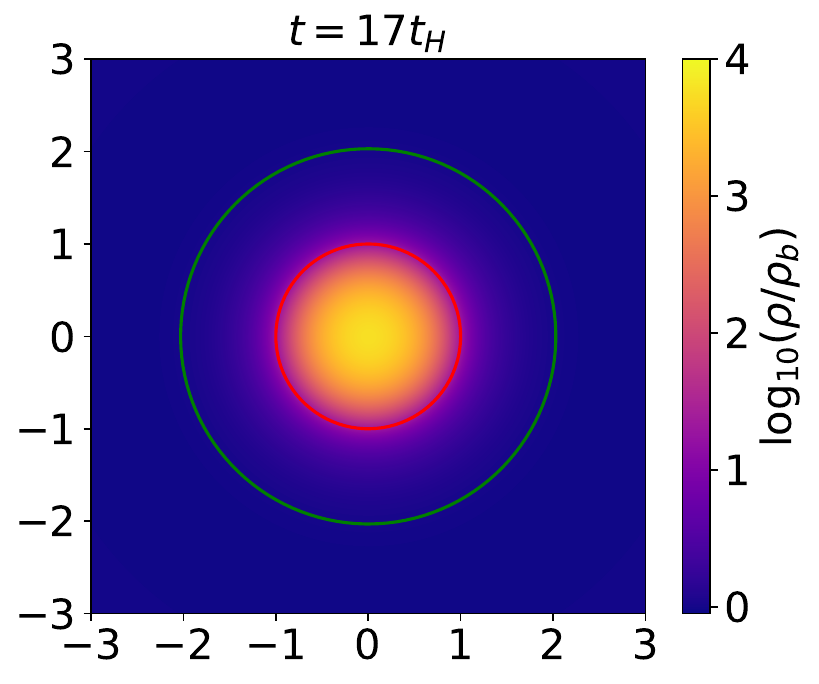}\\[3mm]
    \includegraphics[width = 0.23675\hsize]{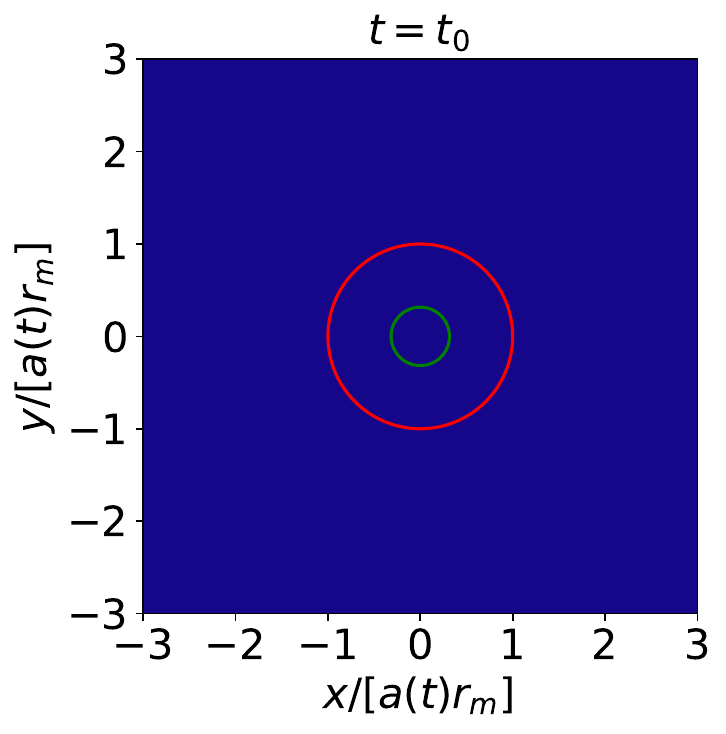}
    \includegraphics[width = 0.22\hsize]{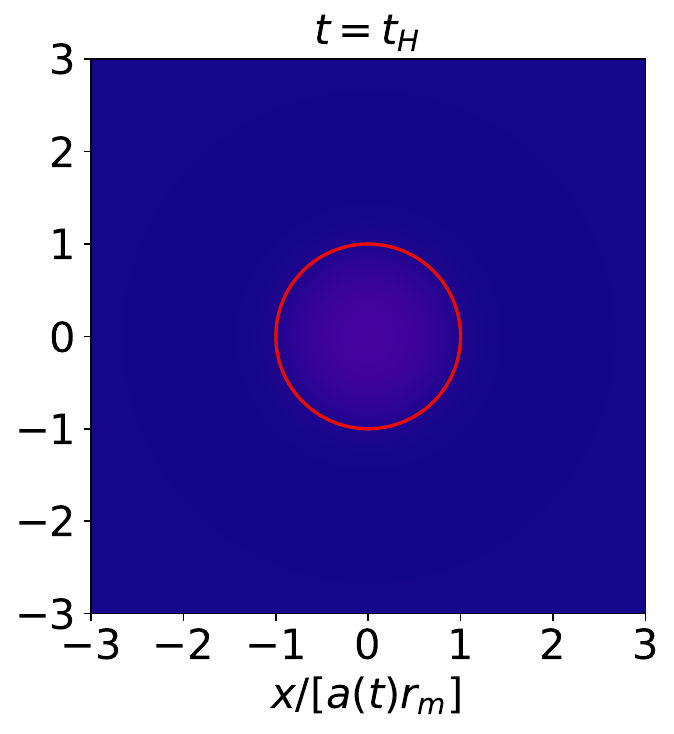}
    \includegraphics[width = 0.22\hsize]{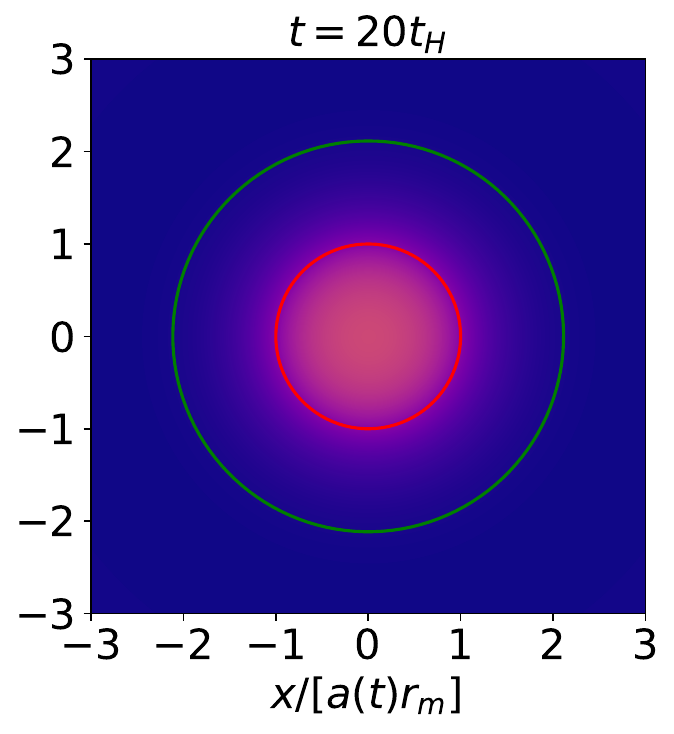}
    \includegraphics[width = 0.2675\hsize]{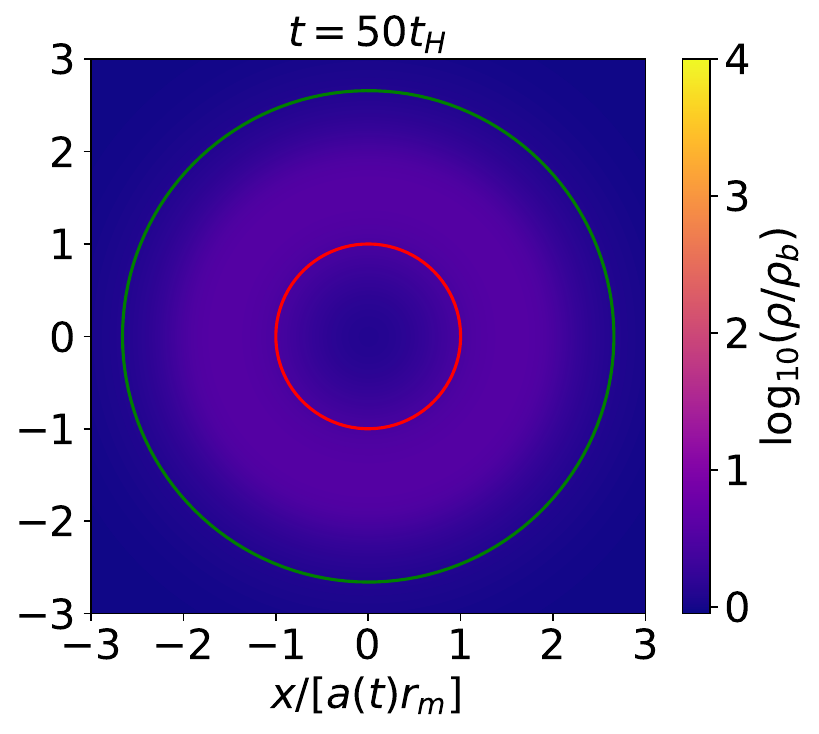}\\
    \caption{
        Snapshots of the time evolution of a Gau{\ss}ian energy-density fluctuation for different times (given in terms of $t_{H}$, which is the time when the fluctuation reenters the cosmological horizon, see Section~\ref{sec:Collapse--Threshold-Definition} for details). The {\it top panels} illustrate this for the case of a supercritical perturbation; the {\it bottom panels} show a subcritical perturbation. The green line represents the cosmological Hubble horizon, while the red one depicts the comoving size $a( t )\.r_{\mrm}$ of the fluctuation [where $r_{\mrm}$ is defined at the initial time $t = t_{0}$, see Equation~\eqref{eq:rm-eq}]. The numerical simulation for this figure utilises the code of Reference~\cite{2020PDU....2700466E}. The plotted magnitude is the ratio between the energy density $\rho$ of the fluid in terms of the energy density $\rho_{\brm}$ of the cosmological background in log scale.
        \vs{3mm}
        }
    \label{fig:collapse-fluc}
\end{figure}

Before entering into details, we illustrate the collapse dynamics in Figure~\ref{fig:collapse-fluc}. As an example, we chose a spherical collapse of a Gau{\ss}ian fluctuation that initially starts on superhorizon scales. Its amplitude determines whether it collapses into a black hole. The top panels of this figure show the case of a {\it supercritical\.} fluctuation (whose amplitude exceeds the threshold). In this case, the energy density continuously increases until formation of an apparent horizon, marking the black hole formation time. In the bottom panels, we depict the opposite case with a {\it subcritical\.} fluctuation (whose amplitude is smaller than the threshold). In this case, the fluctuation seemingly starts a collapse; its energy density is first increasing, but then starts to disperse and smoothens out onto the Friedmann--Lema{\^i}tre--Robertson--Walker (FLRW) background.

Note that in both cases, the length scale of the fluctuation starts on superhorizon scales (red circle), but the Hubble horizon (green circle) increases continuously such that at some point the fluctuation reenters the horizon. 

\newpage

\subsection{Gravitational-Collapse Equations}
\label{sec:Gravitational--Collapse-Equations}
\vs{-1mm}
In this Subsection, we focus on the dynamics of cosmological fluctuations which are initially on superhorizon scales and eventually collapse to black holes.

Following the standard approach, we consider the approximation in which the Universe is filled by a perfect fluid with the equation of state $p = w\mspace{1.5mu}\rho$, with constant parameter $w$, yielding the energy-momentum tensor
\vs{-1mm}
\begin{align}
\label{eq:tensor-energy}
    T^{\mu \nu} 
        = 
            ( p + \rho )\.
            u^{\mu}u^{\nu}
            +
            p\.g^{\mu\nu}
            \, .
\end{align}
Here, $p$ is the pressure, $\rho$ is the energy density, $g^{\mu\nu}$ and $u^{\mu}$ are the components of the spacetime metric and of the four-velocity, respectively. Under the assumption of spherical symmetry, the spacetime metric can be written as 
\begin{align}
\label{eq:2-metricsharp}
    \d s^{2}
        = 
            - A( r,\mspace{1.5mu}t )^{2}\.\d t^{2}
            + B( r,\mspace{1.5mu}t )^{2}\.\d r^{2}
            + R( r,\mspace{1.5mu}t )^{2}\.\d \Omega^{2}
            \, ,
\end{align}
with $R( r,\mspace{1.5mu}t )$ being the {\it areal radius}, $A( r,\mspace{1.5mu}t )$ the {\it lapse function}, and $\d \Omega^{2} \equiv \d \theta^{2} + \sin^{2}(\theta)\.\d\phi^{2}$ the line element of a two-sphere. The definition of the components of the four-velocity $u^{\mu}$ is gauge-dependent. For instance, in {\it comoving gauge} (\cf~Reference~\cite{2015PhRvD..91h4057H, 2022Galax..10..112Y} for other choices), we have $u^{t} = 1 / A$ and $u^{i} = 0$ for $i \in \{ r,\.\theta,\.\phi \}$. Above and below we use units in which $G_{\mspace{-2mu}\Nrm} = c = 1$.

Assuming comoving gauge and solving the Einstein field equations using the energy-momentum tensor given in Equation~\eqref{eq:tensor-energy} together with the spacetime metric as specified in Equation~\eqref{eq:2-metricsharp}, we get the so-called {\it Misner--Sharp} equations, which describe the evolution of a relativistic fluid in curved spacetime~\cite{1964PhRv..136..571M}:
\begin{subequations}
\begin{align}
\label{eq:u-simply}
    \dot{U}
        &= 
            - A\mspace{-1.5mu}
            \left[
                \frac{w}{1+w}
                \frac{\Gamma^{2}}{\rho}\.
                \frac{\rho'}{R\mspace{1mu}'}
                +
                \frac{M}{R^{2}}
                +
                4\mspace{1.5mu}\pi 
                R\.w\mspace{0.5mu}\rho
             \right]
            ,
            \displaybreak[1]
            \\[1.5mm]
            \label{eq:r-simply}
    \dot{R}
        &= 
            A\.U
            \, ,
            \displaybreak[1]
            \\
            \label{eq:rho-simply}
    \dot{\rho}
        &= 
            -\.A\.\rho\.( 1 + w )\!
            \left[
                \frac{2\.U}{R}
                +
                \frac{U'}{R\mspace{1mu}'}
            \right]
            , 
            \displaybreak[1]
            \\[1.5mm]
            \label{eq:m-simply}
    \dot{M}
        &= 
            -\.4\mspace{1.5mu}\pi 
            A\.w \mspace{0.5mu}\rho\,U R^{2}
            \, ,
            \displaybreak[1]
            \\[1.5mm]
            \label{eq:a-lapse}
    A'
        &= 
            -
            A\.\frac{\rho'}{\rho} 
            \frac{w}{1+w}
            \, ,
            \displaybreak[1]
            \\[2.5mm]
    M'
        &= 
            4\mspace{1.5mu}\pi\rho\.
            R^{2}\mspace{1mu}R\mspace{1mu}'
            \, ,
            \label{eq:M-spatial}
\end{align}
\end{subequations}
where a radial derivative is denoted by a prime and a time derivative by a dot. The radial velocity of the fluid, measured from the centre of coordinates (associated with an Eulerian frame), is given by $U$.

A measure of the gravitational and potential energy is provided by the {\it Misner--Sharp mass} $M\mspace{-0.5mu}( \mspace{-1mu}R )$, defined as
\begin{align}
    M\mspace{-0.5mu}( \mspace{-1mu}R )
        \coloneqq
            \int_{0}^{R}\d \tilde{R}\;
            4\mspace{1.5mu}\pi 
            \tilde{R}^{2} \rho
            \, .
\end{align}
The so-called {\it generalised Lorentz} factor $\Gamma$ appearing in Equation~\eqref{eq:u-simply} is obtained by solving the Einstein equations; it relates $M$, $U$ and $R$ through the constraint
\begin{align}
\label{eq:gamma-constraint}
    \Gamma 
        = 
            \sqrt{1 + U^{2} - \frac{2 M}{R}}
            \, .
\end{align}
It is also useful to know that Equation~\eqref{eq:2-metricsharp} implies the relation $B = R\mspace{1mu}' / \Gamma$.

The lapse equation~\eqref{eq:a-lapse} for $A( r,\mspace{1.5mu}t )$ can be solved analytically in the case of constant $w$, yielding
\vs{-1mm}
\begin{align}
    A( r,\mspace{1.5mu}t )
        = 
            \!\left(
                \frac{\rho_{\brm}( t )}
                {\rho\mspace{1mu}( r,\mspace{1.5mu}t )}
            \right)^{\mspace{-6mu}w / ( w + 1 )}
            \, ,
\end{align}
where $\rho_{\brm}( t )$ is the energy density of the FLRW background, and $\rho_{\brm} \coloneqq 3\mspace{1mu}H^{2} / 8\mspace{1mu}\pi$, with $H$ being the Hubble factor. Notice that the solution of $A( r,\mspace{1.5mu}t )$ at very large radii approaches unity, \ie~$A( r \rightarrow \infty,\mspace{1.5mu}t ) = 1$, and hence consistently ensures to recover FLRW background geometry at scales much larger than that of the cosmological fluctuation. In order to fully describe the PBH formation process, we need to introduce realistic initial conditions for the set of Misner--Sharp equations. We will do this in the next Subsection.

It is worth mentioning that the numerical solution of the system of Equations~(\ref{eq:u-simply}--\ref{eq:M-spatial}) (or also equivalently in other gauges~\cite{1999PhRvD..60h4002S, 2015PhRvD..91h4057H}) is essential to capture the highly-nonlinear dynamics of the collapse process. This has been covered by several works~\cite{1999PhRvD..60h4002S, 2015PhRvD..91h4057H, 2020PDU....2700466E, 2013CQGra..30n5009M, 2021JCAP...01..030E, 2009CQGra..26w5001M, 2005CQGra..22.1405M, 2002CQGra..19.3687H, 1998PhRvL..80.5481N, 1978SvA....22..129N, 2015arXiv150402071B}. The numerical implementation for solving these equations goes beyond the scope of this review (see Reference~\cite{2022Univ....8...66E}).

\subsection{Collapse-Threshold Definition}
\label{sec:Collapse--Threshold-Definition}
\vs{-1mm}
Before defining the collapse threshold for PBH formation, we present a basic introduction below. We already remark that its definition is closely related to the problem of formulating consistent initial conditions for Equations~(\ref{eq:u-simply}--\ref{eq:M-spatial}).

In Reference~\cite{1999PhRvD..60h4002S}, it was shown that the metric~\eqref{eq:2-metricsharp} on superhorizon scales (for which $R_{\mrm} \gg R_{H}\mspace{1mu}$, with $R_{H}\!\coloneqq\!H^{-1}$ being the Hubble horizon radius and $R_{\mrm}\!=\!a( t )\.r_{\mrm} = a( t )\.\tilde{r}_{\mrm}\.\erm^{\mspace{1mu}\zeta( \tilde{r} )}$ the comoving length scale of the fluctuation) can be approximated by
\vs{-1mm}
\begin{align}
\label{eq:metrica-222}
    \d s^{2}
        = 
            -\,
            \d t^{2}
            +
            a^{2}( t )\.
            \erm^{2\mspace{1.5mu}
            \zeta( \tilde{r} )}\mspace{-2.5mu}
            \left(
                \d\tilde{r}^{2}
                +
                \tilde{r}^{2}\.
                \d\Omega^{2}
            \right)
            ,
\end{align}
with $a$ being the scale factor. Equation~\eqref{eq:metrica-222} is equivalent to a FLRW metric with a radial curvature dependence $\zeta( \tilde{r} )$ and can be recast as
\begin{align}
\label{eq:2-FLRWmetric5}
    \d s^{2}
        = 
            -\,\d t^{2}
            +
            a^{2}( t )\mspace{-2.5mu}
            \left[
                \frac{\d r^{2}}
                {1 - K( r )\.r^{2}}
                +
                r^{2}\.\d\Omega^{2}
            \right]
            .
\end{align}
Above, $\zeta( \tilde{r} )$ and $K( r )$ are comoving curvature perturbations defined on superhorizon scales. Note that on those scales, and considering adiabatic fluctuations, $\zeta( \tilde{r} )$ and $K( r )$ are frozen (\ie~constant)~\cite{1983PhRvD..28..679B, 2021JCAP...06..012I} and are related by~\cite{2019PhRvD.100l3524M, 2012EPJC...72.2242R, 2009PhRvD..79d4006H, 2015PhRvD..91h4057H}
\begin{align}
    K( r )\.r^{2}
        = 
            -\,\tilde{r}\.
            \zeta'( \tilde{r} )
            \big[
                2
                +
                \tilde{r}\.\zeta'( \tilde{r} )
            \big]
            \, .
\end{align}
Furthermore, the coordinates $r$ and $\tilde{r}$ can be expressed as
\begin{subequations}
\begin{align}
\label{eq:transforms-tilde-K}
    r
        &= 
            \tilde{r}\mspace{4mu}
            \erm^{\mspace{1mu}\zeta( \tilde{r} )}
            \, ,
            \displaybreak[1]
            \\[2mm]
     \tilde{r}
        &= 
            r\mspace{3mu}\exp\!
            \left[
                \int_{\infty}^{r}
                \frac{\d\hat{r}}{\hat{r}}
                \left(
                    \frac{1}
                    {\sqrt{1 - K( \hat{r} )\.
                    \hat{r}^{2}}}
                    -1
                \right)\!
            \right]
            \mspace{-1mu},
            \displaybreak[1]
            \\[2mm]
     \frac{\d r}{\d \tilde{r}}
        &= 
            \erm^{\mspace{1mu}\zeta( \tilde{r} )}
            \big[
                1 + \tilde{r}\.
                \zeta'( \tilde{r} )
            \big]
            \, ,
\end{align}
\end{subequations}
which makes the nonlinear relation between $K( r )$ and $\zeta( \tilde{r} )$ apparent. 

As shown in Reference~\cite{2007CQGra..24.1405P}, using a gradient-expansion approach~\cite{PhysRevD.42.3936, 1998PThPh..99..763S, 1999PhRvD..60h4002S, 1990PhRvD..42.3936S, 1996CQGra..13..705N, 1996PThPh..95..295T} on the Einstein field equations (specifically on the Misner--Sharp equations), one can relate the different hydrodynamic magnitudes of the fluctuations on superhorizon scales to the curvature fluctuation $\zeta$ (as well as to $K$). Therefore, the shape of the cosmological fluctuation is characterised by $K( r )$ or $\zeta( \tilde{r} )$, where the latter can directly be inferred from the power spectrum $\Pcal_{\zeta}( k )$ (see below). We refer the reader to the previously mentioned references for more details about the gradient-expansion approach. Essentially it considers that the characteristic length scale of the inhomogeneity, say $L$, is much larger than the Hubble horizon.
\newpage

As an expansion parameter, we define
\begin{align}
\label{eq:epsilon}
    \epsilon( t )
        \coloneqq
            \frac{ R_{H} }{ R_{\mrm} }
        = 
            \frac{ 1 }
            {a( t )\.r_{\mrm}\.H( t )}
        = 
            \frac{ 1 }
            {a( t )\.\tilde{r}_{\mrm}\.
            \erm^{\mspace{1mu}
            \zeta( \tilde{r}_{\mrm} )}\mspace{1.5mu}
            H( t )}
            \, ,
\end{align}
where we have made the identification $L = R_{\mrm}$. The above parameter fulfils $\epsilon \ll 1$ on superhorizon scales. This consideration is equivalent to saying that the magnitude of spatial gradients of the different fields is proportional to the fields themselves times a term $\Ocal( \epsilon )$ \eg~$\partial_{r} A \sim A \cdot \Ocal( \epsilon )$.

It is useful to define a specific reference time given by the {\it time of horizon crossing}, $t_{H}$, which corresponds to the time at which the cosmological fluctuations reenter the Hubble horizon. This has to be determined by the nonlinear evolution of the gravitational collapse.

Expanding the Misner--Sharp equations in $\epsilon$, yields~\cite{2007CQGra..24.1405P, 2015PhRvD..91h4057H}
\begin{subequations}
\begin{align}
\label{eq:2-expansion}
    A( r,\mspace{1.5mu}t )
        &= 
            1
            +
            \epsilon( t )^{2}\.
            \tilde{A}( r,\mspace{1.5mu}t )\mspace{1.5mu}
            \, ,
            \displaybreak[1]
            \\[1.5mm]
    R( r,\mspace{1.5mu}t )
        &= 
            a( t )\.r\mspace{-1.5mu}
            \left[
                1
                +
                \epsilon( t )^{2}\.
                \tilde{R}\mspace{1mu}( r,\mspace{1.5mu}t )
            \right]
            ,
            \displaybreak[1]
            \\[1.5mm]
    U\mspace{-1mu}( r,\mspace{1.5mu}t )
        &= 
            H( t )\.R\mspace{1mu}( r,\mspace{1.5mu}t )
            \mspace{-1.5mu}
            \left[
                1
                +
                \epsilon( t )^{2}\.
                \tilde{U}\mspace{-1mu}( r,\mspace{1.5mu}t )
            \right]
            ,
            \displaybreak[1]
            \\[1.5mm]
    \rho\mspace{1mu}( r,\mspace{1.5mu}t )
        &= 
            \rho_{\brm}( t )
            \mspace{-1.5mu}
            \left[
                1
                +
                \epsilon( t )^{2}\.
                \tilde{\rho}\mspace{1mu}( r,\mspace{1.5mu}t )
                \vphantom{\tilde{U}}
            \right]
            ,
            \displaybreak[1]
            \\[1.5mm]
    M( r,\mspace{1.5mu}t )
        &= 
            \frac{4\mspace{1.5mu}\pi}{3}\.
            \rho_{\brm}( t )\.
            R( r,\mspace{1.5mu}t )^{3}\mspace{-1.5mu}
            \left[
                1
                +
                \epsilon^{2}( t )\.
                \tilde{M}( r,\mspace{1.5mu}t ) 
            \right]
            .
\end{align}
\end{subequations}
Notice that for $\epsilon \rightarrow 0$, we recover the FLRW solution. The perturbation variables at leading order in the gradient expansion, \ie~at order $\Ocal( \epsilon^{2} )$ (see Reference~\cite{2012JCAP...09..027P} for higher-order calculations) can be found in References~\cite{2007CQGra..24.1405P, 2015PhRvD..91h4057H} and are summarised below:
\vs{-1mm}
\begin{subequations}
\begin{align}
\label{eq:2-perturbations}
    \tilde{\rho}\mspace{1mu}( r )
        &= 
            \frac{ 3\.( 1 + w ) }{ 5 + 3\mspace{1.5mu}w }\mspace{-3mu}
            \left[
                K( r )
                +
                \frac{r}{3}\.K'( r )
            \right]
            r^{2}_{\mrm}
            \, ,
            \displaybreak[1]
            \\[2mm]
    \tilde{U}\mspace{-1mu}( r )
        &= 
            -\frac{1}{5 + 3\mspace{1.5mu}w}\.
            K( r )\.r^{2}_{\mrm}
            \, ,
            \displaybreak[1]
            \\[2mm]
    \tilde{A}\mspace{1.5mu}( r )
        &= 
            -\frac{w}{1 + w}\.
            \tilde{\rho}( r )
            \, ,
            \displaybreak[1]
            \\[2.5mm]
    \tilde{M}\mspace{-0.5mu}( r )
        &= 
            -\.3\.( 1 + w )\.
            \tilde{U}\mspace{-1mu}( r )
            \, ,
            \displaybreak[1]
            \\[1mm]
    \tilde{R}\mspace{1.5mu}( r )
        &= 
            -
            \frac{w}{( 1 + 3\mspace{1.5mu}w )( 1 + w )}\.
            \tilde{\rho}\mspace{1mu}( r )
            +
            \frac{1}{1 + 3\mspace{1.5mu}w}\.
            \tilde{U}\mspace{-1mu}( r )
            \, .
\end{align}
\end{subequations}
\newpage

\noindent The perturbations in terms of the coordinate $\tilde{r}$ are~\cite{2019PhRvD.100l3524M}
\begin{subequations}
\begin{align}
    \tilde{\rho}\mspace{1mu}(\tilde{r})
        &= 
            -
            \frac{2\mspace{1.5mu}( 1 + w )}
            {5 + 3\mspace{1.5mu}w}\.
            \frac{\exp\mspace{-3mu}
            \big[
                    2\.\zeta( \tilde{r}_{\mrm} )
            \big]}
            {\exp\mspace{-3mu}
            \big[
                2\.\zeta( \tilde{r} )
            \big]}\mspace{-2mu}
            \left[
                \zeta''( \tilde{r} )
                +
                \zeta'( \tilde{r} )\mspace{-1.5mu}
                \left(
                    \frac{2}{\tilde{r}}
                    +
                    \frac{\zeta'( \tilde{r} )}{2}
                \right)\mspace{-2mu}
            \right]
            \tilde{r}^{2}_{\mrm}
            \, ,
            \label{eq:rho-tilde-perturb}
            \\[3mm]
    \tilde{U}\mspace{-1mu}(\tilde{r})
        &= 
            \frac{1}{5 + 3\mspace{1.5mu}w}\.
            \frac{\exp\mspace{-3mu}
            \big[
                2\.\zeta( \tilde{r}_{\mrm} )
            \big]}
            {\exp\mspace{-3mu}
            \big[
                2\.\zeta( \tilde{r} )
            \big]}\,
            \zeta'( \tilde{r} )\mspace{-1.5mu}
            \left[
                \frac{2}{\tilde{r}}
                +
                \zeta'( \tilde{r} )
            \right]
            \tilde{r}^{2}_{\mrm} 
            \, .
            \label{eq:U-tilde-perturb}
\end{align}
\end{subequations}
This allows us to express the density contrast 
\begin{align}
    \delta
        \coloneqq
            \frac{ \delta \rho }
            { \rho_{\brm} }
        \equiv
            \frac{\rho - \rho_{\brm} }
            { \rho_{\brm} }
            \, , 
            \\[-9mm]
            \notag
\end{align}
as a function of $r$ and $\tilde{r}$, as
\begin{subequations}
\begin{align}
\label{eq:tilde-perturb-contrast}
    \frac{\delta \rho}{\rho_{\brm}}( r,\mspace{1.5mu}t )
        &= 
            f( w )\mspace{-1.5mu}
            \left(
                \frac{1}{a( t )\mspace{1mu}H( t )}
            \right)^{\mspace{-6mu}2}
            \left[
                K( r )
                +
                \frac{r}{3}\.K'( r )
            \right]
            ,
            \displaybreak[1]
            \\[3.5mm]
\begin{split}
    \frac{\delta \rho}{\rho_{\brm}}( \tilde{r},\mspace{1.5mu}t )
        &= 
            -\.
            f( w )\mspace{-1.5mu}
            \left(
                \frac{1}{a( t )\mspace{1mu}H( t )}
            \right)^{\mspace{-6mu}2}
            \exp\mspace{-3mu}
            \big[\mspace{-1.5mu}
                -
                2\.\zeta( \tilde{r} )
            \big]
            \\[2.5mm]
        &\phantom{=\;}
            \times
            \left[
                \zeta''( \tilde{r} )
                +
                \zeta'( \tilde{r} )\mspace{-1.5mu}
                \left(
                    \frac{2}{\tilde{r}}
                    +
                    \frac{\zeta'( \tilde{r} )}{2}
                \right)\mspace{-2mu}
            \right]
            ,
\end{split}
\end{align}
\end{subequations}
where
\vs{-2.5mm}
\begin{align}
    f( w )
        \coloneqq
            \frac{ 3\.( 1 + w ) }
            { (5 + 3\mspace{1.5mu}w) }
            \, .
            \\[-9mm]
            \notag
\end{align}

If formed from collapse of inflationary perturbations, the PBH abundance is exponentially sensitive to the threshold $\delta_{\crm}$ of the gravitational collapse~\cite{1974MNRAS.168..399C}. Here, $\delta_{\crm}$ is the minimum amplitude of the peak of the gravitational potential, related to the perturbation undergoing gravitational collapse leading to a black hole.

Several approaches have been used to define the PBH formation threshold, and hence several ways for defining the amplitude of the cosmological fluctuations. In this Section, we shall use the definition introduced in Reference~\cite{1999PhRvD..60h4002S} and confirmed in References~\cite{2015PhRvD..91h4057H, 2019PhRvD.100l3524M}, wherein it has been found that a good criterion for PBH formation is to define the associated threshold as the peak value of the so-called {\it compaction function} $\Ccal$. This closely resembles the gravitational Schwarzschild potential and is defined as the average mass excess in a given volume on superhorizon scales. Recently, this has been intensively studied numerically for various types of cosmological fluctuations~\cite{2019PhRvD.100l3524M, 2020PhRvD.101d4022E, 2021JCAP...01..030E, 2019PhRvL.122n1302G, 2021PhRvD.103f3538M, 2019JCAP...11..012Y, 2015PhRvD..91h4057H}.
\newpage

Concretely, the {\it compaction function} can be defined as 
\begin{tcolorbox}
\vs{-3mm}
\begin{align}
\label{eq:2-compactionfunction}
     \com( r,\mspace{1.5mu}t )
        \coloneqq
            2\,
            \frac{M( r,\mspace{1.5mu}t )
            - M_{\brm}( r,\mspace{1.5mu}t )}
            {R( r,\mspace{1.5mu}t )}
            \, ,
\end{align}
\end{tcolorbox}
\noindent with $M_{\brm}( r,\mspace{1.5mu}t ) \coloneqq 4\mspace{1.5mu}\pi \rho_{\brm}\.R^{3} / 3$ being the mass of the FLRW background within a volume $V = 4\mspace{1.5mu}\pi R^{3} /3$. Interestingly, at leading order of the gradient expansion, $\Ocal( \epsilon^{2} )$, the compaction function can be written as
\begin{align}
\label{eq:comaction-function}
    \com( r )
        \simeq
            f( w )\.K( r )\.r^{2}
        = 
            f( w )
            \mspace{-1.5mu}
            \left(
                1 - 
                \big[
                    1
                    +
                    \tilde{r}\.
                    \zeta'( \tilde{r} )
                \big]^{2}
            \right)
            ,
\end{align}
which is a time-independent quantity for superhorizon fluctuations as long as $w$ is constant. We denote the innermost\footnote{\setstretch{0.9}In some cases, it is also possible to have curvature profiles which generate compaction functions with several secondary peaks beyond the innermost one, like when monochromatic power spectra are considered~\cite{2020JCAP...05..022A}. Regardless, such secondary peaks will not substantially contribute to the gravitational collapse as well as to the threshold value for PBH formation as long as those peaks are at least slightly smaller than the main one. See Reference~\cite{2023arXiv231016482E} for a detailed numerical study using compaction functions characterised by two peaks.}. peak of the compaction function (considered also as the fluctuation length scale) by $r_{\mrm}$ (or, equivalently, $\tilde{r}_{\mrm}$ in case of using the $\tilde{r}$-coordinates), which is determined by the first root of Equation~\eqref{eq:comaction-function}, leading to the set of equations
\begin{subequations}
\begin{align}
\label{eq:rm-eq}
     \frac{\d\mspace{1.5mu}\Ccal( r )}
     {\d r}
        &= 
            f( w )
            \big[
                K'( r )\.r^{2}
                +
                2\.r\mspace{1mu}K( r )
            \big] 
        = 
            0
            \, ,
            \displaybreak[1]
            \\[4mm]
            \label{eq:zeta-root}
    \frac{\d\mspace{1.5mu}\Ccal( \tilde{r} )}
    {\d \tilde{r}}
        &= 
            -\.2\.f( w )\mspace{1.5mu}
            \big(
                1
                +
                \tilde{r}\.\zeta'
            \big)
            \big(
                \zeta'
                +
                \tilde{r}\.\zeta''
            \big)
        = 
            0 
            \, .
\end{align}
\end{subequations}
Hence, the root $r_{\mrm}$ is obtained from $K( r_{\mrm} ) + r_{\mrm}\.K'( r_{\mrm} ) / 2 = 0$, according to Equation~\eqref{eq:rm-eq}. On the contrary, Equation~\eqref{eq:zeta-root} yields two types of fluctuations: so-called fluctuations of {\it type I} and of {\it type II}.

The former fulfil $1 + \tilde{r}\mspace{2mu}\zeta' > 0$ for any $\tilde{r}$, being equivalent to the condition that the areal radius $R = a\.\tilde{r}\.\erm^{\mspace{1mu}\zeta}$ (defined with the FLRW background) is a monotonically increasing function, \ie~$R\mspace{1mu}' > 0$. This is evident if we take into account that
\vs{-1mm}
\begin{align}
    R\mspace{1mu}'
        = 
            a\.\erm^{\mspace{1mu}\zeta}
            \big(
                1
                +
                \tilde{r}\.\zeta'
            \big)
            \, .
\end{align}
For this case, the location of the compaction-function peak $\tilde{r}_{\mrm}$ is derived via $\zeta'( \tilde{r}_{\mrm} ) + \tilde{r}_{\mrm}\.\zeta''( \tilde{r}_{\mrm} ) = 0$. The peak $\Ccal( \tilde{r}_{\mrm} )$ is also a monotonic function of the amplitude $\mu$, being a maximum.
\newpage

On the other hand, for very large fluctuations, beyond the critical value (with correspondingly large amplitude $\mu$), the areal radius can be non-monotonic, at least for some values of $\tilde{r}$. We refer to those fluctuations as {\it type-II} fluctuations~\cite{2011PhRvD..83l4025K, 2015PhRvD..91h4048C}. Here, the peak of the compaction function at $\tilde{r}_{\mrm}$ becomes a local minimum. As clarified in Reference~\cite{2011PhRvD..83l4025K}, this peak value can be smaller than the threshold, despite the fact that these fluctuations will always lead to PBH formation. Points for which $R( \tilde{r}_{m,\mspace{1.5mu}{\rm II}} )\mspace{1mu}' = 0$ (\ie~those fulfilling $1 + \tilde{r}_{m,\mspace{1.5mu}{\rm II}}\.\zeta'( \tilde{r}_{m,\mspace{1.5mu}{\rm II}} )$ = 0) imply local maxima in $\Ccal$. These peak values are given by $\Ccal( \tilde{r}_{m,\mspace{1.5mu}{\rm II}} ) = f( w )$~\cite{2022JCAP...05..012E}; the compaction function can be rewritten as
\begin{align}
\label{eq:type2}
    \Ccal( \tilde{r} )
        = 
            f( w )\mspace{-1.5mu}
            \left[
                1
                -
                \left(
                   \frac{R\mspace{1mu}'}
                   {a\,\erm^{\mspace{1mu}\zeta}}
                \right)^{\mspace{-6mu}2}
            \right]
        \leq
            f( w )
            \, .
\end{align}
Therefore, $\Ccal_{\umax} = f( w )$ is the maximum value of the compaction function.

We focus on type-I fluctuations, which are standard in the literature and the ones that characterise the PBH formation threshold, which is the most important quantity in estimating the PBH abundance. Since type-II fluctuations are highly suppressed, they are {\it a priori} not expected to contribute significantly to the abundance of PBHs, except for some particular scenarios associated with non-Gau{\ss}ianities where the situation could be different, as recently discussed in References~\cite{2023EL....14249001G, 2023JCAP...10..035E}. For a recent numerical study on fluctuations of type-II, we refer the reader to Reference~\cite{2024arXiv240106329U}.

Therefore, we can unambiguously define the amplitude of a type-I cosmological fluctuation as the peak value of the compaction function [Equation~\eqref{eq:comaction-function}], defined on superhorizon scales and at leading order in gradient expansion,
\begin{align}
    \delta_{\mrm}
        \coloneqq
            \com( r_{\mrm} )
            \, .
\end{align}
The threshold corresponds to the critical compaction-function peak value $\delta_{\crm} = \Ccal_{\crm}( r_{\mrm} )$. Cosmological fluctuations with an amplitude larger than the threshold value, \ie~with $\delta_{\mrm} > \Ccal_{\crm}( r_{\mrm} )$ will collapse and form black holes. In the opposite case, fluctuations with $\delta_{\mrm} < \Ccal_{\crm}( r_{\mrm} )$ will disperse onto the FLRW background avoiding black hole formation.

\subsection{Threshold Values}
\label{sec:Threshold-Values}
\vs{-1mm}
At this point, we have all the necessary formalism for numerical studies on PBH formation under the assumption of spherically-symmetric perturbations. 

In order to explore the behaviour of the threshold values in terms of different profiles{\,---\,}as mentioned, the thresholds depend substantially on the fluctuation profile{\,---\,}, usually specific families of curvature profiles are considered~\cite{2019PhRvD.100l3524M, 2014JCAP...01..037N, 2009PhRvD..79d4006H, 1999PhRvD..60h4002S, 2021JCAP...01..030E,
2015PhRvD..91h4057H}. Common examples are
\begin{subequations}
\begin{align}
    K_{\rm pol}( r )
        &= 
            \frac{\delta_{\mrm}}{f( w )\.r_{\mrm}^{2}}\.
            \frac{1 + 1/q}
            {1+
            ( r / r_{\mrm} )_{}^{2\mspace{1.5mu}(q+1)} / q}
            \, ,
            \label{eq:K-polynominal}
            \displaybreak[1]
            \\[4mm]
    K_{\rm exp}( r )
        &= 
            \frac{\delta_{\mrm}}{f( w )\.r_{\mrm}^{2}}
            \left(
                \frac{r}{r_{\mrm}}
            \right)^{\mspace{-6mu}2\mspace{1.5mu}\lambda}\,
            \exp\!
            \left[
                \frac{ ( 1 + \lambda )^{2} }{ q }\!
                \left(
                    1 - 
                    \left(
                        \frac{r}{r_{\mrm}}
                    \right)^{\mspace{-6mu}2\mspace{1.5mu}q / ( 1 + \lambda )}
                \right)\mspace{-2mu}
            \right]
            ,
            \label{eq:K-exponential}
\end{align}
\end{subequations}
\ie~polynomial and exponential profiles, respectively. The latter is centrally peaked for $\lambda = 0$ and non-centrally peaked for $\lambda \neq 0$.

Notice that the profiles depend on the ratio $r / r_{\mrm}$. Moreover, as the compaction-function peak is proportional to the amplitude of the curvature fluctuation $K( r )$ [see Equation~\eqref{eq:comaction-function}], we can rewrite these profiles in a convenient way such that when computing $\com( r )$ using Equation~\eqref{eq:comaction-function}, being valid at leading order in gradient expansion, automatically gives $\com( r_{\mrm} ) = \delta_{\mrm}$. A different situation occurs when working with $\zeta( \tilde{r} )$ instead, due to the nonlinear relation between the compaction function $\com$ and $\zeta$.

The shape around the peak of the compaction function is determined by the (dimensionless) parameter $q$, which can be expressed as~\cite{2020PhRvD.101d4022E}
\begin{subequations}
\begin{align}
\label{eq:q-factor}
    q
        &= 
            -
            \frac{\com''( r_{\mrm} )\.r^{2}_{\mrm}}
            {4\.\com( r_{\mrm} )}
            \, .
\intertext{In terms of the $\tilde{r}$-coordinate, Equation~\eqref{eq:q-factor} can be rewritten upon change of variables as
\vs{-1mm}}
\label{eq:q-tilde}
    q
        &= 
            -\frac{\com''( \tilde{r}_{\mrm} )\.
            \tilde{r}^{2}_{\mrm}}
            {4\.\com( \tilde{r}_{\mrm} )
            \big[
                1
                -
                \com( \tilde{r}_{\mrm} ) / f( w )\.
            \big]}
            \, .
\end{align}
\end{subequations}
The parameter $q$ is indeed crucial: As has been shown in Reference~\cite{2020PhRvD.101d4022E} employing detailed numerical simulations, different curvature profiles with the same $q$-parameters have the same threshold $\delta_{\crm}$ upon deviation of $\Ocal( 2\,\text{--}\,3 )\mspace{0.5mu}\%$ in the case of a radiation-dominated universe. We will see in Section~\ref{sec:Analytical-Threshold-Formulae} how to use this result for an analytical estimation of the threshold.
\newpage

\begin{figure}[t]
\centering
    \centering
    \hs{9mm}\includegraphics[width = 0.75\linewidth]{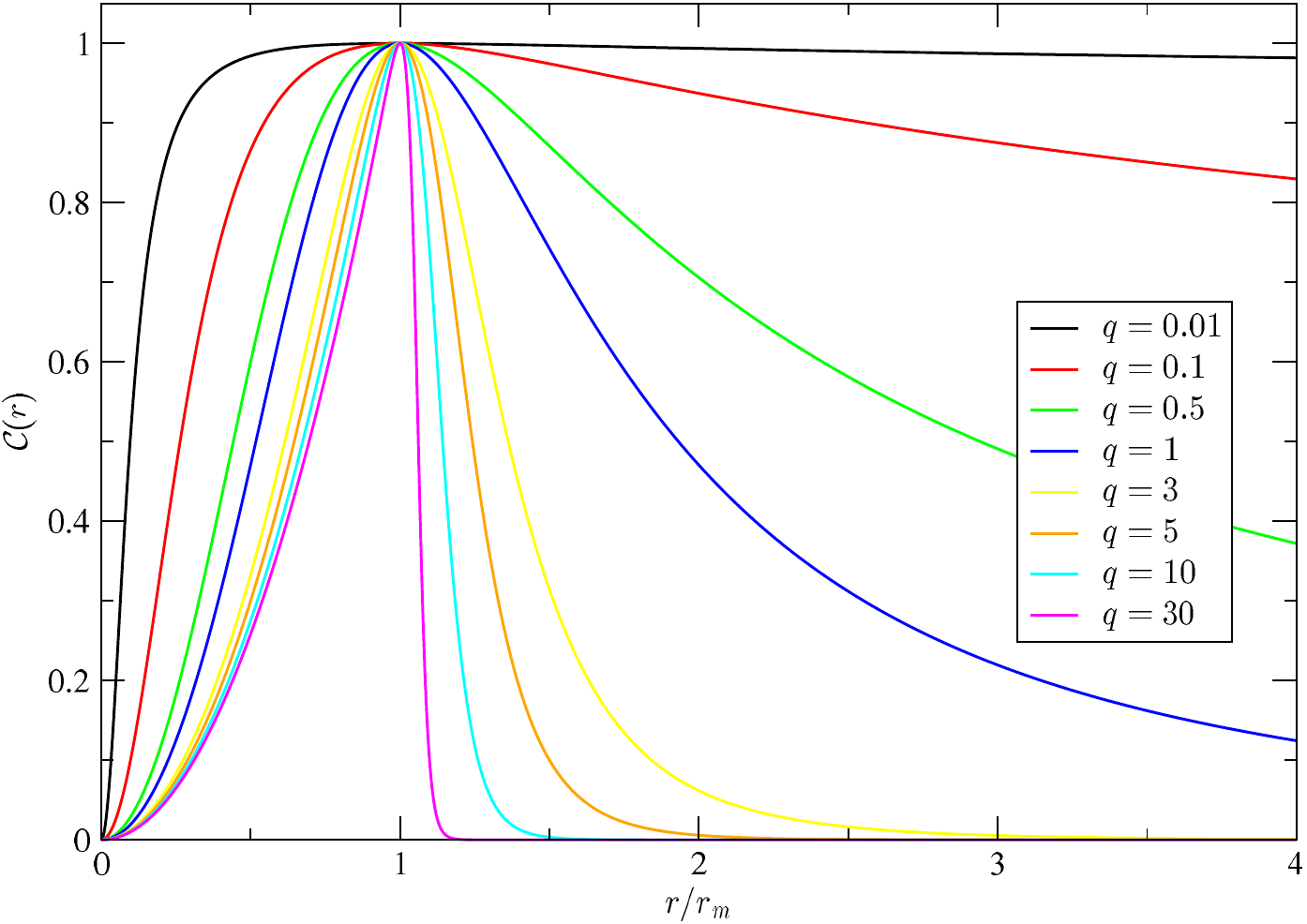}\\[5mm]
    \includegraphics[width = 0.80\linewidth]{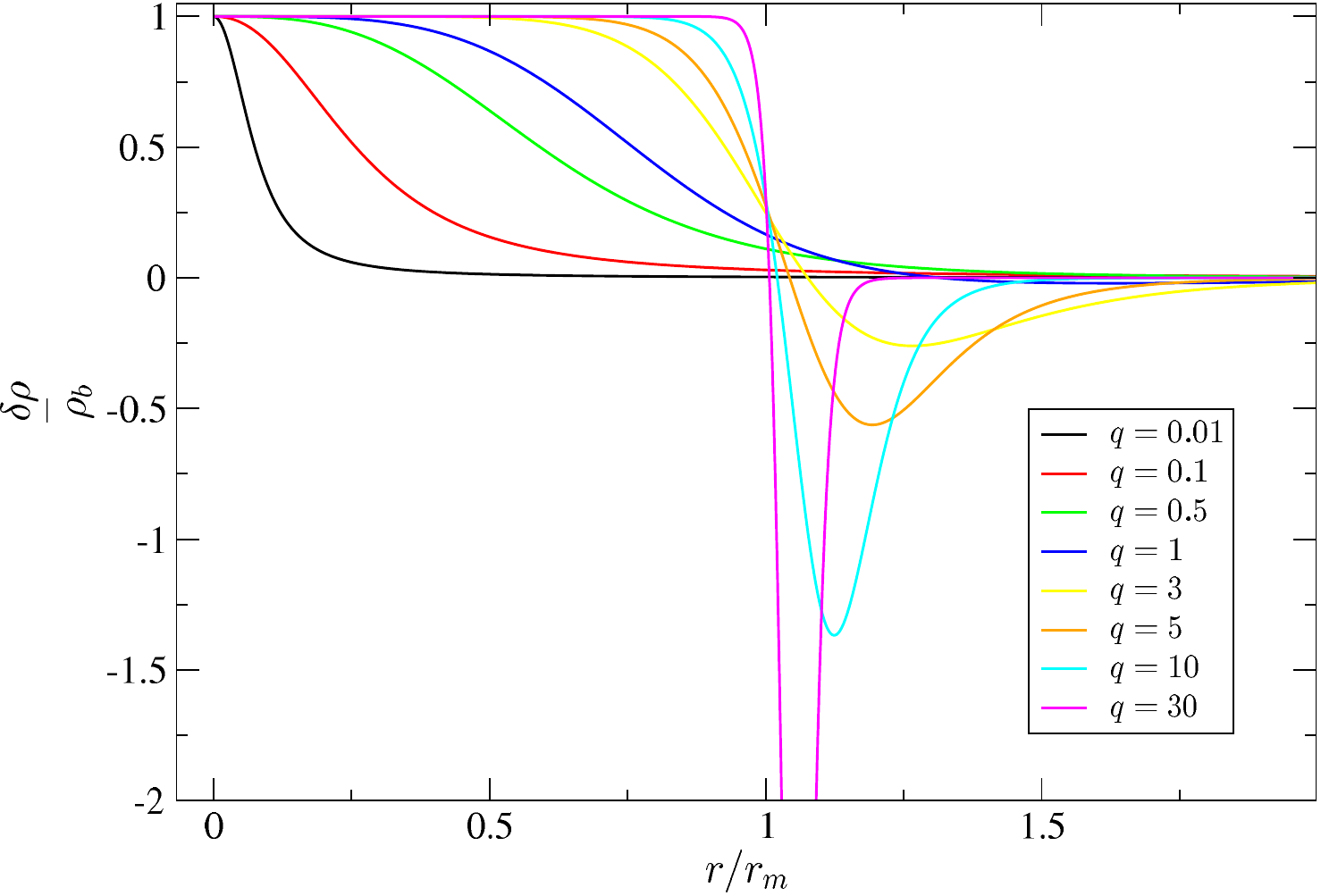} 
    \caption{
        {\it Upper panel}:
            Compaction function as a function of radius for the family of polynomial profiles in Equation~\eqref{eq:K-polynominal}, for various values of the parameter $q$ (see legend). All peak amplitudes $\delta_{\mrm}$ are normalised such that $\delta_{\mrm} = 1$.
        {\it Lower panel}:
            Density contrast for the same profiles as in the upper panel, these being normalised to the peak $\delta \rho\mspace{1mu}(r = 0) / \rho_{\brm}$. Figures from Reference~\cite{2022Univ....8...66E}.
        }
    \label{fig:C-profiles}
\end{figure}

In Figure~\ref{fig:C-profiles} we show the compaction function as well as the density contrast for the two profiles of Equations~(\ref{eq:K-polynominal}--\ref{eq:K-exponential}) for different values of $q$. Notice that for $q \gg 1$, the peak of the compaction function is sharp, while it is broad (which would be an approximate homogeneous sphere) for $q \ll 1$. In the limit $q \rightarrow \infty$, the pressure gradients are maximal and therefore yield the maximally-allowed threshold, which we denote by $\delta_{\crm,\mspace{1.5mu}\umax}$. In the opposite case, when $q$ approaches zero, both the pressure gradients and the threshold are minimised, the latter being denoted by $\delta_{\crm,\mspace{1.5mu}\umin}$.

\begin{figure}
    \includegraphics[width = 0.77\columnwidth]{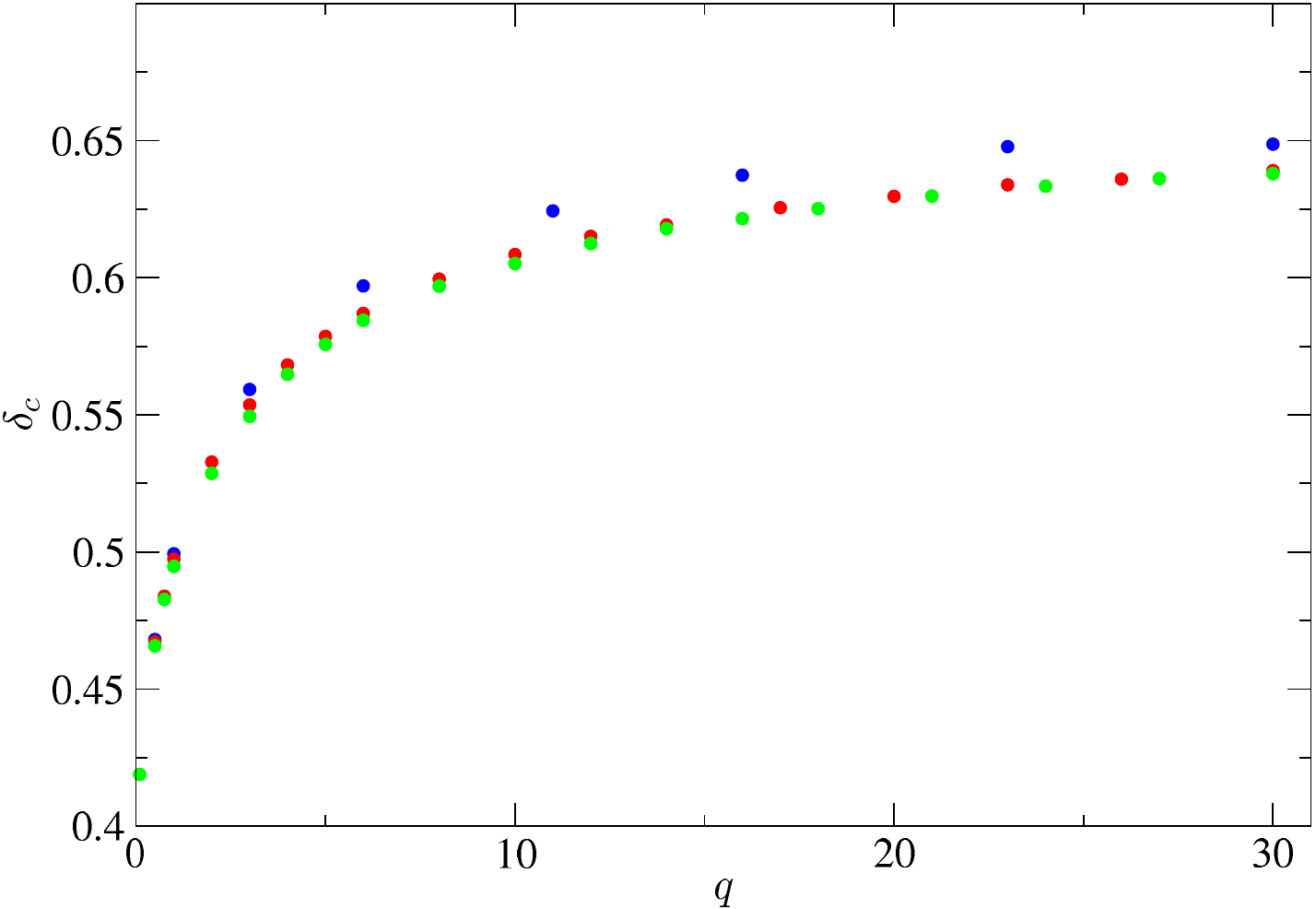}\hs{2mm}\\[4mm]
    \hs{2mm}\includegraphics[width = 0.75\columnwidth]{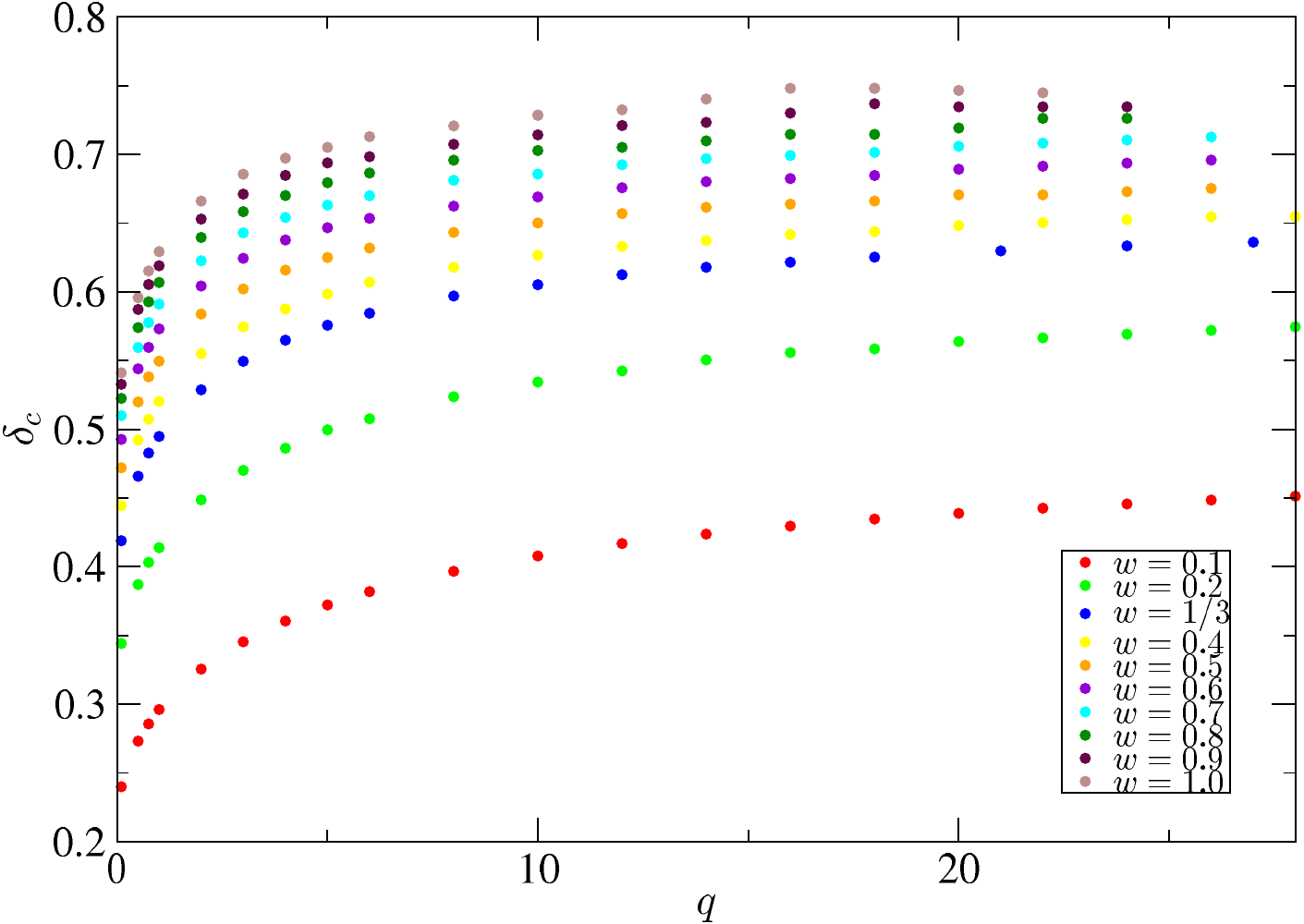}
    \caption{
        {\it Upper panel}: 
            Threshold $\delta_{\crm}$ as a function of $q$ for three different profiles. Blue points denote the polynomial profile of Equation~\eqref{eq:K-polynominal}; the exponential profile of Equation~\eqref{eq:K-exponential} for $\lambda = 0$ and $\lambda = 1$ is indicated by green and red dots, respectively. 
        {\it Lower panel}: 
            Threshold $\delta_{\crm}$ as a function of $q$ for the polynomial profile Equation~\eqref{eq:K-polynominal} and for different values of the equation-of-state parameter $w$.
        Figures from References~\cite{2022Univ....8...66E, 2021JCAP...01..030E}.
        }
    \label{fig:thresholds-plots-1}
\end{figure}

\begin{figure}
    \includegraphics[width = 0.75\columnwidth]{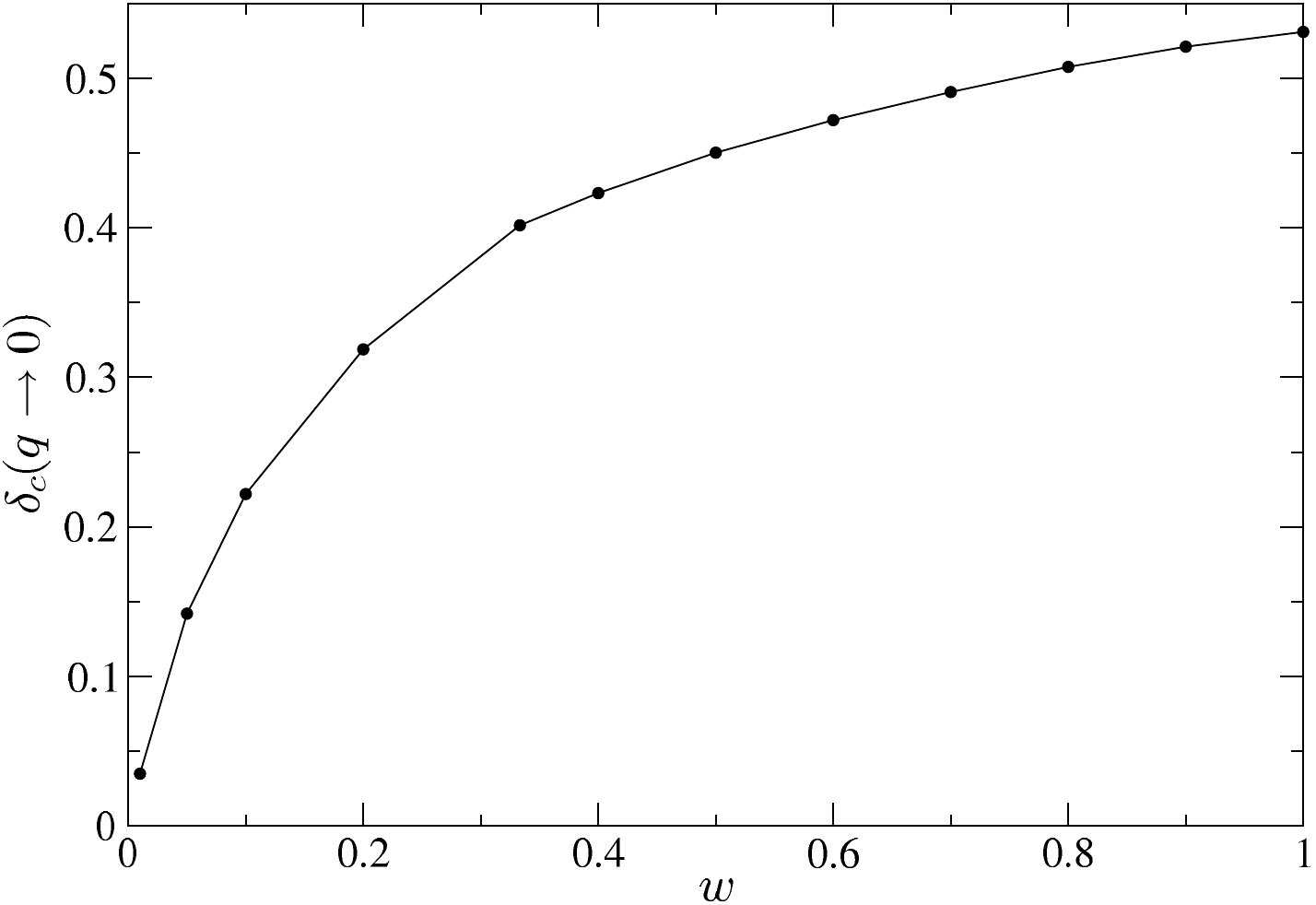}\\[4mm]
    \hs{2mm}\includegraphics[width = 0.75\columnwidth]{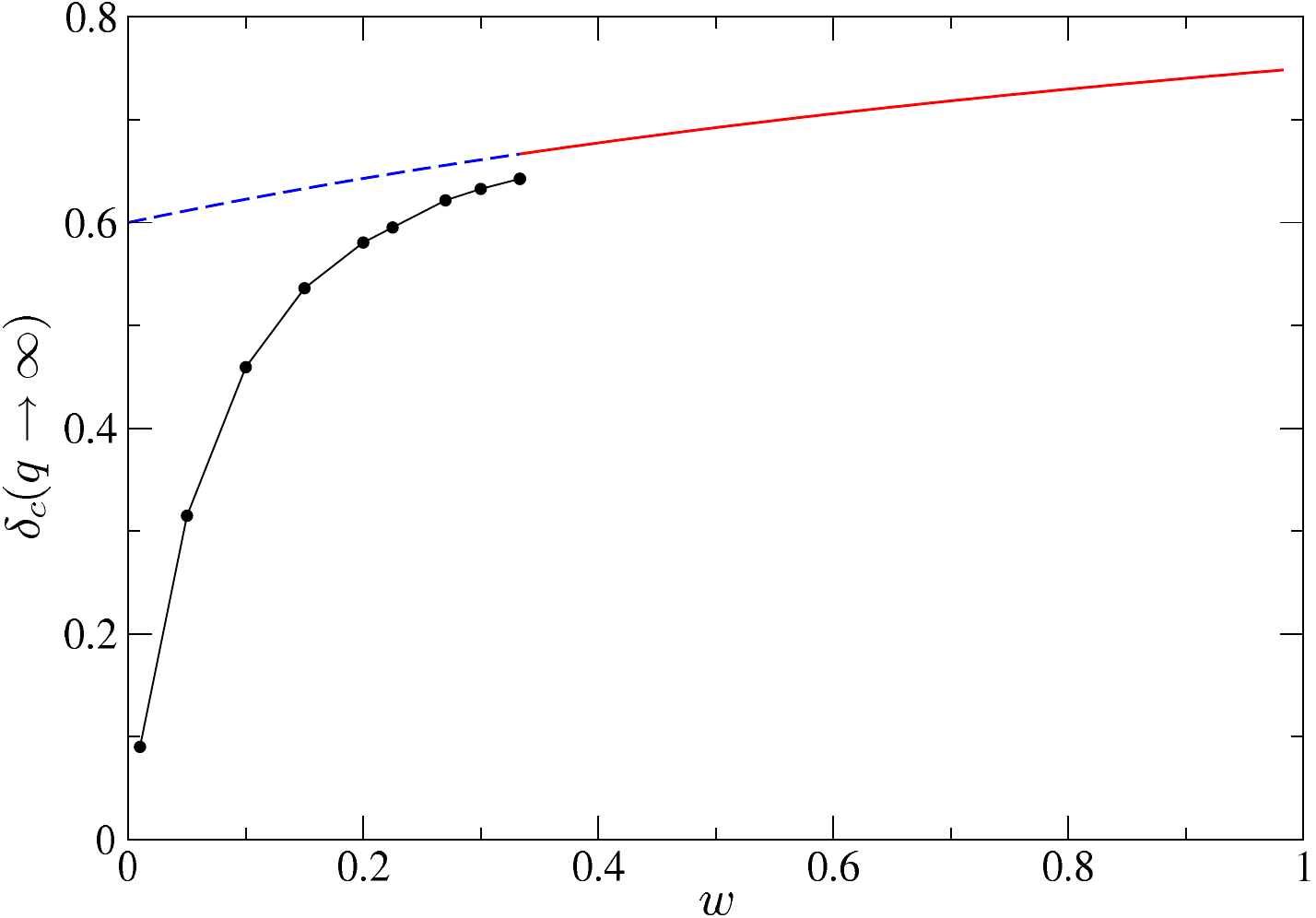}
    \caption{
        {\it Upper panel}:
            Threshold $\delta_{\crm}$ as a function of $w$ for $q \rightarrow 0$. 
        {\it Lower panel}:
            Threshold $\delta_{\crm}$ as a function of $w$ for $q \rightarrow \infty$.
         The red line corresponds to the analytical case $f( w )$, where the black dots represent the numerical simulations in the region where $\delta_{\crm}(q \rightarrow \infty) \neq f( w )$, as shown in Reference~\cite{2021JCAP...01..030E}. The blue dashed line corresponds to $f( w )$ in that region. Figures from References~\cite{2022Univ....8...66E, 2021JCAP...01..030E}.
        }
    \label{fig:thresholds-plots-2}
\end{figure}

Figures~\ref{fig:thresholds-plots-1} and~\ref{fig:thresholds-plots-2} depict numerical results for thresholds of different profiles and equations of states. Concretely, the upper panel of Figure~\ref{fig:thresholds-plots-1} shows threshold values for different profiles in terms of $q$, for a range of choices of $\delta_{\crm} \in [0.4,\mspace{1.5mu}2/3]$ for the case of a radiation-dominated universe~\cite{2019PhRvD.100l3524M, 2020PhRvD.101d4022E}. Observe the comparatively small differences of the threshold values between different profiles with the same $q$-value; the deviation can be quantified to be $\Ocal( 2 )\mspace{0.5mu}\%$. Instead, in the lower panel, we show the threshold for different values of $w$ in terms of $q$, considering the same curvature profile~\cite{2021JCAP...01..030E}. For the same $q$-value, the threshold becomes smaller as $w$ decreases, since pressure gradients are also reduced. Figure~\ref{fig:thresholds-plots-2} shows the minimum threshold [characterised by $q \rightarrow 0$ ({\it upper panel})\hs{-0.3mm}] and maximum threshold [characterised by $q \rightarrow \infty$ ({\it lower panel})\hs{-0.3mm}] as functions of $w$. In particular, $\delta_{\crm,\mspace{1.5mu}\umax}$ matches the analytical estimate for $\delta_{\crm,\mspace{1.5mu}\umax} = f( w )$ if $w \geq 1/3$, but not anymore for softer equations of state for which a numerical approach is required.

We next consider the difference between using $K( r )$ and $\zeta( \tilde{r} )$. As an illustration, we will use a specific example in order to highlight the implications of the nonlinearities. Concretely, we take a Gau{\ss}ian profile, parametrised by the amplitudes $\Acal$ and $\mu$, as $K( r ) = \Acal\,\exp[ - ( r/r_{\mrm} )^{2} ]$ and $\zeta( \tilde{r} ) = \mu\.\exp[ - ( \tilde{r} / \tilde{r}_{\mrm} )^{2} ]$, respectively. As we have seen, on superhorizon scales, the relation between the compaction function and $K$ is linear but nonlinear between $\com$ and $\zeta$. When computing the peak value of the compaction function, $\delta_{\mrm}$, using Equation~\eqref{eq:comaction-function} yields
\begin{subequations}
\begin{align}
    \delta_{m,\mspace{1.5mu}K}
        &= 
            f( w )\.\Acal\mspace{4.5mu}
            r^{2}_{\mrm}\,\erm^{-1}
            \, ,
            \displaybreak[1]
            \\[3mm]
    \delta_{m,\mspace{1.5mu}\zeta}
        &= 
            4\.f( w )\.
            ( \erm - \mu )\.\mu\,\erm^{-2}
            \, ,
            \label{eq:deltam-zeta}
\end{align}
\end{subequations}
where $\delta_{m,\mspace{1.5mu}K} \coloneqq \Ccal( r_{\mrm} )$ and $\delta_{m,\mspace{1.5mu}\zeta} \coloneqq \Ccal( \tilde{r}_{\mrm} )$, which depend nonlinearly on $\mu$. Computing the effective $q$-value for both profiles using Equations~(\ref{eq:q-factor}--\ref{eq:q-tilde}), we see that $q_{K} = 1$, whereas
\vs{-2mm}
\begin{align}
    q_{\zeta}
        = 
            \erm^{2}\.
            \big[
                \erm^{2}
                -
                3\.\erm\mspace{1.5mu}\mu
                +
                2\.\mu^{2}
            \big]^{-1}
            \, .
            \\[-5mm]
            \notag
\end{align}

\begin{figure}[t]
    \centering    
    \includegraphics[width = 0.72\columnwidth]{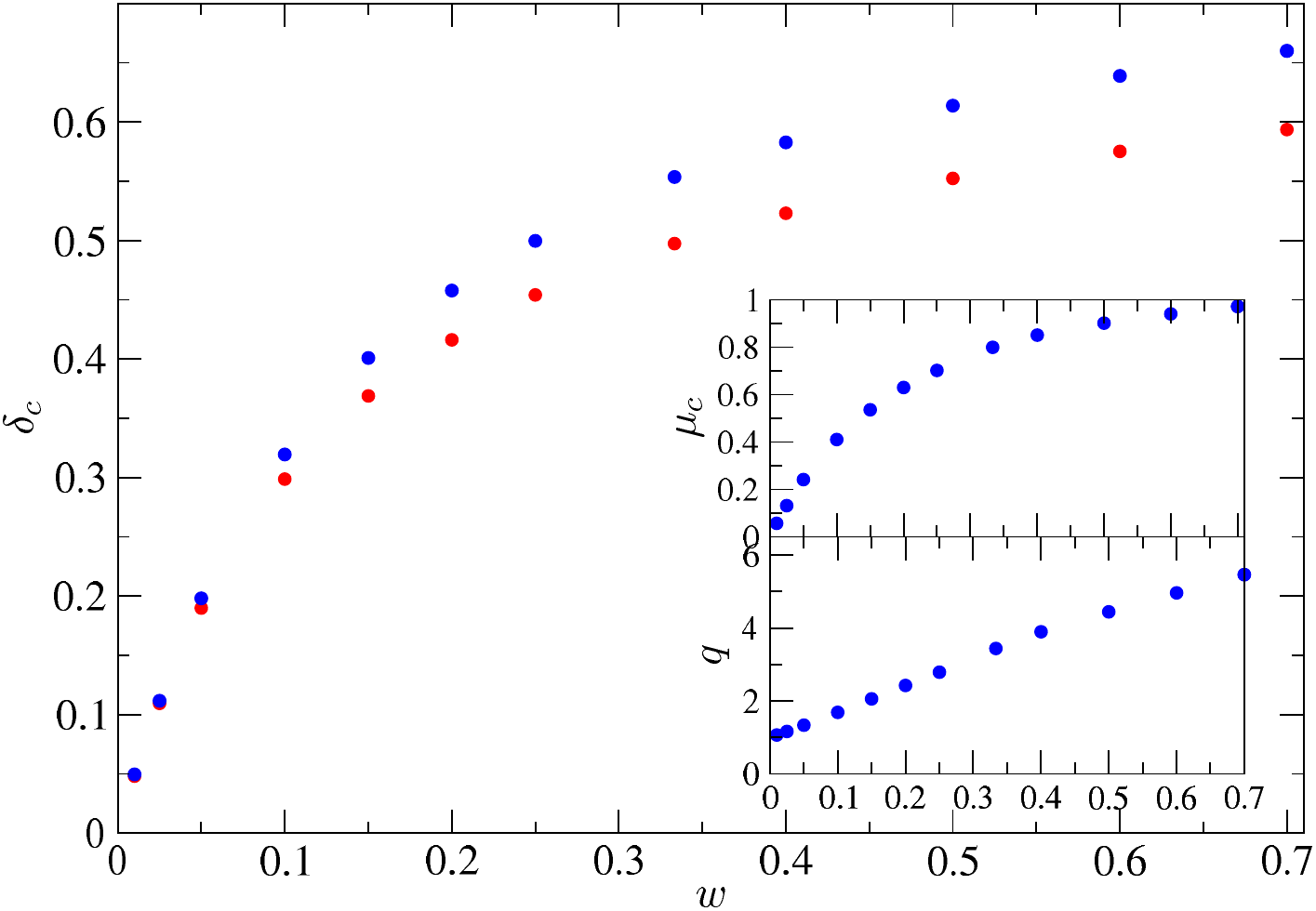}
    \caption{
        Threshold $\delta_{\crm}$ as a function of $w$ for two Gau{\ss}ian profiles: Red dots correspond to using $K( r )$, blue dots to $\zeta( \tilde{r} )$.
        {\it Upper subpanel:} 
            critical amplitude $\mu_{\crm}$ of the profile in $\zeta( \tilde{r} )$.
        {\it Lower subpanel:} 
            dependence of the parameter $q$ of the critical profile $\zeta( \tilde{r} )$ on $w$. 
        Figure (updated) from Reference~\cite{2022Univ....8...66E}.
        \vs{-1mm}
        }
    \label{fig:thresholds-zeta}
\end{figure}

A plot of the threshold $\delta_{\crm}$ for both profiles is shown in Figure~\ref{fig:thresholds-zeta} for different values of $w$. Although both profiles are Gau{\ss}ian, $\delta_{\crm}$ is different, due to the nonlinear relation in $\zeta( \tilde{r} )$. The threshold for $\zeta( \tilde{r} )$ is higher than for $K( r )$ since the shape around the compaction function is sharper (meaning larger $q$), as can be observed in the bottom panel of the subplot in Figure~\ref{fig:thresholds-zeta}. It has additional implications for the PBH mass, as we shall see in Section~\ref{sec:Apparent--Horizon-Formation-and-Primordial-Black-Hole-Mass}. A family of Gau{\ss}ian profiles with $\zeta( \tilde{r} ) = \mu\.\exp[ - ( r / r_{\mrm} )^{2\mspace{1.5mu}p} ]$ was considered in Reference~\cite{2019JCAP...11..012Y} for the estimate of the PBH abundance, taking into account different profiles in $\zeta( \tilde{r} )$ for $w = 1/3$. In this study, a range of thresholds was found, $0.442 < \delta_{\crm} < 0.656$ for $0.34 \lesssim p \lesssim 2$.

As we will see in more detail in Section~\ref{sec:Peak--Theory-Procedure-with-Curvature-Peaks}, the full curvature fluctuation $\zeta( \tilde{r} )$ can be connected to the inflationary power spectrum $\Pcal_{\zeta}( k )$. For now, we just focus on obtaining the radial dependence $\zeta( \tilde{r} ) = \mu\.g( \tilde{r} )$. Therefore, the amplitude $\mu$ of the curvature fluctuation in terms of the compaction-function peak $\Ccal( \tilde{r}_{\mrm} ) = \delta_{\mrm}$ is given by
\vs{-2mm}
\begin{align}
\label{eq:amplitud-mu}
    \mu
        = 
            \frac{\pm \sqrt{1 - \delta_{\mrm}/f( w )} - 1}
            {g'( \tilde{r}_{\mrm} )\.\tilde{r}_{\mrm}}
            \, .
\end{align}
Note that the factor inside the square root is always positive since $\delta_{\crm} \leq f( w )$. The solution with `$+$'-sign gives the $\mu$-value corresponding to fluctuations of type I, whereas taking the `$-$'-sign corresponds to fluctuations of type II. The critical $\mu$-value (denoted by $\mu_{\crm}$) is obtained upon substituting $\delta_{\crm}$ for $\delta_{\mrm}$ in Equation~\eqref{eq:amplitud-mu}.

In addition to the conditions assumed above, the PBH formation threshold has also been estimated by taking into account various additional effects{\,---\,}for instance, concerning its variation in the presence of anisotropic pressure~\cite{2022PhRvD.106h3017M}, or nontrivial velocity dispersions in a matter-dominated era~\cite{2023JCAP...02..038H}. Its modification for a time-dependent equation of state will be discussed in detail in Section~\ref{sec:Thermal--History--Induced-Mass-Function} (see Reference~\cite{2022arXiv220906196E} for a recent numerical study regarding the QCD epoch).

On the other hand, using a phenomenological approach, non-sphericities have been considered in Reference~\cite{2016PhRvD..94f3514K}. By considering ellipsoidal overdensities, an analytical approximation for the collapse threshold, which is larger than in the spherical case, has been derived:
\vs{-1mm}
\begin{align}
\label{eq:delta-ec}
	\frac{ \delta_{\rm ec} }{ \delta_{\crm} }
		&\simeq
			1
			+
			\kappa
			\left(
				\frac{ \sigma^{2} }
				{ \delta_{\crm}^{2} }
			\right)^{\mspace{-6mu}\tilde{\gamma}}
			\, ,
\end{align}
with `ec' denoting `ellipsoidal collapse'. Above, $\delta_{\crm}$ is the threshold value for spherical collapse and $\sigma^{2}$ is the amplitude of the density power spectrum at the given scale. The two phenomenological parameters $\kappa$ and $\tilde{\gamma}$ depend on the characteristics of the scenario under consideration, such as the statistics of the density field.

Note that Reference~\cite{2001MNRAS.323....1S} had already obtained this result for a limited class of cosmologies but this did not include the case of ellipsoidal collapse in a radiation-dominated model. Substantial numerical studies are still needed in order to carefully investigate the effects of the non-sphericities on PBH formation, but a recent work of numerical PBH formation beyond spherical symmetry, for the case of a radiation-dominated universe, has shown that the threshold is not substantially affected by slight deviations from sphericity~\cite{2020PhRvD.102d3526Y}. The situation can be different when non-zero angular momentum is considered, or for a softer equation of state than that of a radiation fluid. Regardless, as far as the ellipsoidal collapse is concerned, Equation~\eqref{eq:delta-ec} determines the form of the collapse threshold.
\vs{2mm}

\subsection{Apparent-Horizon Formation and Primordial Black Hole Mass}
\label{sec:Apparent--Horizon-Formation-and-Primordial-Black-Hole-Mass}
\vs{-1mm}
Sufficiently large cosmological perturbations (with $\delta_{\mrm} > \delta_{\crm}$) will not dissipate after entering the cosmological horizon and continue to grow until the formation of a trapped surface~\cite{1965PhRvL..14...57P}.
\newpage

In order to understand the concept of trapped surfaces in general relativity, we summarise some basic facts. For identifying when trapped surfaces are formed, we need to take into account the expansion $\Theta^{\pm} \coloneqq h^{\mu \nu}\.\nabla^{}_{\mspace{-4mu}\mu}\mspace{1mu}k^{\pm}_{\nu}$ of null geodesics' congruences $k^{\pm}$ orthogonal to a spherical surface $\Sigma$. Here, $h^{\mu \nu}$ is the metric induced on $\Sigma$. We can consider two possible congruences: one, $k_{\mu}^{+}$, inwards and another one, $k_{\mu}^{-}$, outwards whose components are given by $k_{\mu}^{\pm} = (A,\mspace{1.5mu}\pm\mspace{1.5mu}B,\mspace{1.5mu}0,\mspace{1.5mu}0)$ with $k^{+} \mspace{-2.5mu}\cdot k^{-} = -\mspace{1.5mu}2$. Therefore, the induced metric $h_{\mu \nu}$ is given by $h_{\mu \nu} = g_{\mu \nu} + ( k^{+}_{\mu}k^{-}_{\nu} + k^{-}_{\mu}k^{+}_{\nu} ) / 2$.

In the case of flat spacetime, we have so-called {\it normal} surfaces, which are characterised by $\Theta^{-} \!< 0$ and $\Theta^{+} \!> 0$. On the other hand, if both congruences have a positive expansion $\Theta^{\pm} > 0$ the surface is called {\it anti-trapped}, while if both are negative $\Theta^{\pm} < 0$, the surface is {\it trapped}. Specifically, as shown in Reference~\cite{2017CQGra..34m5012H}, taking into account the previous definitions together with Equation~\eqref{eq:gamma-constraint} and using $B = R\mspace{1mu}' / \Gamma$, the expansion of the congruences is given by $\Theta^{\pm} = 2\.( U \pm \Gamma ) / R$.

Under the assumption of spherical symmetry, any point spacetime $( r,\mspace{1.5mu}t )$ (which can be classified as normal, trapped, and anti-trapped) can be considered a closed surface $\Sigma$ with proper radius $R$. Specifically, we define an {\it apparent horizon} (AH) as a marginally-trapped surface, which has a transition from a normal to a trapped surface, characterised by $\Theta^{-}\!< 0$ and $\Theta^{+}\mspace{-1.5mu}= 0$. Taking the identity $\Theta^{+}\.\Theta^{-} \equiv ( U^{2} - \Gamma^{2} )\,4 / R^{2}$ into account, the condition for the existence of an apparent horizon is given by $U^{2} = \Gamma^{2}\,\Rightarrow\,2\mspace{1.5mu}M = R$. For a more detailed discussion about horizons, we refer the reader to References~\cite{2017PhRvD..95h4031M, 2017CQGra..34m5012H, 2005CQGra..22.2221D, 2008AnHP....9.1029W, 2004LRR.....7...10A, 2006CQGra..23..413B, 2013Galax...1..114F, 1994PhRvD..49.6467H, 2012PhRvD..85h4031J, 2014PhRvD..89l3502Y}.

Once an apparent horizon has formed, the initial PBH mass, \ie~the mass of the PBH, $M_{\irm}$, at the moment of formation of the first apparent horizon, $t_{\rm AH}$, will start to grow to a certain stationary value $M_{\frm}$. This situation is different from 
    ({\it i$\mspace{1.5mu}$}) 
        the case of dust collapse $w = 0$, where the mass would continuously increase due to the lack of pressure gradients which could avoid accretion, and 
    ({\it ii$\mspace{1.5mu}$}) 
        from an asymptotically-flat spacetime, where no accretion is expected.
 
\begin{figure}[t]
    \centering
    \includegraphics[width = 0.75\hsize]{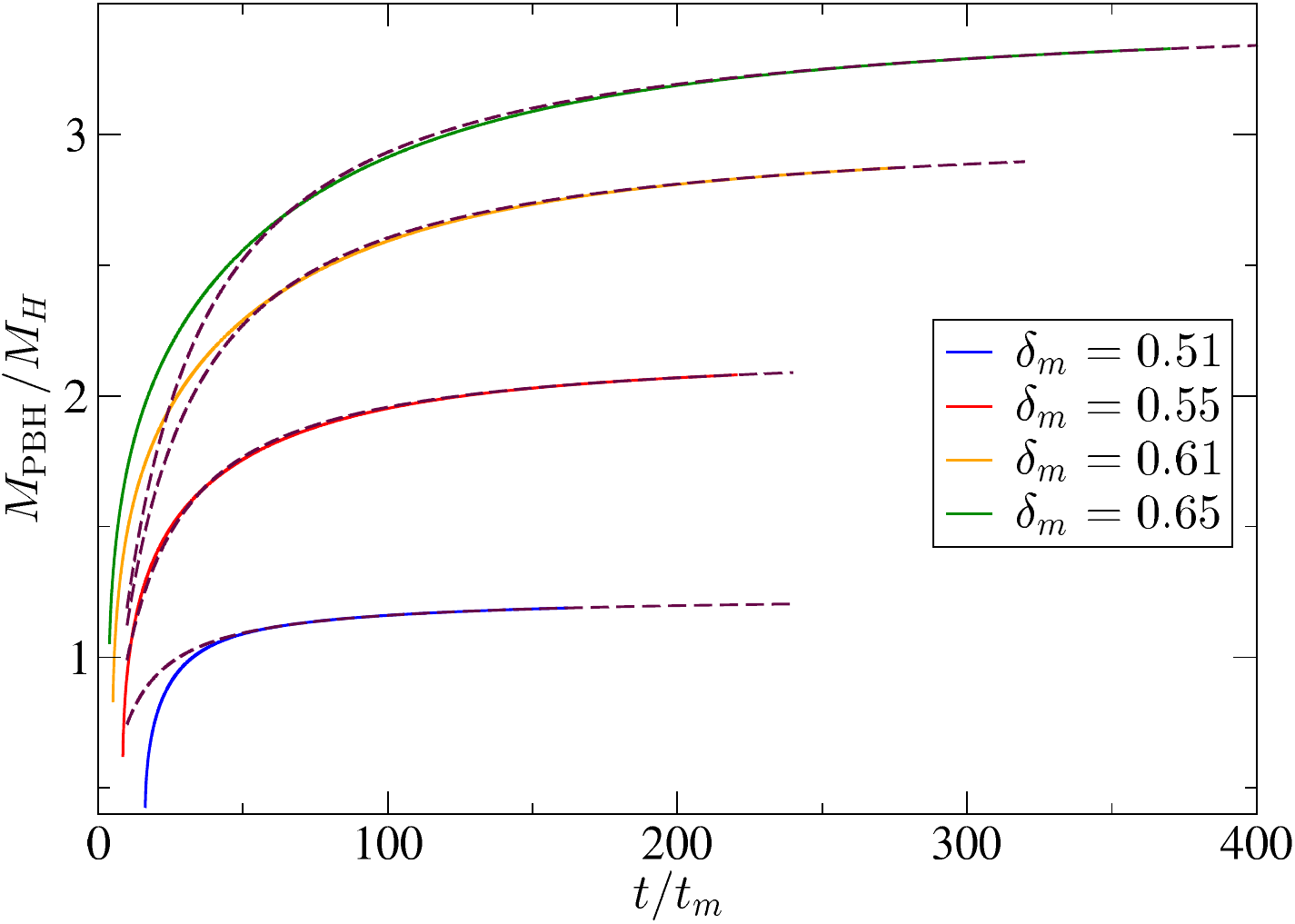}
    \caption{
        Time evolution of the PBH mass for $w = 1/3$ after the formation of the apparent horizon for a Gau{\ss}ian profile. Coloured lines correspond to different values of initial amplitude $\delta_{\mrm}$; the dashed line indicates an analytical fit to the Novikov--Zel'dovich approximation (\cf~Reference~\cite{2020PDU....2700466E}); a threshold value of $\delta_{\crm} \approx 0.498$ has been utilised. Figure from Reference~\cite{2020PDU....2700466E}.
        \vs{3mm}
        }
    \label{fig:mass-pbh-2}
\end{figure}

The process of accretion from a FLRW background has been intensively studied in References~\cite{2005PhRvD..71j4010H, 1967SvA....10..602Z, 2005PhRvD..71j4009H, 1974MNRAS.168..399C, 1998PhRvD..58b3504C}. It has also been shown numerically~\cite{2017JCAP...04..050D, 2020PDU....2700466E, 2022PhRvD.105j3538Y} that accretion from a FLRW background to the apparent horizon can be characterised by Bondi accretion~\cite{1967SvA....10..602Z, 2002PhRvD..66h3509G, 2011Prama..76..173N} at sufficiently late times $t \gg t_{\rm AH}$, which assumes that the energy density right outside the apparent horizon decreases as in a FLRW universe (see Figure~\ref{fig:mass-pbh-2} as an example).

The increase of the PBH mass is not important for fluctuations very close to the critical regime $\delta_{\mrm} \approx \delta_{\crm}$~\cite{2005PhRvD..71j4009H, 1974MNRAS.168..399C}. However, it can be substantial in the case of large fluctuations $\delta_{\mrm} \gg \delta_{\crm}$ as shown numerically in Reference~\cite{2021JCAP...05..066E}, up to $\Ocal( 10 )$, depending on the shape of the fluctuation. It has also been shown that for sharp profiles (\ie~large $q$) that the accretion effect is less important in comparison to broader profiles (\ie~small $q$), the pressure gradients preventing accretion. For those PBHs with a high probability to form, \ie~with $M \approx M_{H}( t_{H} )$ it was found that $M / M_{\irm} \approx 3$. Moreover, the effect of accretion is highly dependent on the specific value of $w$, such that for larger values it is smaller, because the pressure gradients are larger in this case~\cite{2021JCAP...05..066E}.

The final primordial black hole mass, $M_{\frm}$, after the completion of the accretion process from the FLRW background, depends on the specific profile of the fluctuation, its initial amplitude $\delta_{\mrm}$, and the equation-of-state parameter $w$ of the cosmological background fluid. Numerical simulations have shown that the black hole mass follows a scaling law~\cite{2009CQGra..26w5001M, 2013CQGra..30n5009M, 1998PhRvL..80.5481N, 2002CQGra..19.3687H, 1998PhRvL..80.5481N, 1994PhRvL..72.1782E} according to the critical phenomena in gravitational collapse~\cite{2007LRR....10....5G, 1993PhRvL..70....9C} when the amplitude of the perturbation $\delta$ is close to the critical value $\delta_{\mrm} - \delta_{\crm} \lesssim 10^{-2}\.$, \ie
\begin{align}
    M
        = 
            \Kcal\.M_{H}( t_{H} )
            \big(
                \delta_{\mrm}
                -
                \delta_{\crm}
            \big)^{\mspace{-2mu}\gamma}
            \, .
            \label{eq:2-scaling}
\end{align}
\newpage

\begin{figure}[t]
    \centering
    \includegraphics[width = 0.65\hsize]{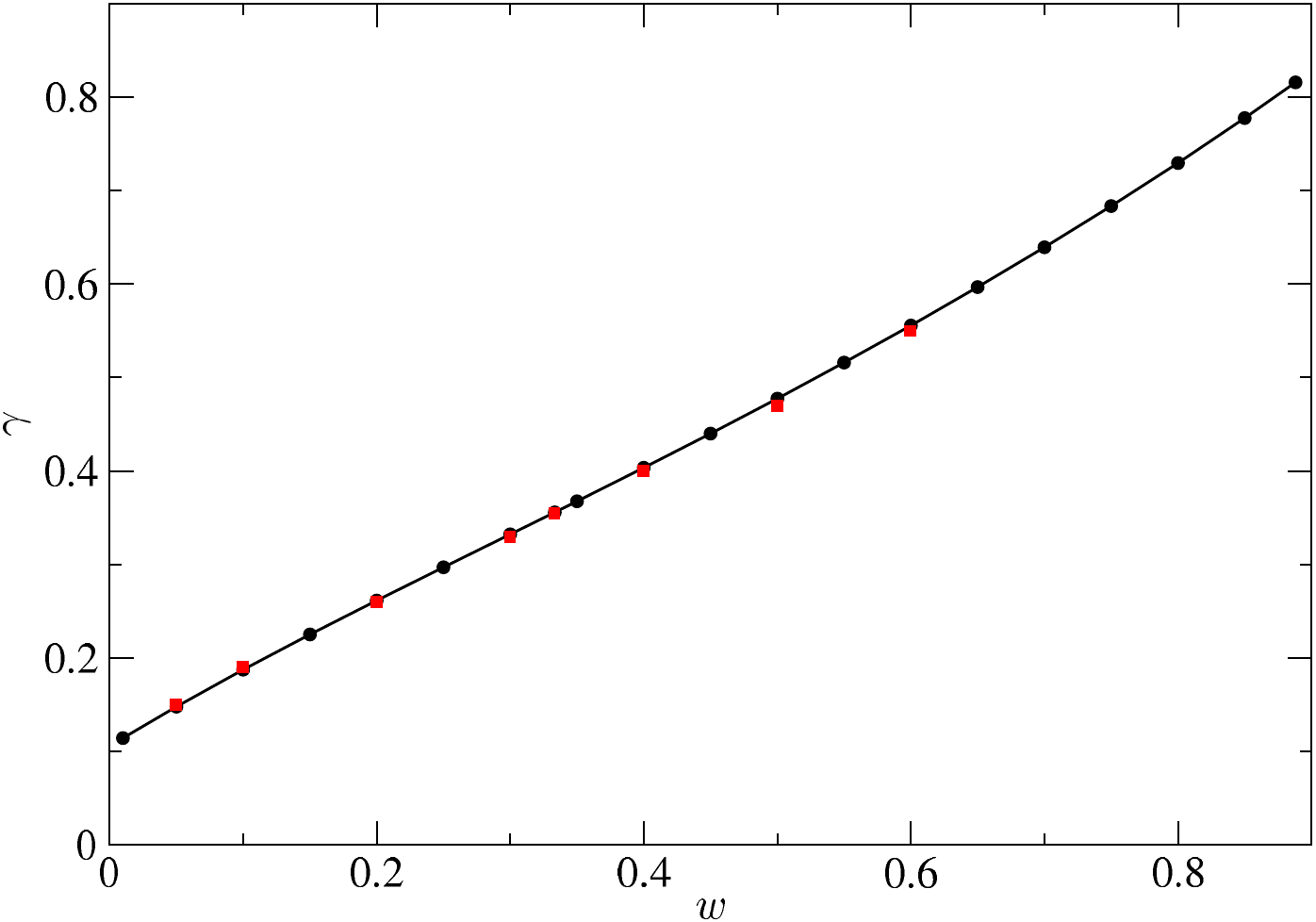}
        \includegraphics[width = 0.65\hsize]{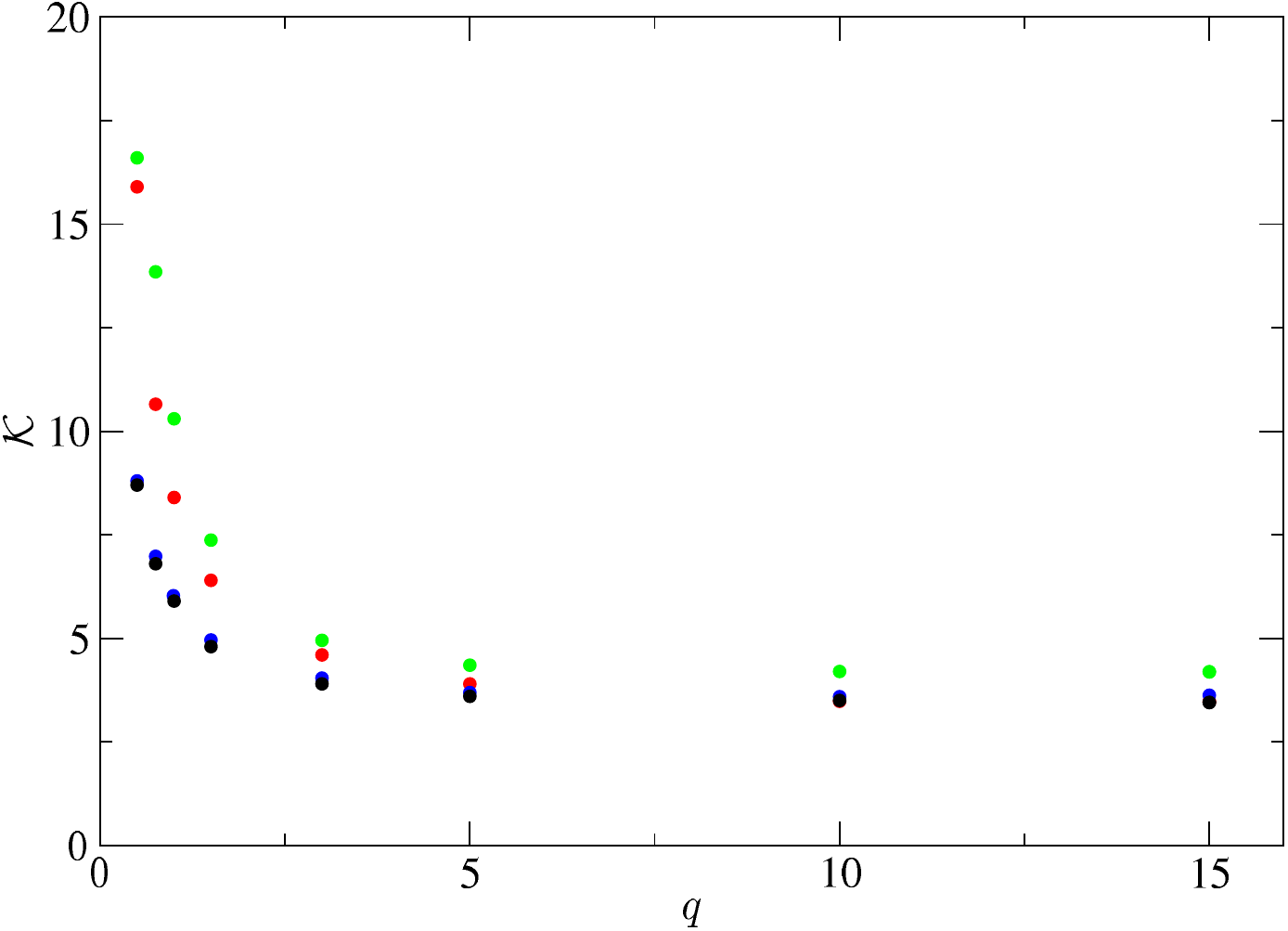}
    \caption{
        {\it Upper panel}:
            Critical exponent $\gamma$ as a function of $w$. Black points indicate the analytical solution from Reference~\cite{1996PhLB..366...82M}; red squares depict the numerical results of         Reference~\cite{2013CQGra..30n5009M}. 
        {\it Bottom panel}:
            Constant $\Kcal$ in terms of the $q$ parameter and         for different curvature profiles (see Reference~\cite{2021JCAP...05..066E} for details) in the case of $w = 1/3$. 
        Figure taken from Reference~\cite{2021JCAP...05..066E}.
        }
    \label{fig:critical-1}
\end{figure}

\noindent Here, $\gamma$ is a universal exponent which only depends on the equation-of-state parameter $w$. The dependence of $\gamma$ on $w$ was found semi-analytically in References~\cite{1996PhLB..366...82M, 1999PhRvD..59j4008K}, which has been numerically confirmed in Reference~\cite{2013CQGra..30n5009M}, see Figure~\ref{fig:critical-1}. The constant $\Kcal$ has a numerical value of $\Ocal( 1 )$ (see bottom panel of Figure~\ref{fig:critical-1}) and depends on both the shape of the curvature fluctuation~\cite{2013CQGra..30n5009M, 2021JCAP...05..066E} as well as on $w$. 

\newpage

\begin{figure}[t]
    \centering
    \includegraphics[width = 0.75\hsize]{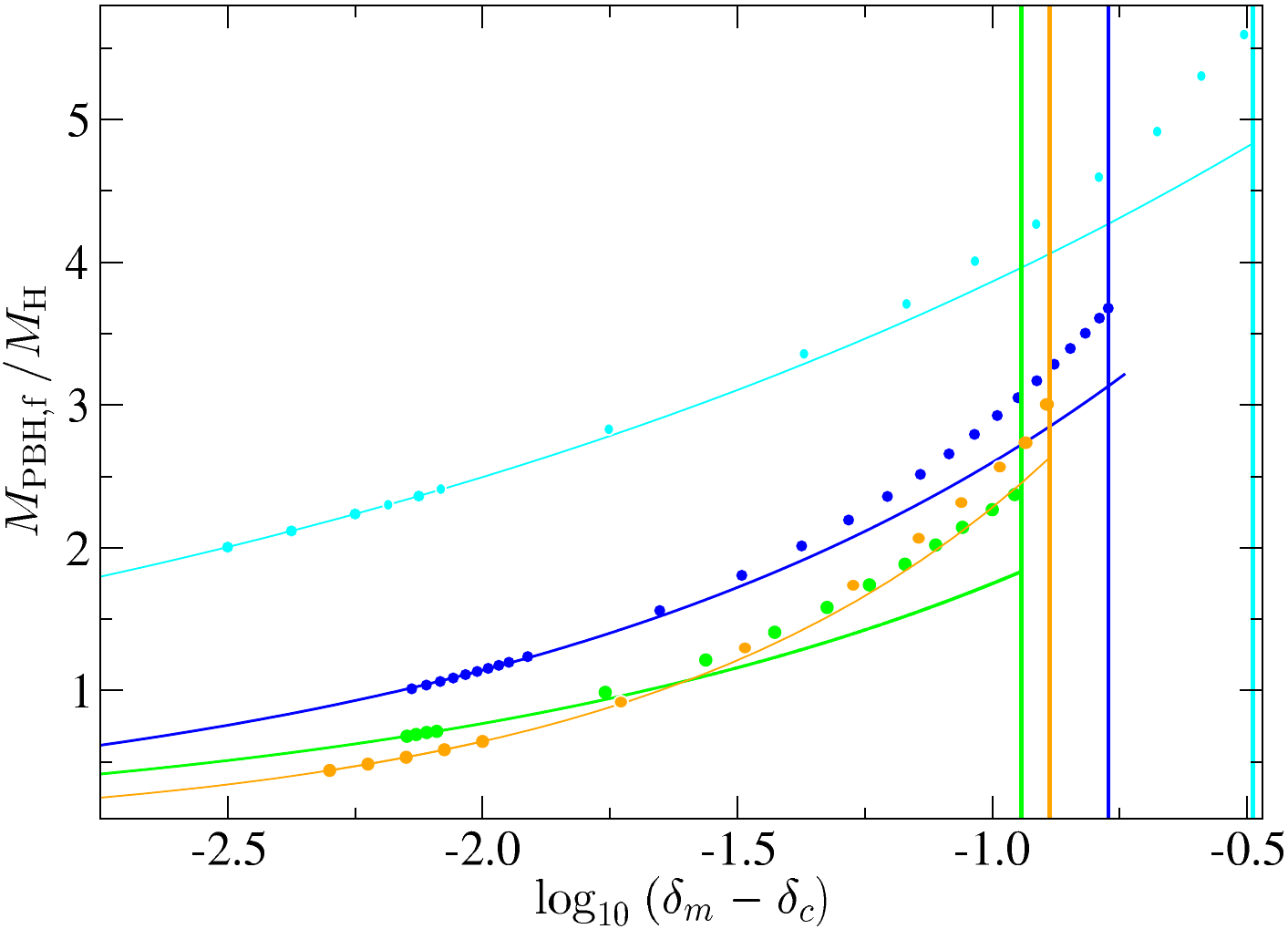}
     \caption{
        Final PBH mass $M_{\rm PBH_{f}} / M_{H}$ as a function of $(\delta_{\mrm} - \delta_{\crm})$ for two Gau{\ss}ian profiles, in $K( r )$ (blue) and $\zeta( \tilde{r} )$ (green) for $w = 1/3$. Also shown are the cases $w = 0.1$ (orange) and $w = 0.6$ (cyan), for a Gau{\ss}ian profile utilising $K( r )$. Points correspond to the numerical values obtained following Reference~\cite{2020PDU....2700466E}. The respective vertical lines set the maximally-allowed value for $\delta_{\umax} - \delta_{\crm}$; solid lines corresponds to the scaling-law behaviour for the different cases. Here, blue corresponds to $\gamma = 0.357$ with $\delta_{\crm} \approx 0.497$ for $\Kcal = 5.91$, green to $\gamma = 0.357$ with $\delta_{\crm} \approx 0.552$ for $\Kcal = 3.98$, cyan to $\gamma = 0.190$ with $\delta_{\crm} \approx 0.299$ for $\Kcal = 5.98$ and orange to $\gamma = 0.550$ with $\delta_{\crm} \approx 0.575$ for $\Kcal  = 8.10$. The results for $( \delta_{\mrm} - \delta_{\crm} ) \ll 10^{-2}$ are not shown since these essentially follow the scaling-law regime.
        }
    \label{fig:critical-2}
\end{figure}

An example of the primordial black hole mass dependence on $\delta_{\mrm}$ is shown in Figure~\ref{fig:critical-2}, wherein the blue and green colours represent results for Gau{\ss}ian profiles in $K( r )$ and $\zeta( \tilde{r} )$, respectively. Although both are Gau{\ss}ian, due to the nonlinear relation in the curvature amplitude~\eqref{eq:deltam-zeta}, these behave differently as a function of the threshold. The critical exponent is the same in both cases, utilising $w = 1/3$. 

Comparing the blue with the cyan and orange cases yields a different situation. For these three cases, the chosen profile is Gau{\ss}ian, but for three different values of the equation-of-state parameter: $w = 1/3$, $0.1$ and $0.6$. In these three cases, the critical exponent $\gamma$ as well as the factor $\Kcal$ are different (see caption of Figure~\ref{fig:critical-2}). Decreasing $w$ increases the PBH mass as pressure gradients become smaller.

The deviation of the scaling law for cosmological perturbations for large amplitudes $\delta_{\mrm}$ beyond the critical regime have been estimated to be $\Ocal( 10\,\text{--}\,15 )\mspace{0.5mu}\%$~\cite{2020PDU....2700466E}, but such cosmological fluctuations are also exponentially rarer than those with smaller deviation. Therefore, the former is usually neglected when estimating the PBH abundance and mass spectra.

\subsection{Analytical Threshold Formul{\ae}}
\label{sec:Analytical-Threshold-Formulae}
\vs{-2mm}
Numerous analytical estimates for $\delta_{\crm}$ have been proposed (see \eg~References~\cite{1974MNRAS.168..399C, 2013PhRvD..88h4051H}), these being based on analytical models with dependence on the equation of state of the cosmological fluid. Specifically, the first analytical approximation has been obtained by Carr~\cite{1975ApJ...201....1C} using a Jeans-length approximation,
\begin{align}
\label{eq:3-CARR}
     \delta_{\crm,\mspace{1.5mu}\rm Carr}
        = 
            w
            \, .
\end{align}
Subsequently, this estimate has been improved~\cite{2013PhRvD..88h4051H} by using a more sophisticated framework (a three-zone model) for the collapse of a homogeneous overdense sphere surrounded by a thin underdense shell, leading to\footnote{\setstretch{0.9}The acronym ``HYK" refers to T.~Harada, C.~Yoo and K.~Kohri{\,---\,}the authors of Reference~\cite{2013PhRvD..88h4051H}.}
\begin{align}
 \label{eq:3-HYK}
     \delta_{\crm,\mspace{1.5mu}\rm HYK}
        = 
            \frac{3\.( 1 + w )}{5 + 3\mspace{1.5mu}w}\,
            \sin^{2}\!
            \left(
                \frac{\pi\.\sqrt{w}}
                {1 + 3\mspace{1.5mu}w} 
            \right)
            .
\end{align}
Numerical simulations have shown that the threshold is furthermore sensitive to the specific shape of the cosmological fluctuations~\cite{1978SvA....22..129N, 1998PhRvL..80.5481N, 2005CQGra..22.1405M, 2015PhRvD..91h4057H, 1999PhRvD..60h4002S, 2019PhRvD.100l3524M, 2021JCAP...01..030E}. 

Recently, a new analytical approach utilising the averaged compact function (the average of $\Ccal$ up to the peak $r_{\mrm}$) has been introduced for an improved analytical threshold estimation~\cite{2020PhRvD.101d4022E}. It takes into account the shape dependence on the curvature fluctuation, which matches that found in simulations to within a few per cent. Specifically, for a radiation-dominated universe, the universal value $\bar{\Ccal}_{\crm} = 2/5$ has been found, independently of the underlying profiles. Here, $\bar{\Ccal}$ is defined as
\begin{align}
\label{eq:averaged-critical-C}
    \bar{\Ccal}_{\crm}
        \coloneqq
            \frac{3}{R^{3}_{\mrm}}\.
            \int_{0}^{R_{\mrm}}
            \d\tilde{R}\;
            \Ccal_{\crm}
            \big(
                \tilde{R}
            \big)
            \tilde{R}^{2}
            \, .
\end{align}
In particular, using the profile of Equation~\eqref{eq:K-polynominal} for Equation~\eqref{eq:averaged-critical-C} yields an analytic threshold formula which only depends on $q$~\cite{2020PhRvD.101d4022E}:
\begin{align}
\label{eq:threshold-analit}
    \delta_{\crm}( q )
        = 
            \frac{4}{15}\,\erm^{-\mspace{1mu}1 / q}\.
            \frac{q^{1 - 5 / 2\mspace{1.5mu}q }}
            {\Gamma( 5 / 2\mspace{1.5mu}q )
            -
            \Gamma( 5 / 2\mspace{1.5mu}q\.,\mspace{1.5mu}1/ q )}
            \, .
\end{align}
The dimensionless parameter $q$ has already been introduced in Equation~\eqref{eq:q-factor} (or, respectively, in Equation~\eqref{eq:q-tilde} as far the $\tilde{r}$-coordinate is concerned).

Using a similar approach as Reference~\cite{2020PhRvD.101d4022E}, Reference~\cite{2021JCAP...01..030E} provides an updated analytical threshold formula which depends on the equation of state and is valid for $w \geq 1/3$,
\vs{-2mm}
\begin{align}
\label{eq:averaged-critical-C-w}
    \bar{\Ccal}_{\crm}( w )
        = 
            \frac{3}{R^{3}_{\mrm}\.V[\alpha( w )]}\,
            \int_{R_{\mrm}[ 1 - \alpha( w ) ]}^{R_{\mrm}} 
            \d\tilde{R}\;
            \Ccal_{\crm}
            \big(
                \tilde{R}
            \big)
            \tilde{R}^{2}
            \, ,
\end{align}
where $V[\alpha( w )]\!\!\coloneqq\!\!\alpha( w )\mspace{1.5mu}[\mspace{1.5mu}3 + ( \alpha( w ) - 3 )\.\alpha( w ) ]$. In turn, for the polynomial profile~\eqref{eq:K-polynominal}, the critical threshold has been found to be
\begin{align}
\label{eq:3-threshold-anal}
     \delta_{\crm}( w,\mspace{1.5mu}q )
        = 
            \frac{ \bar{\com}_{\crm}( w ) }
            { p\mspace{1mu}( w,\mspace{1.5mu}q ) }\,
            \frac{1}{
            \big[
                ( 1 - \alpha )^{3-2\mspace{1.5mu}q}\.
                F_{2}( q,\mspace{1.5mu}\alpha )
                -
                F_{1}( q )
            \big]}
            \, ,
\end{align}
where
\vs{-3mm}
\begin{subequations}
\begin{align}
\label{eq:average-c-fit1}
    \bar{\com}_{\crm}( w )
        &= 
            a
            +
            b\,{\rm Arctan}
            \big( 
                c\,w^{d} 
            \big)
            \, ,
            \\[3mm]
\label{eq:alpha-fit1}
    \alpha( w )
        &= 
            e 
            + 
            f\.{\rm Arctan}
            \big( 
                g\,w^{h} 
            \big)
            \, ,
            \\[-8mm]
            \notag
\end{align}
\end{subequations}
and 
\vs{-1mm}
\begin{align}
    p\mspace{1mu}( w,\mspace{1.5mu}q )
        =
            \frac{3\mspace{1.5mu}( 1 + q )}
            {\alpha( w )\mspace{1.5mu}
            ( 2\mspace{1.5mu}q - 3 )
            \big[\.
                3
                +
                \alpha( w )\mspace{1.5mu}
                \big(
                    \alpha( w ) - 3
                \big)
            \big]} 
            \, ,
            \\[-8mm]
            \notag
\end{align}
and
\begin{subequations}
\begin{align}
    F_{1}( q )
        &= 
            {}_{2}F_{1}\!
            \left[
                1,\mspace{1.5mu}
                1 - \frac{5}{2\mspace{1.5mu}( 1 + q )},\mspace{1.5mu}
                2 - \frac{5}{2\mspace{1.5mu}( 1 + q )},\mspace{1.5mu}
                -\mspace{1.5mu}q\.
            \right]
            ,
            \displaybreak[1]
            \\[3mm]
    F_{2}(q,\mspace{1.5mu}w)
        &=
            {}_{2}F_{1}\!
            \left[
                1,\mspace{1.5mu}
                1
                -
                \frac{5}{2\mspace{1.5mu}( 1 + q )},\mspace{1.5mu}
                2
                -
                \frac{5}{2\mspace{1.5mu}( 1 + q )},\mspace{1.5mu}
                - q\mspace{1.5mu}
                \big[
                    1
                    -
                    \alpha( w )
                \big]^{-2( 1 + q )}
            \right]
            ,
\end{align}
\end{subequations}
where ${}_{2}F_{1}$ is the Gau{\ss} hypergeometric function, and 
$a = -\.0.140381$, 
$b = 0.79538$, 
$c = 1.23593$, 
$d = 0.357491$, 
$e = 2.00804$, 
$f = -\.1.10936$, 
$g = 10.2801$ and 
$h = 1.113$.\footnote{\setstretch{0.9}Another set of fitting values following an equivalent but independent approach has been obtained in the same work (see Reference~\cite{2021JCAP...01..030E} for details).} For the case $w = 1/3$ one finds $\bar{\com}_{\crm}( 1/3 ) = 0.412748$ and $\alpha( w = 1/3 ) = 0.619466$. The estimate using Equation~\eqref{eq:3-threshold-anal} for the same equation-of-state parameter is slightly more accurate than that of Equation~\eqref{eq:threshold-analit}, as shown in Reference~\cite{2021JCAP...01..030E}. Therein, it was found that the analytical estimate for $w \geq 1/3$ gives results within $\approx 6 \mspace{0.5mu}\%$ of the simulated values, and within $\approx 2 \mspace{0.5mu}\%$ for the case $w = 1/3$.
\newpage

The analytical estimates of Equations~\eqref{eq:threshold-analit} and \eqref{eq:3-threshold-anal} are contrasted with numerical simulations. This is particularly important when estimating the abundance of PBHs, which is exponentially sensitive to the threshold of formation. We refer the reader to Reference~\cite{2021JCAP...01..030E} for a comparison and discussion between the different existing analytical estimates.

To finalise, the main conclusion of References~\cite{2020PhRvD.101d4022E, 2021JCAP...01..030E} can be summarised as:
\begin{tcolorbox}
\centering
    {\it The physics of the gravitational collapse of a cosmological perturbation, which affects the threshold value for primordial black hole formation,\\
    mainly depends upon its shape around the compaction-function peak\\
    and the equation of state of the collapsing medium.}
\end{tcolorbox}

\subsection{Threshold-Estimation Scheme}
\label{sec:Threshold--Estimation-Scheme}
\vs{-1mm}
This Section provides a clear and detailed scheme for estimating the PBH threshold using the latest analytical results. We contrast these by numerical simulations quoted in Section~\ref{sec:Collapse--Threshold-Definition}. The presented procedure (utilising the results of Section~\ref{sec:Analytical-Threshold-Formulae}) has already been partly discussed in previous works~\cite{2020JCAP...05..022A, 2022JCAP...05..012E, 2021JCAP...10..053K, 2021PhRvD.103f3538M} for $w = 1/3$; here, we extract its essential aspects in enhanced clarity, such that it can be directly applied by the reader. Generally, there are basically two ways to estimate the PBH formation threshold. Firstly, we can follow the remarkable result of References~\cite{2021JCAP...01..030E, 2020PhRvD.101d4022E} that the averaged critical compaction function is a universal quantity independent on the curvature profile considered (upon deviation at per-cent-level). In turn, from a given profile [in $K( r )$ or $\zeta( \tilde{r} )$] one can compute the compaction function and integrate it iteratively with different amplitude values $\mu$ or $\Acal$ in order to find their critical values such that Equations~(\ref{eq:averaged-critical-C}--\ref{eq:averaged-critical-C-w}) hold. Secondly, we can also take profit of the analytical estimate of Equations~(\ref{eq:threshold-analit}--\ref{eq:3-threshold-anal}) in order to obtain the threshold in a direct way. This last procedure is explained in detail below.

\begin{sidewaysfigure}[t]
     \centering
     \includegraphics[width = .88\hsize]{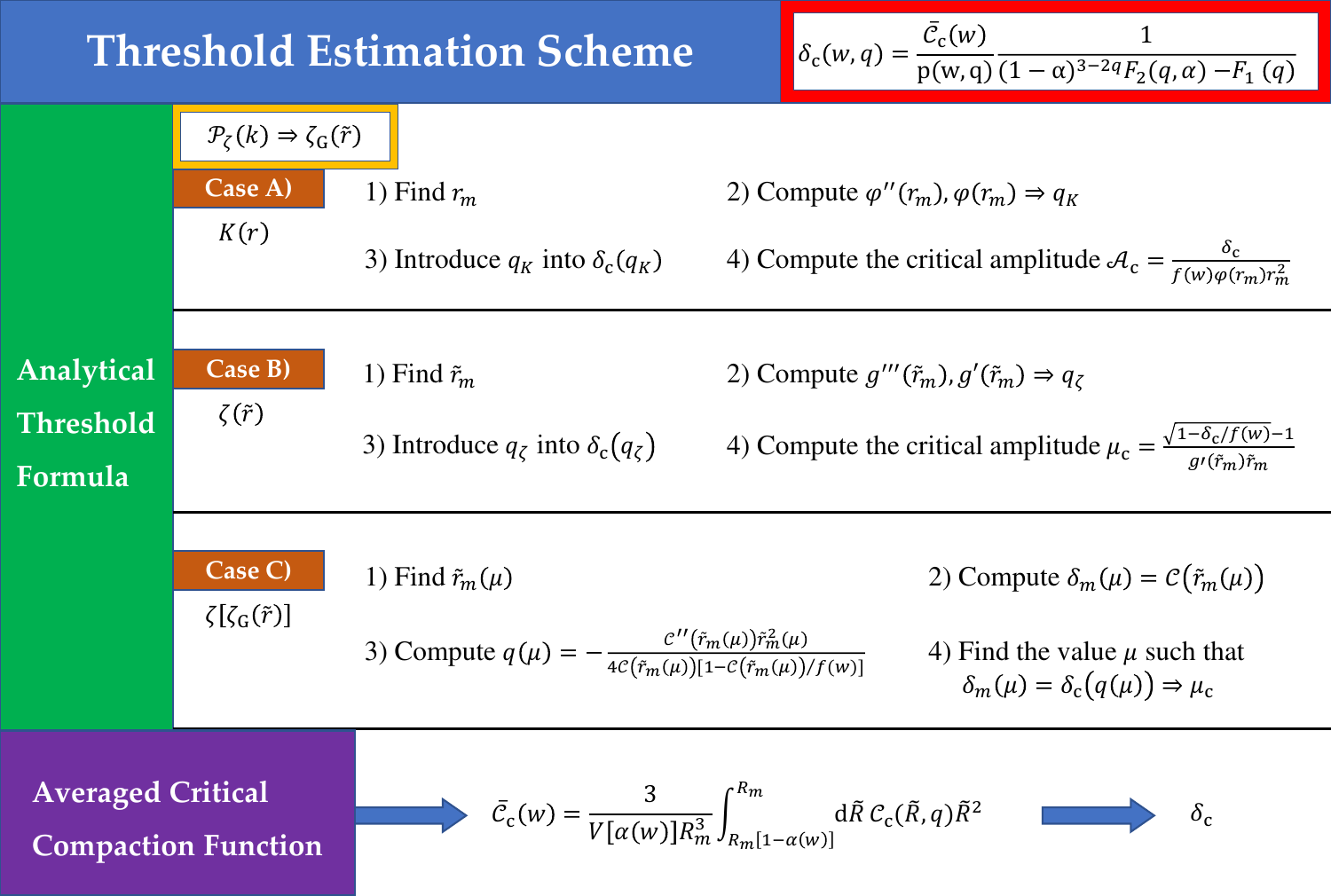}
     \vs{1.5mm}
     \caption{
        Skeleton of the analytical procedure for estimating the PBH formation threshold for $w \geq 1/3$. In the case of $w = 1/3$, the equations for $\bar{\com}_{\crm}( w )$ and $\delta_{\crm}( w,\mspace{1.5mu}q )$ can also be substituted with Equations~(\ref{eq:averaged-critical-C}) and (\ref{eq:threshold-analit}), respectively.
        }
     \label{tab:diagram-procedure}
\end{sidewaysfigure}

We consider three different situations for obtaining the threshold $\delta_{\crm}$:
    ({\it A}) 
        starting from a curvature $K( r )$,
    ({\it B})
        starting from a curvature $\zeta( \tilde{r} )$, and 
    ({\it C}$\mspace{1mu}$) 
        starting from a curvature that has a (local-type) non-Gau{\ss}ian contribution $\zeta( \tilde{r} ) = \zeta\big[ \zeta_{\Grm}( \tilde{r}) \big]$ (see also Section~\ref{sec:Aspects-of-Inflationary-Quantum-Perturbations}). 
A simple skeleton of the procedure is visualised in Figure~\ref{tab:diagram-procedure}. We summarise the main points below, followed by a few illustrative applications.
\newpage

Before proceeding, we make suitable refinements of some utilised quantities. In turn we separate the curvature perturbations into its amplitude $\Acal$ and radial dependence $\varphi( r )$,
\vs{-2mm}
\begin{subequations}
\begin{align}
    K( r )
        &= 
            \Acal\,\varphi( r )
            \, ,
\intertext{likewise,}
\vs{-2mm}
    \zeta_{\crm}( \tilde{r} )
        &= 
            \mu_{\crm}\,g( \tilde{r} )
            \, .
\end{align}
\end{subequations}

Then, the $q$-factors in the terms of the curvature perturbations and their derivatives read
\begin{subequations}
\begin{align}
\label{eq:q-k}
    q_{K}
        &= 
            -\frac{\varphi''( r_{\mrm} )\.r^{2}_{\mrm}
            - 6\.\varphi( r_{\mrm} )}
            {4\.\varphi( r_{\mrm} )}
            \, ,
            \\[3mm]
            \notag
            \displaybreak[1]
            \\[1mm]
\begin{split}
    q_{\zeta}
        &= 
            \frac{2\.\zeta_{\crm}'( \tilde{r}_{\mrm} )
            - \tilde{r}^{2}_{\mrm}\.
            \zeta_{\crm}'''( \tilde{r}_{\mrm} )}
            {\zeta_{\crm}'( \tilde{r}_{\mrm} )}\.
            \frac{1}{
            \big[
                4
                +
                2\.\tilde{r}_{\mrm}\.
                \zeta_{\crm}'( \tilde{r}_{\mrm} )
            \big]\!
            \big[
                1
                +
                \tilde{r}_{\mrm}\.
                \zeta_{\crm}'( \tilde{r}_{\mrm} )
            \big]}
            \\[3mm]
        &= 
            G( \tilde{r}_{\mrm} )\.
            \frac{1}{\sqrt{1 - \delta_{\crm}( q )/f( w )}
            \left[
                1
                +
                \sqrt{ 1- \delta_{\crm}( q )/f( w )}
            \right]}
            \, ,
            \label{eq:q-zeta}
\end{split}
\end{align}
with
\begin{align}
    G( \tilde{r} )
        \coloneqq
            \frac{g'( \tilde{r} )
            - \tilde{r}^{2}\.
            g'''( \tilde{r} ) / 2}
            {g'( \tilde{r} )}
            \, ,
\end{align}
\end{subequations}
where in the second line of Equation~\eqref{eq:q-zeta} we used Equation~\eqref{eq:amplitud-mu} in order to rewrite $\tilde{r}_{\mrm}\.\zeta'( \tilde{r}_{\mrm} )$ in terms of $\delta_{\crm}( q )$, which avoids the $\mu$-dependence of $q_{\zeta}$. Notice that this is only possible when only the Gau{\ss}ian contribution is considered. 
In the non-Gau{\ss}ian case, $q_{\zeta}$ can still depend on $\mu$, as explained in Reference~\cite{2021JCAP...10..053K}. For definiteness, in the subsequent examples, we mainly focus on the case of radiation, \ie~using $w = 1/3$.
\newpage

\subsubsection{Case A: Gau{\ss}ian Curvature $K$}

\vs{-3mm}
\begin{tcolorbox}
\begin{flushleft}
\begin{enumerate}

    \item
        From $K( r )$ find the location $r_{\mrm}$ of the maximum of the compaction function $\Ccal$ using Equation~\eqref{eq:rm-eq}.

    \item
        Compute the values $\varphi''( r_{\mrm} )$ and $\varphi( r_{\mrm} )$ and evaluate Equation~\eqref{eq:q-k} in order to estimate $q_{K}$.

    \item
        Obtain the peak of the critical compaction function by evaluating $\delta_{\crm}( q_{K} )$, using Equation~\eqref{eq:threshold-analit} for a radiation- dominated universe, or Equation~\eqref{eq:3-threshold-anal} for a more general equation of state with $w \geq 1/3$.

    \item
        Compute the corresponding critical amplitude $\Acal_{\crm}$ using\\
        the compaction-function peak $\delta_{\crm}( q_{K} )$,
        \vs{-1mm}
        \begin{align}
            \Acal_{\crm}
                = 
                    \frac{\delta_{\crm}( q_{K} )}
                    {f( w )\.\varphi( r_{\mrm} )\.r^{2}_{\mrm}}
                    \, .
    \end{align}

\end{enumerate}
\end{flushleft}
\end{tcolorbox}
\vs{1mm}

\noindent\textbf{Example A.1}
Consider the curvature
\begin{align}
\label{eq:curvature-K}
    K( r )
        = 
            \Acal\,
            \frac{1}{1 + ( r / B )^{C}}
            \, ,
\end{align}
where the model parameters $B$ and $C$ satisfy $B > 0$ and $C > 2$ [in order to ensure regularity and boundary conditions with $K'( r \rightarrow 0 ) = 0$ and $K''( r \rightarrow 0 ) = 0$]. Applying the first step (1) leads to
\vs{-2mm}
\begin{align}
    r_{\mrm}
        = 
            B\mspace{-1.5mu}
            \left(
                \frac{2}{C - 2}
            \right)^{\mspace{-6mu}1/C}
            \, .
\end{align}
Following step (2), evaluation of $\varphi( r_{\mrm} )$ and $\varphi''( r_{\mrm} )$ yields $q_{K} = ( C - 2 )/2$. In order to get the threshold $\delta_{\crm}$ [\hs{0.2mm}step\,(3)\hs{-0.3mm}], we solve Equation~\eqref{eq:threshold-analit} or~\eqref{eq:3-threshold-anal} with $q = q_{K}$, which yields a numerical value upon choosing a specific value of $C \in \Rbb$. Finally,\, [\hs{0.2mm}step\,(4)\hs{-0.3mm}], we obtain $\Acal_{\crm}$ as
\begin{align}
    \Acal_{\crm}
        = 
            \delta_{\crm}\.
            \frac{4^{-1/C}( C - 2 )^{- 1 + 2/C}\.C}
            {B^{2}\mspace{1.5mu}f( w )}
        = 
            \frac{C}{C - 2}\.
            \frac{\delta_{\crm}}
            {f( w )\.r^{2}_{\mrm}}
            \, .
\end{align}
Notice that Equation~\eqref{eq:curvature-K} can be recast into Equation~\eqref{eq:K-polynominal} upon substituting $\Acal_{\crm}$ and $B$, and using $C = 2\.( q + 1 )$. The respective threshold values are shown in Figures~\ref{fig:thresholds-plots-1} and~\ref{fig:thresholds-plots-2}.
\newpage

\begin{figure}[t]
     \centering
     \hs{2mm}\includegraphics[width = 0.75\hsize]{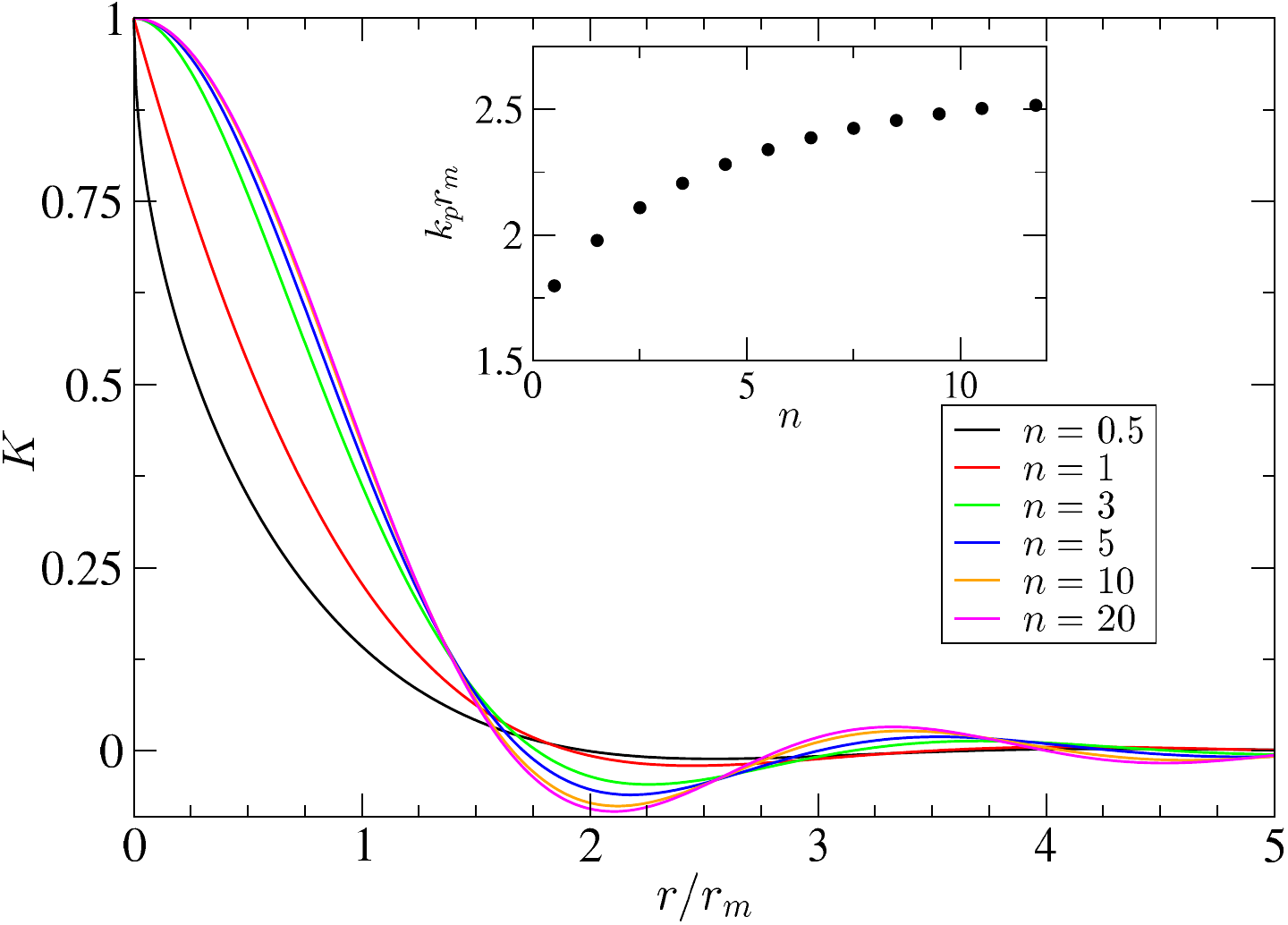}\\[4mm]
     \includegraphics[width = 0.75\hsize]{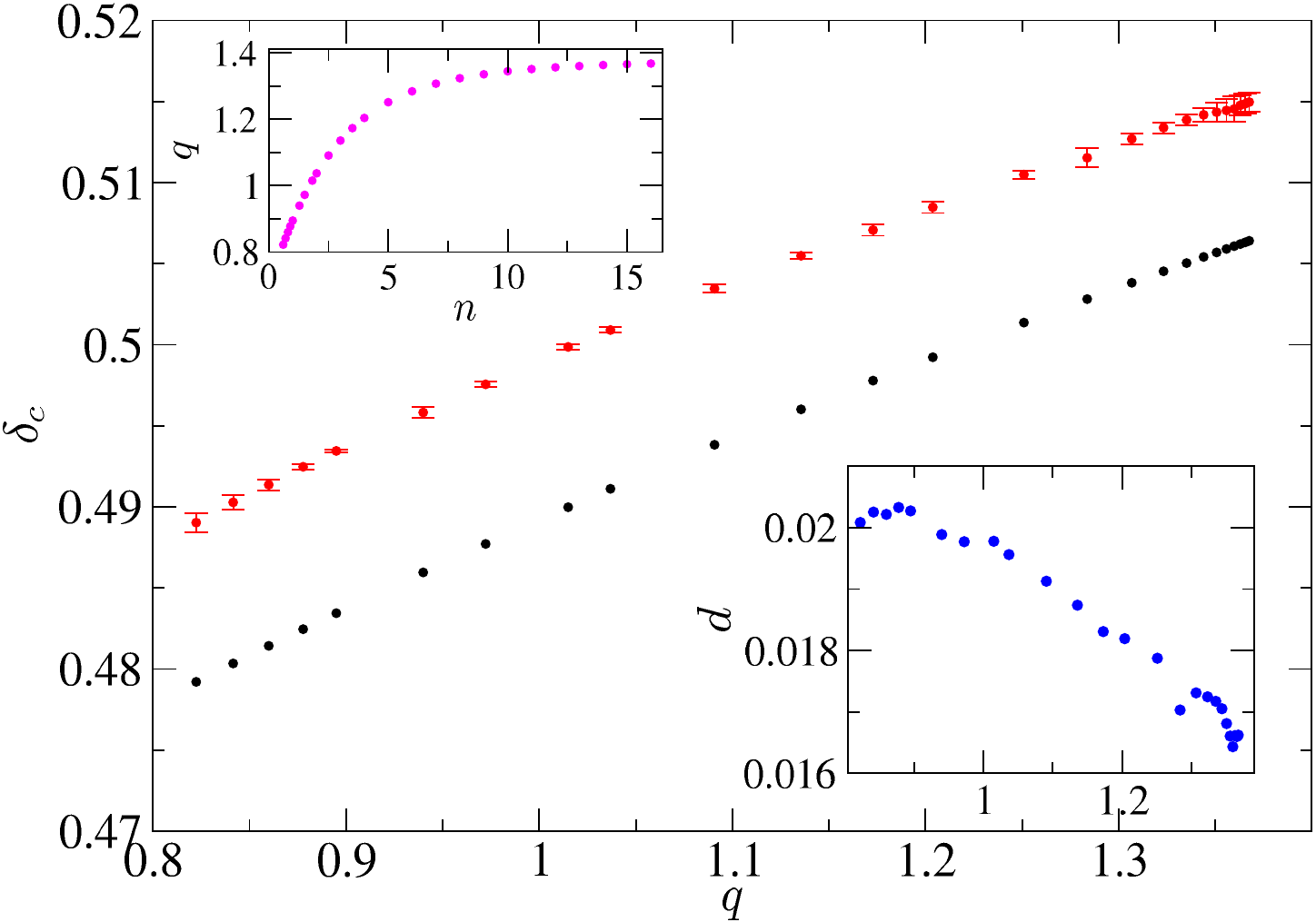}
     \caption{
        {\it Upper panel}: 
            Curvature profiles of Equation~\eqref{eq:K-ps} for different $n$-values as functions of radius $r$. The subplot shows the relation between $n$ and $k_{\prm}\mspace{1.5mu}r_{\mrm}$, obtained from numerically solving Equation~\eqref{eq:num-r-mk-p}.
        {\it Lower panel}: 
            Threshold values for the profile of Equation~\eqref{eq:K-ps}. Numerical results are indicated by red colour; black dots depict the threshold values obtained using the analytical estimation scheme. The relative deviation between both is shown in blue colour. Magenta dots indicate the relation between the $q$ and $n$. 
        The blue dots in the subplot of the lower panel show the relative deviation $d$ between the numerical results (red points) with the analytical estimate (black points). Figures (adapted) from References~\cite{2020PhRvD.101d4022E, 2020PDU....2700466E}.
        }
     \label{fig:plots-k}
\end{figure}

\noindent\textbf{Example A.2}
We now illustrate a more complex example. Consider the curvature
\vs{-5mm}
\begin{subequations}
\begin{align}
\label{eq:K-ps}
    K( r )
        &=
            K_{n}( r )
        = 
            \frac{\delta_{\mrm}}{f( w )\.r^{2}_{\mrm}}
            \frac{r^{3}_{\mrm}}{r^{3}}\.
            \frac{g_{n}( r,k_{\prm},\mspace{1.5mu}r )}
            {g_{n}( n,\mspace{1.5mu}k_{\prm},\mspace{1.5mu}r_{\mrm} )}
            \, ,
            \displaybreak[1]
            \\[4mm]
\begin{split}
    g_{n}( n,\mspace{1.5mu}k_{\prm},\mspace{1.5mu}r )
        &= 
            r\.k_{\prm}
            \big[
                E_{3+n}( -\.i\.k_{\prm}\mspace{1.5mu}r )
                +
                E_{3+n}( i\.k_{\prm}\mspace{1.5mu}r )
            \big]
            \\[2mm]
        &\phantom{=\;}
            +
            i\.
            \big[
                -
                E_{4+n}( i\.k_{\prm}\mspace{1.5mu}r )
                +
                E_{4+n}( -\.i\.k_{\prm}\mspace{1.5mu}r )
            \big]
            ,
\end{split}
\end{align}
\end{subequations}
which has been studied in Reference~\cite{2020PDU....2700466E}, and can be obtained from a power spectrum $\Pcal_{\zeta}( k ) = \Pcal_{0}\.( k/k_{\prm} )^{-n}$ for $k \geq k_{\prm}$ and $\Pcal_{\zeta}( k ) = 0$ for $k \leq k_{\prm}$~\cite{2019PDU....24..275A}, where $k_{\prm}$ is the location of the peak of the power spectrum. The function $E_{n}( x )$ above is defined as $E_{n}( x ) \coloneqq \int_{1}^{\infty}\d t\;\erm^{-x\mspace{1mu}t} / t^{n}$. Notice from Equation~\eqref{eq:K-ps} that $K( r_{\mrm} ) = \delta_{\mrm}/f( w )\.r^{2}_{\mrm}$. Applying step (1), the equation for finding $r_{\mrm}$ can be obtained by solving numerically the following equation, which has been obtained from Equation~\eqref{eq:rm-eq},
\begin{align}
\label{eq:num-r-mk-p}
    2\.
    \big[
        \sin( x )
        -
        x \cos( x )
    \big]
    +
    i\.x^{2}\.(2 + n)
    \big[
        E_{2 + n}(-\.i\.x)
        -
        E_{2 + n}(i\.x)
    \big]
        = 
            0
            \, ,
\end{align}
with $x = k_{\prm}\mspace{1.5mu}r$, which yields $x_{\mrm}( n ) = k_{\prm}( n )\.r_{\mrm}$ as a function of $n$ (see subplot of the upper panel of Figure~\ref{fig:plots-k}). Then from step (2), solving Equation~\eqref{eq:q-k} yields gives a relation between $q$ and $n$ (shown in a subplot of the lower panel of Figure~\ref{fig:plots-k}). Finally step (3): Solving Equation~\eqref{eq:threshold-analit} with $q = q\mspace{0.5mu}( n )$ leads to $\delta_{\crm}$. This is represented as black dots in the lower panel of Figure~\ref{fig:plots-k}. Notice that for this example, we automatically have $\Acal_{\crm} = \delta_{\crm}$.

\subsubsection{Case B: Gau{\ss}ian Curvature $\zeta$}

\vs{-2mm}
\begin{tcolorbox}
\begin{flushleft}
\begin{enumerate}

    \item
        Find the location $\tilde{r}_{\mrm}$ of the maximum of the compaction function $\Ccal$ using Equation~\eqref{eq:rm-eq}. 
    
    \item
        Compute $g'( \tilde{r}_{\mrm} )$ and $g'''( \tilde{r}_{\mrm} )$, and numerically solve  Equation~\eqref{eq:q-zeta} in order to find $q_{\zeta}$.
    
    \item
        Solve Equation~\eqref{eq:threshold-analit} (when considering radiation domination) 
        or Equation~\eqref{eq:3-threshold-anal} (for a more general equation of state with $w \geq 1/3$)
        for $q = q_{\zeta}$ in order to obtain the peak, $\delta_{\crm}$, of the critical compaction function.
    
    \item
        Obtain the corresponding critical amplitude $\mu_{\crm}$ from
        \vs{-1mm}
        \begin{align}
        \label{eq:mu-c-B}
            \mu_{\crm}
                =                
                    \frac{ \sqrt{1 - \delta_{\crm}/f( w )} - 1}
                    {g'( \tilde{r}_{\mrm} )\.\tilde{r}_{\mrm}}
                    \, .
    \end{align}

\end{enumerate}
\end{flushleft}
\end{tcolorbox}
\vs{4mm}

For illustrative proposes, we consider a simplified connection between the power spectrum $\Pcal_{\zeta}$ and the function $g( \tilde{r} )$. Specifically, we assume for the two-point correlation function, 
\begin{align}
\label{eq:two-point-correlation}
    g( \tilde{r} )
        = 
            \frac{1}{\sigma^{2}_{0}}
            \int_{0}^{\infty} \frac{\d k}{k}\;
            \frac{\sin( k\.\tilde{r} )}{k\.\tilde{r}}\,
            \Pcal_{\zeta}( k )
            \, ,
\end{align}
with 
\begin{align}
    \sigma^{2}_{0}
        \coloneqq
            \int_{0}^{\infty}\d k\;
            \frac{\Pcal_{\zeta}( k )}{k}
            \, .
\end{align}
Therefore, we have
\vs{-1mm}
\begin{align}
\label{eq:zeta-simple}
    \zeta( \tilde{r} )
        = 
            \mu\.g( \tilde{r} )
            \, .
\end{align}
It should be noted, that Equations~(\ref{eq:two-point-correlation}--\ref{eq:zeta-simple}) are good approximations (indeed exact) when the power spectrum is monochromatic, but not when the power spectrum is broad. In Section~\ref{sec:Peak--Theory-Procedure-with-Curvature-Peaks} we will see a general and more refined construction of the function $g( \tilde{r} )$.
\newpage

\begin{figure}[t]
     \centering
     \includegraphics[width = 0.63\hsize]{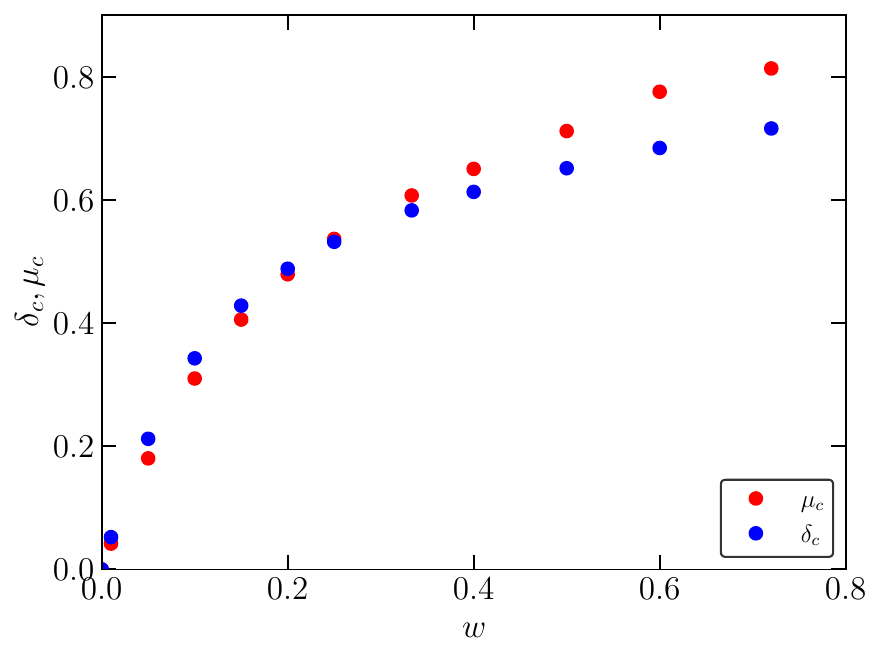}
     \vs{-2mm}
     \caption{
         Threshold values for the curvature profile given by a monochromatic power spectrum. Numerical results (obtained using Reference~\cite{2020PDU....2700466E}) are represented by the dots, with blue corresponding to $\delta_{\crm}$ and red to $\mu_{\crm}$.}
     \label{fig:plots-w-monocromatic}
\end{figure}

\noindent\textbf{Example B.1}
Consider a monochromatic power spectrum given by $\Pcal_{\zeta}( k ) = \Pcal_{0}\,\delta\big( \!\ln[ k / k_{*} ] \big)$ with $\delta$ being the Dirac delta function, $\Pcal_{0} = \sigma_{0}^{2}$, and $k_{*}$ is the location of the peak of the power spectrum. First, making the anti-Fourier transform in order to compute the two-point correlation function from Equation~\eqref{eq:two-point-correlation}, leads to $g( \tilde{r} ) = \sinc( k_{*}\.\tilde{r} )$. Then, following step (1), compute $r_{\mrm}$ by numerically solving
\begin{align}
    \zeta'( \tilde{r} )
    \,\.+\,\.
        &\tilde{r}\.\zeta''( \tilde{r} )
            = 
            0
            \notag
            \\[2mm]
        &\Downarrow
            \\[2mm]
            \sin( \tilde{x} )
            -
            \tilde{x}\mspace{1mu}
            \big[
                \cos
        &( \tilde{x} )
                +
                \tilde{x}\.\sin( \tilde{x} )
            \big]
        = 
            0
            \, ,
            \notag
\end{align}
with $\tilde{x} \coloneqq \tilde{r}\.k_{*}$. This leads to $\tilde{r}_{\mrm} \approx  2.747 / k_{*}$. Applying step (2), evaluating $g'( \tilde{r}_{\mrm} )$ and $g'''( \tilde{r}_{\mrm} )$, yields
\vs{-2mm}
\begin{align}
    G( \tilde{x} )
        = 
            -\.2
            +
            \frac{3\.\tilde{x}^{2}}{2}
            +
            \frac{\tilde{x}^{3} \cos( \tilde{x} )}
            {\sin( \tilde{x} )
            -
            \tilde{x} \cos( \tilde{x} )}
            \, .
\end{align}
Then, taking $\tilde{x}_{\mrm} \mspace{-1mu}= \mspace{-1mu}\tilde{r}_{\mrm}\.k_{*}\!\approx\!2.747\.$ implies $\.G( \tilde{x}_{\mrm} )\!\approx\!2.766$. In turn, solving the transcendental equation~\eqref{eq:q-zeta}, we get $q_{\zeta} \approx 6.3$, which [\hs{0.2mm}step\,(3)\hs{-0.3mm}] from Equation~\eqref{eq:threshold-analit} yields $\delta_{\crm} \approx 0.588$ and [\hs{0.2mm}step\,(4)\hs{-0.3mm}] $\mu_{\crm}\!\approx\!0.628$; the numerical result is $\delta_{\crm,\mspace{1.5mu}\mathrm{num}}\!\approx\!0.589$~\cite{2020JCAP...05..022A}. The numerical values of the thresholds $\delta_{\crm}$ and $\mu_{\crm}$ are shown in Figure~\ref{fig:plots-w-monocromatic} for a selection of equation-of-state parameters $w$.

\newpage

\noindent\textbf{Example B.2}
Consider now the opposite situation: a flat scale-invariant power spectrum defined as $\Pcal_{\zeta}( k ) = \Pcal_{0}\,\Theta( k - k_{\umin} )\.\Theta( k_{\umax} - k )$ with $k_{\umax} \gg k_{\umin}$. This case has been considered in detail in References~\cite{2019PhRvL.122n1302G, 2019JCAP...11..001M, 2020PhLB..80735550D}. Using Equations~(\ref{eq:two-point-correlation}--\ref{eq:zeta-simple}) in order to compute $\zeta$ leads to
\begin{align}
\begin{split}
    \zeta( \tilde{r} )
        &= 
            \frac{\Pcal_{0}}{\sigma^{2}_{0}}\,
            \big[
                {\rm Cos}_{I}( k_{\umax}\.\tilde{r} )
                - 
                {\rm Cos}_{I}( k_{\umin}\.\tilde{r} )
                \\[-1mm]
        &\phantom{= \frac{\Pcal_{0}}{\sigma^{2}_{0}}\,
            \big[}-
                \sinc( k_{\umax}\.\tilde{r} )
                + 
                \sinc( k_{\umin}\.\tilde{r} )
            \big]
            \, ,
\end{split}
\end{align}
with the {\it cosine integral} ${\rm Cos}_{I}( x ) \coloneqq -\int_{x}^{\infty}\d t\,\cos( t ) /t$. In order to find $\tilde{r}_{\mrm}$ [\hs{0.2mm}step\,(1)\hs{-0.3mm}], we solve Equation~\eqref{eq:zeta-root}; in the limit $k_{\umax} \gg k_{\umin}$, this yields
\begin{align}
\label{eq:flat-ps}
    \zeta'( \tilde{r} )
    \,+\,
    &\tilde{r}\.\zeta''( \tilde{r} )
        = 
            0
            \notag
            \\[2mm]
        &\Downarrow
            \\[2mm]
    \tan( \tilde{x} )
    &-
    \tilde{x}
        = 
            0
            \, ,
            \notag
\end{align}
where $\tilde{x} = \tilde{r}\.k_{\umax}$. The numerical solution of Equation~\eqref{eq:flat-ps} gives $\tilde{x}_{\mrm} \approx 4.493$. At this point, the function
\vs{-1mm}
\begin{align}
    G( \tilde{x} )
        = 
            \frac{
            (
                4 
                - 
                \tilde{x}^{2}
            )
            \sin( \tilde{x} )
            - 
            4\.\tilde{x} \cos( \tilde{x} )}
            {2\.\tilde{x} 
            - 
            2 \sin( \tilde{x} )}
\end{align}
assumes the value $G( \tilde{x}_{\mrm} ) \approx 1.802$. Then, numerically solving Equation~\eqref{eq:q-zeta} \;\;\;\;\;[\hs{0.2mm}step\,(2)\hs{-0.3mm}] leads to $q_{\zeta} \approx 3.1$, in agreement the results of Reference~\cite{2021PhRvD.103f3538M}. This yields the threshold value $\delta_{\crm} \approx 0.553$ [\hs{0.2mm}step\,(3)\hs{-0.3mm}]. The corresponding $\mu_{\crm}$ [\hs{0.2mm}step\,(4)\hs{-0.3mm}] is given by Equation~\eqref{eq:mu-c-B} with $\mu_{\crm} \approx 0.486\,\sigma^{2}_{0} / \Pcal_{0}$ with $\sigma^{2}_{0} = \Pcal_{0}\.\log( k_{\umax}/k_{\umin} )$.

\begin{figure}[t]
    \centering
    \includegraphics[width = 0.44\hsize]{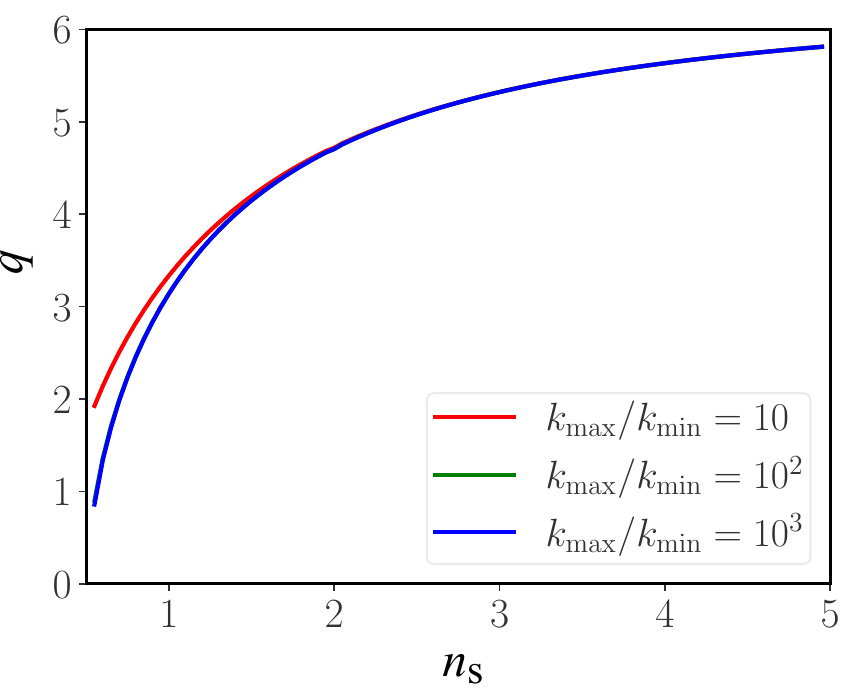}\;
    \includegraphics[width = 0.48\hsize]{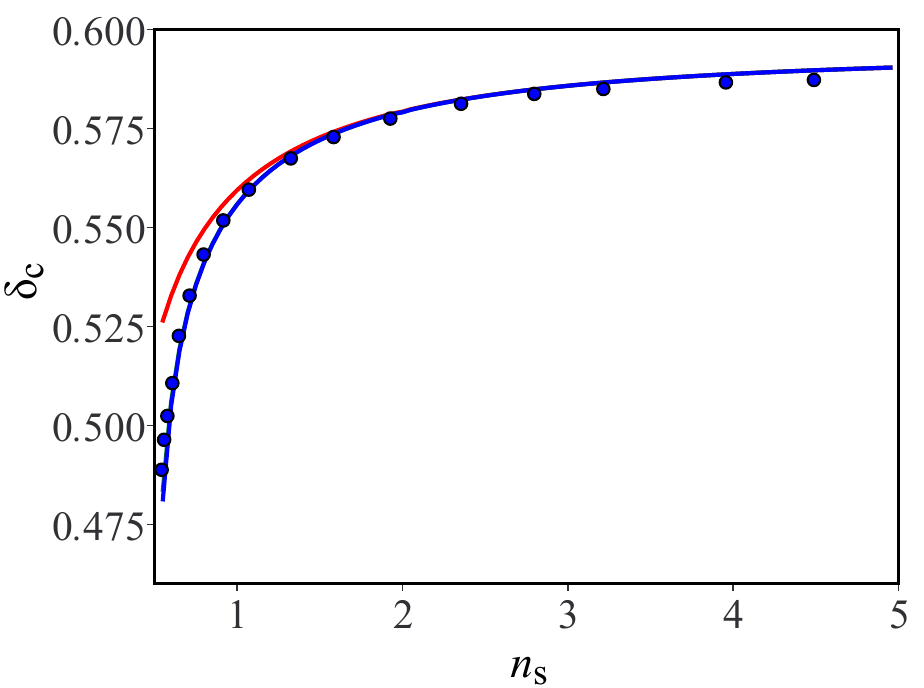}
    \caption{
        {\it Left panel}: 
            Parameter $q_{\zeta}$ for a nearly flat power spectrum as a function of the spectral index $n_{\srm}$ for different values of the ratio $k_{\umax} / k_{\umin}$.
        {\it Right panel}: 
            Threshold $\delta_{\crm}$ as a function of $n_{\srm}\mspace{1mu}$, for the values of $q_{\zeta}$ from the left panel utilising Equation~\eqref{eq:threshold-analit}. Blue points indicate the numerical values for $k_{\umax} / k_{\umin} = 10^{3}$.
        Notice that in both cases, the green line essentially overlaps the blue one.
        \vs{2mm}
        }
    \label{fig:analytical-nearly-flat-PS}
\end{figure}

\noindent\textbf{Example B.3} An example related to the previous one utilises the nearly flat, scale-invariant power spectrum $\Pcal_{\zeta}( k ) = \Pcal_{0}\. ( k/k_{\umin} )^{n_{\srm} - 1}\,\Theta( k - k_{\umin} )\.\Theta( k_{\umax} - k )$ with $k_{\umax} \gg k_{\umin}$, where $n_{\srm}$ is the spectral index. Again, using Equations~(\ref{eq:two-point-correlation}--\ref{eq:zeta-simple}) in order to compute $\zeta$, leads to
\begin{align}
     \zeta( \tilde{r} )
        = 
            \frac{\Pcal_{0}}{2\mspace{1.5mu}\sigma^{2}_{0}}\,
            i\mspace{1.5mu}
            \big(
                k_{\umin}\. \tilde{r}
            \big)^{\mspace{-2mu}1-n_{\srm}}\,
            \erm^{-i\mspace{1mu}n_{\srm} \pi/2}\,
            \big[
                \Gamma_{n_{\srm}(+)}( \tilde{r} )
                -
                e^{-i\mspace{1mu}n_{\srm} \pi/2}\.
                \Gamma_{n_{\srm}(-)}( \tilde{r} )
            \big]
            \, ,
\end{align}
where $\sigma^{2}_{0}\!=\!\Pcal_{0}\mspace{1mu}\big[ k_{\umax} - k_{\umin}\.( k_{\umax}/k_{\umin} )^{n_{\srm}} \big]\./\.[ k_{\umax}( 1 - n_{\srm} ) ]$ and $\Gamma_{n_{\srm}(\pm)}( \tilde{r} )\!\coloneqq\!\Gamma( n_{\srm} - 2,\.\pm\mspace{1.5mu}i\mspace{1.5mu}k_{\umax}\.\tilde{r} )\.-\.\Gamma( n_{\srm} - 2,\.\pm\mspace{1.5mu}i\mspace{1.5mu}k_{\umin}\.\tilde{r} )$. In order to find $\tilde{r}_{\mrm}$ [\hs{0.2mm}step\,(1)\hs{-0.3mm}], solve numerically Equation~\eqref{eq:zeta-root} in the limit $k_{\umax} \gg k_{\umin}$ with different ratios $k_{\umax} / k_{\umin}$. The numerical solution gives $\tilde{x}_{\mrm} \approx [ 5.6,\.3.2 ]$ with  $G( \tilde{x}_{\mrm} ) \approx [ 0.7,\.2.6 ]$ in the range $n_{\srm} \approx [ 0.5,\.5 ]$. Then numerically solving Equation~\eqref{eq:q-zeta} [\hs{0.2mm}step\,(2)\hs{-0.3mm}] leads to $q_{\zeta} \approx [ 1,\.6 ]$. This then yields the threshold values $\delta_{\crm} \approx [ 0.48,\.0.59 ]$ [\hs{0.2mm}step\,(3)\hs{-0.3mm}] for the mentioned range of the spectral index. Figure~\ref{fig:analytical-nearly-flat-PS} shows the value of the $q_{\zeta}$-parameters (\emph{left panel}) and the threshold (\emph{right panel}) for different values of $n_{\srm}$ and ratios $k_{\umax}/k_{\umin}$. Notice that for $n_{\srm} \rightarrow 1$ we recover the case of a flat scale-invariant power spectrum (see previous example).

\subsubsection{Case C: Non-Gau{\ss}ian Contribution to $\zeta$}
\label{sec:Case-C:-Non--Gaussian-Contribution-to-zeta}
\vs{-1mm}
Finally, consider a curvature with non-Gau{\ss}ian contribution, which we modulate with a non-Gau{\ss}ianity parameter $f_{\NL}$ as $\zeta\!= \zeta\!\left[ \zeta_{\Grm}(\tilde{r}, f_{\NL}) \right]$ (see also Section~\ref{sec:Local--type-Non--Gaussianity} for this type of non-Gau{\ss}ianity).\footnote{\setstretch{0.9}We should mention that this corresponds to a limited class of non-Gau{\ss}ianity, where deviations from Gau{\ss}ianity are described by some nonlinearity parameters (such as $f_{\NL}$ or $g_{\NL}$). These are the models which have mostly been considered in the literature when estimating the PBH formation threshold (see \eg~References~\cite{2019JCAP...09..033Y, 2021JCAP...10..053K, 2022JCAP...05..012E, 2020JCAP...05..022A, 2013JCAP...08..052Y, 2019JCAP...09..073A, 2022JCAP...05..037Y, 2007arXiv0708.3875H, 2023PhRvD.107d3520F}).} In this case, the situation is more complicated as compared to case \textit{B}, since the peak $\tilde{r}_{\mrm}$ will not only be depending on the specific parameters of $\zeta$ but also on the amplitude $\mu$. In this case, the reader can find more straightforward the direct integration of the compaction function with Equation~\eqref{eq:averaged-critical-C}, but indeed, as we will comment later, it is more accurate to use again Equation~\eqref{eq:threshold-analit} following now a different procedure than in case {\it B} (which has been introduced in Reference~\cite{2022JCAP...05..012E}):
\newpage

\begin{tcolorbox}
\begin{flushleft}
\begin{enumerate}

    \item
        Find the location $\tilde{r}_{\mrm}$ of the maximum of the compaction function $\Ccal$ using Equation~\eqref{eq:rm-eq}. This will depend on the amplitude $\mu$.
    
    \item
        For $\tilde{r}_{\mrm}$ obtained in the previous step, compute the peak $\delta_{\mrm}$ value of the compaction function following Equation~\eqref{eq:comaction-function}.

    \item
        With the values of $\mu$, $\tilde{r}_{\mrm}$ and $f_{\NL}$, compute $\Ccal( \tilde{r}_{\mrm} )$ and $\Ccal''( \tilde{r}_{\mrm} )$\\
        in order to obtain the parameter $q$ using Equation~\eqref{eq:q-tilde}.

    \item
        Use Equation~\eqref{eq:threshold-analit} to analytically derive $\delta_{\crm}( q )$. In general, this\\
        value will be different from the one obtained in the second step\\
        using Equation~\eqref{eq:comaction-function}. If so, start an iterative process by taking\\
        different $\mu$-values for step (1) until coincidence between $\delta_{\crm}( q )$ [obtained through Equation~\eqref{eq:threshold-analit}] and the peak value of $\Ccal$ [obtained from Equation~\eqref{eq:comaction-function}] is achieved, this yielding the\\
        critical $\mu$-value, denoted by $\mu_{\crm}$.
 
\end{enumerate}
\end{flushleft}
\end{tcolorbox}

\begin{figure}[t]
    \centering
    \hs{2mm}\includegraphics[width = 0.74\hsize]{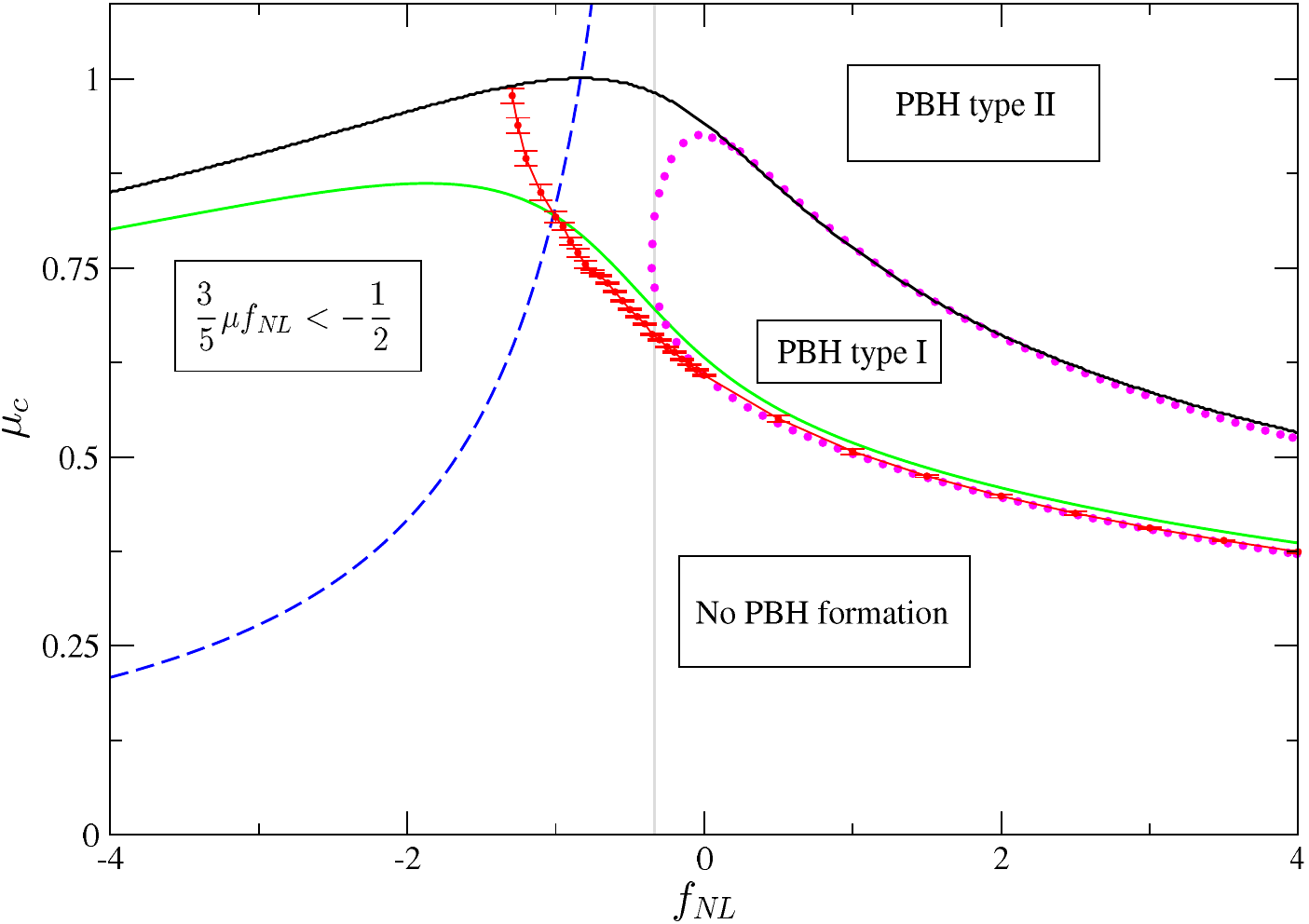}\\[4mm]
    \includegraphics[width = 0.75\hsize]{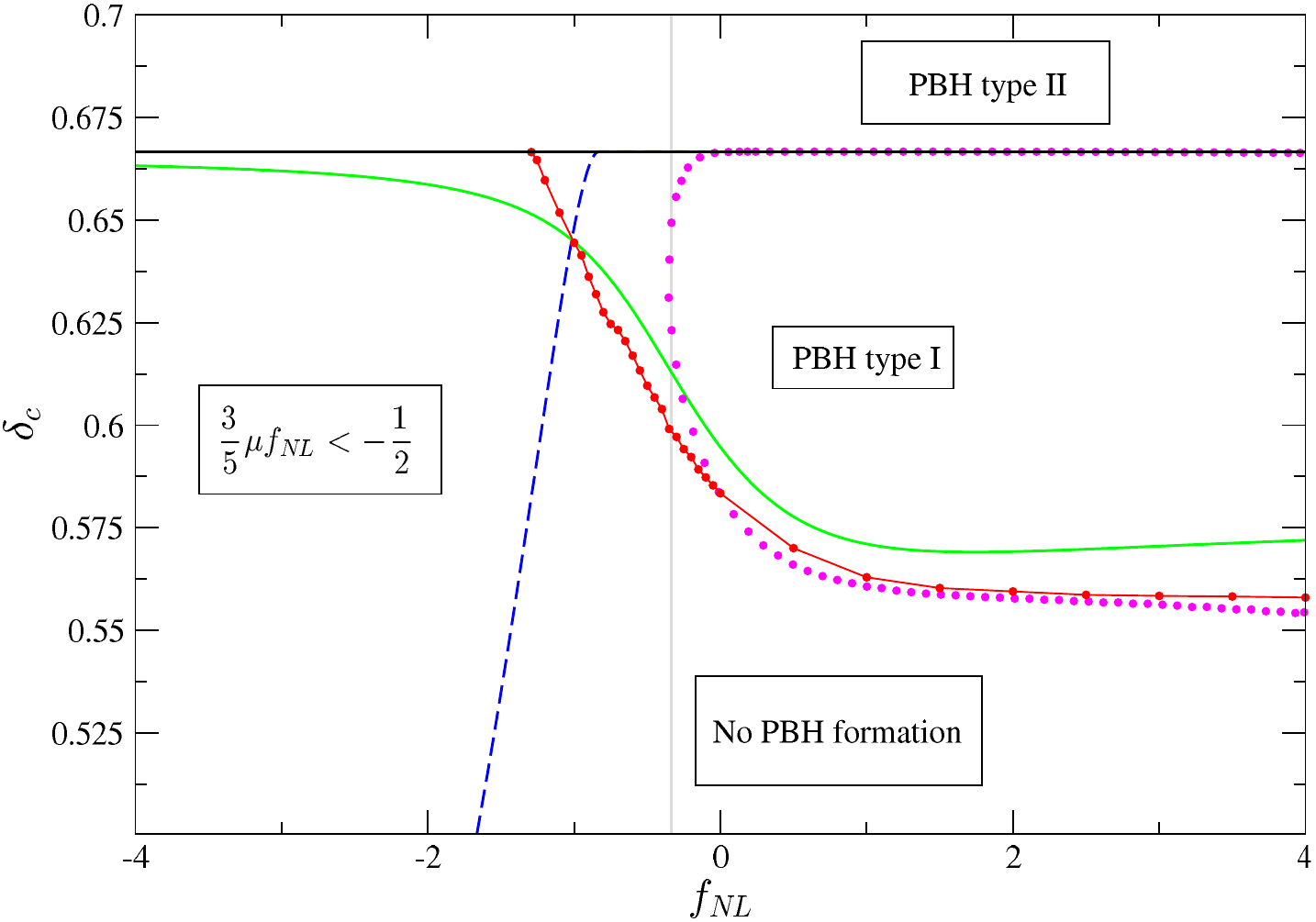}
    \caption{
        Thresholds $\mu_{\crm}$ ({\it upper panel}) and $\delta_{\crm}$ ({\it lower panel}) as functions of the nonlinearity parameter $f_{\NL}$. In both cases: red lines indicate the numerical values, while black lines represent the boundary between PBHs of type I and type II, following Equation~\eqref{eq:type2}. The green line depicts the analytic estimate following case {\it C}. The analytic estimate using the averaged compaction-function approach (also computed in Reference~\cite{2021JCAP...10..053K}) is shown by magenta points. The blue dotted line represents the boundary at which the assumption that peaks of $\zeta$ correspond to peaks of $\zeta_{\Grm}$ breaks down. The vertical grey solid line indicates the value of $f_{\NL}$ for which no PBHs of type I are formed, considering the approach of the averaged critical compaction function. Figures from Reference~\cite{2022JCAP...05..012E}.}
    \label{fig:diagram-ngs}
\end{figure}

\noindent\textbf{Example C.1}
Consider an example with local non-Gau{\ss}ianity (see Section~\ref{sec:Local--type-Non--Gaussianity}):
\vs{-0.5mm}
\begin{align}
\label{eq:zeta-ng}
    \zeta
        = 
            \zeta_{\Grm}
            +
            \frac{3}{5}\.f_{\NL}\.
            \zeta^{2}_{\Grm}
            \, ,
\end{align}
where the parameter $f_{\NL}$ determines the degree of the non-Gau{\ss}ianity. For simplicity, we consider the case of Gau{\ss}ian fluctuation for the monochromatic power spectrum of case {\it B}, given by $\zeta_{\Grm} = \sinc( k_{*}\.\tilde{r} )$. This implies
\vs{-1mm}
\begin{align}
    \zeta
        = 
            \mu\.\sinc( k_{*}\.\tilde{r} )
            +
            \frac{3}{5}\.f_{\NL}\.
            \mu^{2}\.
            \sinc^{2}( k_{*}\.\tilde{r} )
            \, .
\end{align}
Then, in order to find $\tilde{r}_{\mrm}$, following step (1) yields
\begin{subequations}
\begin{align}
    \zeta' + r\.\zeta''
        &= 
            0
            \displaybreak[1]
            \\[2.5mm]
    \begin{split}
        &\mspace{-40mu}\Rightarrow\quad
            \frac{ \mu }{5\.\tilde{x}^{2}\.\tilde{r}}\.
            \Big[\!
                \left(
                    \tilde{x}^{2} - 1
                \right)\!
                \big[\mspace{1mu}
                    6 f_{\NL}\.\mu\mspace{1.5mu}\cos( 2\tilde{x} )
                    -
                    5\.\tilde{x}\mspace{1.5mu}\sin( \tilde{x} )
                \big]
                \\[2mm]
        &\phantom{\mspace{-40mu}\Rightarrow\quad
            \frac{ \mu }{5\.\tilde{x}^{2}\.\tilde{r}}
            \Big[\;\,}
                -
                \tilde{x} \cos( \tilde{x} )
                \big[\mspace{1mu}
                    18\.f_{\NL}\.\mu\mspace{1.5mu}
                    \sin( \tilde{x} )
                    + 5\.\tilde{x}
                \big]
                +
                6\.f_{\NL}\.\mu\,
            \Big]\!
        = 
            0
            \, ,
    \end{split}
\end{align}
\end{subequations}
with $\tilde{x} = \tilde{r}\.k_{*}$ (or, $\tilde{x}_{\mrm} = \tilde{r}_{\mrm}\.k_{*}$). Notice that now $\tilde{r}_{\mrm}$ depends on $k_{*}$, $f_{\NL}$ and most importantly, on $\mu$. With the solution of $\tilde{x}_{\mrm}$ and with the same previous parameters $\mu$ and $f_{\NL}$ compute the compaction-function peak $\Ccal( \tilde{r}_{\mrm} )$ [\hs{0.2mm}step\,(2)\hs{-0.3mm}],
\begin{align}
\label{eq:fnl-c}
    \Ccal( \tilde{x}_{\mrm} )
        = 
            f( w )\!
            \left[
                1
                -
                \frac{
                \Big(
                    5\.\tilde{x}_{\mrm}
                    +
                    \mu\.
                    \big[
                        \tilde{x}_{\mrm}
                        \cos( \tilde{x}_{\mrm} )
                        -
                        \sin( \tilde{x}_{\mrm} )
                    \big]\mspace{-2mu}
                    \big[
                        5
                        +
                        6\.f_{\NL}\.
                        \mu\.\sinc( \tilde{x}_{\mrm} )
                    \big]
                \Big)^{\!2}}
                {25\.\tilde{x}^{2}_{\mrm}}
            \right]
            .
\end{align}
Now estimate $q$ [\hs{0.2mm}step\,(3)\hs{-0.3mm}] by evaluating $\Ccal( \tilde{r}_{\mrm} )$ and $\Ccal''( \tilde{r}_{\mrm} )$, and solving Equation \eqref{eq:q-tilde}. Obtain the corresponding critical threshold $\delta_{\crm}( q )$ [\hs{0.2mm}step\,(4)\hs{-0.3mm}] and compare with the peak value $\Ccal( \tilde{x}_{\mrm} )$ in Equation~\eqref{eq:fnl-c} obtained previously. Then iterate by changing the value of $\mu$ until obtaining coincidence. In Figure~\ref{fig:diagram-ngs} we show the results corresponding to the analytic estimate following this procedure (green line) compared with the numerical results (red points). As was found in Reference~\cite{2021JCAP...10..053K}, this approach gives a more accurate estimate in comparison with the computation of the threshold through the general procedure with the averaged critical compaction function (magenta points) for the case of $f_{\NL} \lesssim 0$. 
\vs{3mm}

\noindent\textbf{Example C.2}
Following the previous example, we now consider a fully nonlinear logarithmic relation between the Gau{\ss}ian curvature fluctuation $\zeta_{\Grm}$ and the non-Gau{\ss}ian one
\begin{equation}
    \label{eq:zeta-log-fnl}
    \zeta
        =
            -
            \ln
            \big(
                1
                -
                \mu_{*}\,\zeta_{\Grm}
            \big) / \mu_{*}
            \, ,
\end{equation}
where $\mu_{*} = 5 / ( 6\.f_{\NL} )$. As above, the parameter $f_{\NL}$ again determines the degree of the non-Gau{\ss}ianity. Such a model has been considered in detail in References~\cite{2019JCAP...09..073A, 2020JCAP...05..022A} in the context of realising an inflationary potential with a bump, where the inflaton field is trapped in a false vacuum for $\mu > \mu_{\star}$ due to large backward fluctuations which prevent horizon-sized regions from overshooting the barrier. For simplicity, we consider, as in the previous example, the case in which the Gau{\ss}ian fluctuation is given by the monochromatic power spectrum $\zeta_{\Grm} = \sinc( k_{*}\.\tilde{r} )$. 

Then, in order to find $\tilde{r}_{\mrm}$, following step (1), yields
\begin{subequations}
\begin{align}
    \zeta' + r\.\zeta''
        &= 
            0
            \displaybreak[1]
            \\[3.5mm]
    \begin{split}
        &\mspace{-40mu}\Rightarrow\quad
            \frac{ 5\.\mu }{\tilde{x}\.\tilde{r}
            \big[\.
                5
                -
                6\.f_{\NL}\.\mu \sinc( \tilde{x} )
            \big]^{2} }\.
            \Big[\.
                3\mspace{1.5mu}f_{\NL}\.\mu
                \big[
                    \sin( 2\.\tilde{x} )
                    -
                    2\.\tilde{x}
                \big]
                \\[2mm]
        &\mspace{110mu}+
                5\.\tilde{x}
                \big[
                    \cos( \tilde{x} )
                    +
                    \tilde{x}\sin( \tilde{x} )
                \big]
                -
                5\.\sin( \tilde{x} )
            \Big]\!
        = 
            0
            \, ,
    \end{split}
            \\[-3mm]
            \notag
\end{align}
\end{subequations}
where $\tilde{x} = \tilde{r}\.k_{*}$ (or $\tilde{x}_{\mrm} = \tilde{r}_{\mrm}\.k_{*}$). Notice that again $\tilde{r}_{\mrm}$ depends on $k_{*}$, $f_{\NL}$ and $\mu$. With the solution of $\tilde{x}_{\mrm}$, and the same previous parameters $\mu$ and $f_{\NL}$, compute the compaction-function peak $\Ccal( \tilde{r}_{\mrm} )$ [\hs{0.2mm}step\,(2)\hs{-0.3mm}],
\begin{align}
\label{eq:fnl-c2}
    \Ccal( \tilde{x}_{\mrm} )
        = 
            f( w )\!
            \left[
                1
                -
                \Bigg(
                    1
                    -
                    \frac{
                        5\.\mu\mspace{1.5mu}
                        \big[\.
                            \tilde{x}_{\mrm}
                            \cos( \tilde{x}_{\mrm} )
                            +
                            \sin( \tilde{x}_{\mrm} )
                        \big]}
                    {6\mspace{1.5mu}f_{\NL}\.\mu \sin( \tilde{x}_{\mrm} )
                    - 5\.\tilde{x}_{\mrm}}
                \Bigg)^{\mspace{-6mu}2}\.
            \right]
            .
\end{align}
As in the previous case, estimate $q$ [\hs{0.2mm}step\,(3)\hs{-0.3mm}] by evaluating $\Ccal( \tilde{r}_{\mrm} )$ and $\Ccal''( \tilde{r}_{\mrm} )$, and then solving Equation \eqref{eq:q-tilde}. Obtain the corresponding critical threshold $\delta_{\crm}( q )$\; [\hs{0.2mm}step\,(4)\hs{-0.3mm}] and compare with the peak value $\Ccal( \tilde{x}_{\mrm} )$ in Equation~\eqref{eq:fnl-c2} derived previously. Then iterate the above steps by changing the value of $\mu$ until coincidence is reached.

\begin{figure}[t]
    \centering
    \includegraphics[width = 0.72\hsize]{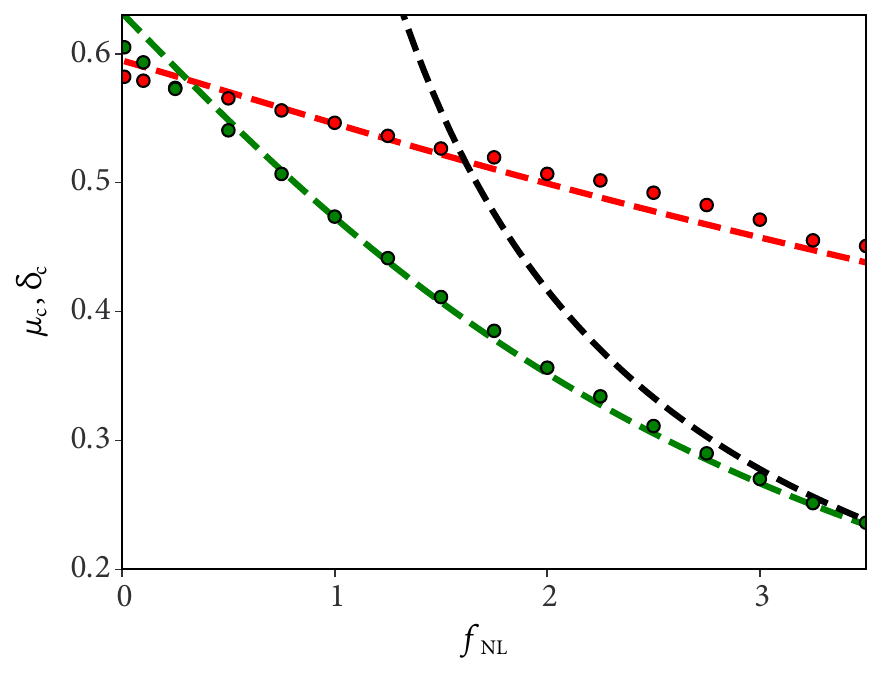}
    \caption{
        Thresholds $\mu_{\crm}$ (green) and $\delta_{\crm}$ (red) as functions of the nonlinearity parameter $f_{\NL}$. Solid dots indicate numerical values; the corresponding green and red dashed lines represent the results obtained using the analytical estimate. The black dashed line corresponds to $\mu= \mu_{*}$. For large values of the nonlinearity parameter, \ie~for $f_{\NL} \gtrsim 3.5$, $\mu$ approaches $\mu_{*}$.
        }
    \label{fig:diagram-ngs-log}
\end{figure}

Figure~\ref{fig:diagram-ngs-log} shows the results corresponding to the analytic estimate following the outlined procedure (dashed lines) compared with the numerical results (points). A similar example in which the same procedure is applied for a power spectrum with spectral index $n_{\srm} = 4$ can be found in Reference~\cite{2020JCAP...05..022A}.

\newpage

\subsection{Generation of Primordial Perturbations in Inflation}
\label{sec:Generation-of-Primordial-Perturbations-in-Inflation} 
\vs{-1mm}
It is widely accepted that our Universe has experienced an accelerated expansion called \emph{cosmic inflation}~\cite{1980PhLB...91...99S, 1981MNRAS.195..467S, 1981PhRvD..23..347G, 1982PhLB..108..389L, 1982PhRvL..48.1220A, 1983PhLB..129..177L} before the standard hot big bang universe. Inflation can make the Universe homogeneous and isotropic globally, and moreover generate local primordial density fluctuations from quantum vacuum fluctuations. They can grow into stars and galaxies in the late Universe by gravitational instability, and therefore provides observability. Sufficiently large primordial fluctuations can also yield primordial black holes. In this Section, we briefly review the production mechanism of primordial perturbations in inflation.

Inflation, the accelerated expansion, can be realised if the Universe is filled with some ``dark energy" component. This should be dynamic such that inflation ends successfully and connects to the hot big bang universe. The simplest realisation of such ``dark energy" is accomplished by a scalar particle $\phi$ with an almost flat potential, called \emph{inflaton}. If the canonical field $\phi$ fills the Universe with the homogeneous mode $\phi_{0}( t )$, its energy density $\rho$ and pressure $p$ are given by
\vs{-1mm}
\begin{subequations}
\begin{align}
    \rho
        = 
            \frac{1}{2}\.\dot{\phi}_{0}^{2}
            +
            V( \phi_{0} )
            \, ,
            \displaybreak[1]
            \\[3.5mm]
    p
        = 
            \frac{1}{2}\.\dot{\phi}_{0}^{2}
            -
            V( \phi_{0} )
            \, ,
\end{align}
\end{subequations}
where dots denote time derivatives, and $V( \phi )$ is the potential of $\phi$. Therefore, if the velocity of the inflaton is sufficiently small, $\dot{\phi}_{0}^{2} \ll V( \phi_{0} )$, the equation-of-state parameter $w = p/\rho$ of the inflaton is well approximated by the numerical value $-1$. Recalling the equation of acceleration of the Universe in terms of its scale factor $a$,
\vs{-2mm}
\begin{align}
    \frac{\ddot{a}}{a}
        = 
            -
            \frac{ 1 + 3\mspace{1.5mu}w }
            { 6\.\Mpl^{2} }\.
            \rho
            \, ,
\end{align}
where $\Mpl = 1/\sqrt{8\mspace{1mu}\pi\mspace{1.5mu}G}$ is the reduced Planck mass, one observes that inflation is indeed realised for $w < -1/3$. Such inflation is known as \emph{slow-roll inflation} because the inflaton velocity is small.

The success of slow-roll inflation is determined by the flatness of the inflaton potential $V( \phi )$. The equation of motion for $\phi_{0}$ reads
\begin{align}
\label{eq: equation of motion for phi0}
    \ddot{\phi}_{0}
    +
    3\mspace{1mu}H\dot{\phi}_{0}
    +
    V^{\prime}( \phi_{0} )
        = 
            0
            \, ,
            \\
            \notag
\end{align}
where $V^{\prime} = \partial_{\phi} V$. The Hubble parameter $H = \dot{a}/a$ satisfies
\vs{-2mm}
\begin{align}
\label{eq:Friedmann}
    3\mspace{1.5mu}\Mpl^{2}\.H^{2}
        = 
            \rho
        = 
            \frac{1}{2}\.\dot{\phi}_{0}^{2}
            +
            V( \phi_{0} )
            \, .
\end{align}
Its time derivative and the inflaton equation of motion~\eqref{eq: equation of motion for phi0} lead to the identity
\begin{align}
\label{eq: SR equation of motion}
    \dot{H}
        = 
            -\.\dot{\phi}_{0}^{2} / 
            \big(
                2\mspace{1.5mu}\Mpl^{2}
            \big)
            \, .
\end{align}

In terms of $\dot{H}$, one finds that the condition of acceleration $\ddot{a} > 0$ is recast into the smallness of the Hubble slow-roll parameter $\epsilon_{H} < 1$ defined by
\vs{-0.5mm}
\begin{align}
\label{eq:eH}
    \epsilon_{H}
        \coloneqq
            -
                \frac{\dot{H}}{H^{2}}
            = 
                \frac{3}{2}\.
                \frac{\dot{\phi}^{2}}
                {
                \dot{\phi}_{0}^{2}/2
                +
                V( \phi_{0} )
                }
                \, .
\end{align}
This gives a more rigorous expression of the small-velocity condition. In order for inflation to continue sufficiently, not only the velocity but also the acceleration is assumed negligible as $| \ddot{\phi}_{0} | \ll | H\dot{\phi}_{0} |$. One hence finds the slow-roll equation
\vs{-0.5mm}
\begin{align}
    3\mspace{1mu}H\dot{\phi}_{0}
        \simeq
        -
        \mspace{1mu}V^{\prime}( \phi_{0} )
        \, .
\end{align}
Through this equation of motion, one can find that the condition of acceleration can be further recast into the smallness of the first potential slow-roll parameter $\epsilon_{V} < 1$ defined by
\vs{-2.5mm}
\begin{align}
    \epsilon_{V}
        \coloneqq
            \frac{\Mpl^{2}}{2}
            \pqty{
                \frac{V^{\prime}}{V}
            }^{\mspace{-6mu}2}
            \, .
\end{align}
That is, this slow-roll parameter is proven to be equivalent to the Hubble one, $\epsilon_{H}$, in the slow-roll limit as
\vs{-0.5mm}
\begin{align}
    \epsilon_{V}
        = 
            \frac{\Mpl^{2}}{2}
            \pqty{
                \frac{V^{\prime}}{V}
            }^{\mspace{-6mu}2}
        \simeq
            \frac{\Mpl^{2}}{2}
            \pqty{
            \frac{3\mspace{1mu}H\dot{\phi}}
            {3\mspace{1.5mu}\Mpl^{2}\.H^{2}}
            }^{\mspace{-6mu}2}
        = 
            \frac{\dot{\phi}^{2}}
            {2\mspace{1.5mu}\Mpl^{2}\.H^{2}}
        = 
            -\frac{\dot{H}}{H^{2}}
        = 
            \epsilon_{H}
            \, .
\end{align}
The time derivative of the slow-roll equation of motion~\eqref{eq: SR equation of motion} leads to another condition,
\vs{-2.5mm}
\begin{align}
    \ddot{\phi}_{0}
        \simeq
            -
            \frac{\dot{H}}{H}\.
            \dot{\phi}_{0}
            -
            \frac{V^{\prime\prime}}
            {3\mspace{1mu}H}\.
            \dot{\phi}_{0}
            \, .
\end{align}
\newpage

\noindent Combining it with the negligible-acceleration condition $\big| \ddot{\phi}_{0} \big| \ll \big| H\dot{\phi}_{0} \big|$, one finds
\vs{-1mm}
\begin{align}
    \abs{\epsilon_{H}
    - \frac{V^{\prime\prime}}
    {3\mspace{1mu}H^{2}}}
        \ll
            1
            \, .
\end{align}
The smallness of the first term $\epsilon_{H}$ yields
\vs{-2mm}
\begin{align}
    \abs{\eta_{V}}
        \ll
            1
            \, ,\quad
    \text{where\; $\displaystyle\eta_{V}
        \coloneqq 
            \Mpl^{2}\.
            V^{\prime\prime} / V$}
            \. .
\end{align}
Therefore, sufficiently long-lasting slow-roll inflation can be realised by a flat-enough potential such that $\epsilon_{V} \ll 1$ and $\abs{\eta_{V}} \ll 1$.

This is not the end of the story for inflation. Taking quantum effects into account, one finds that the inflaton receives a fluctuation $\delta\phi$ every Hubble time due to the ``temperature" $T_{\dS} \coloneqq H/2\mspace{1.5mu}\pi$ of the horizon of the (quasi) de Sitter universe~\cite{1977PhRvD..15.2738G}. It is almost Gau{\ss}ian with the variance $\big\langle \delta\phi^{2} \big\rangle \simeq T_{\dS}^{2}$ and its wavelength is given by the Hubble scale $H^{-1}$ at that time. How does such a fluctuation affect the late Universe?

To this end, it is useful to introduce the {\it number of e-folds} $N$ as a novel time variable. It is defined by $\dd{N} \coloneqq H\dd{t}$, that is, it is a time variable normalised by the Hubble scale. The considered fluctuation is Hubble-sized, so even the fluctuating universe can be locally seen as almost homogeneous. Therefore, the fluctuation is understood as a shift of initial condition $\phi_{0} \to \phi_{0} + \delta\phi$ for a late universe. Furthermore, the slow-roll equation of motion~\eqref{eq: SR equation of motion} for a homogeneous universe can be rewritten in terms of $N$ as
\vs{-1mm}
\begin{align}
\label{eq: SR equation of motion 2}
    \dv{\phi_{0}}{N}
        \simeq 
            -\.\Mpl^{2}\.
            \frac{V^{\prime}}
            {V}( \phi_{0} )
            \, ,
\end{align}
which can be determined only by the current field value $\phi_{0}$. Therefore, the fluctuation $\delta\phi$ results in the change of the inflation duration $\delta N$ given by
\vs{-2.5mm}
\begin{align}\label{eq: SR linear dN}
    \delta N
        \simeq
            \frac{1}
            {\sqrt{2\.\epsilon_{V}}}\,
            \delta\phi
            \, ,
\end{align}
at leading order in $\delta\phi$. Note that the coefficient $1 / \sqrt{2\.\epsilon_{V}}$ should be evaluated at the time when the considered perturbation scale is equivalent to the Hubble scale. Using $\big\langle \delta\phi^{2} \big\rangle = ( H/2\mspace{1.5mu}\pi )^{2}$, the time fluctuation is expected as
\vs{-3mm}
\begin{align}
\label{eq: delta N squared}
    \big\langle
        \delta N^{2}
    \big\rangle
        \simeq
            \frac{1}
            {2\.\epsilon_{V}}
            \pqty{
                \frac{H}
                {2\mspace{1.5mu}
            \pi}}^{\mspace{-6mu}2}
            \, ,
\end{align}
which is consistent with standard linear-perturbation quantum field theory~\cite{1985PhRvD..31.1792L}.
\newpage

As the e-folds are normalised by the Hubble parameter, if the dynamics of the late Universe is locally determined only by the energy density [or the Hubble parameter through the Friedmann equation~\eqref{eq:Friedmann}], in which case the Universe is called \emph{adiabatic}, it can be understood that each local Universe behaves completely in the same manner and the difference in the number of e-folds $\delta N$ is conserved until the perturbation scale becomes shorter than the Hubble scale again, \ie~the time when the local Universe cannot be seen homogeneous any longer. $\delta N$ can always be converted to the perturbation of the energy density $\delta\rho$ (on a flat slice, the choice of the spacetime coordinate such that the spatial curvature is uniform, strictly speaking) as
\vs{-2mm}
\begin{align}
    \delta\rho
        \simeq
            -\mspace{1.5mu}
            \frac{\dot{\rho}_{0}}{H}\,
            \delta N
\end{align}
at linear order, where $\rho_{0}$ is the average density. Therefore, cosmic inflation can generate density perturbations for the late Universe as ``time fluctuations". Such a formulation of the primordial perturbation is known as the \emph{$\delta N$ formalism}~\cite{Starobinsky:1985ibc, 1990PhRvD..42.3936S, 1996PThPh..95...71S, 1998PThPh..99..763S, 2000PhRvD..62d3527W, 2005JCAP...05..004L, 2005PhRvL..95l1302L}. The conservation of $\delta N$ on a super-Hubble scale in the adiabatic Universe is proven in Reference~\cite{2005JCAP...05..004L} in a rigorous manner. Therein, it is shown that $\delta N$ is equivalent to the curvature perturbation $\zeta$ on a uniform density slice.

\subsection{Inflation Models for Primordial Black Hole Production}
\label{sec:Inflation-Models-for-Primordial-Black-Hole-Production}
\vs{-1mm}
The perturbation scale plays an important r{\^o}le for the PBH mass: a large overdense region yields a more massive black hole than a small one (see Section~\ref{sec:Statistics} for details). The scale of the curvature perturbations is often expressed in terms of its power spectrum, the (comoving) Fourier mode of its two-point function,
\begin{align}
    \Pcal_{\zeta}( k )
        = 
            \frac{k^{3}}
            {2\mspace{1.5mu}\pi^{2}}
            \int\dd[3]{\xbm}\;
            \erm^{-i\mspace{1.5mu}\kbm\cdot\xbm}
            \Braket{
                \zeta\!
                \left(
                    -
                    \frac{ \xbm }{ 2 }
                \right)\mspace{-2mu}
                \zeta\!
                \left(
                    \frac{ \xbm }{ 2 }
                \right)
            }
            \, .
\end{align}
In order to obtain a non-negligible PBH abundance, this power spectrum should be as large as $\Pcal_{\zeta} \sim 10^{-2}$ (see Section~\ref{sec:Statistics}) on a small scale ($\lesssim {\rm Mpc}$). However, observations on cosmological scales such as of the cosmic microwave background temperature perturbation or of large-scale structure (LSS) have already revealed that primordial perturbations are as small as $\Pcal \sim 10^{-9}$ on large scales ($\gtrsim{\rm Mpc}$)~(see \eg~Reference~\cite{2015JCAP...12..052H}). This is actually the main constraint on a PBH-realising inflation model. 
\newpage

\begin{figure}[t]
	\centering
	\vs{-1mm}\hs{-10mm}\includegraphics[width = 0.75\hsize]{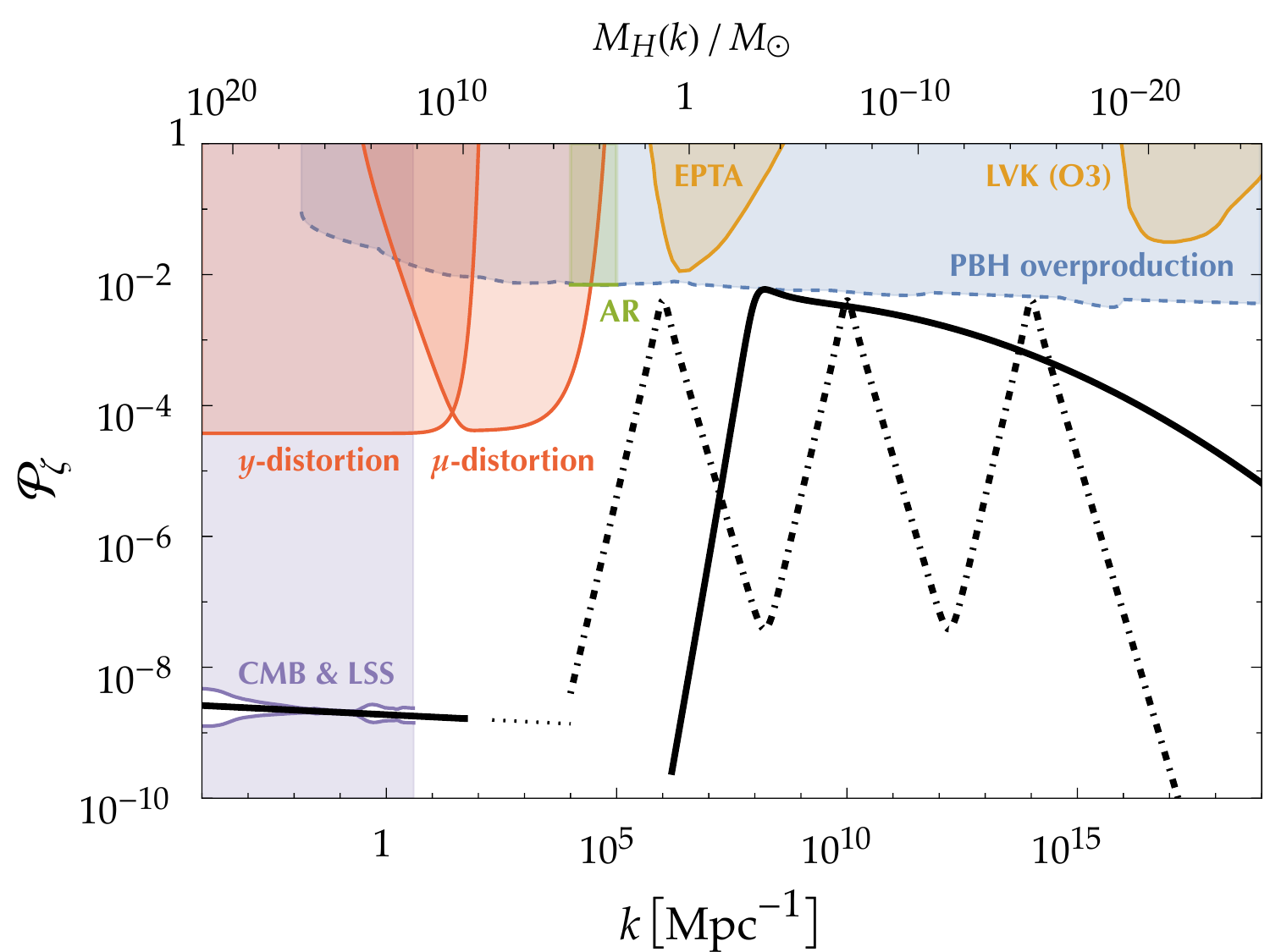}
	\caption{
        Observationally excluded region (shaded) and schematic pictures of amplification of the primordial power spectrum $\Pcal_{\zeta}$ (black). The individual constraints are due to CMB and LSS observations (purple~\cite{2015JCAP...12..052H}), CMB spectral $\mu$- and $y$-distortion (red~\cite{2014PhRvD..90h3514K, 2012ApJ...758...76C}) {acoustic reheating} (AR) (green~\cite{2016PhRvD..94d3527I}; see also References~\cite{2014PhRvL.113f1301J, 2014PhRvL.113f1302N}), induced gravitational waves (orange~\cite{2019PhRvD..99d3511I,2021ApJ...910L...4K,2022PhRvL.128e1301R}; see also Section~\ref{sec:Primordial-Black-Hole-Formation-Time}), and PBH overproduction (blue dotted; see also Section~\ref{sec:Constraints-on-Primordial-Perturbations}). The power spectrum is well determined on large scales ($\sim{\rm Mpc}$) but the constraints on smaller scales are weak and perturbations can be enhanced as represented by black plain or dot-dashed lines (these are schematic forms in the flat-inflection and multi-phase inflation models, respectively). The upper ticks indicate the horizon mass~\eqref{eq: horizon mass} at horizon reentry of the corresponding comoving mode as a rough indicator of the primordial black hole mass.
        }
	\label{fig: amplification power}
\end{figure}

The single-field $\delta N$ formula~\eqref{eq: delta N squared} is understood as a power spectrum,
\begin{align}
\label{eq: Pcal zeta formula}
    \Pcal_{\zeta}( k )
        = 
            \eval{\frac{1}{2\.\epsilon_{V}\mspace{0.5mu}\Mpl^{2}}
            \pqty{
                \frac{H}{2\mspace{1.5mu}\pi}
            }^{\mspace{-6mu}2}}_{a H\mspace{1mu}=\.k}
            \, .
\end{align}
During slow-roll inflation, both $H$ and $\epsilon_{V}$ vary only slowly and thus the power spectrum becomes almost scale-invariant. In fact, making use of the approximation $\dd{\ln k} = \dd{\ln( a H )} \simeq \dd{N}$ and the slow-roll equation of motion~\eqref{eq: SR equation of motion 2}, one finds that the scale-dependence of the power spectrum, dubbed \emph{spectral index $\ns$}, is given by
\vs{-2mm}
\begin{align}
\label{eq: ns SR}
    \ns
        \coloneqq
            \dv{\ln\Pcal_{\zeta}}{\ln k}
            + 1
        = 
            -\.6\.\epsilon_{V}
            + 2\.\eta_{V}
            + 1
            \, ,
\end{align}
which is almost unity during slow-roll inflation. Therefore, in order to obtain a large enough perturbation on small scales consistent with cosmological observations (see Figure~\ref{fig: amplification power}), one has to violate some of the simplifying assumptions: The inflaton is a 
    ({\it i$\mspace{1.5mu}$}) 
        canonical, 
    ({\it ii$\mspace{1.5mu}$}) 
        single and 
    ({\it iii$\mspace{1.5mu}$}) 
        slow-rolling 
    ({\it iv}) 
        scalar field.
In this Section, we review several models which accomplish the required amplification.

\subsubsection{Flat Inflection}
\begin{figure}[t]
    \centering
    \includegraphics[width = 0.75\hsize]{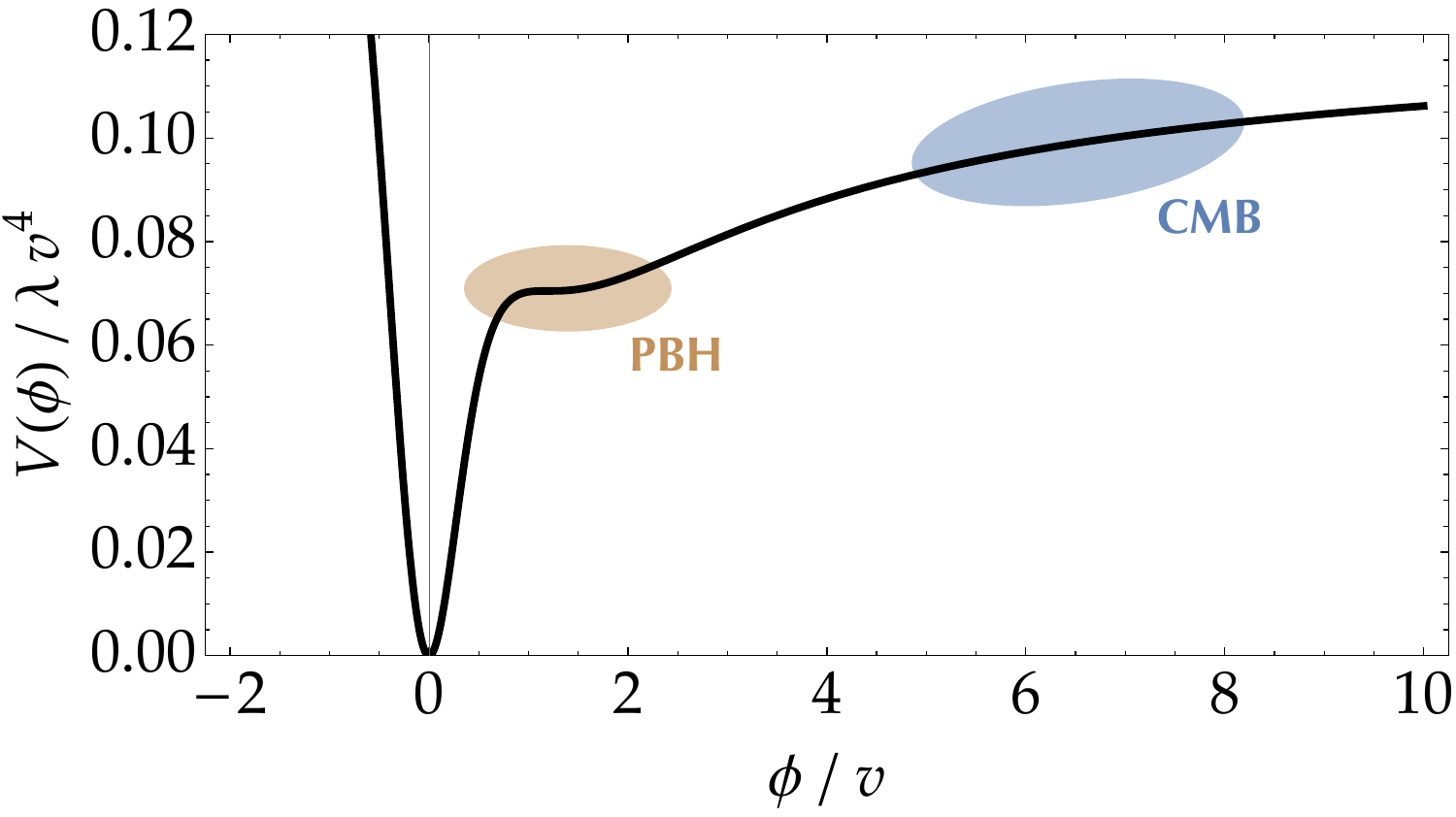}
    \caption{
        Flat-inflection potential~\eqref{eq: Vinflection} with $a = 1$ and $\alpha = 10^{-4}$. The upper part (blue) corresponds to CMB-scale perturbations and the flat inflection region (orange) can source PBHs.
        }
    \label{fig: Vinflection}
\end{figure}

\vs{-1mm}
One simple way to amplify the power spectrum~\eqref{eq: Pcal zeta formula} is to rapidly reduce $\epsilon_{V}$, violating the second slow-roll condition, \ie~allowing $\eta_{V} \sim \Ocal( 1 )$. That is, we suppose a nearly-flat-inflection point in the inflaton potential around which the potential tilt $V^{\prime}$ is rapidly changed due to a large enough curvature $\abs{V^{\prime\prime}}$ (see Figure~\ref{fig: Vinflection}). An inflaton coming from the upper side of the potential can overshoot this inflection point and the power spectrum of perturbations can be quickly amplified. Such a system with a too-small potential tilt is called \emph{ultra slow-roll} system. If the second slow-roll condition is not violated, the inflaton cannot quickly pass through the inflection point and fails to make large-scale perturbations which are small enough ($\sim 10^{-5}$).

A toy model of such an inflection model has first been studied by Starobinsky~\cite{1992PZETF..55..477S}, where two or three linear potentials are connected non-smoothly (see Figure~\ref{fig: USR} for a sketch). This model has soon been applied for primordial black hole formation in Reference~\cite{1994PhRvD..50.7173I}, and the application of the stochastic formalism (see Section~\ref{sec:Stochasti-Approach}) for this model is discussed in Reference~\cite{1998PhRvD..57.7145I}. More realistic flat-inflection models have originally been introduced in order to obtain a flat-enough potential for slow-roll inflation in the context of the minimal supersymmetric extension of the Standard Model of particle physics for example (see the first papers~\cite{2006PhRvL..97s1304A, 2007JCAP...01..015B}). There, the inflaton is understood as a supersymmetric flat direction and its potential is only lifted by a non-renormalisable term in the superpotential,
\vs{-1mm}
\begin{align}
    W
        = 
            \frac{ \lambda_{n} }{ n }
            \frac{ \Phi^{n} }{ \Mpl^{n-3} }
            \, ,
\end{align}
where $\Phi$ is the superfield of the inflaton.
This leads to the potential,
\begin{align}
    V
        = 
            \frac{1}{2}\.m^{2}\mspace{1.5mu}\phi^{2}
            -
            A\.
            \frac{\lambda_{n}\.\phi^{n}}
            {n\.\Mpl^{n-3}}
            +
            \lambda_{n}^{2}\.
            \frac{\phi^{2\mspace{1mu}( n - 1 )}}
            {\Mpl^{2\mspace{1mu}( n - 3 )}}
            \, ,
\end{align}
where the first two terms are a consequence of soft supersymmetry breaking. If the supersymmetry breaking is gravity-mediated, the dimensionful coefficient $A$ becomes of order $m$ and can hence have a flat inflection point. This potential is however too steep at the upper part and then the inflaton coming from there cannot keep inflation around the inflection point: the velocity is too large to satisfy the accelerated-expansion condition $\epsilon_{H} < 1$ [\cf~Equation~\eqref{eq:eH}].

Then, in order to have another flat part in the upper side of the potential, the ratio-of-polynomials model has phenomenologically been proposed as~\cite{2017PDU....18...47G, 2017PhRvD..96f3503M}
\begin{align}
\label{eq: Vinflection}
    V
        = 
            \frac{\lambda\.v^{4}}{12}\.
            \frac{x^{2}\mspace{2mu}
            \big(
                6
                -
                4\.a\.x
                +
                3\.x^{2}
            \big)}
            {
            \big(
                1
                +
                b\.x^{2}
            \big)^{\mspace{-2mu}2}}
            \, ,
\end{align}
where $x = \phi/v$. If the parameter combination
\begin{align}
    \alpha
        \coloneqq
            \!\bqty{1 - \frac{a^{2}}{3}
            +
            \frac{a^{2}}{3}
            \pqty{\frac{9}{2\.a^{2}} - 1}^{\mspace{-6mu}2/3}\.}
            -
            b
            \, ,
\end{align}
is positive and small, the potential has an almost-flat inflection point as shown in Figure~\ref{fig: Vinflection}. This model can successfully produce perturbations on scales of the cosmic microwave background as well as on much scales associated with PBHs, at the upper flat region and the flat inflection point, respectively.

\newpage

Respective realisations have been also proposed in Higgs inflation for example~\cite{2018PhLB..776..345E}. Taking unitary gauge, we only consider its radial mode $\phi$ for simplicity. We assume that it has a non-minimal coupling to gravity as
\begin{align}
    \Scal
        = 
            \int\dd[4]x\,
            \bqty{
                \frac{1}{2}\mspace{1.5mu}
                \pqty{
                    \Mpl^{2}
                    +
                    \xi\.\phi^{2}
                }
                R
                -
                \frac{1}{2}\.
                \partial_{\mu}\phi\.
                \partial^{\mspace{1mu}\mu}\phi
                -
                \frac{\lambda}{4}\.
                \phi^{4}
            }
            \, ,
\end{align}
where we neglected the small mass term as it is irrelevant to inflation dynamics. By conformal transformation of the metric,
\begin{align}
    g_{\mu\nu}
        \to
            \Omega^{2}( \phi )\.
            g_{\mu\nu}
            \, ,
            \quad
    \Omega^{2}( \phi )
        = 
            1
            +
            \xi\.
            \phi^{2} / \Mpl^{2}
            \, ,
\end{align}
the gravity part reduces the ordinary Einstein--Hilbert term as
\begin{align}
\label{eq: Higgs Einstein frame}
    \Scal
        = 
            \int\dd[4]{x}\,
            \bqty{\frac{1}{2}\.\Mpl^{2}\.R
            -
            \frac{\Omega^{2}( \phi )
            +
            6\.\xi^{2}\phi^{2}/\Mpl^{2}}
            {2\.\Omega^{4}( \phi )}\,
            \partial_{\mu}\phi\.\partial^{\mspace{1mu}\mu}\phi
            -
            \frac{\lambda}{4}
            \frac{\phi^{4}}
            {\Omega^{4}( \phi )}}
            \, .
\end{align}
Hence, the potential converges to the constant value $( \lambda/4\.\xi^{2} )\.\Mpl^{4}$ in the large-field limit $\phi \gg 1$. The inflection point can be also realised by taking account of the renormalisation group of the couplings $\lambda$ and $\xi$. The Higgs self-coupling $\lambda$ depends on the energy scale of the system through the quantum loop effects of the Standard Model particles. Depending on the Standard Model parameters (in particular, the top-quark mass), $\lambda$ can have an almost vanishing minimum value $\lambda_{0}$ at very high energy $\mu \sim 10^{17}\,\text{--}\,10^{18}\.{\rm GeV}$. In this case, the approximated expressions of $\lambda$ and $\xi$ around the critical point $\mu$ are functions of the Higgs field value $\phi$
\begin{subequations}
\begin{align}
    \lambda( \phi )
        &\simeq
            \lambda_{0}
            +
            b_{\lambda} 
            \ln^{2}( \phi/\mu )
            \, ,
            \displaybreak[1]
            \\[2mm]
    \xi( \phi )
        &\simeq
            \xi_{0}
            +
            b_{\xi} \ln( \phi/\mu )
            \, ,
\end{align}
\end{subequations}
with two parameters $b_{\lambda}$ and $b_{\xi}$. Therefore, the effective potential reads
\begin{align}
    V
        &= 
            \frac{1}{4}\.
            \frac{
                \big[
                    \lambda_{0}
                    +
                    b_{\lambda}
                    \ln^{2}( \phi/\mu )\mspace{1mu}
                \big]\.\phi^{4}}
            {\Big[
                1
                +
                \big[
                    \.\xi_{0}
                    +
                    b_{\xi}\ln( \phi/\mu )\mspace{1mu}
                \big]\.\phi^{2}/\Mpl^{2}
            \Big]^{2}\.}
            \, ,
\end{align}
which can have an inflection point around $\mu$. Note that the detailed calculation needs care, though we do not show it explicitly here, because $\phi$ has non-canonical kinetic term~\eqref{eq: Higgs Einstein frame} in this model.\footnote{\setstretch{0.9}See \eg~Reference~\cite{2021JCAP...01..032C} for recent progress on PBH formation in Higgs inflation. Note also that there are many other approaches to realise an inflection point in the potential (see \eg~References~\cite{2018PhRvD..97b3501B, 2018JCAP...06..034C, 2020JCAP...07..025B, 2022PhRvD.106f3535G}). There, the inflection point is not even ``flat" but the potential has a local shallow minimum. In such a case, the slow-roll approximation fails even for the maximal amplitude of the power spectrum and hence one has to resort to a full numerical analysis. It is also implied that stochastic effects (see Section~\ref{sec:Stochasti-Approach}) are significant in these models~\cite{2018JCAP...07..032B}.}

Phenomenologically, in these flat-inflection models, the power spectrum is rapidly amplified with the maximal spectral index at the beginning, $\ns - 1 = 4$ (in the standard scenario; see Reference~\cite{2019JCAP...06..028B}), and then slowly decreases with a convex cubic potential (see thick black line in Figure~\ref{fig: amplification power}). Even though a precise prediction requires a full analysis beyond the slow-roll approximation, the maximal power spectrum amplitude is often well-approximated by Formula~\eqref{eq: Pcal zeta formula}.

\subsubsection{Uphill Inflation}
\begin{figure}[t]
    \centering
    \includegraphics[width = 0.7\hsize]{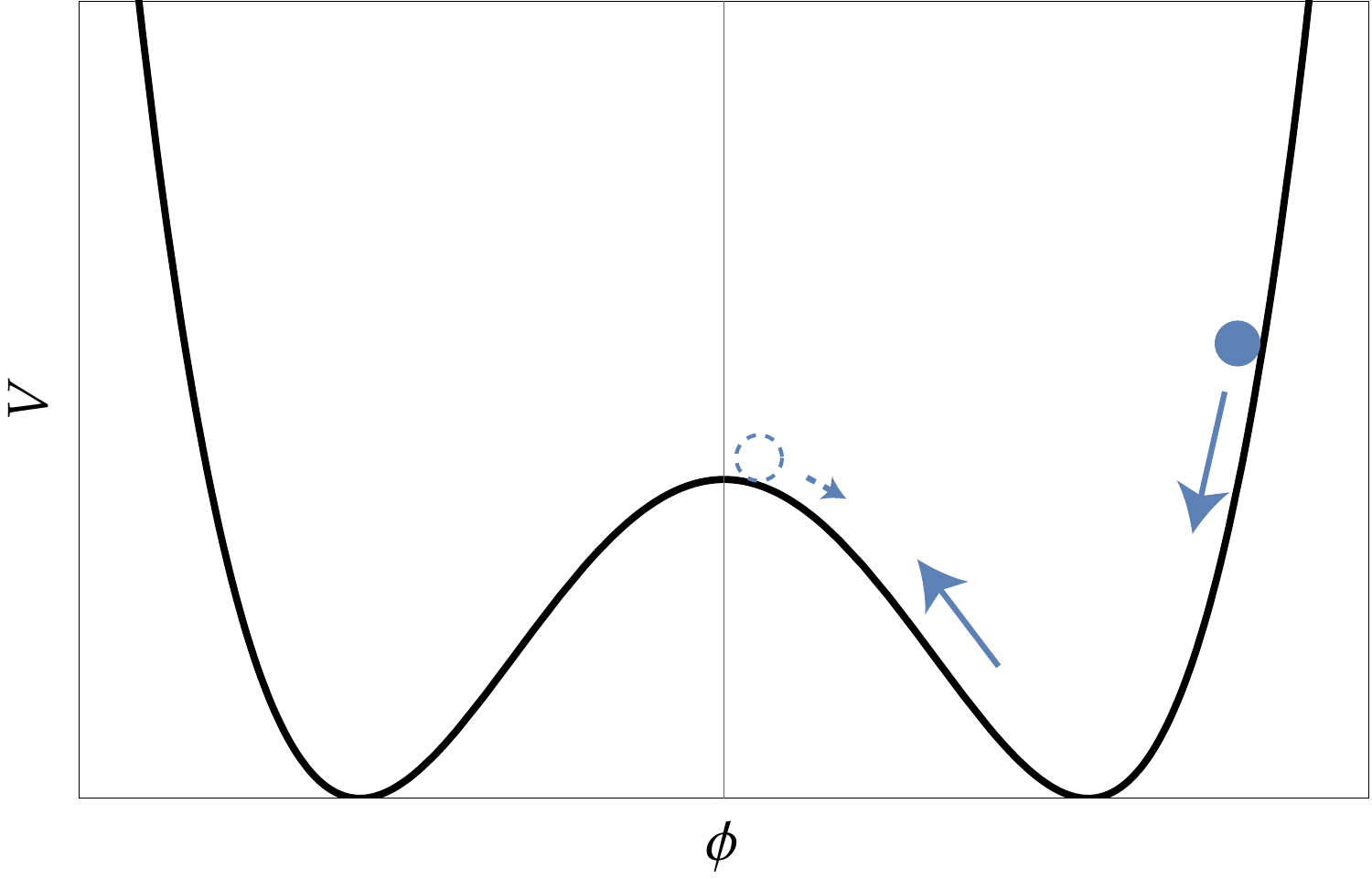}
    \caption{
        In the ``wine-bottle"-potential model, the inflaton can climb up the local maximum and return to the minimum, if the potential parameters are fine-tuned accordingly. As the inflaton velocity is reduced (or even becomes exactly zero) through this process, the curvature perturbation can be significantly enhanced.
        \vs{5mm}
        }
    \label{fig: Vuphill}
\end{figure}

\vs{-1mm}
Other than the through a flat inflection point, the reduction of the inflaton velocity can be realised by climbing up a potential hill, dubbed \emph{chaotic new inflation}~\cite{1998PhRvD..58h3510Y} or \emph{uphill inflation}~\cite{2023JCAP...06..029B}. It can be realised in the ``wine-bottle" potential as shown in Figure~\ref{fig: Vuphill}. If the potential parameters are fine tuned, the inflaton coming from the outer part can climb up the potential local maximum, stop around the hilltop, and turn back to the potential minimum. At the returning point, the inflaton velocity becomes exactly zero and the curvature perturbation is expected to be significantly enhanced. In this model, the inflaton stops at the point where the slow-roll linear relation~\eqref{eq: SR linear dN} leads to the curvature perturbation greater than unity, which indicates the breakdown of the perturbative approach. One hence has to resort to some non-perturbative way, for instance the stochastic formalism of inflation (see Section~\ref{sec:Stochasti-Approach}).

\subsubsection{Multi-Phase Inflation}
\vs{-1mm}
The slow-roll condition can be violated in a bolder way. That is, we can assume that inflation takes place more than once. In these cases, cosmic microwave background scale perturbations are disconnected from the primordial black hole scale ones, with the latter being large enough to generate PBHs, while the former being sufficiently small in accordance with CMB observations.

One possible scenario is motivated by supersymmetry/supergravity~\cite{1998PhRvD..57.6050K, 1999PhRvD..59d3512K}. If one assumes multiple scalar fields $\phi_{i}$ ($i = 1$, $2$, $\ldots$), even if they have their own intrinsic potential $V_{i}(\phi_{i})$, supergravity generally yields a Planck-suppressed coupling between them:\footnote{\setstretch{0.9}Note that the self-coupling $( \phi_{i}^{2}/\Mpl^{2} )\.V_{i}(\phi_{i})$ breaks the (second) slow-roll condition, and should hence be prohibited. This is known as the $\eta$ problem (see \eg~Reference~\cite{1999PhR...314....1L} for a review).}
\vs{-3mm}
\begin{align}
\label{eq: Planck-suppressed int}
    V_{i}( \phi_{j} )\.\phi_{i}^{2} / \Mpl^{2}
    \, .
\end{align}
If \eg~the typical energy scales of $\phi_{1}$ and $\phi_{2}$ are hierarchical as $V_{2}( \phi_{2} ) \ll V_{1}( \phi_{1} )$, this coupling acts as a large effective mass for $\phi_{2}$ during the first inflation by $\phi_{1}$.

\begin{figure}[t]
    \centering
    \includegraphics[width = 0.7\hsize]{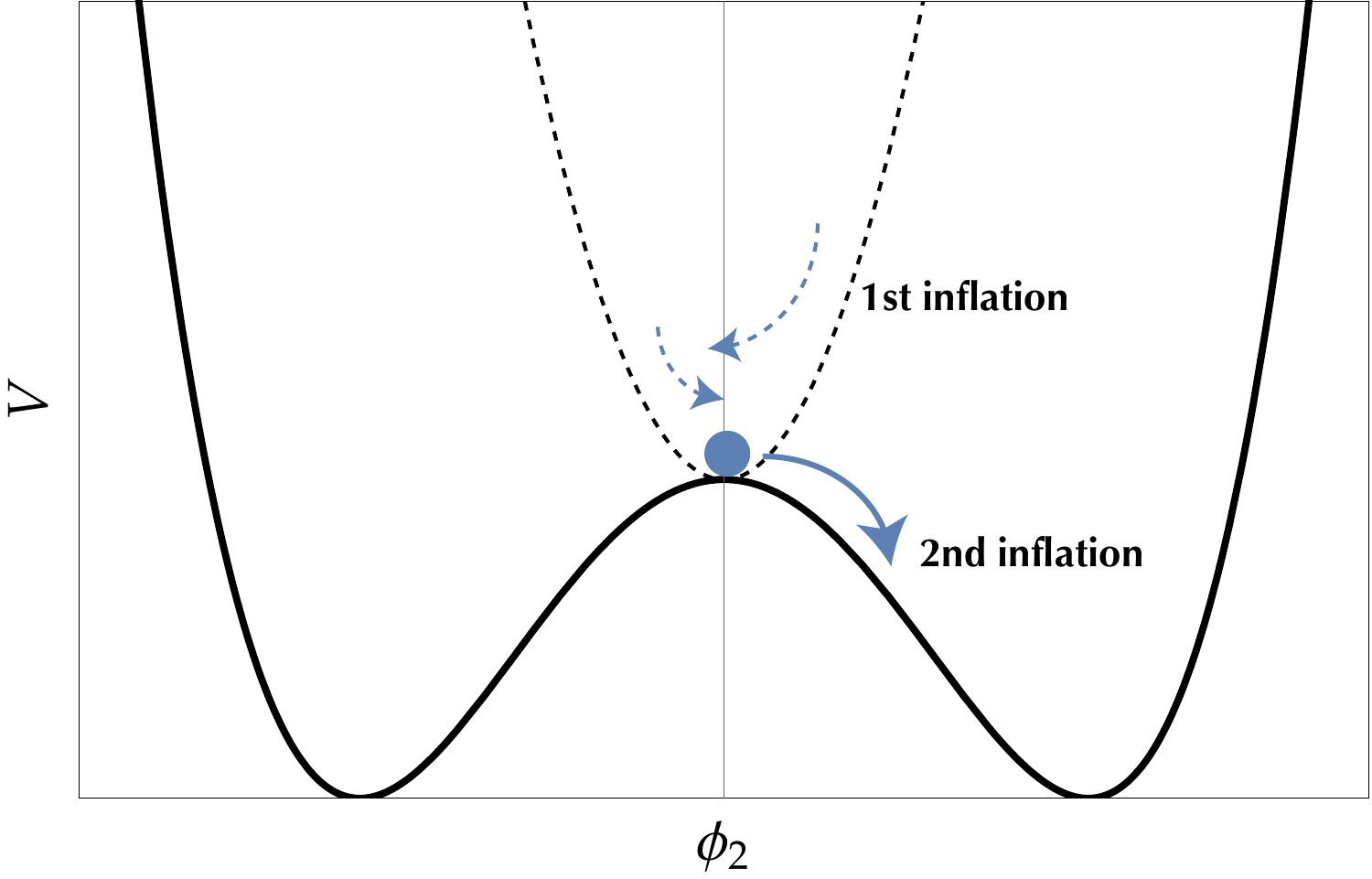}
    \caption{
        During the first phase of inflation, the potential of the second inflaton $\phi_{2}$ is uplifted by the interaction~\eqref{eq: Planck-suppressed int} and $\phi_{2}$ is stabilised on the potential top. After the first inflation, $\phi_{1}$'s energy is diluted and $\phi_{2}$ is released to yield the second period of inflation. In principle, this process can be repeated many times. At the onset of each phase of inflation, perturbations grow large because the potential is nearly flat around its top.
        }
    \label{fig: Vmulti-phase}
\end{figure}

In the case $\phi_{2}$ has a "wine-bottle"-type potential, it can drive the second phase of inflation after $\phi_{1}$'s energy is sufficiently diluted (see Figure~\ref{fig: Vmulti-phase}). References~\cite{2006PhRvD..74d3525K, 2008MNRAS.388.1426K, 2010JCAP...04..023F, 2012PhLB..711....1K} assume so-called smooth hybrid inflation for the first period of inflation. The large amplification of perturbations is realised by the preheating at the end of this period. The second instance of inflation simply makes the perturbation scale larger so that the corresponding PBH mass becomes sizeable enough. References~\cite{2016PhRvD..94f3509K, 2016PhRvD..94h3523K, 2017PhRvD..95l3510I, 2017PhRvD..96d3504I, 2017PhRvD..96l3527I} investigate simple polynomial potentials. There, the second phase of inflation generates large perturbations. Furthermore, Reference~\cite{2019PhRvD.100b3537T} shows that more than double inflation works and even three peaks in the PBH mass function can be realised in a quadruple inflation model, corresponding to the second, third, and fourth phases of inflation.
\newpage

For the quadratic hilltop potential, each peak of the power spectrum follows a broken power law given by
\vs{-2mm}
\begin{align}
    \Pcal_{\zeta}( k )
        \sim
            \begin{cases}
                k^{3}
                    & \text{for $k < k_{\prm}$}
                \, ,
                \\[1mm]
                k^{3-2\mspace{1.5mu}\nu}
                    & \text{for $k > k_{\prm}$}
                \, ,
    \end{cases}
\end{align}
with the peak scale $k_{\prm}$ and $\nu = \big[ 9 / 4 - V^{\prime\prime} / ( 3\mspace{1.5mu}H^{2} ) \big]^{1/2}$. This is illustrated by the black dot-dashed line in Figure~\ref{fig: amplification power}.

We note that there are several other mechanisms for a multi-spiked PBH mass spectrum{\,---\,}for instance from oscillations of the sound speed which lead to parametric amplification of the curvature perturbation. This yields significant peaks in the power spectrum of the density perturbations~\cite{2018PhRvL.121h1306C}. Another mechanism for generating multi-spiked PBH mass spectra~\cite{2019PhRvD..99j3535C} relies on the choice of non-Bunch--Davies vacua. These lead to oscillatory features in the inflationary power spectrum, which in turn generates oscillations in the PBH mass function with exponentially enhanced spikes. Furthermore, even if the primordial power spectrum is featureless, the thermal history of the Universe naturally generates spikes at various masses as we will see in Section~\ref{sec:Thermal--History--Induced-Mass-Function}.
\newpage

\subsubsection{Mild Waterfall Hybrid Inflation}
\vs{-1mm}
If the two energy scales of the above double inflation are not hierarchical, it is called \emph{hybrid inflation}~\cite{1994PhRvD..49..748L}.
The hybrid inflation potential is often parametrised as
\vs{-1.5mm}
\begin{align}
    V( \phi,\mspace{1.5mu}\psi )
        = 
            V( \phi )
            +
            \Lambda^{\mspace{-2mu}4}
            \Bigg[
                \pqty{
                    1
                    -
                    \frac{\psi^{2}}{M^{2}}
                }^{\mspace{-6mu}2}
                +
                2\.\frac{\phi^{2}\mspace{1.5mu}\psi^{2}}
                {\phi_{\crm}^{2}\.M^{2}}
            \Bigg]
            \. ,
\end{align}
with two scalars $\phi$ and $\psi$ as well as three dimensionful parameters $\Lambda$, $M$ and $\phi_{\crm}$.
The field $\psi$ has a "wine-bottle"-type potential; for $\phi > \phi_{\crm}$, one finds that it is stabilised to the origin due to the coupling $2\mspace{1.5mu}\Lambda^{\mspace{-2mu}4}\mspace{1.5mu}\phi^{2}\mspace{1.5mu}\psi^{2} / \phi_{\crm}^{2}\mspace{1.5mu}M^{2}$. Inflation is mainly driven by its false-vacuum energy $\Lambda^{\mspace{-2mu}4}$. If $\phi$ slowly rolls down due to its potential $V( \phi )$ and gets smaller than $\phi_{\crm}$, $\psi$ becomes unstable and inflation ends when $\psi$ falls to $M$ or $-\mspace{1.5mu}M$, which is the reason why $\psi$ is often called \emph{waterfall} field.

The waterfall phase is typically assumed to be very rapid, but depending on the parameters, it can work as a second-phase slow-roll inflation dubbed \emph{mild-waterfall hybrid inflation}. The transition from the valley phase ($\phi\!>\!\phi_{\crm}$) to the waterfall phase and the later dynamics are determined by $\psi$'s fluctuations about the origin just before the transition. In other words, the perturbation determines the whole dynamics in this system, hence being completely beyond perturbative physics, similarly to other second-order phase transitions. It is then na{\"i}vely expected that the corresponding curvature perturbations are much amplified, and if mild-waterfall inflation sufficiently expands such perturbations, sizeable PBHs can be formed~\cite{1996PhRvD..54.6040G}.

Clesse and Garc{\'i}a-Bellido~\cite{2015PhRvD..92b3524C} follow a semi-perturbative approach. By taking account of the quantum fluctuation as Brownian noise{\,---\,}a procedure known as {\it stochastic formalism} (see Section~\ref{sec:Stochasti-Approach}){\,---\,}they estimate the typical fluctuation amplitude of $\psi$ just before the critical point $\phi_{\crm}$ as
\vs{-1mm}
\begin{align}
    \sigma_{\psi}^{2}
        \coloneqq
            \big\langle
                \psi^{2}
            \big\rangle_{\!\phi_{\crm}}
        = 
            \frac{\sqrt{2}
            \Lambda^{\mspace{-2mu}4}
            \mspace{1.5mu}M
            \sqrt{\phi_{\crm}\.\mu_{1}}}
            {96\mspace{1.5mu}\pi^{3/2}\Mpl^{4}}
            \, ,
\end{align}
where $1/\mu_{1} = \eval{V'( \phi )/\Lambda^{\mspace{-2mu}4}}_{\phi_{\crm}}$. They then compute the curvature perturbation in the waterfall phase using the standard $\delta N$ formalism. The power spectrum reads
\vs{-0.5mm}
\begin{align}
    \Pcal_{\zeta}( k )
        \simeq
            \frac{M\sqrt{\phi_{\crm}\.\mu_{1}}}
            {2\.\sqrt{2\mspace{1.5mu}\pi}
            \.\chi_{2}\.\Mpl^{2}}
            \exp\mspace{-3mu}
            \bqty{-\frac{4\.\Mpl^{4}\mspace{1mu}
            ( N_{k} - N_{\rm water} )^{2}}
            {M^{2}\.\phi_{\crm}\.\mu_{1}}}
            \. ,
            \\[-3mm]
            \notag
\end{align}
where $N_{k}$ is the number of backward e-folds from the end of inflation to the time when $k = a H$, $N_{\rm water}$ is the number of e-folds estimated for the waterfall phase,
\vs{-0.5mm}
\begin{align}
    N_{\rm water}
        \simeq
            \pqty{
                \frac{ \sqrt{\chi_{2}} }{ 2 }
                +
                \frac{1}{4\.\sqrt{\chi_{2}}}
            }
            \frac{M\sqrt{\phi_{\crm}\.\mu_{1}}}
            {\Mpl^{2}}
            \, ,
\end{align}
and the parameter $\chi_{2}$ is defined as
\vs{-1.5mm}
\begin{align}
    \chi_{2}
        \coloneqq
            \ln\mspace{-1.5mu}
            \pqty{
                \frac{M\sqrt{\phi_{\crm}}}
                {2\.\sqrt{\mu_{1}}\.\sigma_{\psi}}
            }
            \, .
\end{align}
Interestingly, except for a weak parameter-dependence of $\chi_{2}$, any of those quantities is entirely characterised by the combination
\begin{align}
    \Pi
        \coloneqq            
            \frac{ \sqrt{\phi_{\crm}\.\mu_{1}}\.M }{ \Mpl^{2} }
            \, .
\end{align}

Furthermore, one should note that the curvature perturbation generated during the valley phase must be small to be consistent with the cosmic microwave background observation, which leads to the condition
\vs{-1.5mm}
\begin{align}
    \frac{\Lambda^{\mspace{-2mu}4}\.\mu_{1}^{2}}
    {12\mspace{1.5mu}\pi^{2}\.\Mpl^{6}}
        \simeq
            \Pcal_{\zeta}( k_{\CMB} )
        \sim
            10^{-9}
            \, .
\end{align}
Under this condition, $\chi_{2}$ can be expressed as
\vs{-1mm}
\begin{align}
    \chi_{2}
        = 
            \ln\mspace{-1.5mu}
            \bqty{
                \pqty{
                    \frac{2}{\pi}
                }^{\mspace{-6mu}1/4}\.
            \sqrt{\Pi / \Pcal_{\zeta}( k_{\CMB} )}\,}
            ,
\end{align}
and one finds $\chi_{2} \sim 10$ for typical values $10 \lesssim \Pi^{2} \lesssim 1000$. Therefore, the number of peak e-folds $N_{\rm water} \simeq \big[ \sqrt{\chi_{2}}/2 + 1/( 4\sqrt{\chi_{2}} ) \big]\mspace{1.5mu}\Pi$ and the maximum value which the power spectrum can assume, $\Pcal_{\zeta,\mspace{1.5mu}\max} = \eval{\Pcal_{\zeta}}_{N_{k} = N_{\rm water}} \simeq \Pi / \big( 2\.\sqrt{2\mspace{1.5mu}\pi}\.\chi_{2} \big)$, satisfy the relation
\vs{-2mm}
\begin{align}
\label{eq: Pzmax and Nwater}
    \Pcal_{\zeta,\mspace{1.5mu}\max}
        \simeq
            \frac{1}
            {\sqrt{2\mspace{1.5mu}\pi\.\chi_{2}^{3}}}\,
            N_{\rm water}
        \simeq
            0.01\.N_{\rm water}
            \, .
\end{align}
If one assumes $N_{\rm water} \sim 10$, the power spectrum inevitably becomes as large as $\Pcal_{\zeta,\mspace{1.5mu}\max} \sim 0.1$. However, as we will see below, this value is too large and PBHs are overproduced. Therefore, successful (massive) primordial black hole production cannot be realised in this setup.\footnote{\setstretch{0.9}Even though Equation~\eqref{eq: Pzmax and Nwater} shows that massive PBHs cannot be realised in mild-waterfall hybrid inflation, Reference~\cite{2015PhRvD..92b3524C} has erroneously reached this conclusion, despite containing the correct formul{\ae}. This has also been confirmed using the full stochastic approach~\cite{2016JCAP...08..041K}.}

\subsubsection{Non-Canonical Inflaton}
\vs{-1mm}
The scalar inflaton is often assumed to have a canonical kinetic term, \ie
\vs{-1mm}
\begin{align}
    \Scal
        = 
            \int\dd[4]x\;\sqrt{- g}\.
            \big[
                X
                -
                V( \phi )
            \big]
            \, ,
\end{align}
with gradient term $X = -\frac{1}{2}\.g^{\mu\nu}\.\partial_{\mu}\phi\.\partial_{\nu}\phi$ and potential energy $V( \phi )$. In this setup, the sound speed $\cs^{2} = \partial p / \partial\rho$ of the inflaton is found equal to the speed of light: $\cs = 1$. However, one may consider a more general form of the action. For example, in the model known as \emph{k-inflation}~\cite{1999PhLB..458..209A, 1999PhLB..458..219G}, the action reads
\begin{align}
    \Scal
        = 
            \int\dd[4]{x}\;\sqrt{- g}\,
            K( \phi,X )
            \, ,
\end{align}
with an arbitrary function $K$ of $\phi$ and $X$. In this case, the sound speed is given by
\begin{align}
    \cs^{2}
        = 
            \frac{K_{X}}
            {K_{X} + 2\.X K_{XX}}
            \, ,
\end{align}
where $K_{X} = \partial_{X}K$ and $K_{XX} = \partial_{X}^{2}K$, and thus it is not necessarily unity, beyond the canonical case $K = X - V( \phi )$.

The general sound speed modifies Formula~\eqref{eq: Pcal zeta formula} for the power spectrum as
\begin{align}
    \Pcal_{\zeta}( k )
        = 
            \eval{
                \frac{1}
                {2\.\epsilon_{H}\.\cs\.\Mpl^{2}}
                \pqty{\frac{H}{2\mspace{1.5mu}\pi}}^{\mspace{-6mu}2}\.
            }_{a H\mspace{1mu} = \.k}
            \, .
\end{align}
Therefore, the reduction of the sound speed can give rise to an amplification of the power spectrum. The PBH production due to such a small sound speed has been studied in References~\cite{2019JCAP...06..016B, 2022JHEP...01..074B, 2022JCAP...02..030G} in terms of \emph{effective field theory}, where all possible relevant terms are included in the action under the assumption that a single scalar field drives the quasi de Sitter universe.

A much more exotic change of sound speed has been proposed in the context of multi-field inflation. Consider the action of multiple scalars $\phi^{I}$ ($I = 1,\mspace{1.5mu}2,\mspace{1.5mu}\ldots$) with a field-dependent kinetic term:
\begin{align}
    \Scal
        = 
            \int\dd[4]{x}\,\sqrt{- g}\.
            \bqty{
                -\.G_{IJ}( \phi )\.
                g^{\mu\nu}\.
                \partial_{\mu}\phi^{I}\mspace{1.5mu}
                \partial_{\nu}\phi^{J}
                -
                V( \phi )
            }
            \, .
\end{align}
The field values $\phi^{I}$ are now understood as \emph{coordinates} of a curved manifold (called \emph{target manifold} or \emph{field-space manifold}) whose metric is given by $G_{IJ}( \phi )$. The function $\phi^{I}( x )$ is thus the mapping from the spacetime manifold to the target manifold. In this case, the inflatons $\phi^{I}$ are driven not only by the potential force but also by the ``geodesic'' force (without the potential force, the inflatons go ``straight'' along the curved target manifold).

This mechanism or inflation model is dubbed in several ways, for instance, \emph{Geometrical Destabilisation}~\cite{2016PhRvL.117n1301R} (where the trajectory of the inflatons is destabilised from the potential minimum by the geometrical force) \emph{Hyperbolic Inflation}~\cite{2018PhRvL.121y1601B} (where the negative curvature of the hyperbolic target manifold yields a centrifugal force), or \emph{Orbital Inflation}~\cite{2020PhRvD.102b1302A} (in which the inflatons are enabled to be in an orbital motion with the radial effective mass arbitrarily controlled). The sound speed of the perturbations in this system is affected by the ``bending'' rate of the trajectory from the geodesic motion; strong bending can even cause an \emph{imaginary} sound speed (\ie~$\cs^{2} < 0\mspace{0.5mu}$; see \eg~References~\cite{2018JCAP...07..057G, 2019PhRvL.123t1302F}). As discussed in References~\cite{2020PhRvL.125l1301P, 2020arXiv200408369F}, the curvature perturbation can be significantly amplified in this way, implying sizeable primordial black hole formation.

\begin{figure}
    \centering
    \includegraphics[width=0.75\hsize]{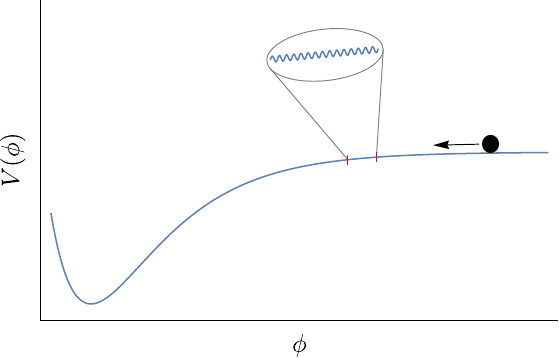}
    \caption{
        Fine features in the inflaton potential can yield parametric-resonance amplification to the curvature perturbation. Figure from Reference~\cite{2020JCAP...06..013C}.
        \vs{3mm}
        }
    \label{fig: resonance}
\end{figure}

\newpage

\subsubsection{Parametric Resonance due to Potential Features}

Fine features in the inflaton potential may lead to significant amplification of the curvature perturbation. For example, one may introduce a small oscillating modulation as illustrated in Figure~\ref{fig: resonance}. In this case, the square of the effective inflaton mass rapidly oscillates as time goes on, and if this oscillation frequency matches the wave oscillation of the mode perturbation, the corresponding perturbation can be amplified as a ``swing". This phenomenon is known as \emph{parametric resonance} (see Reference~\cite{2020JCAP...06..013C} for PBH formation in this class of scenarios).



\subsubsection{Curvaton}
\vs{-1mm}
Not only do the inflatons induce primordial perturbation, but also a (scalar) field which is negligible during inflation and sources curvature perturbations later{\,---\,}a \emph{curvaton}~\cite{1997PhRvD..56..535L, 2002NuPhB.626..395E, 2002PhLB..524....5L, 2001PhLB..522..215M}. The minimal curvaton mechanism can be realised by the simplest mass term potential:
\begin{align}
    V(\sigma)
        = 
            \frac{1}{2}\.
            m_{\sigma}^{2}\.
            \sigma^{2}
            \, ,
\end{align}
where $\sigma$ represents the curvaton field.
If the curvaton mass is light enough, $m_{\sigma} \ll H_{\inf}$, and its field value is sub-Planckian, $\sigma \ll \Mpl$, the curvaton energy density is negligible during inflation. As it is a light scalar, the curvaton acquires fluctuations $\delta \sigma \sim H/2\mspace{1.5mu}\pi$ from the zero-point quantum fluctuation as well as the inflatons; these are conserved. After inflation has ended, the field value $\sigma = \sigma_{0} + \delta\sigma$ ($\sigma_{0}$ is the background value) is still frozen until the background energy density (from oscillating inflatons or radiation) becomes small enough, $H \sim m_{\sigma}$.

The curvaton then starts to oscillate around its potential minimum and its energy density decreases as a matter component $\rho_{\sigma} \propto a^{-3}$. On the other hand, the background density damps as radiation $\rho_{\rrm} \propto a^{-4}$ once reheating is completed. Hence the relative density of the curvaton to the background grows as $\rho_{\sigma} / \rho_{\rrm} \propto a$, and can be significant even though the curvaton is initially negligible. If the curvaton then decays into radiation by some interaction, the curvaton density fluctuation $\delta\rho_{\sigma}$ is converted into the radiation fluctuation (and hence adiabatic curvature perturbation). The sourced curvature perturbation is proportional to the curvaton density contrast, $\zeta \propto \delta\rho_{\sigma} / \rho_{\sigma_{0}} \simeq 2\.\delta\sigma/\sigma_{0}$, where the coefficient is determined by the relative energy density $\rho_{\sigma}/\rho_{\rrm}$ at curvaton decay. As the curvaton is free from inflation dynamics, the power spectrum of the sourced curvature perturbation is not necessarily restricted by the inflatons' potential [like in Equation~\eqref{eq: ns SR}]. Kasuya {\it et al.}~\cite{2009PhRvD..80b3516K} showed that an axion-like curvaton can produce a strongly blue-tilted curvature perturbation, and Kawasaki {\it et al.}~\cite{2013PhRvD..87f3519K} investigated PBH formation in this model. The original model is supersymmetry-inspired, with three chiral superfields $\Phi$, $\bar{\Phi}$ and $S$, and their superpotential given by
\begin{align}
    W
        = 
            h\.S\.
            \Big(
                \Phi\bar{\Phi}
                -
                f_{a}^{2}
            \Big)
            \, ,
\end{align}
Here, $h$ is a coupling and $f_{a}$ a decay constant. The corresponding scalar potential reads
\vs{-1mm}
\begin{align}
    V
        = 
            h^{2}\.
            \abs{
                \Phi\bar{\Phi}
                -
                f_{a}^{2}
            }^{2}
            +
            h^{2}\.\abs{S}^{2}\mspace{2mu}
            \Big(
                \abs{\Phi}^{2}
                +
                \abs{\bar{\Phi}}^{2}
            \Big)
            \, .
\end{align}
Writing the complex scalar components explicitly as
\begin{subequations}
\begin{align}
    \Phi
        &= 
            \varphi\,
            \erm^{i\mspace{1.5mu}\theta}
            \, ,
            \displaybreak[1]
            \\[2mm]
    \bar{\Phi}
        &= 
            \bar{\varphi}\,
            \erm^{i\mspace{1.5mu}\bar{\theta}}
            \, ,
\end{align}
\end{subequations}
the potential minimum is reached for
\begin{align}
    \varphi\.\bar{\varphi}
        = 
            f_{a}^{2}\,\qc
    \theta
        = 
            -\.\bar{\theta}\,\qc
    S
        = 
            0
            \, .
\end{align}
Assuming $h\mspace{1.5mu}f_{a} \gg H_{\inf}$, all scalars are strongly stabilised along this constraint. Conversely, $\varphi$ and $\bar{\varphi}$, and $\theta$ and $\bar{\theta}$ can freely move as long as this constraint is satisfied. These flat-direction features are common in supersymmetry contexts.

During inflation, supergravity generically uplifts the potential as
\begin{align}
    V_{H}
        = 
            c\.H^{2}\.\abs{\Phi}^{2}
            +
            \bar{c}\.H^{2}\.
            \abs{\bar{\Phi}}^{2}
            +
            c_{S}\.H^{2}\.\abs{S}^{2}
            \, ,
\end{align}
with order-unity coefficients $c$, $\bar{c}$, and $c_{S}$. The $\theta$-direction is still flat. If $\varphi$ or $\bar{\varphi}$ have large initial field values $\varphi \ll f_{a}$ or $\bar{\varphi} \ll f_{a}$, they slowly roll down to the potential minimum
\vs{-2mm}
\begin{subequations}
\begin{align}
    \varphi_{\umin}
        \simeq&
            \pqty{\bar{c}/{c}}^{1/4}\.f_{a}
            \, ,
            \displaybreak[1]
            \\[3mm]
    \bar{\varphi}_{\umin}
        \simeq&
            \pqty{c/\bar{c}}^{1/4}\.f_{a}
            \, ,
            &
            \notag
\end{align}
\end{subequations}
due to the {\it Hubble-induced mass terms}. Without loss of generality, let us assume that $\varphi$ has a large value. Then, $\bar{\varphi} = \varphi / f_{a}$ can be neglected. In the slow-roll limit, in which $H$ is approximately constant, $\varphi$'s equation of motion,
\begin{align}
    \ddot{\varphi}
    +
    3\mspace{1.5mu}H\mspace{1.5mu}\dot{\varphi}
    +
    c\.H^{2}\mspace{1.5mu}
    \varphi
        = 
            0
            \, ,
\end{align}
can be analytically solved as
\begin{align}
    \varphi
        \propto
            \erm^{-\lambda\mspace{1mu}Ht}
            \quad
            \text{with $\displaystyle\!
                \quad
                \lambda
                    = 
                        \frac{3}{2}
                        -
                        \frac{3}{2}\.
                        \sqrt{1 - \frac{4}{9}\mspace{2mu}c}$
            }
            \, .
\end{align}

Therefore, $\varphi$'s value at $k = a H$, $\varphi( k )$, has the dependence
\vs{-0.5mm}
\begin{align}
    \varphi( k )
        \propto
            k^{-\lambda}
            \, .
\end{align}

The $\theta$-direction is massless during this process and hence acquires fluctuations. Note here that $\theta$ itself is not a canonical field but $\varphi\.\theta$ is approximately one. Therefore, the perturbation corresponding to the wavenumber $k$ is evaluated as
\vs{-0.5mm}
\begin{align}
    \varphi( k )\,\delta\theta( k )
        \sim
            H / 2\mspace{1.5mu}\pi
            \, .
\end{align}
Even though $H$ is almost constant, $\theta$'s perturbation is blue-tilted,
\vs{-0.5mm}
\begin{align}
    \Pcal_{\delta\theta}( k )
        \propto
            \varphi^{-2}( k )
        \propto
            k^{2\mspace{1.5mu}\lambda}
            \, ,
\end{align}
for $2\.\lambda \sim 1\,\text{--}\,3$ and order-unity $c$. After $\varphi$ reaches $\varphi_{\umin}$, the power spectrum becomes almost scale-invariant.

\begin{figure}
    \centering
    \includegraphics[width = 0.8\hsize]{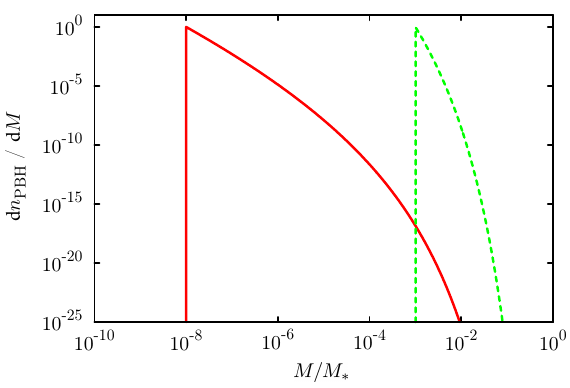}
    \caption{
        Exemplary primordial black hole mass spectra $\dd{n_{\PBH}} / \dd{M}$, utilising the minimum mass (corresponding to the curvaton decay time) $M_{\umin}/M_{*} = 10^{-8}$ (red solid) and $10^{-3}$ (green dashed). Here, $M_{*}$ is the peak scale of the power spectrum (at the time when $\varphi$ reaches $\varphi_{\umin}$). Figure from Reference~\cite{2013PhRvD..87f3519K}.
        \vs{3mm}
        }
    \label{fig: curvaton PBH}
\end{figure}

The field $\theta$ is massless at high energies, but may acquire mass at low energies well after inflation due to some non-perturbative mechanism like for the QCD axion. Then, $\theta$ can play the r{\^o}le of the curvaton, and sourced curvature perturbations are also blue-tilted with the same spectral index. By tuning the decay time of the curvaton, one can put a lower limit on the PBH mass because no PBH is expected to form before curvaton decay. The overproduction of light PBHs can be avoided in this way. Kawasaki {\it et al.}~\cite{2013PhRvD..87f3519K} illustrates exemplary PBH mass spectra of this model, which are shown in Figure~\ref{fig: curvaton PBH}.

Another interesting amplification of the curvaton perturbation can be accomplished by kinetic mixing between the inflaton $\phi$ and the curvaton $\sigma$~\cite{2023PhRvD.108j1301P}. There, the kinetic term of the curvaton in the Lagrangian $\Lcal$ is modulated by some function $f( \phi )$ of the inflaton, while the inflaton has a minimal kinetic term as
\vs{-0.5mm}
\begin{align}
	\Lcal_\mathrm{kin}
        =
            -
            \frac{1}{2}\.
            \partial_{\mu}\phi\.\partial^{\mspace{1mu}\mu}\phi
            -
            \frac{1}{2}\.f( \phi )\.
            \partial_{\mu}\sigma\.\partial^{\mspace{1mu}\mu}\sigma
            \, .
\end{align}
In this case, the curvaton perturbation is modified by $f$ at the time $t_{k}$ when the mode of interest $k$ exits the horizon:
\vs{-0.5mm}
\begin{align}
	\Pcal_{\delta\sigma/\sigma}( k )
        =
            \eval{\frac{1}{\sigma^{2}}
            \pqty{\frac{H}{2\mspace{1.5mu}\pi f}}^{\mspace{-6mu}2}}_{t_{k}}
            \, .
\end{align}
Therefore, if $\eval{f( \phi )}_{t_{k}}$ has a sharp dip around some field value, the curvaton, and hence curvature perturbation, can easily be significantly amplified.

Beyond linear perturbation theory, the curvaton naturally has a nonlinear relation between its fluctuation $\delta\sigma$ and the curvature perturbation $\zeta$, because, in general, the curvaton energy density $\rho_{\sigma}$ is a nonlinear function of $\sigma$. Consequently, the curvature perturbations imply so-called \emph{local-type non-Gau{\ss}ianity}, which we discuss in the next Subsection. Hence, PBH production in the curvaton scenario requires a non-Gau{\ss}ian treatment, such as the one elaborated below.
\newpage


\subsection{Aspects of Inflationary Quantum Perturbations}
\label{sec:Aspects-of-Inflationary-Quantum-Perturbations}
\vs{-1mm}
So far, we have only considered the power spectrum (\ie~the two-point function) as a property of curvature perturbations. However, these are not necessarily Gau{\ss}ian, and many kinds of \emph{non-Gau{\ss}ianities} can be relevant. In particular, as PBHs are related to large curvature perturbations, $\zeta \sim 1$, it has been pointed out that nonlinearity/non-Gau{\ss}ianity significantly affects the prediction for the PBH abundance (see \eg~Reference~\cite{2012PhRvD..86d3512B}). Here, we review this important aspect of primordial black holes.

\subsubsection{Local-type Non-Gau{\ss}ianity}
\label{sec:Local--type-Non--Gaussianity}
\vs{-1mm}
The \emph{local-type} non-Gau{\ss}ianity is one of the simplest types of non-Gau{\ss}ianity. In this model, the full curvature perturbation $\zeta( \xbm )$ is approximated by a nonlinear mapping $\Fcal_{\NL}\big[ \zeta_{\Grm}( \xbm ) \big]$ of the Gau{\ss}ian field $\zeta_{\Grm}( \xbm )$ at the same spatial point $\xbm$. In particular, the series expansion up to quadratic order,
\vs{-1mm}
\begin{align}
    \zeta( \xbm )
        = 
            \zeta_{\Grm}( \xbm )
            +
            \frac{3}{5}\.f_{\NL}\.
            \zeta_{\Grm}^{2}( \xbm )
            \, ,
\end{align}
is commonly used as a minimal non-Gau{\ss}ian correction. The coefficient $f_{\NL}$ is called \emph{nonlinearity parameter}; the factor $3/5$ is just a convention. 

The quadratic expansion is inversely solved as\footnote{\setstretch{0.9}We neglect the other solution, which is actually suppressed probabilistically.}
\begin{align}
    \zeta_{\Grm}( \xbm )
        = 
            \frac{\sqrt{60\.f_{\NL}\.
            \zeta( \xbm ) + 25} - 5\,}
            {6\.f_{\NL}}
            \, .
\end{align}
For very small values of the curvature perturbation, $\zeta( \xbm ) \ll 1$, as a typical perturbation, $\zeta_{\Grm}( \xbm )$ is almost equivalent to $\zeta( \xbm )$ for not-so-large values of the nonlinearity parameter, \ie~for $f_{\NL} \lesssim \Ocal( 1 )$. Hence, this non-Gau{\ss}ianity can be seen as a small correction to the Gau{\ss}ian approximation. However, PBH formation associated with $\zeta( \xbm ) \sim 1$ is a different story. In fact, for $\zeta( \xbm ) = 1$ and $f_{\NL} = 1$ as an example, $\zeta_{\Grm}( \xbm )$ is non-negligibly shifted to $\zeta_{\Grm}( \xbm ) \simeq 0.7$. If $\zeta_{\Grm}( \xbm )$ follows a Gau{\ss}ian probability distribution,
\vs{-1mm}
\begin{align}
    P_{\Grm}( g )
        = 
            \frac{1}
            {\sqrt{2\mspace{1.5mu}\pi\sigma_{0}^{2}}}\mspace{1mu}
            \exp\!\pqty{-\frac{g^{2}}{2\.\sigma_{0}^{2}}}
            \. ,
\end{align}
with $\sigma_{0}^{2} = 10^{-2}$ for example, the probability difference between $\zeta_{\Grm} = 1$ and $\zeta_{\Grm} = 0.7$ is huge:
\vs{-2.5mm}
\begin{align}
    \frac{P_{\Grm}(\zeta_{\Grm} = 0.7)}
    {P_{\Grm}(\zeta_{\Grm} = 1)}
        \simeq
            1.2 \times 10^{11}
            \, .
\end{align}
Therefore, even small amounts of non-Gau{\ss}ianity should carefully be taken into account in PBH abundance calculations (see Reference~\cite{2012PhRvD..86d3512B}).

\begin{figure}
    \centering
    \includegraphics[width = 0.72\hsize]{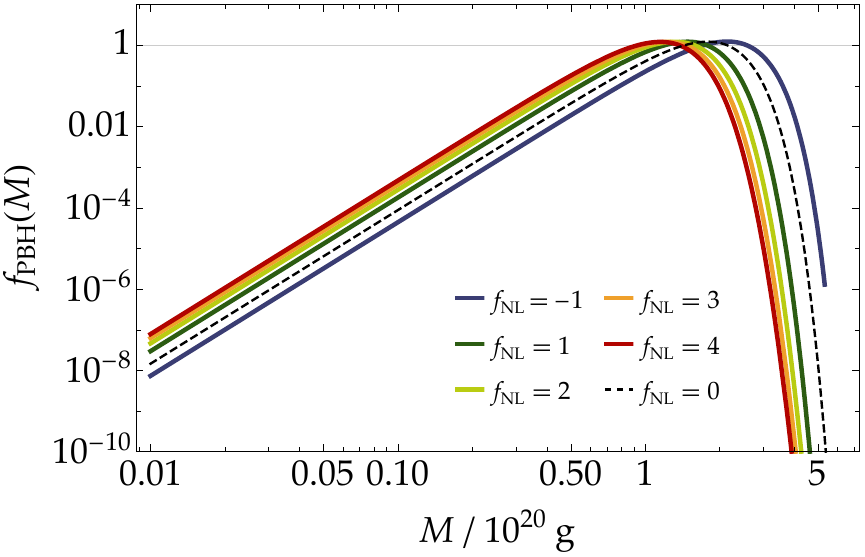}\\[6mm]
    \hs{1.68mm}\includegraphics[width = 0.749\hsize]{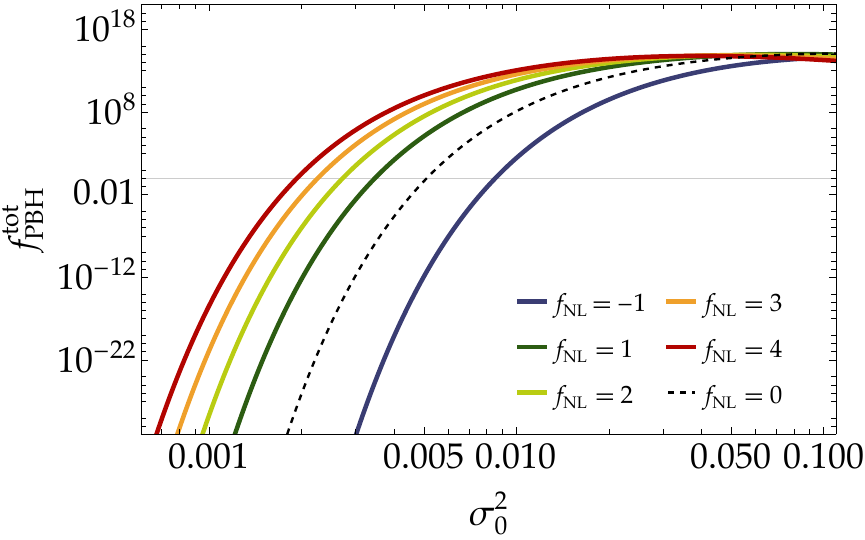}
    \caption{
        Mass spectra $f_{\PBH}( M )$ ({\it upper panel}) and total abundance $f^{\rm tot}_{\PBH}( \sigma_{0}^{2} )$ ({\it lower panel}) of primordial black holes for various values of the nonlinearity parameter $f_{\NL}$ under the monochromaticity assumption~\eqref{eq: monochromatic power} with $k_{*} = 1.56 \times 10^{13}\.{\rm Mpc}^{-1}$ corresponding to $M_{H} = 10^{20}\.\grm$~\eqref{eq: horizon mass}. The {\it upper panel}, uses different values of $\sigma_{0}^{2}$ such that $f_{\NL}^{\rm tot} = 1$. Upper and lower panels correspond to Figure~5 of Reference~\cite{2021JCAP...10..053K} and Figure~9 of Reference~\cite{2022JCAP...05..012E}, respectively.
        \vs{3mm}
        }
    \label{fig: fPBHfNL}
\end{figure}

The strong point of the local-type non-Gau{\ss}ianity approximation is that its statistics is fully determined as it is just a mapping of a Gau{\ss}ian field. Therefore, the peak-theory approach for the PBH abundance (see Section~\ref{sec:Peak--Theory-Procedure-with-Curvature-Peaks}), which is usually only used for Gau{\ss}ian fields, can be straightforwardly applied to this type of non-Gau{\ss}ianity~\cite{2019JCAP...09..033Y, 2021JCAP...10..053K, 2022JCAP...05..012E}, with the PBH abundance being estimated most precisely to our current knowledge. In Figure~\ref{fig: fPBHfNL}, we show examples of the $f_{\NL}$-dependence of both the mass function of primordial black holes and their total abundance as a function of $\sigma_{0}^{2}$ for a monochromatic power spectrum of $\zeta_{\Grm}$: 
\begin{align}
\label{eq: monochromatic power}
    \Pcal_{\Grm}( k )
        = 
            \sigma_{0}^{2}\,
            \delta
            \big(
                \!\ln[ k / k_{*} ]
            \big)
            \, .
\end{align}
Whilst the shape of the mass function is not very sensitive to changes in $f_{\NL}$, one sees that the total abundance indeed strongly depends on the nonlinearity parameter. In order to obtain the mass function, we adopted the $q$-parameter criterion (Section~\ref{sec:Case-C:-Non--Gaussian-Contribution-to-zeta}) and the peak-theory procedure (Section~\ref{sec:Peak--Theory-Procedure-with-Curvature-Peaks}).

\subsubsection{Exponential/Heavy Tail}
\label{sec:Exponential/Heavy-Tail}
\vs{-1mm}
The local-type non-Gau{\ss}ianity can be naturally understood in terms of the $\delta N$ formalism. Though the inflaton perturbations $\delta\phi( \xbm )$ originating from quantum zero-point fluctuations can be well viewed as a Gau{\ss}ian field, the curvature perturbation $\zeta$ and the time difference $\delta N$ as functions $\delta\phi( \xbm )$, are nonlinear mappings in general. If one assumes that a single noise $\delta\phi( \xbm )$ (with a monochromatic spectrum) becomes quite large by chance, the curvature perturbation is given by $\zeta( \xbm ) = \delta N\big[ \delta\phi( \xbm ) \big]$, which is exactly the form of local-type non-Gau{\ss}ianity.

Recently, some non-perturbative features of this mapping function are attracting attention in the context of ultra-slow-roll behaviour in flat-inflection models (see, \eg, References~\cite{2017JCAP...10..046P, 2019JCAP...09..073A, 2020JCAP...05..022A, 2020JCAP...03..029E, 2021PhRvL.127j1302F, 2021JCAP...04..080P, 2021PhLB..82036602B, 2022JCAP...05..027F, 2022PhLB..83437400H, 2022PhLB..83437461C, 2022JCAP...12..034C, 2023PhRvL.131a1002P}). There, the probability density of $\zeta$ decays only exponentially $P_{\zeta} \!\propto \!\exp\mspace{1mu}(-\Lambda\.\zeta)$ ($\Lambda$ being a decay constant) in the large-$\zeta$ regime, contrary to the Gau{\ss}ian decay $\propto \exp\!\big( \!-\mspace{-1mu}\zeta^{2}/2\.\sigma_{\zeta}^{2} \big)$, and is hence called \emph{exponential-tailed} curvature perturbation. The exponential tail significantly amplifies the probability density of large $\zeta$ and is expected to completely alter the PBH abundance prediction even compared with the simple $f_{\NL}$ correction.

\begin{figure}
    \centering
    \includegraphics[width = 0.75\hsize]{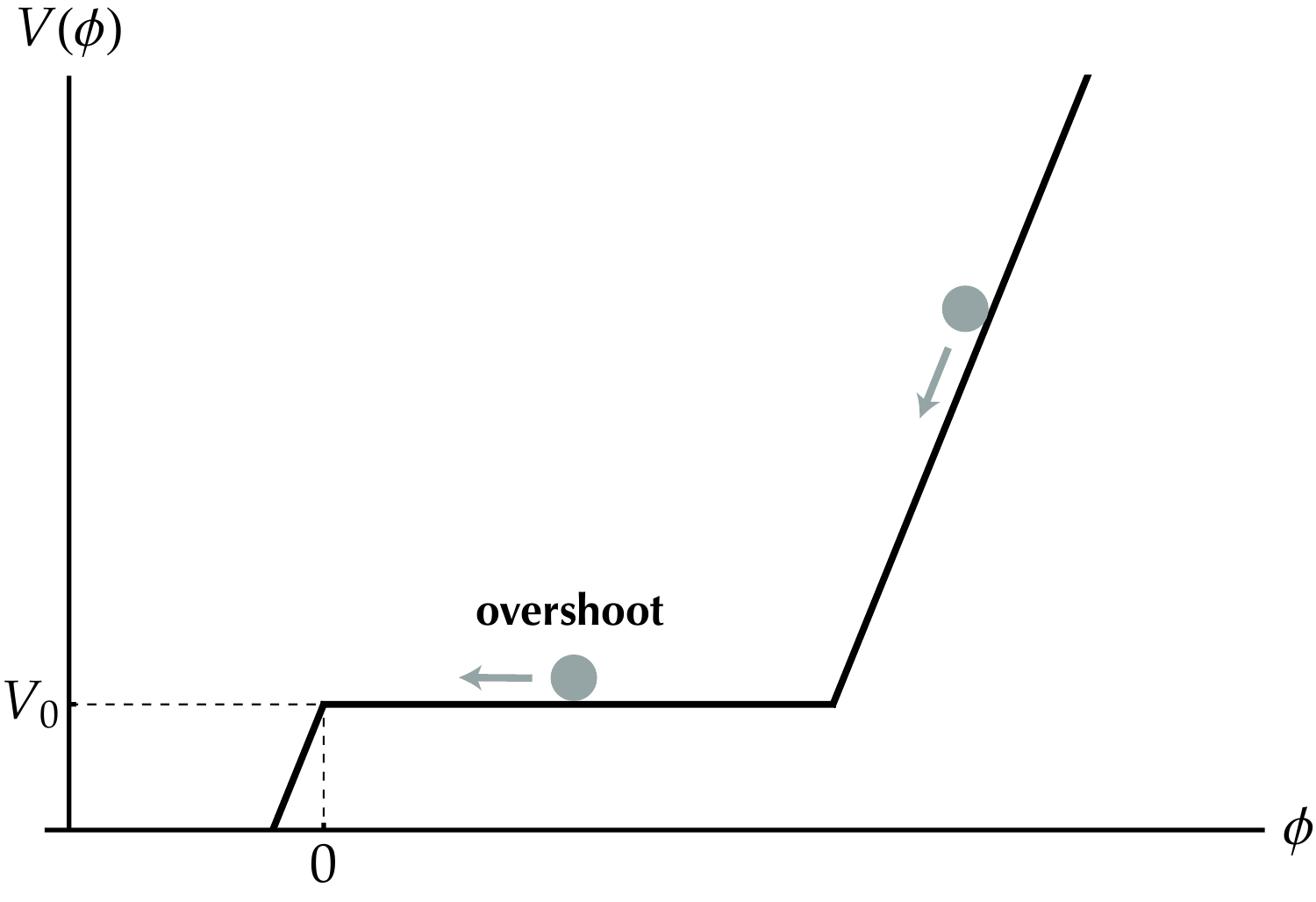}\\[12mm]
    \includegraphics[width = 0.75\hsize]{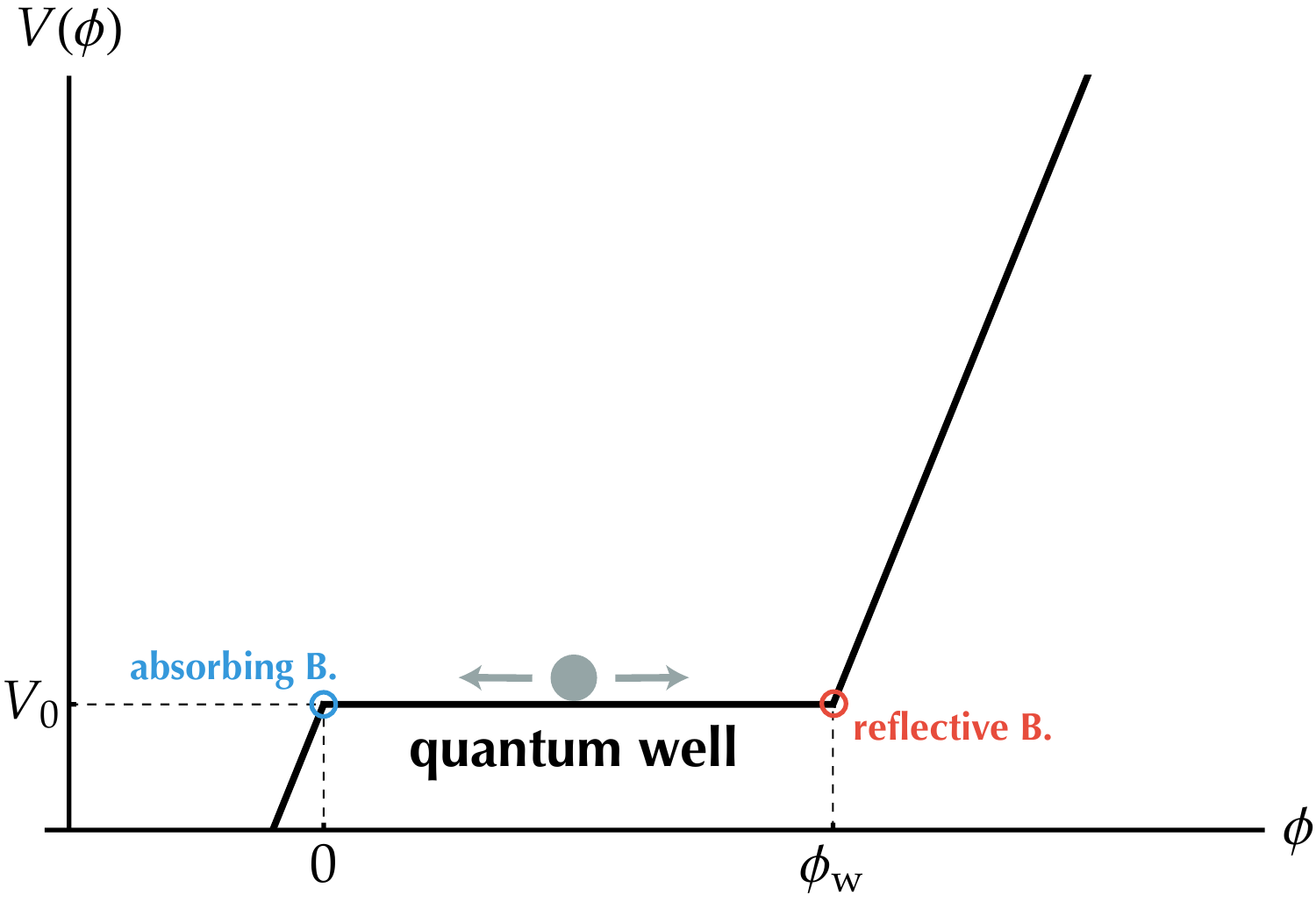}
    \caption{
        Toy model of a flat-inflection potential.
        \emph{Upper panel}: 
            The inflaton is assumed to overshoot the flat region by the initial velocity given in the upper part of the potential as an ultra-slow-roll situation. Single noise $\delta\phi$ is added during the flat-potential phase.
        \emph{Lower panel}: 
            The inflaton velocity is neglected but it is assumed to exit the flat region by stochastic noise as a quantum well situation.
        }
    \label{fig: USR}
\end{figure}

\newpage

Let us study the exponential-tail probability in the simplest toy model of the inflection potential illustrated in the upper panel of Figure~\ref{fig: USR}. The potential has an exactly-flat region and the inflaton overshoots it by an initial velocity given in the precedent steep-slope part. In the flat region, the inflaton background equation of motion reads
\begin{align}
    \dv[2]{\phi_{0}}{N}
    +
    3\.\dv{\phi_{0}}{N}
        \simeq
            0
            \, ,
\end{align}
where the Hubble parameter is almost constant, given by $H \simeq \big[ V_{0} / ( 3\mspace{1.5mu}\Mpl^{2} ) \big]^{1/2}$. Such a situation in which the potential tilt is too small to neglect the second time derivative of the inflaton is called \emph{ultra slow-roll}.

This equation of motion is readily solved as
\begin{align}
    \phi_{0}( n )
        &= 
            \phi_{\irm}
            +
            \frac{\pi_{\irm}}
            {3\mspace{1mu}H}\.
            \Big(
                1
                -
                \erm^{-3N}
            \Big)
            \notag
            \\[2mm]
        &\Updownarrow
            \\[2mm]
    N( \phi_{0}\,|\,\phi_{\irm} )
        &= 
            -
            \frac{1}{3}\ln\mspace{-2mu}
            \pqty{
                1
                -
                3\mspace{1mu}H\.
                \frac{\phi_{0} - \phi_{\irm}}
                {\pi_{\irm}}
            }
            \, ,
            \notag
\end{align}
where $\phi_{\irm} \equiv \phi_{0}( N_{\irm} )$ and $\pi_{\irm} \equiv H\eval{\dd{\phi_{\irm}} / \dd{N}}_{N_{\irm}}$ denote the inflaton field value and its momentum at some time $N_{\irm}$ during the flat regime, respectively. The number of e-folds taken from $\phi_{\irm}$ to $\phi_{0}$ is denoted by $N(\phi_{0}\,|\,\phi_{\irm})$. This exact solution allows to obtain the time difference to the end edge $\phi_{\frm} = 0$ due to the fluctuation $\delta\phi_{\irm}$ at $\phi_{\irm}$:
\begin{align}
\begin{split}
    \zeta( \xbm )
        &= 
            \delta N( \xbm )
        = 
            N
            \big[
                \phi_{\frm}\,
                \big|\,
                \phi_{\irm}
                +
                \delta\phi_{\irm}( \xbm )
            \big]
            -N
            \pqty{
                \phi_{\frm}\,|\,\phi_{\irm}
            }
            \\[3mm]
        &= 
            -
            \frac{1}{3}\.
            \ln\mspace{-1.5mu}
            \pqty{
                1
                +
                3\mspace{1mu}H\mspace{1.5mu}
                \frac{\delta\phi_{\irm}( \xbm )}{\pi_{\frm}}
            }
            \, ,
\end{split}
\end{align}
where $\pi_{\frm} \mspace{-1mu}\equiv \mspace{-1mu}\pi_{\irm}\,\erm^{-3N( \phi_{\frm}\,|\,\phi_{\irm} )} \mspace{-1mu}= \mspace{-1mu}\pi_{\irm} - 3\mspace{1mu}H (\phi_{\frm}-\phi_{\irm})$ is the (global) momentum at $\phi_{\frm}$. Therefore, defining the Gau{\ss}ian part via $\zeta_{\Grm}( \xbm ) \mspace{-1mu}\coloneqq \mspace{-1mu}-\mspace{1.5mu}H \.\delta\phi_{\irm}( \xbm )/\pi_{\frm}$, the curvature perturbation is understood to follow the nonlinear mapping given by
\vs{-1.5mm}
\begin{align}
\label{eq: exp-tail zeta}
    \zeta( \xbm )
        = 
            -
            \frac{1}{3}\.
            \ln\mspace{-1.5mu}
            \big[
                1
                -
                3\mspace{1.5mu}\zeta_{\Grm}( \xbm )
            \big]
            \, .
\end{align}
\newpage

One can check that the curvature perturbation is almost Gau{\ss}ian [$\zeta( \xbm ) \sim \zeta_{\Grm}( \xbm )$] in the small region $\abs{\zeta( \xbm )} \ll 1$ by its series expansion:
\begin{align}
    \zeta( \xbm )
        = 
            -\frac{1}{3}
            \ln\mspace{-1.5mu}
            \big[
                1
                -
                3\.\zeta_{\Grm}( \xbm )
            \big]
        = 
            \zeta_{\Grm}( \xbm )
            +
            \frac{3}{2}\.
            \zeta_{\Grm}^{2}( \xbm )
            +
            \Ocal\qty[ \zeta_{\Grm}^{3}( \xbm ) ]
            \, .
\end{align}
Note that it corresponds to the numerical value $5/2$ of the nonlinearity parameter $f_{\NL}$, this being a well-known result for the ultra-slow-roll scenarios (see \eg~Reference~\cite{2013EL....10139001N}). The series expansion, however, obviously fails for a large perturbation $\zeta_{\Grm}( \xbm ) \sim 1/3$. There, the probability-density function of $\zeta$ shows in fact an exponential tail as
\begin{align}
    P_{\zeta}( \zeta )
        = 
            \abs{\dv{\zeta_{\Grm}}{\zeta}}\.
            P_{\Grm}( \zeta_{\Grm} )
        = 
            \erm^{-3\mspace{1.5mu}\zeta}\.
            P_{\Grm}( \zeta_{\Grm} )
        \underset{\zeta_{\Grm}\.\to\.1/3}{\sim}
            \erm^{-3\mspace{1.5mu}\zeta}\.
            P_{\Grm}\qty(\zeta_{\Grm} = 1/3)
            \, .
\end{align}

\begin{figure}
    \centering
    \includegraphics[width = 0.75\hsize]{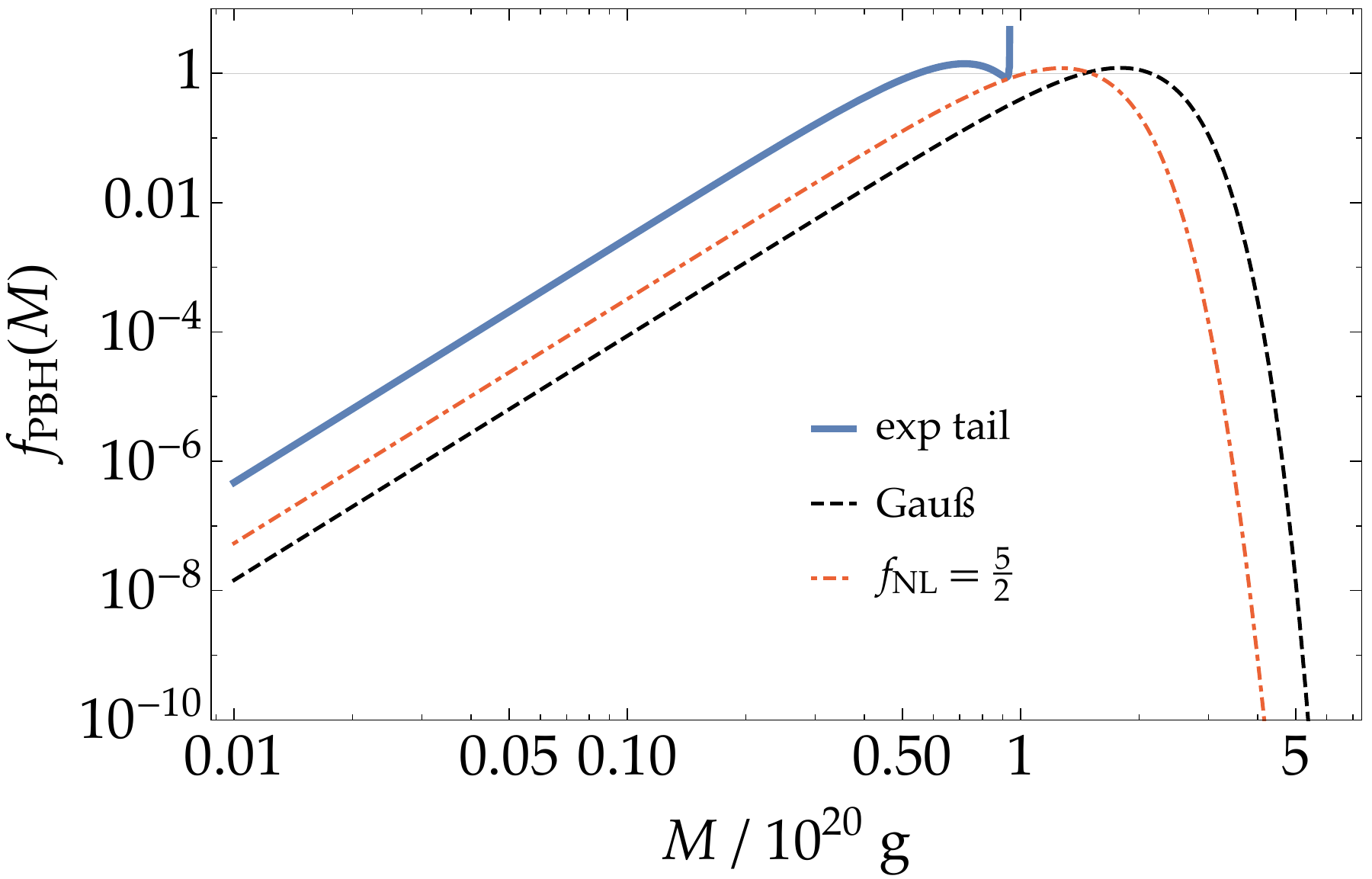}\\[6mm]
    \hs{2mm}\includegraphics[width = 0.78\hsize]{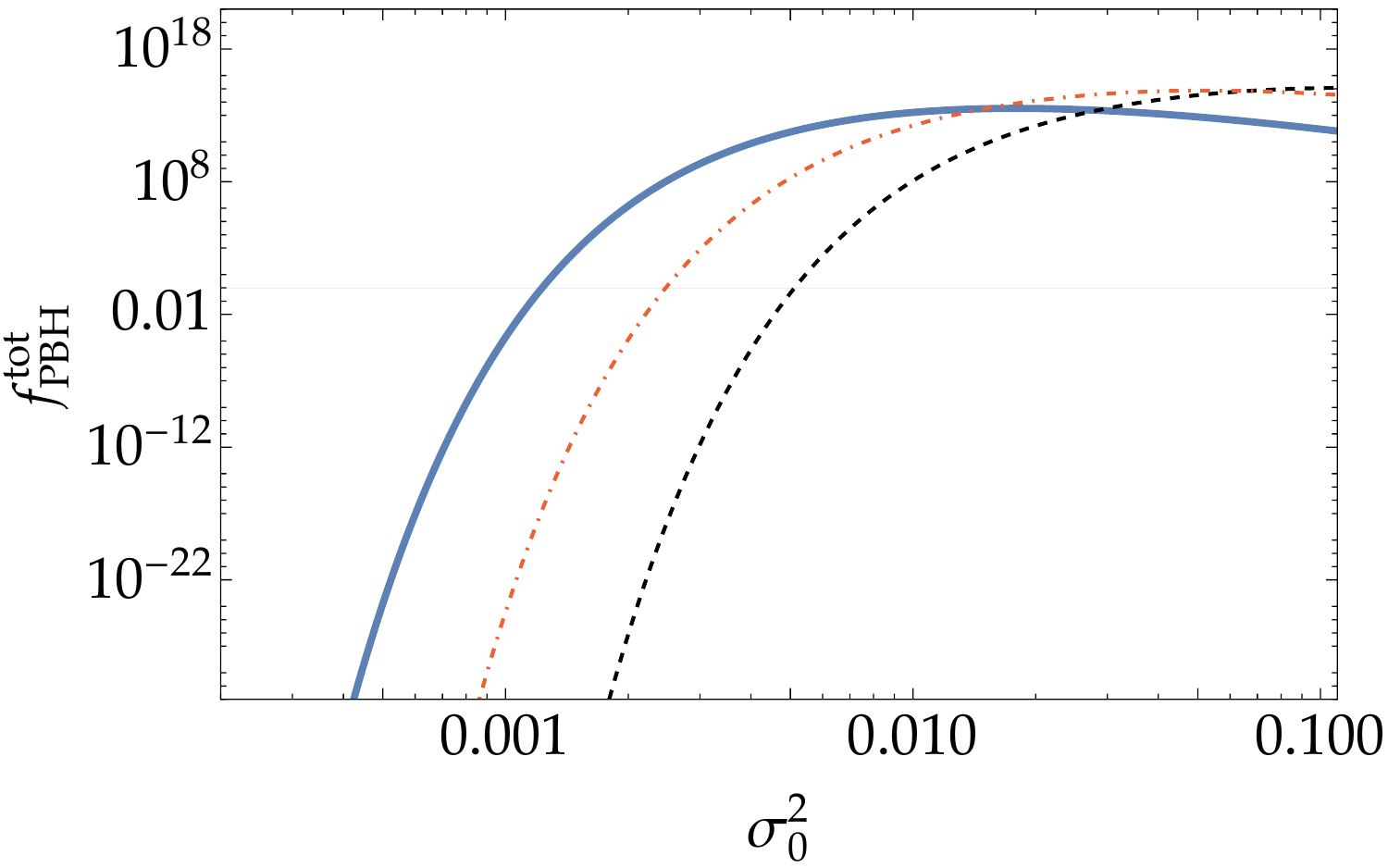}
    \caption{
        Similar plots to Figure~\ref{fig: fPBHfNL} under the exponential-tail assumption~\eqref{eq: exp-tail zeta}. One sees that neither the Gau{\ss}ian nor the quadratic expansion ($f_{\NL} = 5/2$) is enough at all. Figures correspond to the panels of Figure~7 in Reference~\cite{2021JCAP...10..053K}.
        }
    \label{fig:fPBHexp}
\end{figure}

Since $P_{\Grm}( \zeta_{\Grm} = 1/3 )$ is merely a constant, one finds that $P_{\zeta}( \zeta )$ decays only as $\erm^{-3\mspace{1.5mu}\zeta}$. A respective example mass function is shown in Figure~\ref{fig:fPBHexp}. The shape is distinctive as it has a hard cut at the maximum primordial black hole mass. Furthermore, the total PBH abundance is much amplified even compared to the quadratic expansion, \ie~for $f_{\NL} = 5/2$. These results demonstrate the necessity of a non-perturbative treatment.

\subsubsection{Stochastic Approach}
\label{sec:Stochasti-Approach}
\vs{-1mm}
So far, the curvature perturbation has been calculated by considering only single-impulsive noise for the inflaton fields. However, since small-scale perturbations actually continuously exit the horizon, the inflatons should in turn receive continuous noise. Hence, the dynamics of the inflatons are understood as \emph{stochastic processes} (or Brownian motions). This picture is well-sophisticated in the context of the stochastic formalism of inflation (see References~\cite{1982PhLB..117..175S, 1986LNP...246..107S, 1988PhLB..205..441N, 1989PhLB..219..240N, 1989PhRvD..39.2245K, 1988PThPh..80.1041N, 1989PThPh..81.1037N, 1991PhRvD..44.1670M, 1994PhRvD..49.1783L, 1994PhRvD..50.6357S} for the very first works).

The stochastic formalism is an effective theory for coarse-grained fields. Starting from the full action and integrating out sub-Hubble perturbations, one obtains an effective action of stochastic processes for the super-Hubble coarse-grained fields (see \eg~References~\cite{1990PhRvD..42.1027M, 1995PThPh..93..685M, 2013PhRvD..88h3537L, 2016arXiv160401015M, 2018JCAP...02..014T, 2018JCAP...11..022T} for such an effective-action formulation). One can also heuristically find the stochastic differential equation from the original equation of motion. For simplicity, let us consider a single-field case (see Reference~\cite{2021JCAP...04..048P} for an extension to multiple fields):\footnote{\setstretch{0.9}We have neglected metric perturbations. The rigorous formulation including those can be found in Reference~\cite{2021JCAP...04..048P}.}
\begin{subequations}
\begin{align}
    \dv{\phi( N,\mspace{1.5mu}\xbm )}{N}
        &= 
            \frac{\pi( N,\mspace{1.5mu}\xbm )}{H}
            \, ,
            \label{eq: single scalar EoM - phi}
            \displaybreak[1]
            \\[4mm]            
    \dv{\pi( N,\mspace{1.5mu}\xbm )}{N}
        &= 
            -\.3\mspace{1.5mu}\pi( N,\mspace{1.5mu}\xbm )
            -
            \mspace{-1.5mu}
            \frac{V^{\prime}\big[\phi( N,\mspace{1.5mu}\xbm )\big]}{H}
            +
            \mspace{-1.5mu}
            \frac{\Delta^{\!2}}{a^{2}H}\,\phi( N,\mspace{1.5mu}\xbm )
            \, .
            \label{eq: single scalar EoM - pi}
\end{align}
\end{subequations}
Hereafter, we will assume that the Hubble parameter $H = \big[ ( \pi^{2} / 2 + V ) / ( 3\mspace{1.5mu}\Mpl^{2} ) \big]^{1/2}$ is almost constant as a slow-roll approximation. Let us then split the inflaton and its canonical momentum into the coarse-grained parts [called {infrared} (IR) {parts}] and the remaining fluctuations [called {ultraviolet} (UV) {parts}] as
\begin{subequations}
\begin{align}
    \mspace{-40mu}X_{\IR}( N,\mspace{1.5mu}\xbm )
        &\coloneqq
            \int\frac{\dd[3]{k}}
            {( 2\mspace{1.5mu}\pi )^{3}}\;
            X_{\kbm}( N )\.
            \Theta\pqty{\sigma a H - k}\.
            \erm^{-i\mspace{1.5mu}\kbm\cdot\xbm}
            \, ,
            \displaybreak[1]
            \\[4mm]
    X_{\UV}( N,\mspace{1.5mu}\xbm )
        &\coloneqq
            X( N,\mspace{1.5mu}\xbm )
            -
            X_{\IR}( N,\mspace{1.5mu}\xbm )
            \, ,
\end{align}
\end{subequations}
where $X$ represents $\phi$ or $\pi$, $X_{\kbm}$ is its Fourier mode; the step function $\Theta( \sigma a H - k )$ picks up only the long-wavelength modes $k^{-1} > ( \sigma a H )^{-1}$. The parameter $\sigma$ is a tiny constant, $\sigma \ll 1$, ensuring that the infrared modes are well super-Hubble. Substituting this decomposition into the original equations of motion~(\ref{eq: single scalar EoM - phi}--\ref{eq: single scalar EoM - pi}), omitting the spatial derivative of the infrared modes, and keeping only linear terms for the ultraviolet modes, one obtains equations of motion for the infrared and ultraviolet modes as
\begin{subequations}
\begin{align}
    \dv{\phi_{\IR}}{N}
        &= 
            \frac{\pi_{\IR}}{H}
            +
            \xi_{\phi}
            \, ,
            \displaybreak[1]
            \\[3mm]
    \dv{\pi_{\IR}}{N}
        &= 
            -\.3\mspace{1.5mu}\pi
            -
            \frac{V^{\prime}( \phi_{\IR} )}{H}
            +
            \xi_{\pi}
            \, ,
\end{align}
\end{subequations}
and 
\begin{align}
    \ddot{\phi}_{\kbm}
    +
    3\mspace{1.5mu}H\dot{\phi}_{\kbm}
    +
    \mspace{-1.5mu}\left[
        \frac{k^{2}}{a^{2}}
        +
        V^{\prime\prime}( \phi_{\IR} )
    \right]\mspace{-1.5mu}
    \phi_{\kbm}
        = 
            0
    \qquad
    \text{for $k > \sigma a H$}
    \, .
\end{align}
The last one is the standard linear equation of motion. On the other hand, as an interesting point of the coarse-grained theory, new terms $\xi_{\phi}$ and $\xi_{\pi}$ appear in the infrared equation of motion, these being defined by
\begin{align}
    \xi_{X}( N,\mspace{1.5mu}\xbm )
        \coloneqq
            \int\frac{\dd[3]{k}}
            {( 2\mspace{1.5mu}\pi )^{3}}\;
            X_{\kbm}( N )\.
            \delta
            \big[
                \mspace{-1mu}\ln( k / \sigma a H)
            \big]\.
            \erm^{-i\mspace{1.5mu}\kbm\cdot\xbm}
            \, .
\end{align}
This is because the coarse-graining scale $\sigma a H \propto \erm^{N}$ is time-dependent and hence one has to include the time-derivative of the window function $\Theta(\sigma a H - k)$. Physically speaking, these terms represent the transit modes from the ultraviolet parts to the infrared ones. As $X_{\kbm}$ originates from the quantum zero-point fluctuation, $\xi_{X}$'s specific value cannot be determined {\it a priori}. However, its statistics can be inferred from the quantum expectation values:
\begin{subequations}
\begin{align}
    \big\langle
        \xi_{X}( N,\mspace{1.5mu}\xbm )
    \big\rangle
        &= 
            0
            \, ,
            \displaybreak[1]
            \\[2mm]
    \big\langle
        \xi_{X}( N,\mspace{1.5mu}\xbm )\.
        \xi_{Y}( N^{\prime},\mspace{1.5mu}\ybm )
    \big\rangle
        &= 
            \Pcal_{XY}( k = \sigma a H )\.
            \delta( N - N^{\prime} )\,
            \frac{\sin\!
            \big(
                \sigma a H\mspace{1.5mu}\abs{\xbm - \ybm}
            \big)}
            {\sigma a H\mspace{1.5mu}\abs{\xbm - \ybm}}
            \, .
\end{align}
\end{subequations}
The infrared modes are well super-Hubble and are viewed as classical fields. Therefore, $\xi_{X}$ is also understood as a classical Gau{\ss}ian random variable satisfying the above correlations. The spatial correlation $\sinc( \sigma a H\mspace{1.5mu}\abs{\xbm - \ybm} )$ gives almost full correlation ($\sim 1$) within the coarse-graining patch [\ie~for $\abs{\xbm - \ybm} < ( \sigma a H )^{-1}$] and nearly no correlation ($\sim 0$) beyond the coarse-graining scale [$\abs{\xbm - \ybm} > ( \sigma a H )^{-1}$]. In the slow-roll limit, the power spectra read
\begin{subequations}
\begin{align}
    \Pcal_{\phi\phi}
        &\simeq
            \pqty{
                H / 2\mspace{1.5mu}\pi
            }^{2}
            \, ,
            \displaybreak[1]
            \\[2.5mm]
    \Pcal_{\phi\pi}
        &\simeq
            \Pcal_{\pi\phi}
        \simeq
            \Pcal_{\pi\pi}
        \simeq
            0
            \, ,
\end{align}
\end{subequations}
being valid for for $k = \sigma a H$, so that $\xi_{\pi}$ can be neglected. Therefore, making use of the slow-roll approximation, $\dd{\pi}/\dd{N} \simeq 0$, each (super-)Hubble patch is understood to evolve \emph{independently} through the stochastic differential equation
\begin{align}
\label{eq: Langevin}
    \dv{\phi_{\IR}}{N}
        = 
            -
            \frac{V^{\prime}( \phi_{\IR} )}
            {3\mspace{1mu}H^{2}}
            +
            \frac{H}{2\mspace{1.5mu}\pi}\.\xi
            \, ,
\end{align}
where $\xi$ is a normalised Gau{\ss}ian \emph{white} stochastic noise:
\begin{subequations}
\begin{align}
    \big\langle
        \xi( N )
    \big\rangle
        &= 
            0
            \, ,
            \displaybreak[1]
            \\[2.5mm]
    \big\langle
        \xi( N )\.\xi(N^{\prime})
    \big\rangle
        &= 
            \delta( N - N^{\prime} )
            \, .
\end{align}
\end{subequations}
That is, $\phi_{\IR}$ receives random noise every moment with typical amplitude $H / ( 2\mspace{1.5mu}\pi )$. Here, the Hubble parameter can be replaced by $H \simeq \big[ V( \phi_{\IR} ) / ( 3\mspace{1.5mu}\Mpl^{2} ) \big]^{1/2}$.

In the stochastic formalism, each coarse-grained patch behaves as an independent stochastic process. Therefore, the e-fold number from some initial value $\phi$ to the end of inflation, denoted by $\Ncal( \phi )$, is not deterministic but understood as a stochastic variable.$\vphantom{_{_{_{_{_{_{_{_{_{_{_{_{_{_{_{_{_{_{_{_{_{_{_{_{_{_{_{_{_{_{_{_{_{_{1}}}}}}}}}}}}}}}}}}}}}}}}}}}}}}}}}}}$ The $\delta N$ approach readily claims that the corresponding curvature perturbation can be defined by its deviation from the average:
\vs{-1mm}
\begin{align}
    \zeta
        \coloneqq
            \Ncal( \phi_{\irm} )
            -
            \big\langle
                \Ncal( \phi_{\irm} )
            \big\rangle
            \, ,
\end{align}
where the initial value $\phi_{\irm}$ represents the possible initial condition of our observable Universe. The computational approach to the curvature perturbation in this way is called \emph{stochastic-$\delta N$} formalism~\cite{2008JCAP...04..025E, 2013JCAP...12..036F, 2014JCAP...10..030F, 2015EPJC...75..413V, 2016JCAP...06..043A, 2017PhRvL.118c1301V, 2017JCAP...10..046P, 2021JCAP...04..057A}. Of central interest here is not the probability distribution of the inflaton fields themselves, but rather that one of the e-fold number $\Ncal( \phi )$. Stochastic calculus interestingly shows that the stochastic differential equation~\eqref{eq: Langevin} is equivalent to the Fokker--Planck equation for the probability density of the inflaton $\phi$ at the time $N$,
\begin{align}
\begin{split}
    \partial_{N} p\mspace{1mu}( \phi\,|\, N )
        &= 
            \Lcal_{\FP} \cdot p\mspace{1mu}( \phi\,|\, N )
            \\[3mm]
        &= 
            \Mpl^{2}
            \left\{
                \partial_{\phi}
                \bigg[
                    \frac{v^{\prime}( \phi )}
                    {v( \phi )}\.p\mspace{1mu}( \phi\,|\, N )
                \bigg]
                +
                \partial_{\phi}^{2}
                \big[
                    v( \phi )\.p\mspace{1mu}( \phi\,|\, N )
                \big]
                \right\}
                ,
\end{split}
\end{align}
and also to the adjoint Fokker--Planck equation for the probability density of the \emph{first-passage time} (FPT) $\Ncal$ from the initial value $\phi$~\cite{2015EPJC...75..413V},\footnote{\setstretch{0.9}These Fokker--Planck and adjoint Fokker--Planck equations become partial-derivative equations in the multi-field case (see again Reference~\cite{2021JCAP...04..048P} for such a generalisation).}
\begin{align}
\label{eq: adjoint FP}
\begin{split}
    \partial_{\Ncal} P_{\FPT}( \Ncal\,|\,\phi )
        &= 
            \Lcal_{\FP}^{\dagger}
            \cdot P_{\FPT}( \Ncal\,|\,\phi )
            \\[3mm]
        &= 
            \Mpl^{2}
            \bigg[\!
                -
                \frac{v^{\prime}}{v}\.\partial_{\phi}
                +
                v\.\partial_{\phi}^{2}\.
            \bigg]\.
            P_{\FPT}( \Ncal\,|\,\phi )
            \, .
\end{split}
\end{align}
Here, $v( \phi )$ is a renormalised potential
\begin{align}
    v( \phi )
        = 
            \frac{V( \phi )}{24\mspace{1.5mu}\pi^{2}
            \mspace{1.5mu}\Mpl^{4}}
            \, ,
\end{align}
and we omitted the subscript `IR' for brevity. One actually sees the exponential tail in this first-passage-time probability. Let us first look at the simplest example, shown in the lower panel of Figure~\ref{fig: USR}, where we consider the same flat potential as shown in the upper panel of this figure but neglect the inflaton velocity for simplicity. The inflaton still can exit the flat region (called \emph{quantum well}) by stochastic fluctuations. The other parts of the potential are steep enough, so two edges of the quantum well are understood as an absorbing and a reflective boundary, respectively. The boundary conditions on the first-passage-time probability are hence given by
\vs{-2mm}
\begin{subequations}
\begin{align}
    P_{\FPT}( \Ncal\,|\,\phi = 0 )
        &= 
            \delta( \Ncal )
            \, ,
            \displaybreak[1]
            \\[2.5mm]
    \eval{\partial_{\phi} P_{\FPT}( \Ncal\,|\,\phi )}_{\phi_{\wrm}}
        &= 
            0
            \, .
\end{align}
\end{subequations}
The adjoint Fokker--Planck equation~\eqref{eq: adjoint FP} reads in this case
\begin{subequations}
\begin{align}
    \partial_{\Ncal} P_{\FPT}( \Ncal\,|\,\phi )
        &= 
            v_{0}\.\Mpl^{2}\.
            \partial_{\phi}^{2}P_{\FPT}( \Ncal\,|\,\phi )
            \, ,
\intertext{with\vs{-1mm}}
    v_{0}
        &= 
            \frac{V_{0}}{24\mspace{1.5mu}\pi^{2}\Mpl^{4}}
            \, .
\end{align}
\end{subequations}
This equation with the above boundary conditions can be easily solved in generalised ``Fourier" space, where the {\it characteristic function} $\chi_{\Ncal}( t\,|\,\phi )$ is defined by~\cite{2017JCAP...10..046P}
\begin{align}
    \chi_{\Ncal}( t\,|\,\phi )
        \coloneqq
            \Big\langle
                \erm^{i\mspace{1.5mu}t\mspace{1.5mu}\Ncal( \phi )}
            \Big\rangle
        = 
            \int_{-\infty}^{\infty}
            \dd{\Ncal}\;
            \erm^{i\mspace{1.5mu}t\mspace{1.5mu}\Ncal}
            P_{\FPT}( \Ncal\,|\,\phi )
            \, ,
\end{align}
with $t$ being a dummy parameter. The probability density is its own inverse transformation:
\begin{align}
\label{eq: inverse Fourier of PFPT}
    P_{\FPT}( \Ncal\,|\,\phi )
        = 
            \frac{1}{2\mspace{1.5mu}\pi}
            \int_{-\infty}^{\infty}\dd{t}\;
            \erm^{-i\mspace{1.5mu}t\mspace{1.5mu}\Ncal}
            \chi_{\Ncal}( t\,|\,\phi )
            \, .
\end{align}
In terms of this characteristic function, the adjoint Fokker--Planck equation and the boundary conditions read
\begin{subequations}
\begin{align}
    \mspace{-60mu}\partial_{\phi}^{2}\chi_{\Ncal}( t\,|\,\phi )
    +
    \frac{i\mspace{1mu}t}{v_{0}\.\Mpl^{2}}\.
    \chi_{\Ncal}( t\,|\,\phi )
        &= 
            0
            \, ,
            \displaybreak[1]
            \\[2mm]
    \chi_{\Ncal}( t\,|\,0 )
        &= 
            1\, ,
            \displaybreak[1]
            \\[2mm]
    \eval{\partial_{\phi}\chi_{\Ncal}( t\,|\,\phi )}_{\phi_{\wrm}}
        &= 
            0
            \, .
\end{align}
\end{subequations}
They can be solved as
\begin{align}
    \chi_{\Ncal}( t\,|\,\phi )
        = 
            \frac{\cos
            \bqty{
                \sqrt{i\mspace{1mu}t}\.\mu\.( x - 1 )\,
                }
            }
            {\cos\bqty{\sqrt{i\mspace{1mu}t}\.\mu}}
            \, ,
\end{align}
where $x = \phi/\phi_{\wrm}$ and $\mu = \phi_{\wrm} / \big( \sqrt{v_{0}}\.\Mpl \big)$. The important point is that the above solution has a pole structure:
\begin{align}
    \chi_{\Ncal}( t\,|\,\phi )
        = 
            \sum_{n}\frac{a_{n}( \phi )}
            {\Lambda_{n} - i\mspace{1mu}t}
            +
            (\text{terms regular in $t$})
            \, ,
            \\[-8mm]
            \notag
\end{align}
with
\begin{subequations}
\begin{align}
    \Lambda_{n}
        &= 
            \frac{\pi^{2}}{\mu^{2}}
            \pqty{n + \frac{1}{2}}^{\mspace{-6mu}2}
            \, ,
            \displaybreak[1]
            \\[3mm]
    a_{n}( \phi )
        &= 
            (-1)^{n}\frac{\pi}{\mu^{2}}\.( 2\.n + 1 )
            \cos\bqty{\frac{\pi}{2}( 2\.n + 1 )( x - 1 )}
            \, .
\end{align}
\end{subequations}
In such a case, the inverse transformation~\eqref{eq: inverse Fourier of PFPT} can be computed by considering the $t$-integration of $\erm^{-i\mspace{1.5mu}t\mspace{1.5mu}\Ncal}\chi_{\Ncal}( t\,|\,\phi )$ along the negative imaginary hemisphere according to the residual theorem as
\begin{tcolorbox}
\vs{-2mm}
\begin{align}
    P_{\FPT}( \Ncal\,|\,\phi )
        = 
            \sum_{n}a_{n}( \phi )\,
            \erm^{-\Lambda_{n}\mspace{1.5mu}\Ncal}
            \, .
\end{align}
\end{tcolorbox}
\noindent Hence, {\it the first-passage-time probability has exponential tails}; the negative imaginary poles $\Lambda_{n}$ of the characteristic function are viewed as the decay constants.

In a more general potential case, the adjoint Fokker--Planck equation reads
\begin{align}
    \bqty{
        \partial_{\phi}^{2}
        -
        \frac{v^{\prime}}{v}\.
        \partial_{\phi}
        +
        \frac{i\mspace{1mu}t}
        {v\mspace{1.5mu}\Mpl^{2}}
    }\.
    \chi_{\Ncal}( t\,|\,\phi )
        = 
            0
            \, ,
\end{align}
which can be recast into
\begin{align}
    \bqty{
        \partial_{z}^{2}
        +
        \frac{1}{\sqrt{i\mspace{1mu}t}}
        \pqty{
            \Mpl\mspace{1.5mu}
            \abs{\frac{v^{\prime}}{v^{3/2}}}
            +
            \Mpl\mspace{1.5mu}
            \abs{
                \frac{v^{\prime}}{
                2\mspace{1.5mu}v^{1/2}}
            }
        }
        \partial_{z}+1}\.
        \chi_{\Ncal}
            = 
                0
                \, ,
\end{align}
with the field redefinition
\begin{align}
    \dd{z}
        \equiv
            -\,
            \sign( v^{\prime} )\.
            \sqrt{
                i\mspace{1mu}t / v\mspace{1.5mu}\Mpl^{2}
            }\.
            \dd{\phi}
            \, ,
\end{align}
where we assumed that $v$ is monotonic in $\phi$ during inflation. In this definition, the inflaton always rolls so that $z$ increases. This is a damped oscillatory system. As it is a linear equation, once one finds a specific solution $\tilde{\chi}_{\Ncal}$, any constant multiplication $C\.\tilde{\chi}_{\Ncal}$ also gives a solution. The boundary condition at the end of inflation, $\chi_{\Ncal}( t\,|\,\phi_{\frm} ) = 1$, is hence easily satisfied: Find first a specific solution $\tilde{\chi}_{\Ncal}$, and then the true solution is its renormalisation as $\chi_{\Ncal}( t\,|\,\phi ) = \tilde{\chi}_{\Ncal}( t\,|\,\phi ) / \tilde{\chi}_{\Ncal}( t\,|\,\phi_{\frm} )$. The origin of the pole can be also understood in this way. That is, if $\tilde{\chi}_{\Ncal}( t\,|\,\phi_{\frm} ) = 0$ for some $t$, the renormalised one $\chi_{\Ncal}( t\,|\,\phi_{\frm} )$ is diverging. In the case in which the friction coefficient is larger than $2$, the system is overdamped, and $\tilde{\chi}_{\Ncal}$ does neither change sign nor pass through the zero point. For having a pole, the friction coefficient should be smaller than $2$ at least in some field-space regions. The size of the second term of the friction coefficient is of order $v\.\sqrt{\epsilon_{V}}$, which should be small enough for ordinary slow-roll inflation, but the magnitude of the first term is order $\Pcal_{\zeta}^{-1/2}$, where the linear-theory formula~\eqref{eq: Pcal zeta formula} has been used. Therefore, for a non-negligible decay scale $\Lambda_{n} = i\mspace{1mu}t$, the potential must be so flat that the linear-theory formula is violated as $\Pcal_{\zeta} \sim 1$. This condition can be changed in multi-field cases and/or beyond the slow-roll approximation.

Note that the coarse-graining scale of the curvature perturbation obtained in the direct stochastic-$\delta N$ approach is of order $H_{\rm inf}^{-1}$, \ie~the Hubble scale at the end of inflation. Therefore, even if one finds large perturbations, it does not necessarily correspond to massive-enough PBHs. In order to obtain information about the PBH mass, one should take arbitrarily large coarse-graining scales. In the stochastic picture, two spatial points start their independent evolutions at the time when their distance becomes equivalent to the Hubble scale. Therefore, if one wants to consider a coarse-graining scale $R$, one should average over the stochastic processes deviating from each other at $N_{\bw}( \mspace{-1mu}R ) \coloneqq \ln( \sigma a H R )$ e-folds before the end of inflation. Accordingly, the probability density of $\zeta_{R}$ coarse-grained on $R$ is formulated as~\cite{2022JCAP...02..021T}
\begin{align}
\begin{split}
    p\mspace{1mu}(\zeta_{R})
        &= 
            \int_{\Omega}\dd{\bm{\Phi}_{*}}\;
            P_{\bw}\big[\bm{\Phi}_{*}\,\big|\, N_{\bw}( \mspace{-1mu}R )\big]
            \\
        &\phantom{ = \;\int_{\Omega}\dd{\bm{\Phi}_{*}}\;}
        \times
        P
        \Big[
            \Ncal( \bm{\Phi}_{0} \to \bm{\Phi}_{*} )
            = 
            \zeta_{R}
            -
            \big\langle
                \Ncal( \bm{\Phi}_{*} )
            \big\rangle
            +
            \big\langle
                \Ncal( \bm{\Phi}_{0} )
            \big\rangle \,\Big|\, \bm{\Phi}_{*}
        \Big]
        \, ,
\end{split}
\end{align}
which describes a general multi-field and beyond-slow-roll case. The symbol $\bm{\Phi}$ formally indicates a multi-dimensional phase-space point, and $\Omega$ is the inflationary region in phase space. Above, $P_{\bw}\big[ \bm{\Phi}_{*}\,\big|\, N_{\bw}( \mspace{-1mu}R ) \big]$ is the backward probability that the inflaton was at the point $\bm{\Phi}_{*}$ at the time $N_{\bw}( \mspace{-1mu}R )$ e-folds before the end of inflation, and $P$ is the probability that the e-folds from the initial value $\bm{\Phi}_{0}$ of our observable Universe to $\bm{\Phi}_{*}$ are equivalent to $\zeta_{R} - \big\langle \Ncal( \bm{\Phi}_{*} ) \big\rangle + \big\langle \Ncal( \bm{\Phi}_{0} ) \big\rangle$ under the condition that the inflaton passes through the point $\bm{\Phi}_{*}$ at least once.
\newpage

\begin{figure}
    \centering
    \includegraphics[width = 0.62\hsize]{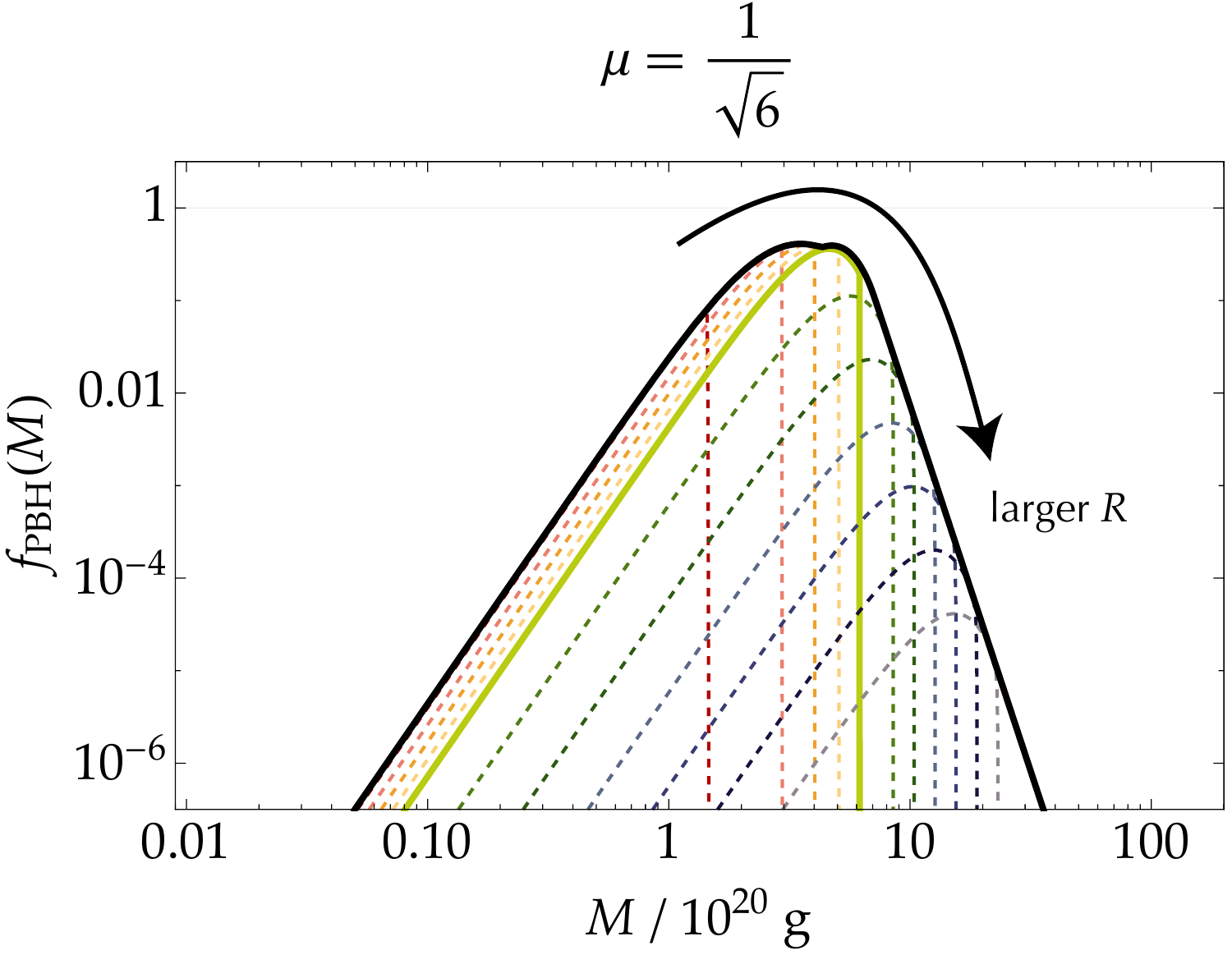}
    \caption{
        Example primordial black hole mass function taking account of stochastic effect in the quantum-well toy model (lower panel of Figure~\ref{fig: USR}). Each coloured (dashed or thick) line corresponds to a different coarse-graining scale $R$ in the Press--Schechter approach and the black thick line represents the total mass function as their envelope curve. Figure from Reference~\cite{2022JCAP...02..021T}.
        }
    \label{fig: stochastic fPBH}
\end{figure}

As this is no longer a simple mapping of the Gau{\ss}ian field, the rigorous peak theory cannot be applied. We therefore follow the Press--Schechter approach in terms of the linear density contrast (see Section~\ref{sec:Press---Schechter-Formalism}). The linear density contrast $\delta^{\lin}$ is related to the curvature perturbation via
\begin{align}
    \delta^{\lin}
        \propto
            \Delta^{\!2}\zeta
            \, .
\end{align}
Though the spatial derivative is not directly calculated in the stochastic approach, it can be approximated by the difference in the curvature perturbations when one changes the coarse-graining scale slightly:
\vs{-2mm}
\begin{align}
    \eval{\Delta^{\!2}\zeta}_{R}
        \approx
            \zeta_{R}
            -
            \zeta_{R\mspace{1.5mu}
            +
            \mspace{1.5mu}\Delta R}
            \, .
\end{align}
In this way, one can calculate the PBH mass function. In Figure~\ref{fig: stochastic fPBH}, we show an example mass function in the quantum-well model presented in the lower panel of Figure~\ref{fig: USR}. Contrary to the single-noise approach (Figure~\ref{fig:fPBHexp}), one can clearly see the heavy-mass tail. This is because the stochastic noise can continue inflation longer by chance and enlarge the scale of the generated perturbation. We finally remark that, for illustrative purposes, we neglected the velocity of the inflaton. However, it must certainly be included in more realistic treatments (see \eg~References~\cite{2021PhRvD.104h3505P, 2021JCAP...04..080P}). See also recent attempts to directly simulate the compaction function~\eqref{eq:comaction-function} in the stochastic formalism~\cite{2023arXiv231212911R, STOLAS}.

\subsubsection{Quantum Loop Corrections onto CMB-scale Perturbations?}

Large stochastic effect, heavy-tail feature, etc.~would be understood as significant loop corrections even dominating tree-level evaluations in a quantum approach (see \eg~References~\cite{2023JCAP...04..011I, 2023arXiv230708358F}). Recently, the discussion whether PBH-scale perturbations could or could not backreact CMB-scale ones at the quantum level has attracted much attention~\cite{2022PhLB..82736956C, 2022arXiv221103395K, 2023arXiv230100599R, 2023arXiv230110000C, 2023PhLB..84538123C, 2023arXiv230300341K, 2023PhRvD.108d3526T, 2023JCAP...11..066C, 2023JCAP...10..006F, 2023arXiv230407801F, 2023arXiv230516810C, 2023arXiv230519263F, 2023arXiv230713636M, 2023arXiv231104080F, 2023arXiv231203498T, 2023arXiv231205694D, 2023arXiv231212424I}.\footnote{See also References~\cite{2023MPLA...3850063O, 2023PhRvD.108d3542O, 2023arXiv231019071O} for similar discussions on tensor perturbations.} If the PBH-scale perturbations lead to non-negligible corrections and spoil the success of large-scale cosmology, the PBH scenario would be ruled out in a general sense. One may recall that the superhorizon curvature perturbations are proven to be conserved non-perturbatively at the classical level irrespective of the small-scale dynamics~\cite{2005JCAP...05..004L}. This fact is understood as a Nambu--Goldstone soft theorem associated with the dilation symmetry~\cite{2012JCAP...07..052C, 2012JCAP...08..017H, 2012JCAP...11..047A, 2014JCAP...01..039H, 2013PhRvD..87j3520G, 2015JHEP...04..061K, 2016JHEP...01..046K}. Can it be violated quantum-mechanically?

The long- and short-wavelength coupling is often characterised by the squeezed bispectrum. According to Maldacena's consistency relation~\cite{2003JHEP...05..013M}, long- and short-wavelength modes seem to have non-vanishing coupling even in the simplest single-field slow-roll inflation scenario as
\begin{align}
	B_\zeta( k_{\Lrm},\.k_{\Srm},\.k_{\Srm} )
        \underset{k_{\Lrm} \ll k_{\Srm}}{=}
            -\,P_\zeta( k_{\Lrm} )\.P_\zeta( k_{\Srm} )\.
            \dv{\ln\Pcal_\zeta( k_{\Srm} )}{\ln k_{\Srm}}
            \, .
\end{align}
However, its physical meaning is just an apparent scale shift by the renormalisation of the local scale factor due to the long-wavelength curvature perturbation $\zeta_{\Lrm}$ as another example of the dilatation soft theorem. In particular, the shift of the loop momentum in the loop integration can be renormalised into the shift of the integration variable, and hence the loop correction is expected to be strongly suppressed~\cite{2012JHEP...07..166P}. Reference~\cite{2023arXiv230804732T} studies a toy model of flat-inflection inflation, and shows that the squeezed bispectrum between the PBH scale and a large-enough (CMB) scale satisfies Maldacena's consistency relation. In turn, the loop backreaction onto such a large-scale mode is negligible. It is, however, still under discussion whether loop corrections remain negligible when including quartic or higher couplings in a more general model.

\subsection{Thermal-History-Induced Mass Function}
\label{sec:Thermal--History--Induced-Mass-Function}
\begin{figure}[t]
	\centering
	\vs{-3mm}
	\includegraphics[width = 0.75\textwidth]{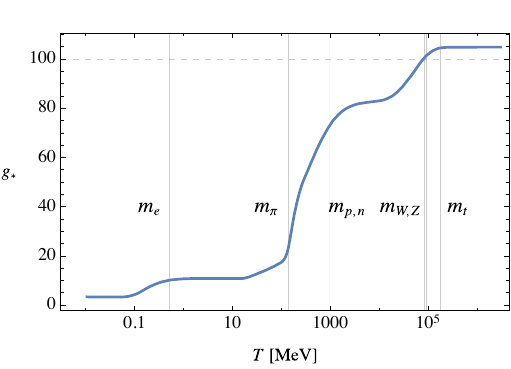}\hs{3mm}
	\includegraphics[width = 0.75\textwidth]{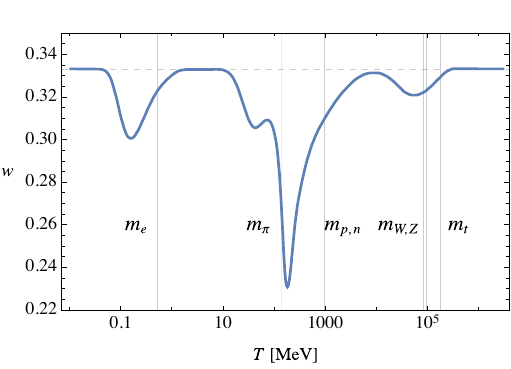}
	\caption{
	    Relativistic degrees of freedom $g_{*}$ ({\it upper panel}) and equation-of-state parameter $w$ ({\it lower panel}), as a function of temperature $T$ (in ${\rm MeV}$). The grey dashed horizontal lines indicate values of $g_{*} = 100$ and $w = 1/3$, respectively. The grey vertical lines indicate the masses of the electron, pion, proton/neutron, $W,\mspace{1.5mu}Z$ bosons and top quark, respectively. Figure from Reference~\cite{Carr:2019kxo}.
        }
	\label{fig:g-and-w-of-T}
\end{figure}

\vs{-1mm}
The thermal evolution of the Universe has been far from uniform. It underwent multiple events which dramatically changed its equation of state. This happened particularly in instances in which the number of relativistic degrees of freedom changed significantly (see Figure~\ref{fig:g-and-w-of-T}). The most pronounced event of these is clearly the QCD phase transition/crossover at around $10^{-5}\.\srm$. Its nature is not yet exactly clear; most authors now believe this to be a crossover~\cite{2006Natur.443..675A} (see also Reference~\cite{2021EPJA...57..136G} for a review), however, a second- or even a first-order transition can currently not be excluded.

\begin{figure}[t]
    \centering
    \includegraphics[width = 0.68\hsize]{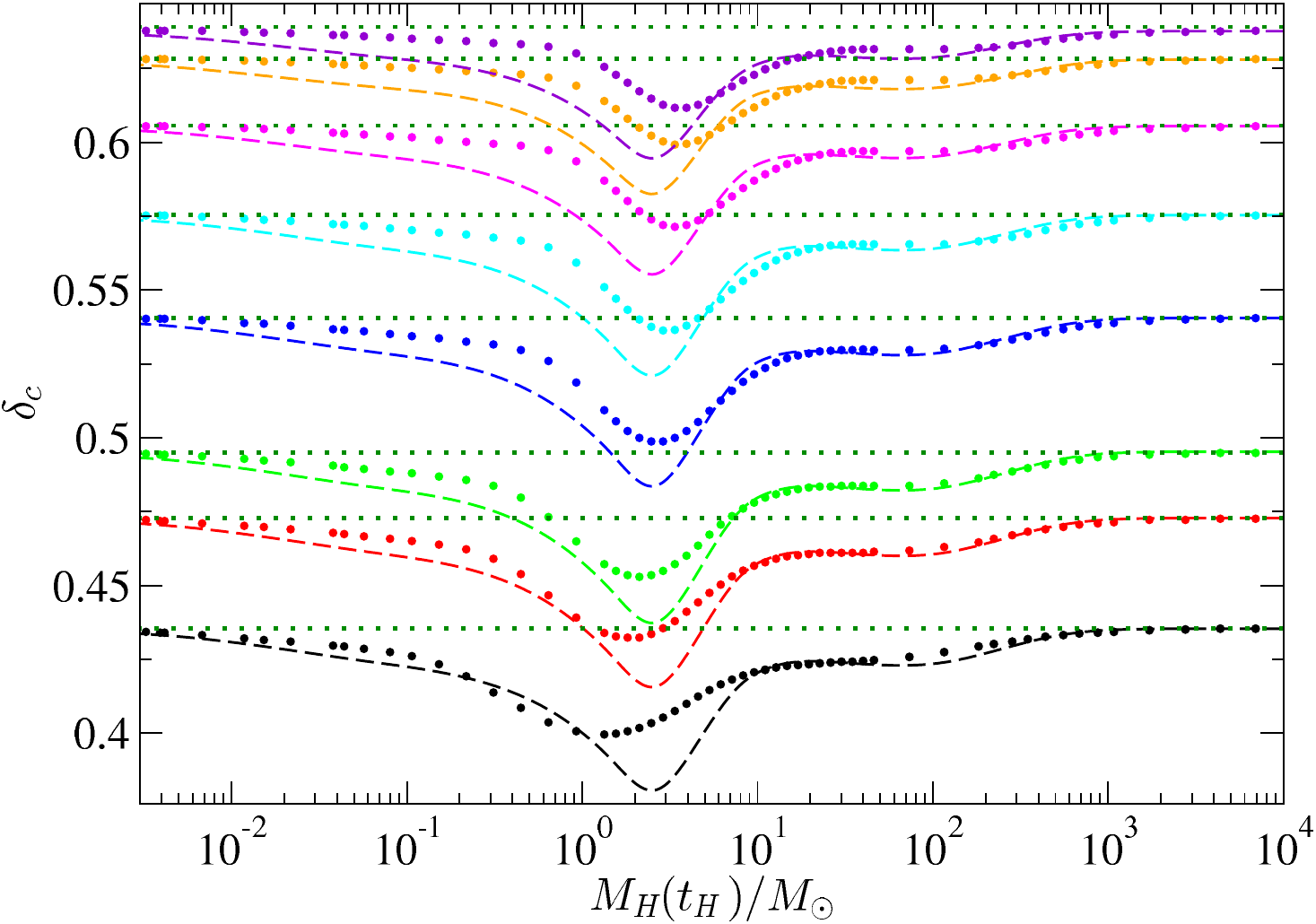}
    \caption{
        Numerical results for the PBH formation threshold $\delta_{\crm,\.{\rm QCD}}$ for the case of a QCD crossover in terms of the horizon mass $M_{H}( t_{H} )/M_{\odot}$ (dots). The dashed lines correspond to thresholds $\delta_{\crm,\.w\.=\.{\text{const}}}\.$, considering a constant value $w$ for a given horizon-mass value. The dark green dotted lines represent the threshold $\delta_{\crm,\.{\rm rad}}$ for a pure radiation era for each profile $q = 0.2,\,0.6,\,1,\,2.5,\,5,\,10,\,20,\,30$ (from black to magenta colour in increasing order). Figure from Reference~\cite{2022arXiv220906196E}.
        \vs{-2mm}
        }
    \label{fig:deltac_qcd}
\end{figure}

The latter is particularly attractive since it leads to the largest known pressure reduction, and hence to the largest{\,---\,}and exponential{\,---\,}enhancement of any PBH mass function which has support during this range. A possible first-order phase transition might be realised in scenarios with a large lepton-flavour asymmetry~\cite{2022PhRvL.128m1301G}. The latter could also provide an even better explanation for the LIGO/Virgo events~\cite{2021PhRvD.103f3506B}. A first-order QCD phase transition would mean that the quark-gluon plasma and hadron phases could coexist, with the cosmic expansion proceeding at constant temperature by converting the quark-gluon plasma to hadrons. The sound speed would then vanish, with the effective pressure being reduced, thus significantly lowering the collapse threshold $\delta_{\crm}$. PBH production during a first-order QCD phase transition was first suggested by Crawford \& Schramm~\cite{1982Natur.298..538C} and later revisited by Jedamzik~\cite{1997PhRvD..55.5871J}. The amplification of density perturbations due to the vanishing of the sound speed during the QCD transition was also considered by Schmid and colleagues~\cite{1999PhRvD..59d3517S, 1998astro.ph..8142W}, while Cardall \& Fuller developed a semi-analytic approach for PBH production during the transition~\cite{1998astro.ph..1103C}. More recently, PBH formation during the cosmic QCD transition has come into recent focus of attention (\cf~References~\cite{2016MNRAS.463.2348S, 2018JCAP...08..041B, 2019PhRvL.123j1102D, 2020JCAP...09..022J, 2021PhRvL.126e1302J, Carr:2019hud, Carr:2019kxo, 2020arXiv200706481C, 2021JCAP...12..023G, 2022PhRvD.106l3526F, 2022arXiv220906196E, 2021PhRvD.103f3506B, 2022PhRvD.105l4055P, 2022PhRvD.106l3519F, 2022JCAP...07..009I}).\footnote{\setstretch{0.9}Recent literature has also considered the consequences of PBH formation from a strongly-coupled crossover at temperatures beyond the electroweak scale. This can lead to a substantially enhanced production of PBHs with masses between $10^{-16}\.\Msun$ and $10^{-6}\.\Msun$~\cite{2022arXiv221115674E}.} In particular, the effects of a QCD crossover on the PBH formation threshold have been computed in Reference~\cite{2022arXiv220906196E}, and are shown in Figure~\ref{fig:deltac_qcd}, this demonstrating the reduction of the threshold values around solar-mass scale for different curvatures profiles following Equation~\eqref{eq:K-polynominal}.

Besides making lesser imprints, also events in the thermal history of the Universe before and after the QCD transition can vitally impact PBH formation. In general, assuming the absence of extensions beyond the Standard Model of particle physics, as the Universe cools down after reheating, the number of relativistic degrees of freedom $g_{*}$ remains constant at a value $g_{*} = 106.75$ until around $200\.{\rm GeV}$, when the temperature of the Universe starts to subsequently fall below the masses of Standard Model particles. As visualised in the upper panel of Figure~\ref{fig:g-and-w-of-T}, the first particle which becomes non-relativistic is the top quark at $T \simeq m_{\trm} = 172\.{\rm GeV}$, which is followed by the Higgs boson at $125\.{\rm GeV}$, and then the $Z$ and $W$ bosons at $92$ and $81\.{\rm GeV}$, respectively. Later, at the QCD transition (around $160\.{\rm MeV}$) the number of relativistic degrees of freedom then falls to $g_{*} = 17.25$, after which first the pions and then the muons become non-relativistic, yielding $g_{*} = 10.75$. Lastly, at $e^{+}e^{-}$ annihilation and neutrino decoupling (around one $\text{MeV}$), it drops to $g_{*} = 3.36$. As mentioned, those changes are reflected in a corresponding variation of both the energy density and the pressure, which impact the equation-of-state parameter $w$ (see lower panel of Figure~\ref{fig:g-and-w-of-T}) as well as the sound speed $c_{\srm}$.

\begin{figure}[t]
	\vs{-3.5mm}
	\centering
	\includegraphics[width = 0.69\textwidth]{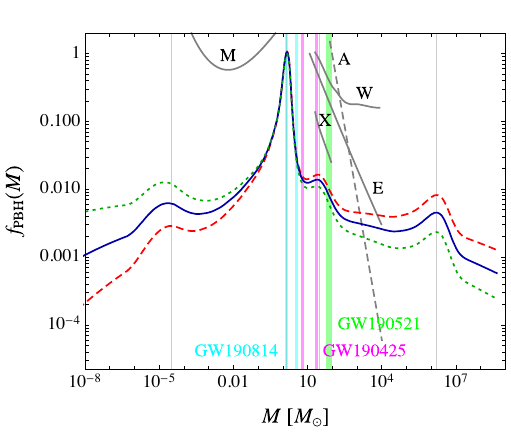}
	\vs{-2mm}
	\caption{
        Thermal-history-induced mass spectrum of PBHs as a function of mass, $f_{\PBH}( M ) \coloneqq {\rho_{\rm CDM}}^{-1}\.\drm\.\rho_{\PBH} ( M ) / \drm \ln M$, for a power-law primordial power spectrum with spectral index $n_{\srm} = 0.955$ (red, dashed), $0.960$ (blue, solid), $0.965$ (green, dotted). Shown are the electroweak and QCD phase transitions (grey vertical lines) and $e^{+}e^{-}$ annihilation. Exemplary, three recent LIGO/Virgo events, which have been {\it predicted} by this mass function, are included. Also shown (grey curves) are constraints from microlensing (M), ultrafaint dwarf galaxies and Eridanus II (E)~\cite{2017ApJ...838....8L}, X-ray/radio counts (X)~\cite{2017PhRvL.118x1101G}, halo wide binaries (W)~\cite{2009MNRAS.396L..11Q}, and accretion (A)~\cite{2017PhRvD..95d3534A}. Note that these constraints are for monochromatic mass functions; they do not apply for the extended ones shown, and have only been included for illustrative purpose. Figure from Reference~\cite{Carr:2019kxo}.
        }
	\label{fig:fPBH}
\end{figure}

The striking observation made by the authors of Reference~\cite{Carr:2019kxo} was that, when using a simple primordial power spectrum of the form $p\mspace{1mu}( k ) = A\.k^{n_{\srm} - 1}$ with {\it the very same spectral index as measured by Planck}, $n_{\srm} = 0.96$~\cite{2020A&A...641A...6P}, the thermal history of the Universe imprints peaks onto the PBH mass function not only at four outstanding masses (\ie~planetary mass, solar mass, of order ten solar masses and about $10^{6}$ solar masses) but also at a relative height such that they can naturally explain numerous cosmic conundra, while providing at the same time the entirety of the dark matter. Figure~\ref{fig:fPBH} depicts this class of models for three exemplary values of the spectral indices. The amplitude $A$ has been chosen to yield $100\mspace{0.5mu}\%$ of dark matter in each of the cases. Using a spectral running, in fact the same as suggested by the Planck collaboration~\cite{2020A&A...641A...6P}, the thermal-history model~\cite{Carr:2019kxo} can be further refined. This has originally been worked out by Garc{\'i}a-Bellido and Hasinger (see Reference~\cite{2020JCAP...07..022H} for the first implementation). It should be stressed that this model requires extremely little input, \ie~merely the value of the power-spectrum amplitude at those small PBH scales, in order to resolve {\it all} of the conundra mentioned in the Introduction, besides providing a, or possibly {\it the} most, natural explanation for the origin of the dark matter.

\newpage

\begin{figure}
	\centering 
    \includegraphics[width = 0.9\textwidth]{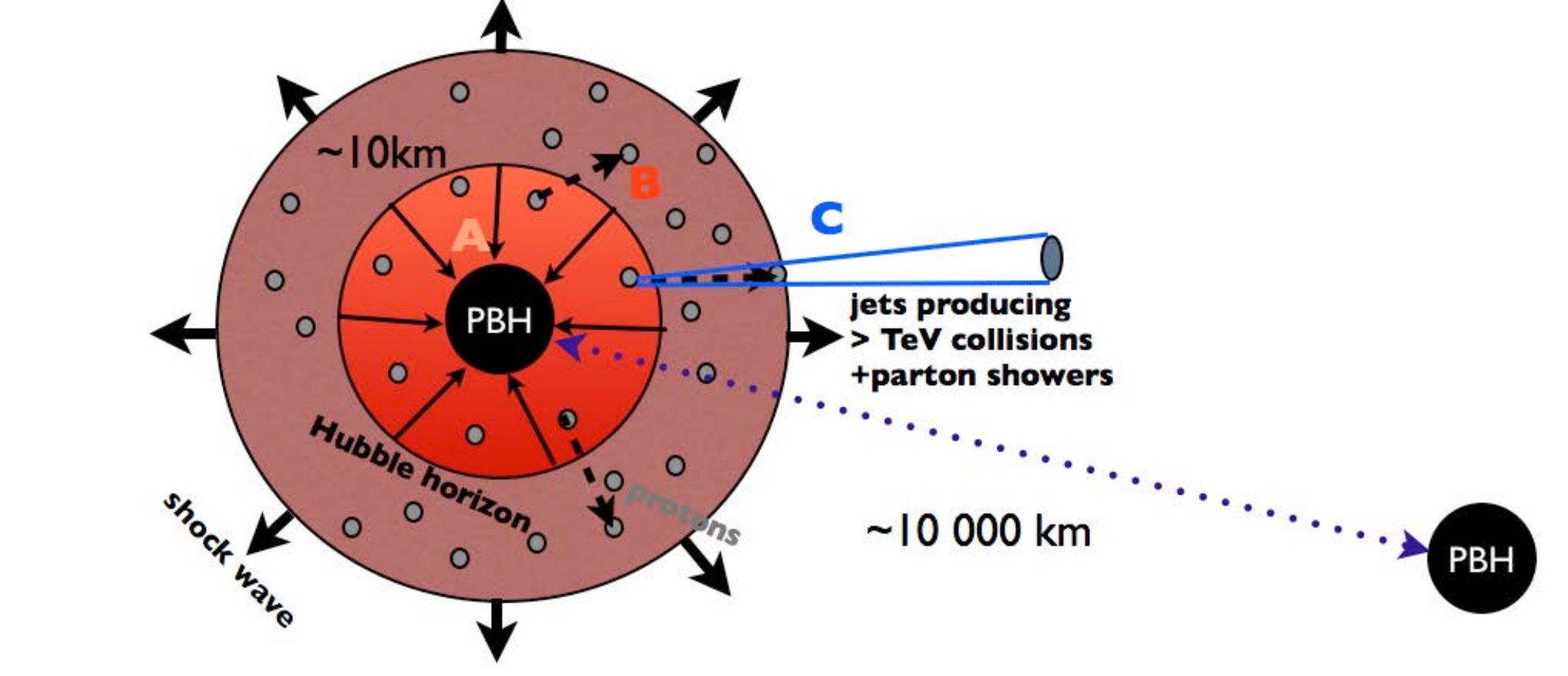}
	\caption{
		Sketch of the steps in the {\it primordial supernova} scenario. 
		(A) Gravitational collapse to a PBH; 
		(B) sphaleron transition 
			around the PBH, these producing $\eta \sim \Ocal( 1 )$ locally through electroweak baryogenesis; 
		(C) propagation of baryons outwards, 
			leading to the observed baryon asymmetry of the Universe $\eta \sim 10^{-9}$.
		Figure (adapted) from Reference~\cite{Garcia-Bellido:2019vlf}.
        \vs{2mm}
        }
	\label{fig:BAU}
\end{figure}

Primordial black hole formation during the QCD epoch can not only be linked to dark matter, but also to baryogenesis as shown in Reference~\cite{Carr:2019hud}. Therein, the authors investigate in detail particle-physics processes connected with the gravitational collapse of a horizon-sized overdensity of radiation into a solar-mass black hole. This is a particularly interesting setting as during this extremely violent process, from the gravitational potential energy released by the collapse, the involved elementary particles acquire energies over three orders of magnitude above their rest mass. This results in shock waves similarly to those ejected by the outer layers of a star when it explodes as a supernova. In turn, this has been termed {\it primordial supernova}. However, in the case of PBHs, the surrounding plasma is much denser, yielding interactions at higher energies, which can exceed that of electroweak sphaleron transitions, and lead in turn to the creation of baryon number~\cite{Shaposhnikov:1998ewc}. Importantly, further sphaleron transitions cannot wash out the local baryon asymmetry $\eta$, since the surrounding plasma is much cooler. Note that since these ``hot spots" are separated by many horizon scales, the propagation of the outgoing baryons dilutes significantly from the initially large local baryon asymmetry. Strikingly, this yields the observed global one~\cite{Carr:2019hud}. Figure~\ref{fig:BAU} shows an illustration of this scenario's processes.

\newpage

\subsection{Quark Confinement}
\label{sec:Quark-Confinement}
\vs{-1mm}
By far most discussed mechanisms for PBH formation utilise the generation of large density perturbations of inflationary origin which collapse to PBHs. This has two significant shortcomings: 
    ({\it i$\mspace{1.5mu}$}) 
        the formation usually happens in the strong-coupling regime, and 
    ({\it ii$\mspace{1.5mu}$}) 
        the PBH abundance is  extremely (often exponentially) sensitive on the choice of the model parameters, implying significant fine-tuning.

\begin{figure}[t]
    \centering
    \includegraphics[width=0.75\textwidth]{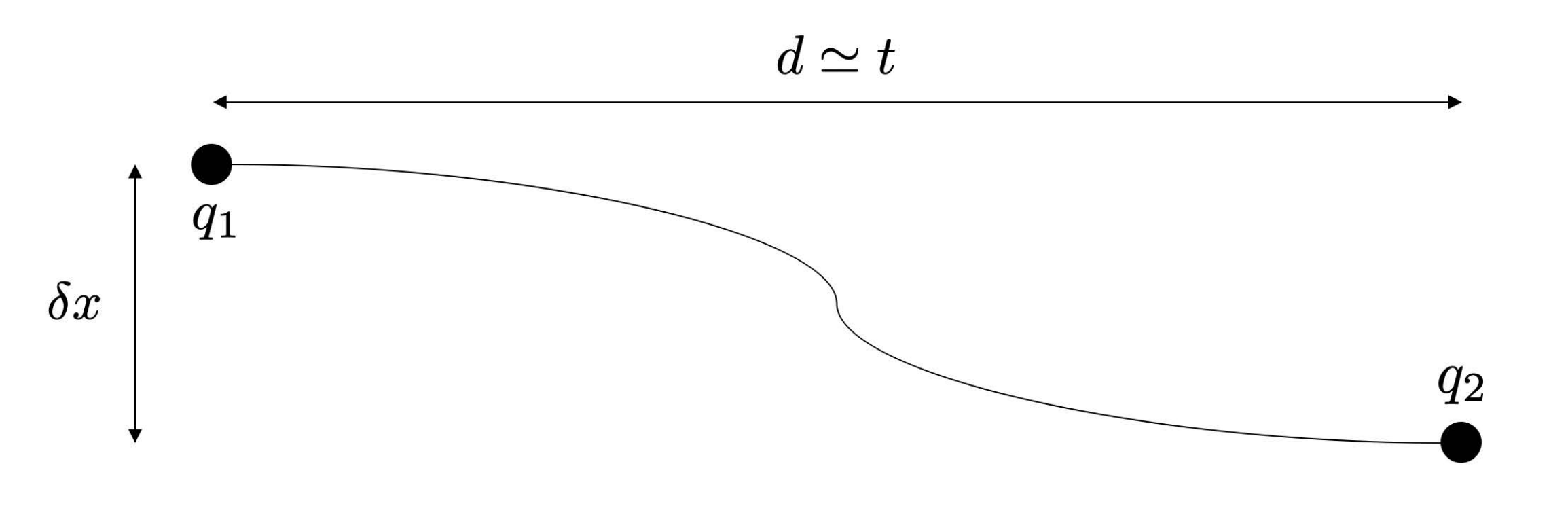}
    \caption{
        Sketch of the initial configuration, consisting of two quarks, $q_{1}$ and $q_{2}$, at collapse; their distance $d$ is assumed to be much larger than their impact parameter $\delta x \ll d$. Figure from Reference~\cite{2021PhRvD.104l3507D}.
        }
    \label{fig:stringbh}
\end{figure}

\begin{figure}[t]
    \centering
    \includegraphics[width = 0.8\textwidth]{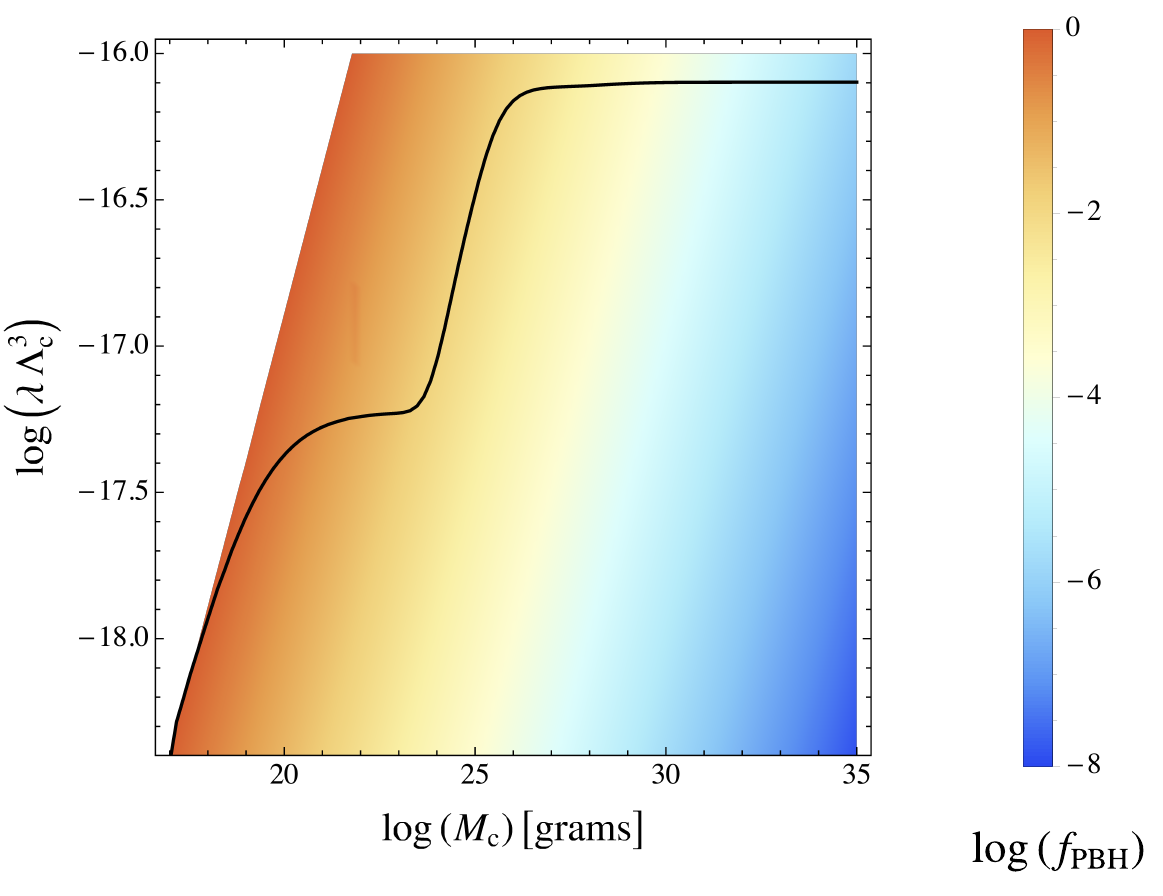}
    \vs{1mm}
    \caption{
        Dependence of $f_{\PBH}$ on the highest value of the mass spectrum, $M_{\crm}$, and the combination $\lambda\mspace{1mu}\Lambda_{\crm}^{3}$, where $\lambda$ is the nucleation rate of the quark/anti-quark pair and $\Lambda_{\crm}$ is the confinement scale. The black line corresponds to the saturation of present constraints (conservatively taken at face value), below which the scenario admits $100\mspace{0.5mu}\%$ PBH dark matter. Figure from Reference~\cite{2021PhRvD.104l3507D}.
        \vs{3mm}
        }
    \label{fig:fpbhext}
\end{figure}

There is, however, a recently developed mechanism~\cite{2021PhRvD.104l3507D}, which is free from those problems and is merely based on quark confinement. Here, heavy quarks of a confining gauge theory produced by de Sitter fluctuations are pushed apart by inflation and then get confined after horizon re-entry. The large amount of energy stored in the colour flux tubes connecting the quark pair leads to the formation of PBHs (see Figure~\ref{fig:stringbh} for a sketch). These are much lighter and can be of significantly higher spin than those produced by standard collapse of horizon-size inflationary overdensities. Furthermore, PBHs formed by the confinement mechanism can account for the entirety of the dark matter in the mass range $10^{17}\,\text{--}\,10^{19}\.\grm$ (see Figure~\ref{fig:fpbhext}). As a by-product, the slowly-decaying mass spectrum, scaling as $M^{-1/2}$, could {\it at the same time} provide seeds for the supermassive black holes observed in the galactic centres.

This mechanism can work with ordinary QCD, being possible by time-variation of physical parameters, such as the QCD scale and quark masses. This is rather generic in inflationary cosmology~\cite{1995hep.ph....5253D}. Correspondingly, the values favourable for the presented mechanism of PBH formation could have easily been attained at early times. It is actually possible to implement this mechanism into a string-theoretic framework of inflation driven by $D$-branes~\cite{1999PhLB..450...72D, 2001hep.th....5203D}. Here, the r{\^o}le of heavy ``quarks'' connected by colour flux tubes is assumed by compact $D$-branes connected by $D$-strings. Interestingly, for rather conservative values of the string-theoretic parameters, $M_{\srm} \sim 10^{16\text{ -- }17}\.{\rm GeV},\;g_{\srm} \sim 10^{-2}$, where $M_{\srm}$ and $g_{\srm}$ are the fundamental string scale and the string coupling constant, respectively, the obtained gravitational-wave signal from PBH formation in this mechanism has an amplitude which is compatible with that recently detected by the {\it North American Nanohertz Observatory for Gravitational Waves} (NANOGrav)~\cite{2020ApJ...905L..34A}. This realisation also includes the possibility to account for potential scalar contributions.

\subsection{Clustering of Primordial Black Holes}
\label{sec:Clustering-of-Primordial-Black-Holes}
\vs{-1mm}
Most constraints on PBHs are based on the assumption that their spatial distribution is homogeneous. However, due to their discrete nature, PBHs would {\it unavoidably} undergo Poisson clustering~\cite{1975A&A....38....5M, 1977A&A....56..377C, 1983ApJ...275..405F, 1983ApJ...268....1C, 2003ApJ...594L..71A, 2006PhRvD..73h3504C, 2015PhRvD..91l3534T, 2016ApJ...823L..25K, 2018MNRAS.478.3756C, 2018PhRvL.121h1304A, 2019PTEP.2019j3E02S, 2019PhRvD.100l3544M, 2022PhRvL.129s1302D, 2023PhRvL.130q1401D} (see also Reference~\cite{Carr:2023tpt} for a recent discussion on this matter, and references therein). Closely related to this is the so-called ``seed'' effect, in which a single black hole generates cosmic structure~\cite{1966RSPSA.290..177H, 1972ApJ...177L..79R, 1984MNRAS.206..801C, 2018MNRAS.478.3756C}, dominating over Poisson clustering on small scales (\cf~Reference~\cite{2018MNRAS.478.3756C}). Both effects are of particular relevance for an extended PBH mass function, which is generic, and may encompass a mass range spanning several orders of magnitude, such as the natural thermal-history-induced mass function discussed in the previous Subsection.

Kashlinski~\cite{2016ApJ...823L..25K} has pointed out that Poisson fluctuations dominate the matter power spectrum on small scales. For stellar-mass PBHs, it has been shown to lead to high-redshift gravitational collapse of almost all small-scale perturbations into clusters of masses up to $10^{6}\,\text{--}\,10^{7}\,\Msun$. As shown in Reference~\cite{2003ApJ...594L..71A}, the Poisson contribution to the matter power spectrum is scale-invariant and well approximated by
\begin{equation}
	P_{\rm Poisson}
		\simeq
				2 \times 10^{-3}\,
				\frac{ \fPBH }{ g( z )^{2} }
				\mspace{-1.5mu}
				\left(
					\frac{ M }{ 3\mspace{1mu}\Msun }
				\right)
				{\rm Mpc}^{3}
				\, ,
\end{equation}
with $g( z )$ being the growth factor for isocurvature fluctuations, having a linear growth when clusters are formed during the matter-dominated era.

\begin{figure}[t]
	\includegraphics[width=0.85\textwidth]{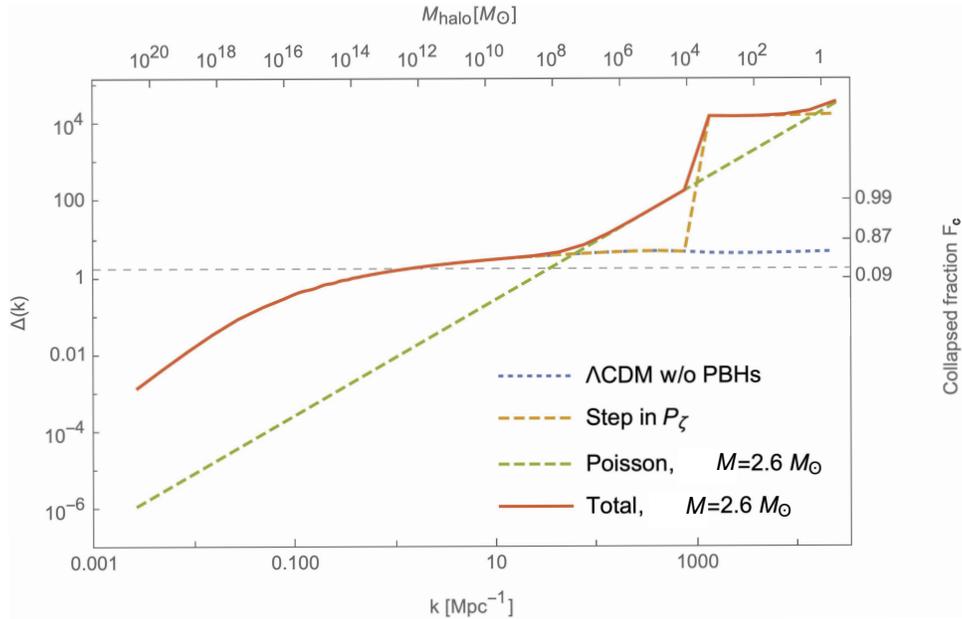}
	\caption{
        Present dimensionless (linear) 
        density contrast $\Delta( k )$ in 
        the standard $\Lambda$CDM model 
        without PBHs (dotted blue line) as 
        well as and with them included with 
        $\fPBH = 1$ at 
        $M = 2.6\,\Msun$ (solid 
        red line). This includes the 
        inevitable Poisson term in the 
        matter power spectrum (dashed green 
        line). Also shown is an exemplary
        power-spectrum enhancement yielding 
        PBH formation for a transition scale 
        at 
        $k_{\rm trans} = 10^{3}\,{\rm Mpc^{-1}}$. 
        The critical threshold for halo formation 
        $\delta_{\rm clust} \simeq 1.686$ is 
        represented by a horizontal dashed 
        line. The upper $x$-axis is the halo 
        mass, and the right $y$-axis 
        indicates the collapsed halo 
        fraction (see 
        Reference~\cite{Carr:2023tpt} 
        for details). Figure adapted 
		from 
        Reference~\cite{Clesse:2020ghq}.} 
	\label{fig:deltas}
\end{figure}

Figure~\ref{fig:deltas} depicts the current (dimensionless) density perturbation $\Delta( k )$, related to the power spectrum through $\Delta( k )^{2} = P( k )\.k^{3} / ( 2 \pi^{2} )$, neglecting non-linear effects. Also shown is the expected spectrum for the standard $\Lambda$CDM scenario and for a primordial spectrum with a sharp enhancement at $k_{\rm trans} = 10^{3}\,{\rm Mpc^{-1}}$ in order to generate PBHs at small scales. This effect clearly modifies the power spectrum below the scale of Poisson dominance, but could also be important to clustering for lower values of $M$ or $\fPBH$.

Clustering of PBHs can strongly affect their merger rate (\cf~References~\cite{2011PhRvD..84l4031C, 2017PDU....15..142C, 2018PhRvD..98l3533D, 2020JCAP...03..004Y, 2020JCAP...11..036A, 2020JCAP...11..028D}). This inevitably impacts the associated gravitational-wave emission in two major ways: 
    ({\it i$\mspace{1.5mu}$}) 
        an enhancement of the merging rates of binaries formed by tidal capture within PBH clusters; 
    ({\it ii$\mspace{1.5mu}$}) 
        a reduction of the merger rate of early binaries due to tidal disruption of binary systems of PBHs if their density/dark matter fraction is sufficiently large.
Overall, the merger rate is reduced, implying a reduction of the stochastic gravitational-wave background~\cite{2021Univ....7...18T}, in turn evading the bounds from LIGO/Virgo~\cite{2021PhRvD.104b2004A}. PBH clustering may also explain the detection of massive black holes in the pair-instability mass gap (\cf~Section~\ref{sec:Non--Stellar-Black-Hole-Merger}), having larger spins through the increased PBH interaction rate in their clusters.\footnote{\setstretch{0.9}Recently, the authors of Reference~\cite{2023Symm...15..637E} have pointed out the importance of inter-cluster tidal interactions, which can lead to the dissipation of orbital energy, resulting in the ejection of constituent PBHs from the clusters. This has been shown to yield consistency of even an order-one dark matter fraction of clustered PBHs (with, exemplary, one central $\sim 30\.\Msun$ PBH surrounded by smaller ones, giving a cluster mass around $100\.\Msun$) with LIGO/Virgo observational data.} This in turn also affects the stochastic gravitational-wave background from close hyperbolic PBH encounters in dense clusters~\cite{2018PDU....21...61G} which might be detectable by third-generation ground-based observatories such as Einstein Telescope and Cosmic Explorer~\cite{2022PDU....3601009G} (see discussion in Section~\ref{sec:Hyperbolic-Encounters}).

\subsection{Other Formation Scenarios}
\label{sec:Other-Formation-Scenarios}
\vs{-1mm}

\paragraph*{Collapse in a Matter-Dominated Era}

It is well possible that the cosmic evolution of the Universe deviates from the standard one, in which the (exponential) inflationary phase is succeeded by periods of first reheating, then matter domination, followed by radiation domination, with a final re-transition into a era of accelerated expansion. If, for instance, the inflaton decays into a heavy metastable particle, this would induce an additional era of matter domination (see \eg~References~\cite{1985MNRAS.215..575K, 2017PhRvD..96f3504H, 2017PhRvD..96f3507C, 2021JCAP...07..001C, 2022PhRvD.106b3519P}). Of course, there are many other ways to generate such a phase of pressurelessness, which provides an environment of significantly enhanced PBH formation. In fact, assuming an infinitely long phase of matter domination, the PBH formation threshold formation is zero if non-spherical effects are neglected. However, as was pointed out in Reference~\cite{2017PhRvD..96h3517H}, such a dust phase could induce angular momentum to the collapsing perturbation, which would effectively increase the threshold value to a nonzero value. Therefore, the abundance of PBHs during a finite matter dominated era differs greatly from the standard case of a radiation-dominated epoch.

It has been shown that then the PBH mass function acquires a power-law rather than an exponential dependence, $\d n / \d M \propto M^{-2}\.\delta_{H}( M )_{}^{5}$~\cite{2016ApJ...833...61H}. For an extended duration of the matter-dominated phase between the times $t_{1}$ to $t_{2}$, the formation of primordial black holes is amplified over the mass range $M_{\umin} \sim M_{H}( t_{1} ) < M < M_{\umax} \sim M_{H}( t_{2} )\,\delta_{H} ( M_{\umax} )_{}^{3/2}$~\cite{2017PhRvD..96f3507C}. In this context, PBH formation has been studied using the Lema{\^i}tre--Tolman--Bondi model~\cite{2016PTEP.2016i3E04H, 2002PhRvD..66j4023H}. Furthermore, Kokubu {\it et al.}~\cite{2018PhRvD..98l3024K} have investigated the effect of inhomogeneity on PBH production during a matter-dominated era.
\vs{1mm}

\paragraph*{Collapse of Cosmic Loops}

Cosmic strings might also be a source of PBH formation if they self-intersect and form closed loops which fall within their Schwarz-schild radii~\cite{1989PhLB..231..237H, 1991PhRvD..43.1106P, 1993PhRvD..48.2502G, 1996PhRvD..53.3002C, 1998PhRvD..57.2158M, 2021arXiv210105040B}. The associated probability depends upon both the string length $\ell$, the string mass per unit length, $\mu$, and the string-correlation scale $s$. Note that the black holes form with equal probability at every epoch, so they should have an extended mass spectrum with $\beta \sim ( G\mspace{1mu}\mu )^{2\mspace{1.5mu}\ell / s - 4}$, where $2 < \ell / s < 4$; avoiding overproduction of PBHs requires $G\mspace{1mu}\mu < 10^{-7}$~\cite{1989PhLB..231..237H}.
\vs{1mm}

\paragraph*{Collapse of Domain Walls}

Second-order phase transition leading to collapsing domain walls could ignite the formation of PBHs~\cite{2000hep.ph....5271R, 2001JETP...92..921R, 2005GrCo...11...99D}. Their masses might span a wide range depending on the specific scenario, which could lead to clustering (even with fractal structure of smaller PBHs around larger ones) and masses significantly below the horizon mass~\cite{2001JETP...92..921R, 1998hep.ph....7343K, 1999PAN....62.1593K, 1999hep.ph...12422K, 2020PDU....2700440G}.
\vs{1mm}

\paragraph*{Collapse through Bubble Collisions}

When bubbles nucleate as a result of first-order phase transitions, they may collide and yield enough energy density to form PBHs~\cite{1982Natur.298..538C, 1982PhRvD..26.2681H, 1982PThPh..68.1979K, 1998hep.ph....7343K, 2000PhRvD..62d3516L, 1994PhRvD..50..676M, 1999hep.ph...12422K, 2017JCAP...04..050D, 2017JCAP...12..044D, 2020JCAP...11..060K, 2021arXiv211201505N}. For this to happen, the bubble-formation rate per Hubble volume must be finely tuned, not to be much larger than the Hubble rate, because then the entire Universe undergoes the phase transition essentially immediately, leaving no time to form black holes. It can also not be much less than the Hubble rate, because then the bubbles are rare and practically never collide.
\newpage

\paragraph*{Collapse of Isocurvature Fluctuations}

While large-scale primordial fluctuations are adiabatic, this need not necessarily be the case on smaller scales. Recently, it has been shown that primordial black holes could have been formed from the collapse of large isocurvature fluctuations of cold dark matter~\cite{2022PhRvD.105j3530P}. This has been confirmed by numerical simulations in Reference~\cite{2022PhRvD.105j3538Y} by considering an isocurvature perturbation of a massless scalar field. Indeed in Reference~\cite{2022JCAP...03..023D}, it has been demonstrated that large isocurvature fluctuations could yield an observable gravitational-wave signal, with a spectrum distinct from the one induced by adiabatic perturbations.
\vs{1mm}

\paragraph*{Particle Trapping by Bubble Walls}

The collision of phase-transition bubbles is not the only mechanism to form PBHs out of them. In some particle models, bubble walls can "sweap" a specific kind of particles and collect them inside the false vacuum regions, which eventually causes PBH formation~\cite{2021arXiv210507481B, 2021arXiv211000005B, 2022PhLB..82436791K, 2022JCAP...10..030K}. This is realised as follows. Let us first assume that some scalar field $\phi$ has zero vacuum expectation value $\braket{\phi}$ in the false vacuum but acquires a non-zero value $\braket{\phi} \neq 0$ in the true vacuum as commonly happens in many phase-transition scenarios. Suppose then that some fermion $\chi$ is coupled to $\phi$ through the Yukawa interaction $y\.\phi\.\bar{\chi}\chi$ and hence acquires an effective mass $m_{\chi} = y\braket{\phi}$ in the true vacuum. If this mass is much larger than the typical kinetic energy of $\chi$ during the phase transition, $\chi$ cannot enter the true vacuum because of kinetic blocking. Hence, $\chi$ is trapped in the false vacuum, which in turn can lead to PBH formation.
\vs{1mm}

\paragraph*{Scalar Fifth Force}

Overdensities may collapse not only by the gravitational force but also by a hypothetical \emph{fifth force}. Reference~\cite{2019PhRvD.100h3518S} shows that if some massive fermion $\psi$ is coupled to a light scalar particle through the Yukawa interaction, this scalar long-range force can lead to the $\psi$-halo formation even during the radiation-dominated era without any enhancement of primordial perturbations. Flores \& Kusenko~\cite{2021PhRvL.126d1101F} further revealed that the same scalar fifth force yields radiative cooling to $\psi$-halos and eventually compresses them into PBHs. Their resulting mass function can account for the entirety of the dark matter as well as for the events detected by LIGO/Virgo. The masses of the resulting black holes are expected to be much less than the horizon mass at their formation time, which can be much later than usually expected (even in the future)~\cite{2022ApJ...932..119C, 2022arXiv221016462L}.
\vs{1mm}

\paragraph*{Scalar-Field Fragmentation}

Scalar fields, particularly those predicted by supersymmetric extensions of the Standard Model, might dynamically develop into a condensate, which can subsequently fragment into certain non-topological solitons, called {\it $Q$-balls}~\cite{Coleman:1985ki} (see Reference~\cite{1992PhR...221..251L} for a review, and also References~\cite{1998PhLB..418...46K, 2016PhRvD..94l3006C}). These can grow until they collapse into black holes~\cite{2017PhRvL.119c1103C, 2017PhRvD..96j3002C, 2019JCAP...10..077C, 2023JCAP...05..013F}. In the case of supersymmetry, where a number of scalar fields develop a towards the end of inflation along the flat directions of their potential, the mass of the produced PBHs can be expressed as $M \sim \Mpl^{3} / \Lambda^{2}$, with $\Lambda$ being the supersymmetry-breaking scale. A value of $\Lambda = 100\.{\rm TeV}$ leads to $M \sim 10^{23}\.\grm$.

As has been shown in References~\cite{2018PhRvD..98h3513C, 2019JCAP...10..077C}, the inflaton can fragment into localised, metastable, pseudosolitonic configurations, called {\it oscillons}~\cite{1976JETPL..24...12B, 1994PhRvD..49.2978G, 1995PhRvD..52.1920C, 2006PhRvD..74l4003F, 2003PhLB..559...99K, 2002PhRvD..65h4037H, 2012PhRvL.108x1302A, 2018PhRvD..98d3531H, 1994PhRvD..49.5040K, 2018JHEP...01..083A}. These can in turn collapse to PBHs whose masses can in principle span a larger range{\,---\,}from approximately $10^{17}\.\grm$ up to $10^{35}\.\grm$, while constituting a significant fraction (possibly all) of the dark matter. Here, sublunar masses can be attained in even simple single-field inflation models; PBH formation at solar-mass scales requires more elaborate scenarios since the inflaton mass needs to be very small.
\vs{0.7mm}

\paragraph*{Metric Preheating}

Oscillations of the inflaton around a local minimum of its potential energy at the end of inflation trigger {\it resonant instabilities} in the equation of motion for scalar perturbations~\cite{2010JCAP...09..034J, 2011JCAP...04..027E}. More specifically, if the inflaton potential is expanded according to $V \simeq m^{2}\.\phi^{2}/2\.$ around its minimum, the instability band consist of comoving scales $k$ satisfying the condition $a H < k < a\.\sqrt{3\mspace{1.5mu}H m}$. For these modes, the density contrast grows linearly with the scale factor. If the oscillations last long enough, \ie~if perturbative reheating occurs with a sufficiently low decay rate, this constitutes an efficient production mechanism of ultralight PBHs~\cite{2020JCAP...01..024M, 2021JCAP...02..038A}. This mechanism has been generalised to multiple-field setups~\cite{2001PhRvD..64b1301G}, and shown to be immune to perturbative inflaton decay~\cite{2020JCAP...05..003M}.
\vs{0.7mm}

\paragraph*{Asynchronous First-Order Phase Transition}

It has been suggested~\cite{2022PhRvD.105b1303L} that first-order phase transitions can form black holes even without bubble collision or particle trapping. As a phase transition occurs {\it stochastically} in each Hubble patch, one may have regions where the transition is significantly delayed by chance. These delayed-decay regions keep the false-vacuum energy whilst the energy density in other regions decreases as the Universe expands, finally leading to black hole formation. Reference~\cite{2022PhRvD.105b1303L} demonstrates that this can generally yield sizeable primordial black hole abundances (see also References~\cite{2022arXiv221014094H, 2023PhRvD.108j3531K}).
\vs{0.7mm}

\paragraph*{Baby Universes}

Primordial black holes can also form from false-vacuum bubbles generated during inflation which continue inflating in an ambient radiation-dominated universe, and eventually pinch off from it. This results in black holes which separate the ambient universe from an inflating ``baby" universe~\cite{2016JCAP...02..064G, 2017JCAP...04..050D, 2019JCAP...09..073A, 2020JCAP...05..022A, 2020JCAP...09..023D}. Note, however, that this baby universe is neither in the trapped region nor in the interior of the black hole. Rather, the trapped region separates two normal regions{\,---\,}one in the baby universe and the other in the parent ambient universe, which were originally causally connected but are not anymore.
\newpage

In particular, in the scenarios considered in References~\cite{2017JCAP...04..050D, 2016JCAP...02..064G, 2020JCAP...09..023D}, tunneling is assumed to be a Poisson process which can happen with nearly constant probability per unit time and volume. Once formed, walls and vacuum bubbles are stretched to large sizes by the inflationary expansion, leading to a nearly scale-invariant distribution of relics. But vacuum bubbles may also arise naturally in models in which PBHs form from large adiabatic fluctuations. This happens in single-field models with a small barrier in the slope of the potential~\cite{2019JCAP...09..073A, 2020JCAP...04..007M, 2021arXiv211110028W, 2022EPJC...82..758R, 2022JCAP...06..007I}. The primary r{\^o}le of the barrier is to slow down the inflaton, thus producing an enhancement of curvature perturbations. Then, large backward fluctuations can prevent localised domains from overshooting the barrier, resulting in vacuum bubbles in which the field remains trapped behind the barrier~\cite{2019JCAP...09..073A, 2020JCAP...05..022A}.

As a consequence, two different channels for PBH production coexist~\cite{2019JCAP...09..073A, 2020JCAP...05..022A}: the conventional one from large adiabatic fluctuations and the bubble channel. The relative importance of both is determined by the sharpness of the barrier, which is in correspondence with the degree of non-Gau{\ss}ianity of curvature perturbations at those scales. This leads to sharply peaked co-moving size distribution~\cite{2023JCAP...10..035E} of bubbles with a near-monochromatic primordial black hole mass function.
\newpage
{\color{white}.}
\newpage

\section{Statistics}
\label{sec:Statistics}
\vs{-3.5mm}
\lettrine[lines=3, slope=0em, findent=0em, nindent=0.2em, lhang=0.1, loversize=0.1]{T}{} he abundance of primordial black holes is usually estimated using a statistical scheme. In this Section, we review the most commonly used ones: the {\it Press--Schechter}~\cite{1974ApJ...187..425P} and the {\it peak-theory}~\cite{1986ApJ...304...15B} approaches. For further reading, also about different approaches, we refer the reader to References~\cite{2019JCAP...07..048D, 2021MNRAS.505.1787E, 2020PDU....3000654W, 2020PhLB..80735550D, 2021PTEP.2021a3E02Y, 2021JCAP...02..002G, 2020JCAP...11..022Y, 2020IJMPD..2930002Y, 2019JCAP...11..012Y, 2013JCAP...08..052Y, 2014JCAP...07..045Y, 2019PhRvL.122n1302G, 2020PTEP.2020b3E03S, 2018PhRvD..97j3528A, 2007JCAP...03..010Z, 1998PhRvD..58j7502Y, 2022JCAP...02..021T, 2019JCAP...10..031K,2023Univ....9..421G} as well as to articles cited therein. For definiteness, and since this is the most relevant case, we focus on the formation primordial black holes during the radiation-dominated era.

In both approaches, one considers the curvature perturbation $\zeta$ and its relation to the power spectrum,
\begin{align}
    \Big\langle
        \tilde{\zeta}^{*}(\kbm)\.
        \tilde{\zeta}(\kbm')
    \Big\rangle
        \equiv
            \frac{2\mspace{1.5mu}\pi^{2}}{k^{3}}\,
            \Pcal_{\zeta}( k )\.
            ( 2\mspace{1.5mu}\pi )^{3}\.
            \delta^{(3)}
            \big(
                \kbm - \kbm'
            \big)
            \, ,
\end{align}
where $\tilde{\zeta}(\kbm)$ is the Fourier transform of $\zeta$, and the bracket $\langle \ldots \rangle$ denotes the ensemble average. Furthermore, we note that the gradient moments $\sigma_{n}$ of $\Pcal_{\zeta}$ can be calculated via
\vs{-1mm}
\begin{align}
\label{eq:sigma-22}
    \sigma_{n}^{2}
        = 
            \int \frac{ \d k}{k}\;
            k^{2\mspace{1.5mu}n}\,
            \Pcal_{\zeta}( k )
            \, .
\end{align}
Note that, on superhorizon scales, the power spectrum of $\zeta$ can be related to that of the density contrast $\delta$ as\footnote{\setstretch{0.9}The factor $16/81$ comes from the term $4\mspace{1.5mu}( 1 + w )^{2} / \mspace{1.5mu}(  3\mspace{1.5mu}w )^{2}$ evaluated during radiation domination, where $w = 1/3$.}
\vs{-1mm}
\begin{align}
    \Pcal_{\delta}
        = 
            \frac{16}{81}\.
            \big(
                k R_{H}
            \big)^{\mspace{-1.5mu}4}\,
            \Pcal_{\zeta}\, .
\end{align}
Therefore, the gradient moments $\sigma_{\delta,\mspace{1.5mu}n}$ of $\Pcal_{\delta}$ become
\begin{align}
    \sigma_{\delta,\mspace{1.5mu}n}^{2}
        = 
            \frac{16}{81}
            \int \frac{ \d k}{k}\;
            \big(
                k\mspace{1.5mu}R_{H}
            \big)^{\!4} \,
            k^{2\mspace{1.5mu}n}\,
            \Pcal_{\zeta}( k )
            \, .
            \\[-4mm]
            \notag
\end{align}

\subsection{Press--Schechter Formalism}
\label{sec:Press---Schechter-Formalism}
\vs{-1mm}
The Press--Schechter formalism~\cite{1974ApJ...187..425P} essentially makes two assumptions: 
    ({\it i$\mspace{1.5mu}$})
        the peak value of the nonlinear volume-averaged density perturbation (\ie~the compaction-function peak) $\delta_{\mrm}$ has a Gau{\ss}ian distribution $p\mspace{1mu}( \delta_{\mrm} )$;
        $\vphantom{_{_{_{_{_{_{_{_{_{1}}}}}}}}}}$
    ({\it ii$\mspace{1.5mu}$}) 
        perturbations with compaction-function peaks $\delta_{\mrm} > \delta_{\crm}$ will collapse into a black hole.
One basically integrates the probability distribution $P$ over the range $\delta_{\crm} \leq \delta < \delta_{\umax}$, where $\delta_{\umax}$ is the maximally-allowed value. In practice, one integrates up to $( \delta_{\umax} \rightarrow )\,\infty$ since the probability distribution is rapidly decreasing above $\delta_{\crm}$, and therefore does not change the result, allowing to simplify the computation. 

Considering a Gau{\ss}ian probability distribution,
\begin{align}
    p\mspace{1mu}( \delta_{\mrm} )
        = 
            \frac{1}{\sqrt{2 \pi}\.
            \sigma_{0}}\.
            \erm^{-\.\delta^{2}_{\mrm} / 
            ( 2\mspace{1.5mu}\sigma_{0}^{2} )}
            \, ,
\end{align}
and taking into account the scaling law for the mass of the primordial black holes (as discussed in Section~\ref{sec:Apparent--Horizon-Formation-and-Primordial-Black-Hole-Mass}), their abundance can be computed as
\begin{align}
    \beta
        &= 
            \frac{\rho_{\PBH}}{\rho_{\rm tot}}
        = 
            2 \int_{\delta_{\crm}}^{\infty}\d \delta_{\mrm}\;
            \frac{M}{M_{H}}\.
            p\mspace{1mu}( \delta_{\mrm} )
            \notag
            \\[4mm]
        &= 
            2 \int_{\delta_{\crm}}^{\infty}
            \d \delta_{\mrm}\;
            \Kcal\.
            ( \delta_{\mrm} -\delta_{\crm} )^{\gamma}\.
            p\mspace{1mu}( \delta_{\mrm} )
            \label{eq:abundance}
            \\[3mm]
        &= 
            \frac{1}{\sqrt{\pi}}\,
            2^{- ( 1 + \gamma ) / 2}\.
            \Kcal\,\delta_{\crm}\.
            \sigma^{-1 + \gamma}_{0}\,
            \Gamma( 1 + \gamma )\,
            U\!
            \left(
                1
                +
                \frac{\gamma}{2},\mspace{1.5mu}
                \frac{3}{2},\mspace{1.5mu}
                \frac{\delta^{2}_{\crm}}
                {2\.\sigma^{2}_{0}}
            \right)
            \erm^{-\.\delta^{2}_{\mrm} / 
            ( 2\mspace{1.5mu}\sigma_{0}^{2} )}
            \, .
            \notag
\end{align}
Here, $U(a,\mspace{1.5mu}b,\mspace{1.5mu}z) \coloneqq \Gamma( a )^{-1} \int_{0}^{\infty}\d t\;\erm^{-z\mspace{1.5mu}t}\.t^{a - 1}\.( 1 + t )^{b - a - 1}$ is the confluent hypergeometric function, and the factor $2$ at the beginning of the integral is introduced to avoid the well-know under-counting in Press--Schechter theory (known as ``Press--Schechter swindle"). We have assumed that the scaling law for the PBH mass is always accurate even when $\delta_{\mrm} \gg \delta_{\crm}$, which is actually not true as can be observed in Figure~\ref{fig:critical-2}. Since the probability distribution $p\mspace{1mu}( \delta_{\mrm} )$ depends exponentially on $\delta_{\mrm}$, it is indeed a good approximation to extend the integral of Equation~\eqref{eq:abundance} up to $( \delta_{\umax} \rightarrow )\,\infty$.

Assuming that all PBHs are formed with the same mass (thereby ignoring the critical regime), \ie~having a monochromatic mass spectrum peaked at an $\Ocal( 1 )$ fraction of the horizon mass $M_{H}$, their abundance is given by 
\begin{align}
    \beta
        = 
            \bar{\Kcal}\.\erf\!
            \left(
                \frac{\delta_{\crm}}
                {\sqrt{2}\sigma_{0}}
            \right)
            ,
            \label{eq:psekeres}
\end{align}
where the {\it error function} `$\erf$' is defined as $\erf( z ) \coloneqq 2 \int_{0}^{z}\d t\,\mspace{1.5mu}\erm^{-t^{2}} / \sqrt{\pi}$. The above equation is obtained by taking the limit $\gamma \rightarrow 0$ and setting $\Kcal = \bar{\Kcal}$ in Equation~\eqref{eq:abundance}.
\newpage

\subsection{Peak-Theory Procedure with Curvature Peaks}
\label{sec:Peak--Theory-Procedure-with-Curvature-Peaks}
\vs{-1mm}
The approach to PBH statistics using peak theory is different from the Press--Schechter formalism presented above. It introduces statistics for counting the number of over-threshold peaks. Several variants based on peak theory, which specifically focus on counting peaks of the overdensity perturbation or also of the compaction function, have been proposed~\cite{2019JCAP...11..012Y, 2019PhRvL.122n1302G, 2020JCAP...11..022Y, 2020PhRvD.101f3520G}. Here, for illustrative purposes, we focus on one specific method, in particular for counting peaks on the comoving curvature fluctuation as developed originally in References~\cite{2018PTEP.2018l3E01Y, 2021PTEP.2021a3E02Y}. This is precisely the method used in other parts of this review in order to account for various aspects of PBH statistics{\,---\,}in particular to explore the effect of non-Gau{\ss}ianities. Utilising the procedure developed in Reference~\cite{2018PTEP.2018l3E01Y}, we first focus on the standard approach using peak theory and consider peaks in the curvature perturbation $\zeta$. Following References~\cite{1986ApJ...304...15B, 2021PTEP.2021a3E02Y, 2018PTEP.2018l3E01Y, 2019JCAP...09..033Y}, the typical (\ie~{\it mean}) profile $\bar{\zeta}$ of a given Gau{\ss}ian random field $\zeta$ with a high peak is given by
\vs{-1mm}
\begin{align}
\begin{split}
\label{eq:psi-tilde}
    \frac{\bar{\zeta}( \tilde{r} )}{\mu_{0}}
        &= 
            \frac{1}{\big( 1 - \gamma^{2}_{1} \big)}
            \left[
                \psi_{0}( \tilde{r} )
                +
                \frac{1}{3}\.R^{2}_{1}\.
                \Delta \psi_{0}( \tilde{r} )
            \right]
            \\[3mm]
        &\phantom{=\;}
            -
            \frac{k^{2}_{1}}
            {\gamma_{1}\mspace{1.5mu}
            \big(
                1
                -
                \gamma^{2}_{1}
            \big)}\.
            \frac{\sigma_{0}}
            {\sigma_{2}}
            \left[
                \gamma^{2}_{1}\.
                \psi_{0}( \tilde{r} )
                +
                \frac{1}{3}\.R^{2}_{1}\.
                \Delta \psi_{0}( \tilde{r} )
            \right]
            ,
\end{split}
\end{align}
with $\mu_{0} \coloneqq \zeta( \tilde{r} = 0 )$ (the height of the peak) and $k^{2}_{1} \coloneqq -\.\Delta \zeta( \tilde{r} = 0 ) / \mu_{0}$ (the width of the peak) as two random variables which characterise the mean profile. Other statistical parameters introduced in Equation~\eqref{eq:psi-tilde} are defined as
\vs{-2mm}
\begin{subequations}
\begin{align}
\label{eq:parameters-statistics}
    \gamma_{n}
        &\coloneqq
            \frac{\sigma_{n}^{2}}
            {\sigma_{n-1}\.\sigma_{n+1}}
            \, ,
            \displaybreak[1]
            \\[2.5mm]
    R_{n}
        &\coloneqq
            \frac{\sqrt{3}\,\sigma_{n}}
            {\sigma_{n+1}}
            \, ,
            \displaybreak[1]
            \\[2.5mm]
    \psi_{n}( \tilde{r} )
        &\coloneqq
            \frac{1}{\sigma_{n}^{2}}
            \int\frac{\d k}{k}\;
            k^{2\mspace{1.5mu}n}\,
            \frac{\sin( k\mspace{1.5mu}\tilde{r} )}
            {k\mspace{1.5mu}\tilde{r}}\.
            P_{\zeta}( k )
            \, ,
\end{align}
\end{subequations}
where the quantities $\sigma_{n}$ are given in Equation~\eqref{eq:sigma-22}. Note that the equations for $\gamma_{n}$ and $R_{n}$ are only valid for odd $n$. It is important to mention that the mean value of $k_{1}$ (considered as a random variable) is given by $k_{1} = k_{\crm} = \sigma_{1} / \sigma_{0}$, which is the value simplifying Equation~\eqref{eq:psi-tilde} in such way that $\bar{\zeta}( \tilde{r} ) = \mu_{0}\.\psi_{0}( \tilde{r} )$. For the illustrative example of the monochromatic power spectrum introduced in Section~\ref{sec:Threshold--Estimation-Scheme}, one has $\sigma_{n} = \sigma_{0}\.k_{*}^{n}$ and therefore $k_{1} = k_{\crm}$.

Following the peak-theory procedure (see Reference~\cite{2018PTEP.2018l3E01Y} for further details), the number of peaks in terms of the parameters $\mu_{0}$ and $k_{1}$ reads
\vs{-1mm}
\begin{align}
\label{eq:peak-number}
\begin{split}
    n^{( \mu_{0},\mspace{1.5mu}k_{1} )}_{\rm peaks}\,
    \dd{\mu_{0}}\mspace{1mu}
    \dd{k_{1}}
        &= 
            \frac{2 \cdot 3^{3/2}}
            {( 2\mspace{1.5mu}\pi )^{3/2}}\,
            \mu_{0}\.k_{1}\.
            \frac{\sigma_{2}^{2}}
            {\sigma_{0}\.\sigma_{1}^{3}}\;
            f\!
            \left(
                \frac{\mu_{0}\.k_{1}^{2}}{\sigma_{2}}
            \right)
            \\[3mm]
        &\phantom{=\;}
            \times
            P_{1}^{(1)}\!
            \left(
                \frac{\mu_{0}}{\sigma_{0}},\mspace{1.5mu}
                \frac{\mu_{0}\.k_{1}^{2}}{\sigma_{2}}
            \right)
            \dd{\mu_{0}}\.\dd{k_{1}}
            \. ,
\end{split}
            \\[-9mm]
            \notag
\intertext{where}
\vs{-1mm}
\label{eq:f-peak-theory}
    f( \xi )
        &= 
            \frac{1}{2}\.\xi
            \big(
                \xi^{2} - 3
            \big)\!
            \left(
                \erf\!\bqty{\frac{1}{2}\.\sqrt{\frac{5}{2}}\,\xi}
                +
                \erf\!\bqty{\sqrt{\frac{5}{2}}\,\xi}
            \right)
            \notag
            \displaybreak[1]
            \\[4mm]
        &\phantom{=\;}
            +
            \sqrt{\frac{2}{5\pi}}\,
            \Bigg\{\mspace{-6mu}
                \left(
                    \frac{8}{5}
                    +
                    \frac{31}{4}\.\xi^{2}\mspace{-3mu}
                \right)
                \exp\!\bqty{-\frac{5}{8}\.\xi^{2}}
            \displaybreak[1]
            \\[1mm]
        &\mspace{100mu}
                +\!
                \left(
                    \frac{1}{2}\.\xi^{2}
                    -
                    \frac{8}{5}
                \right)
                \exp\!\bqty{-\frac{5}{2}\.\xi^{2}}
            \Bigg\}
            \notag
            \, ,
            \\[-9mm]
            \notag
\intertext{and}
    P_{1}^{( n )}( \nu,\mspace{1.5mu}\xi )
        &= 
            \frac{1}{2\mspace{1.5mu}\pi\sqrt{1 - \gamma_{n}^{2}}}\.
            \exp\!
            \Bigg[
                -
                \frac{1}{2}\!
                \left(
                    \nu^{2}
                    +
                    \frac{(\xi - \gamma_{1}\mspace{1.5mu}\nu)^{2}}
                    {1 - \gamma_{n}^{2}}
                \right)\!
            \Bigg]
            .
\end{align}
The procedure of Reference~\cite{2018PTEP.2018l3E01Y} was updated in Reference~\cite{2021PTEP.2021a3E02Y} in order to be able to compute the PBH abundance for an arbitrary power spectrum (including broad shapes). The difference comes from counting peaks of the Laplacian of the curvature perturbation, $\Delta \zeta$, in comparison with peaks of $\zeta$ itself. It allows us to characterise the typical profile of the curvature perturbation $\zeta$ around the peak by using the values of $\Delta \zeta$ and $\Delta^{\!2} \zeta$. As suggested in Reference~\cite{2021PTEP.2021a3E02Y}, this provides the possibility to decouple the effects of the environment from the absolute value of $\zeta$, which can be contaminated by long-wavelength perturbations.

The update of the previous equations, taking into account the counting of peaks of $\Delta \zeta$, is in fact simple, as shown in Reference~\cite{2021PTEP.2021a3E02Y}. Therefore, it is only needed to replace the terms $n \rightarrow n + 2$ in such a way that
\vs{-1mm}
\begin{subequations}
\begin{align}
    \mu_{2}
        &\coloneqq
            -\.\Delta \zeta( \tilde{r} = 0 )
            \, ,
            \displaybreak[1]
            \\[2mm]
    k^{2}_{3}
        &\coloneqq
            \frac{\Delta^{\!2} \zeta( \tilde{r} = 0 )}{\mu_{2}}
            \, .
            \\[-15mm]
            \notag
\end{align}
\end{subequations}
\newpage

\noindent Then, Equation~\eqref{eq:psi-tilde} becomes 
\begin{subequations}
\begin{align}
\begin{split}
    \frac{\bar{\zeta}_{2}( \tilde{r} )}{\mu_{2}}
        &= 
            \frac{1}
            {\.\big( 1 - \gamma^{2}_{3} \big)}
            \left(
                \psi_{2}( \tilde{r} )
                +
                \frac{1}{3}\.R^{2}_{3}\.
                \Delta \psi_{2}( \tilde{r} )
            \right)
            \\[2mm]
        &\phantom{=\;}
            -
            \frac{\sigma_{2}\.k^{2}_{3}}
            {\sigma_{4}\.\gamma_{3}\.
            \big(
                1
                -
                \gamma^{2}_{3}
            \big)}
            \left(
                \gamma^{2}_{3}\.
                \psi_{2}( \tilde{r} )
                +
                \frac{1}{3}\.R^{2}_{3}\.
                \Delta \psi_{2}( \tilde{r} )
            \right)
            .
\end{split}
\end{align}
\end{subequations}
The typical profile with $\hat{\zeta}$ can be obtained by integrating $\bar{\zeta_{2}}$ and considering the regularity condition at the centre, $\partial_{r}\hat{\zeta}( \tilde{r} ) = 0$, which yields
\begin{subequations}
\begin{align}
\begin{split}
\label{eq:zeta-laplacian-integrated}
    \frac{\hat{\zeta}( \tilde{r} )}
    {\tilde{\mu}_{2}}
        &= 
            \frac{1}{\.
            \big(
                1
                -
                \gamma^{2}_{3}
            \big)}
            \left(
                \psi_{1}( \tilde{r} )
                +
                \frac{1}{3}\.R^{2}_{3}\.
                \Delta \psi_{1}( \tilde{r} )
            \right)
            \\[2mm]
        &\phantom{=\;}
            -
            \frac{\tilde{\kappa}^{2}_{3}}
            {\gamma_{3}\.
            \big(
                1
                -
                \gamma^{2}_{3}
            \big)}
            \left(
                \gamma^{2}_{3}\.
                \psi_{1}( \tilde{r} )
                +
                \frac{1}{3}\.
                R^{2}_{3}\.
                \Delta \psi_{1}( \tilde{r} )
            \right)
            +
            \zeta_{\infty}
            \, ,
\end{split}
\end{align}
\end{subequations}
where the integration constant $\zeta_{\infty}$ is a new random variable that can be set to zero as shown in Reference~\cite{2021PTEP.2021a3E02Y}. Notice that the parameters $\tilde{\mu}_{2} = \sigma^{2}_{1}\.\mu_{2} / \sigma^{2}_{2}$ and $\tilde{\kappa}_{3} = k_{3}\.\sqrt{\sigma_{2}/\sigma_{4}}$ are dimensionless.

\begin{figure}[t]
    \centering
    \hs{5mm}\includegraphics[width = 0.8624\hsize]{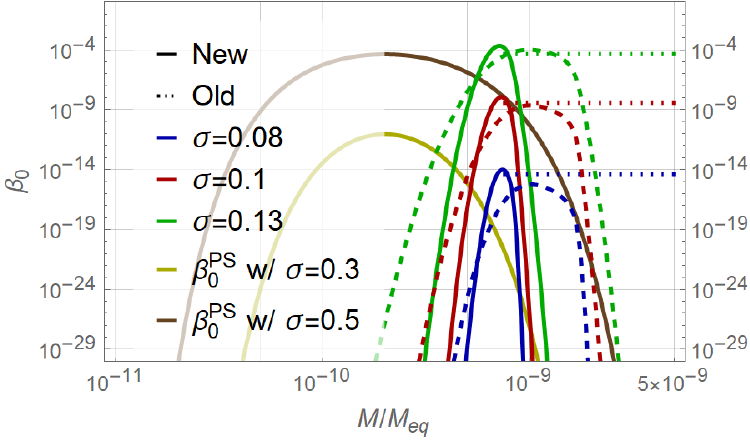}\\[5mm]
    \hs{-4mm}\includegraphics[width = 0.78\hsize]{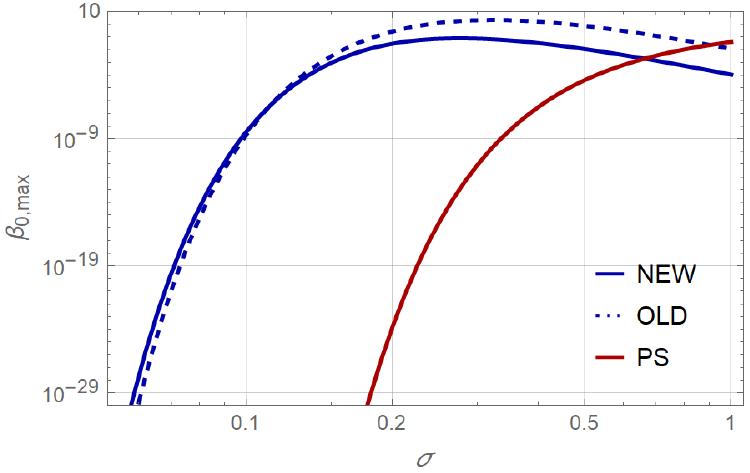}
    \caption{
        {\it Upper panel}: 
            Fraction $\beta_{0}$ of collapsed horizon patches as a function of mass in units of $M_{\eq}$. The label ``NEW" refers to the approach of considering peaks of $\Delta \zeta$; ``OLD" denotes results considering peaks on $\zeta$. Different colours indicate different values of $\sigma$. The quantity $\beta^{\rm PS}_{0}$ has been obtained using the Press--Schechter estimate. The horizontal dotted lines indicate the corresponding maximum values $\beta_{0,\mspace{1.5mu}\mathrm{max}}$ of the ``NEW" approach estimated in Reference~\cite{2021PTEP.2021a3E02Y}.
        {\it Lower panel}: 
            Estimated maximum value $\beta_{0,\mspace{1.5mu}\mathrm{max}}$ as a function of $\sigma$.
        In both cases $k_{*} = 10^{5}\.k_{\eq}$, using $M = M_{H}$. Figures from Reference~\cite{2021PTEP.2021a3E02Y}.
        }
    \label{fig:abundance-peak-zeta}
\end{figure}

The number of peaks as a function of the new variables is given by
\begin{align}
\label{eq:number-peaks}
    n^{( k_{2} )}_{\rm peaks}(\mu_{2},\mspace{1.5mu}k_{2})\,
    \d \mu_{2}\mspace{2mu}\d k_{3}
        &= 
            n^{( \tilde{\kappa}_{2} )}_{\rm peaks}
            \big(
                \tilde{\mu}_{2},\mspace{1.5mu}
                \tilde{\kappa}_{3}
            \big)\,
            \d \tilde{\mu}_{2}\.\d \tilde{\kappa}_{3}
            \notag
            \displaybreak[1]
            \\[2.5mm]
        &= 
            \frac{2 \cdot 3^{3/2}}{( 2\mspace{1.5mu}\pi )^{3/2}}\.
            \frac{ \sigma^{2}_{2}\.\sigma^{3}_{4} }
            { \sigma^{4}_{1}\.\sigma^{3}_{3} }\,
            \tilde{\mu}_{2}\.\tilde{\kappa}_{3}\.
            f\!
            \left(
                \frac{ \sigma_{2} }{ \sigma^{2}_{1} }\.
                \tilde{\mu}_{2}\.
                \tilde{\kappa}^{2}_{3}
            \right)
            \displaybreak[1]
            \\[1.5mm]
        &\phantom{=\;}
            \times
            P_{1}^{(3)}\!
            \left(
                \frac{\sigma_{2}}{\sigma^{2}_{1}}\.
                \tilde{\mu}_{2},\mspace{1.5mu}
                \frac{\sigma_{2}}{\sigma^{2}_{1}}\.
                \tilde{\mu}_{2}\.
                \tilde{\kappa}^{2}_{3}
            \right)
            \d \tilde{\mu}_{2}\mspace{2mu}
            \d \tilde{\kappa}_{3}
            \, .
            \notag
\end{align}
For estimating the current PBH dark matter fraction, we still need to make a change of variables in Equation~\eqref{eq:number-peaks} in order to relate the number of peaks solely in terms of mass. Therefore, consider the threshold value~\eqref{eq:amplitud-mu},
\begin{align}
    \tilde{\mu}_{2,\mspace{1.5mu}\crm}( \tilde{\kappa}_{3} )
        = 
            -
            \frac{1 - \sqrt{1 - 3\.\delta_{\crm}/2}}
            {\hat{r}_{\mrm}\.
            \hat{g}\mspace{1mu}'_{\mrm}( \hat{r}_{\mrm},\tilde{\kappa}_{3} )}
            \, ,
\end{align}
with $\hat{g}( \hat{r}_{\mrm},\mspace{1.5mu}\tilde{\kappa}_{3} ) = \hat{\zeta}( \hat{r}_{\mrm},\tilde{\kappa}_{3} ) / \tilde{\mu}_{2}$. Therefore, we can relate $\hat{r}_{\mrm}( \tilde{\kappa}_{3} )$ to the profile of $\hat{\zeta}$. In terms of the new variables, the PBH mass can be expressed as
\begin{align}
    M( \tilde{\mu}_{2},\mspace{1.5mu}\tilde{\kappa}_{3} )
        &= 
            \frac{1}{2\mspace{1.5mu}H( t_{H} )}\.
            \Kcal( \tilde{\kappa}_{3} )\.
            (
                \tilde{\mu}_{2}
                -
                \tilde{\mu}_{\crm}
            )^{\gamma}
            \notag
            \displaybreak[1]
            \\[2mm]
        &= 
            \frac{1}{2}\.a\,\Kcal( \tilde{\kappa}_{3} )\.
            (
                \tilde{\mu}_{2}
                -
                \tilde{\mu}_{\crm}
            )^{\gamma}\.
            \hat{r}_{\mrm}\,
            \erm^{\mspace{1.5mu}\tilde{\mu}_{2}\.\hat{g}_{\mrm}}
            \displaybreak[1]
            \\[2mm]
        &= 
            M_{\eq}\.k^{2}_{\eq}\.
            \hat{r}^{2}_{\mrm}\.
            \Kcal( \tilde{\kappa}_{3} )\.
            (
                \tilde{\mu}_{2}
                -
                \tilde{\mu}_{\crm}
            )^{\gamma}\.
            \erm^{\mspace{1.5mu}\tilde{\mu}_{2}\.\hat{g}_{\mrm}}
            \, ,
            \notag
\end{align}
where we have used that $a = a^{2}_{\eq}\.H_{\eq} \.\hat{r}_{\mrm}\,\erm^{\mspace{1.5mu}\tilde{\mu}_{2}\.g_{\mrm}}$, $H \sim a^{-2}$ and $k_{\eq} = a_{\eq}\.H_{\eq}$ with $M_{\eq} = 2.8 \times 10^{17}\.\Msun$. Upon change of variables, and utilising Equation~\eqref{eq:number-peaks}, the number of peaks per logarithmic mass interval is given by
\begin{align}
\label{eq:peaks-2}
    n_{\PBH}( M )\,
    \d \ln M
        = 
            \left\{
                \int_{\tilde{\mu}_{2}( M,\mspace{2mu}\tilde{\kappa}_{3} )
                    \mspace{2mu}\geq\mspace{1mu}
                        \tilde{\mu}_{2,\mspace{1.5mu}\crm}
                        ( \tilde{\kappa}_{3} )}
                \frac{\d \ln M}{\d \tilde{\mu}}\.
                \d \tilde{\kappa}_{3}\;
                n^{(\tilde{\mu}_{2},\mspace{1.5mu}
                \tilde{\kappa}_{3})}_{\rm peaks}\!
                \big[
                    \tilde{\mu}
                    ( M,\mspace{1.5mu}\tilde{\kappa}_{3} ),\mspace{1.5mu}
                    \tilde{\kappa}_{3}
                \big]
            \right\}
            \d\ln M
            \, .
\end{align}
Using the previous equation, the current PBH abundance, denoted by $\beta_{0}$, can readily be computed. It is defined as the relative density of PBHs (as compared to the dark matter density), which would still exist today,
\vs{-1mm}
\begin{align}
\label{eq:beta0}
    \beta_{0}\.\d \ln M
        = 
            \frac{M n_{\PBH}}
            {\rho\.a^{3}}\,\d \ln M
            \, .
\end{align}
This implies for the PBH dark matter fraction
\vs{-1mm}
\begin{align}
    f_{\PBH}^{\rm tot}
        = 
            \int \d \ln( M )\;
            f_{\PBH}( M ) 
            \, ,
\end{align}
with $f_{\PBH}( M )$ being the mass function, which using Equation~\eqref{eq:peaks-2}, can be expressed as
\vs{-2mm}
\begin{align}
\begin{split}
    f_{\PBH}( M )\.
    \dd{\ln M}
        &= 
            \frac{M\.n_{\PBH}( M )}
            {3\mspace{1.5mu}\Mpl^{2}\.
            H_{0}^{2}\.\Omega_{\DM}}\;
            \d \ln M
            \\[1mm]
        &= 
            \frac{\rho\.a^{3}}
            {3\mspace{1.5mu}\Mpl^{2}\.
            H_{0}^{2}\.\Omega_{\DM}}\,
            \beta_{0}\,\d \ln M
            \, .
\end{split}
\end{align}
Here, $\Omega_{\DM} = \rho_{\DM} / ( 3\mspace{1.5mu}\Mpl^{2}\.H_{0}^{2} )$ and $H_{0}$ and $\rho_{\DM}$ are current values of the dark matter (DM) energy density and the Hubble constant, respectively. An example of the abundance estimate which follows this approach is visualised in Figure~\ref{fig:abundance-peak-zeta}, using the power spectrum
\begin{equation}
\label{eq:ps-peak-paper}
    \Pcal_{\zeta}( k )
        = 
            3\.
            \sqrt{\frac{6}{\pi}}\,
            \sigma^{2}\mspace{-2mu}
            \left(
                \frac{k}{k_{*}}
            \right)^{\mspace{-6mu}3}
            \erm^{-3\mspace{1.5mu}k^{2} / 2\mspace{1.5mu}k_{*}^{2}}
            \, .
\end{equation}

\newpage

Although it is not explicit in Equation~\eqref{eq:beta0}, the fraction of collapsed horizon patches, $\beta_{0}$, being a measure for the PBH abundance, has an exponential dependence on the PBH formation threshold, as shown by the Press--Schechter formalism [\cf~Equation~\eqref{eq:psekeres}]. This is the reason why an accurate numerical determination of the threshold is important.

\begin{figure}[p]
    \centering
    \includegraphics[width = 0.82\hsize]{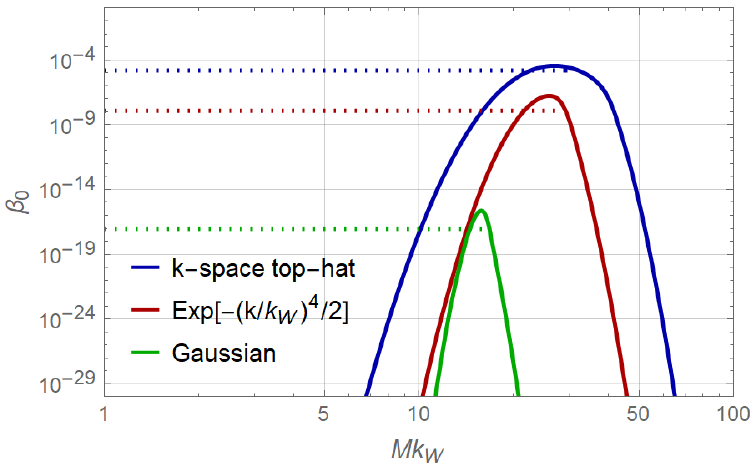}\\[5mm]
    \includegraphics[width = 0.80\hsize]{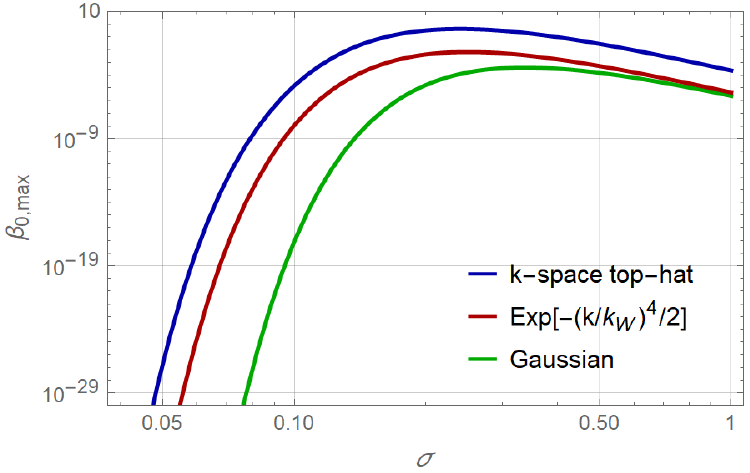}
    \caption{
        {\it Upper panel}: 
            Primordial black hole mass spectrum $\beta_{0}$ (for $\sigma = 0.1$) as a function of the mass, $M_{k_{\Wrm}}$, associated to the mode $k_{W}$. The dotted horizontal lines represent the respective values of $\beta_{0,\mspace{1.5mu}\umax}$ estimated in Reference~\cite{2021PTEP.2021a3E02Y}. 
        {\it Lower panel}: 
            Estimations of $\beta_{0,\mspace{1.5mu}\umax}$ as function of $\sigma$.
        Both cases show a comparison of the results for different window functions (see plot legends). Figures from Reference~\cite{2021PTEP.2021a3E02Y}.
        }
    \label{fig:abundance-peak-zeta-window}
\end{figure}

We would like to mention that a window function has to be used in order to correctly estimate the PBH abundances. This leads to $P_{\zeta,\mspace{1.5mu}\Wrm} \rightarrow P_{\zeta} \,W( k,\mspace{1.5mu}k_{\Wrm} )^{2}$ where $W( k,\mspace{1.5mu}k_{\Wrm} )$ is a window function satisfying $W( k,\mspace{1.5mu}k_{\Wrm} ) \leq 1$ and $W( k,\mspace{1.5mu}k_{\Wrm} ) \equiv 0$ for $k \gg k_{\Wrm}$. In particular, for broad power spectra (where several wavelength scales $k$ are involved), the use of a window function is important. The main reason is that correct counting of the peak number is invalidated by contamination of small scales which would dominate without a window function. Such a function allows us to study different scales since smaller-scale inhomogeneities can be smoothed out. Despite of this fact, the choice of the window function is not unique, and the results depend upon its choice, see Figure~\ref{fig:abundance-peak-zeta-window} for illustrative examples. However, we should emphasise that although the mentioned choice is not unique, the smoothing procedure is physically meaningful. Pragmatically, the freedom of choosing different window functions is related to our incomplete knowledge of how to statistically account for a broad power spectrum where a substantial number of scales (compared to the simplistic case of a monochromatic power spectrum) are involved. For further details, we refer the reader to References~\cite{2020IJMPD..2930002Y, 2018PhRvD..97j3528A, 2020JCAP...12..038T}.
\vfill

\newpage
\vphantom{.}
\newpage

\section{Spin}
\label{sec:Spin}
\vs{-3mm}
\lettrine[lines=3, slope=0em, findent=0em, nindent=0.2em, lhang=0.1, loversize=0.1]{S}{} \!o far, we have focused on the {\it mass} of the primordial black holes. However, a black hole is also characterised by its {\it charge} and {\it spin}. Although PBHs as dark matter candidates are basically expected to be charge-neutral because of the global neutrality of the Universe, their spins can be significant. From the viewpoint of the gravitational-wave detection of binary black hole mergers, spins are important as they can be measured through the effective inspiral spin. In this Section, we review the spin-statistics of primordial black holes.

The spin distribution of primordial black holes has been extensively studied in the literature~\cite{2017PTEP.2017h3E01C, 2017PhRvD..96h3517H, 2020JCAP...03..017M, 2019PhRvD.100f3520H, 2021PhRvD.104f3008F, 2021PhRvD.104h3018C, 2021JCAP...12..041E, 2021PDU....3100791G}. De Luca {\it et al.}~\cite{2020JCAP...04..052D} and Harada {\it et al.}~\cite{2021ApJ...908..140H} performed recent studies on PBH spin using peak theory, with the density contrast being assumed to follow a Gau{\ss}ian distribution. Peaks of these density contrast typically have a spherically-symmetric profile, particularly in the case of monochromatic power spectra, but can have small anisotropy through a deviation from exact monochromaticity. Furthermore, the critical behaviour~\eqref{eq:2-scaling} indicates that only a relatively small fraction of the overdense region collapses into a black hole. In such a case, the black hole spin is further enhanced. Harada {\it et al.}~\cite{2021ApJ...908..140H} found the average PBH Kerr parameter $a$ [$\coloneqq S / ( G M^{2} )$, with $S$ being the spin amplitude] to be proportional to $( M / M_{H} )^{-1/3}$, implying that the low-mass tail yields large spins.

Including the leading-order anisotropy around the density contrast peak, peak theory yields the probability distribution of the tidal torque. Heavens and Peacock~\cite{1988MNRAS.232..339H} found a fitting formula for the resulting probability density function of the normalised spin parameter $h$ (see below) as\footnote{\setstretch{0.9}De Luca {\it et al.}~\cite{2020JCAP...04..052D} found another fitting formula:
\begin{align}
    P_{H}( h )
        \simeq
            \exp
            \big[
                -
                2.37
                -
                4.12\.\ln h
                -
                1.53\.( \ln h )^{2}
                -
                0.13\.( \ln h )^{3}
            \big]
            \, .
\end{align}
However, as it is singular for $h \to 0$, we adopt the Heavens--Peacock formula hereafter.
}
\begin{align}
    P_{H}( h )
        \simeq
            563\mspace{1.5mu}H^{2}
            \exp\mspace{-2mu}
            \bqty{
                -
                12\.h
                +
                2.5\.h^{1.5}
                +
                8
                -
                3.2\.
                \big(
                    1500
                    +
                    h^{16}
                \big)^{\mspace{-2mu}1/8}
            }
            \, .
\end{align}
In the case of PBHs, this $h$-parameter is related to the Kerr parameter $a$ as~\cite{2021ApJ...908..140H}
\begin{subequations}
\begin{align}
    h
        &\coloneqq
            a / C( M,\mspace{1.5mu}\nu )
            \, ,
\intertext{where}
    C( M,\mspace{1.5mu}\nu )
        &\equiv 
            3.25\times10^{-2}\.
            \sqrt{1 - \gamma_{1}^{2}}\,
            \sigma_{0}\.
            \pqty{\frac{M}{M_{H}}}^{\mspace{-6mu}-1/3}\.
            \pqty{\frac{\nu}{10}}^{\mspace{-5mu}-1}
            \, .
\end{align}
\end{subequations}
Here, we assume an almost monochromatic power spectrum for the density contrast as $\Pcal_{\delta}( k ) \approx \sigma_{0}^{2}\.\delta( \ln k/k_{*} )$. The parameter $\gamma_{1} \coloneqq \sigma_{1}^{2} / ( \sigma_{0}\.\sigma_{2} ) \lesssim 1$, with $\sigma_{i}^{2} \coloneqq \int\dd{\ln k}\,k^{2i}\.\Pcal_{\delta}( k )$, characterises the width of the power spectrum (where $\gamma_{1} = 1$ for an exactly monochromatic spectrum).\footnote{\setstretch{0.9}Note that peak theory is adopted for $\delta$, and $\sigma_{i}$ is defined for $\Pcal_{\delta}$, while Section~\ref{sec:Peak--Theory-Procedure-with-Curvature-Peaks} is for $\Pcal_{\zeta}$.} The peak value $\nu$ of the density contrast in unit of the standard deviation $\sigma_{0}$ is given by $\nu = \delta_{\rm peaks} / \sigma_{0}$.

\begin{figure}
    \centering
    \includegraphics[width = 0.68\hsize]{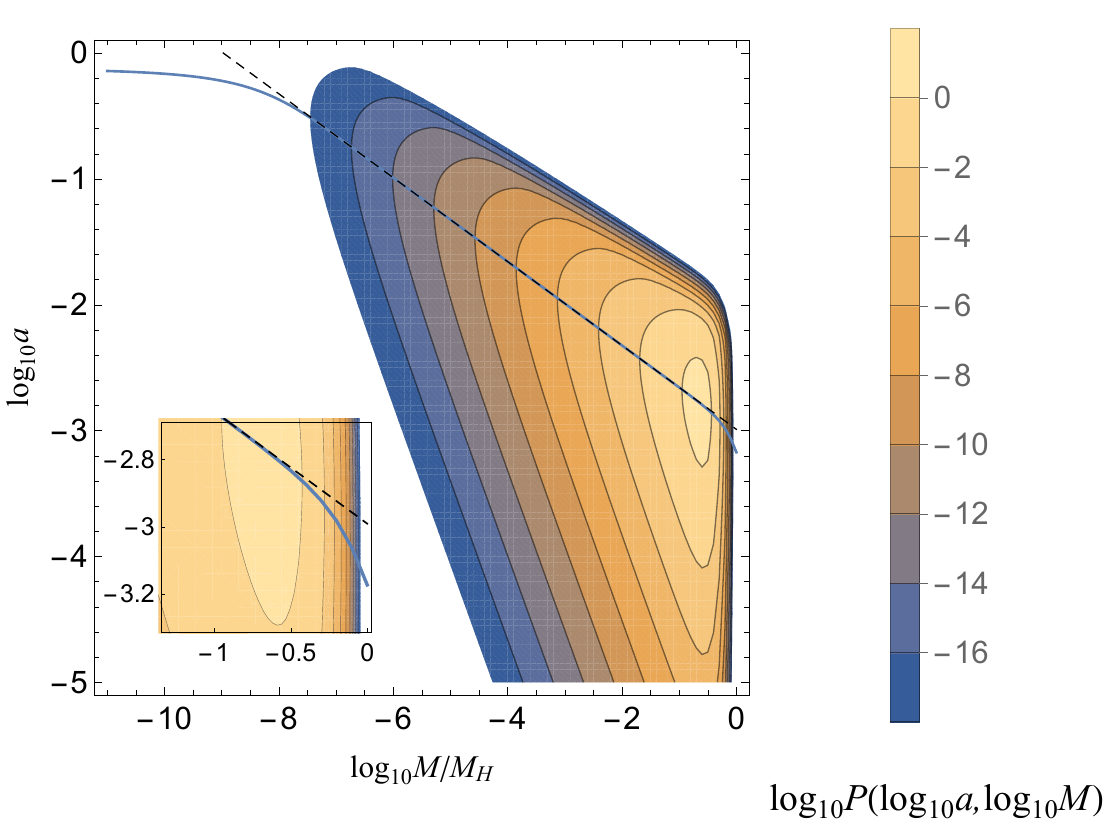}
    \caption
        {Contour plot of $\log_{10}p\mspace{1mu}( \log_{10}a,\mspace{1.5mu}\log_{10}M )$ for $\nu_{\uth} = 10$ and $\sigma_{0} = 0.192$, corresponding to $f_{\PBH} \sim 0.1\mspace{0.5mu}\%$ for $M_{H} \sim \Msun$, setting $\gamma = 0.85$. The solid blue line is the expectation value $\braket{a}$ for given mass $M$, and the dashed black line is its power-law fit $\sim ( M / M_{H} )^{-1/3}$. Figure from Reference~\cite{2022ApJ...939...65K}.}
    \label{fig: single PBH spin}
\end{figure}

Although $h$ can in principle take an arbitrarily large value, the Kerr parameter cannot be larger than unity for a black hole. In other words, a density peak with $h > 1 / C( M,\mspace{1.5mu}\nu )$ does not form a black hole. Therefore, the conditional probability of $a$ for PBHs is restricted as
\begin{align}
    p\mspace{1mu}
    \big(
        a\,\big|\, M,\mspace{1.5mu}\nu
    \big)\.
    \dd{a}
        = 
            \frac{P_{H}
            \big[
                ( a / C( M,\mspace{1.5mu}\nu )
            \big]}
            {C( M,\mspace{1.5mu}\nu )
            \int_{0}^{1/C( M,\mspace{2.5mu}\nu )}\dd{h}\.
            P_{H}( h )}
            \;\dd{a}
            \, .
\end{align}
Furthermore, the critical behaviour determines the PBH mass $M$ as a function of $\nu$. Therefore, the joint probability of $a$ and $M$ is formulated as
\begin{align}
    p\mspace{1mu}( a,\mspace{1.5mu}M )\,
    \dd{a}\dd{M}
        = 
            P
            \big[
                a\,\big|\, M( \nu ),\mspace{1.5mu}\nu
            \big]\.
            P_{\nu}( \nu )\,
            \dd{a}\dd{\nu}
            \, ,
\end{align}
where
\vs{-1mm}
\begin{align}
    P_{\nu}( \nu )
        = 
            \frac{\erm^{-\nu^{2}/2}}
            {\int_{\nu_{\uth}}^{\infty}\dd{\tilde{\nu}}\,
            \erm^{-\tilde{\nu}^{2}/2}}
        = 
            \sqrt{\frac{2}{\pi}}\.
            \frac{\erm^{-\nu^{2}/2}}
            {\erfc\mspace{-2mu}
            \big(
                \nu_{\uth}/\sqrt{2}\mspace{1mu}
            \big)}
            \, ,
\end{align}
being a restricted Gau{\ss}ian distribution, and $\nu_{\uth} = \delta_{\uth} / \sigma_{0}$ is the threshold value for the density contrast in unit of $\sigma_{0}$. An example of this joint probability is shown in Figure~\ref{fig: single PBH spin}. Typically, the PBH spin is as small as $a \sim 10^{-3}$ with $M \sim M_{H}$. However, a large spin $a \sim 1$ is allowed for very small PBHs $M \ll M_{H}$ although the probability is strongly suppressed.

Stellar-mass black holes have been extensively searched for using gravitational waves from mergers of binary black holes. Gravitational waves of any binary system can be characterised by its chirp mass $\Mcal$, mass ratio $q$, and {\it effective inspiral spin} $\chi_{\eff}$ defined by
\vs{-1mm}
\begin{subequations}
\begin{align}
    \Mcal
        &\coloneqq
            \frac{( M_{1}\.M_{2} )^{3/5}}
            {( M_{1} + M_{2} )^{1/5}}
            \in( 0,\mspace{1.5mu}\infty )
            \, ,
            \displaybreak[1]
            \\[3mm]
    q
        & \coloneqq 
            \frac{ M_{2} }{ M_{1} }
            \in( 0,\mspace{1.5mu}1 ]
            \, ,
            \displaybreak[1]
            \\[3mm]
    \chi_{\eff}
        &= 
            \frac{a_{1}\mspace{-0.5mu}\cos\theta_{1}
            +
            q\.a_{2}\cos\theta_{2}}
            {1 + q}
            \in[ -1,\mspace{1.5mu}1 ]
            \, ,
\end{align}
\end{subequations}
respectively, with the subscript $1$ indicating the primary black holes, and $2$ stands for the secondary one. By $\theta_{i}$ we denote the angles between the respective black hole spins and the orbital angular momentum of the binary.

\begin{figure}
   $\mspace{30mu}$\includegraphics[width = 1.45\hsize]{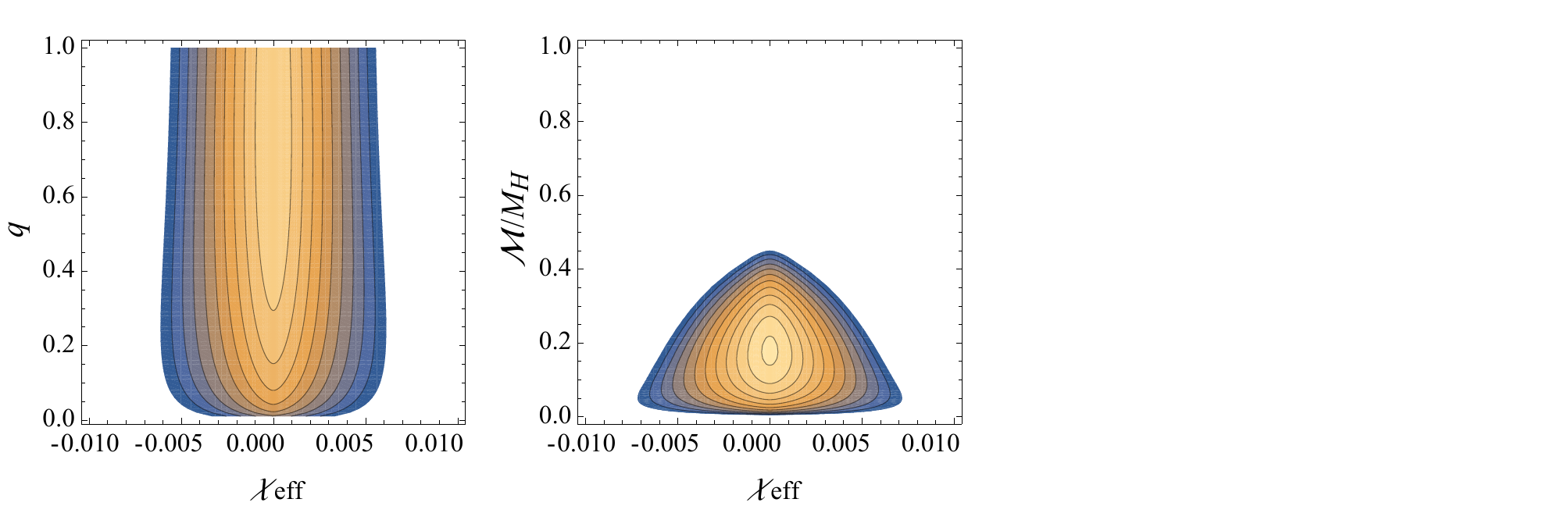}\\
    $\mspace{-50mu}$\includegraphics[width = 1.45\hsize]{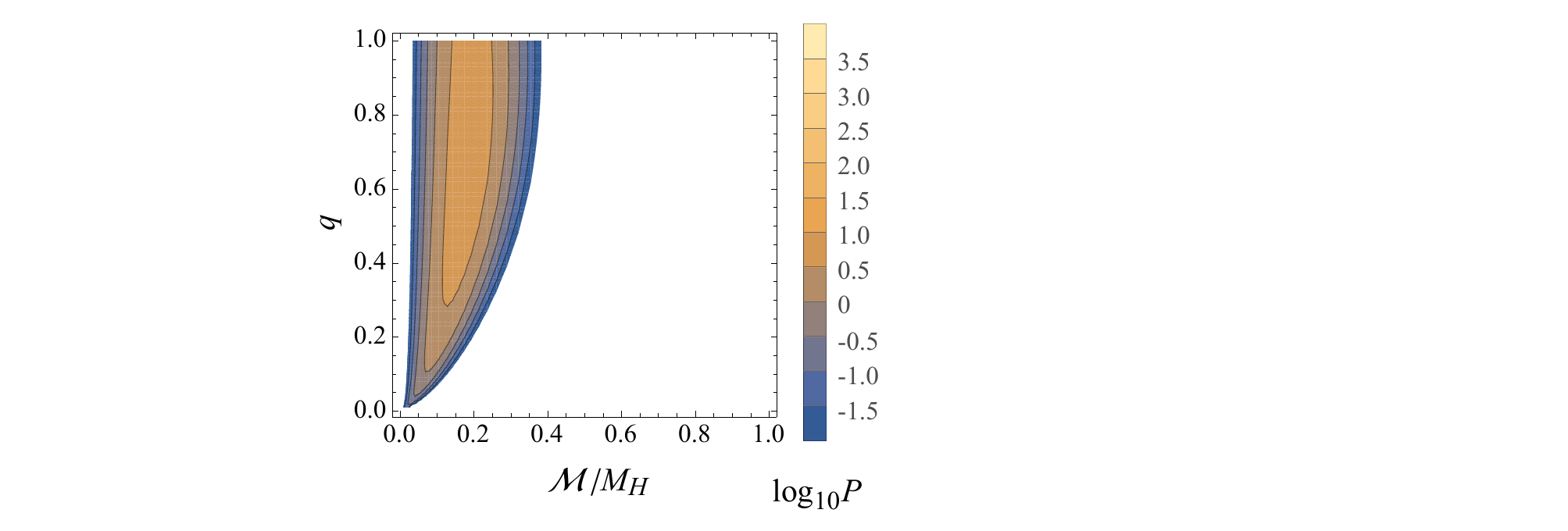}
    \caption{
        Two-variable probabilities of binary primordial black holes{\,---\,}$p\mspace{1mu}(q,\mspace{1.5mu}\chi_{\eff})$ ({\it top-left panel}), $p\mspace{1mu}(\Mcal,\mspace{1.5mu}\chi_{\eff})$ ({\it top-right panel}) and $p\mspace{1mu}(q,\mspace{1.5mu}\Mcal)$ ({\it bottom panel}), for $\nu_{\uth} = 10$, $\sigma_{0} = 0.192$, and $\gamma_{1} = 0.85$.
        }
    \label{fig: binary PBH spin}
\end{figure}

Primordial black holes can also form binary systems (see Section~\ref{sec:Binary-Mergers}), where basically two constituent PBHs are chosen randomly. Assuming that PBH formation is statistically isotopic, the joint probability of their intrinsic parameters $\wbm \coloneqq ( a_{1},\mspace{1.5mu}a_{2},\mspace{1.5mu}M_{1},\mspace{1.5mu}M_{2},\mspace{1.5mu}\cos\theta_{1},\mspace{1.5mu}\cos\theta_{2},\mspace{1.5mu}\phi_{1},\mspace{1.5mu}\phi_{2} )$ is hence given by the direct product of the single PBH distribution as
\vs{-1mm}
\begin{align}
    p\mspace{1mu}(\wbm)\dd{\wbm}
        = 
            \frac{2}{( 4\mspace{1.5mu}\pi )^{2}}
            \prod_{i = 1}^{2}
            p\mspace{1mu}( a_{i},\mspace{1.5mu}M_{i} )
            \dd{a_{i}}
            \dd{M_{i}}
            \dd{\cos\theta_{i}}
            \dd{\phi_{i}}
            \, .
\end{align}
With an appropriate Jacobian, it can be easily recast into the following probability of $\Mcal$, $q$, and $\chi_{\eff}$:
\vs{-1mm}
\begin{align}
\label{eq:p(M-q-chi)}
    p\mspace{1mu}( \Mcal,\mspace{1.5mu}q,\mspace{1.5mu}\chi_{\eff} )
        &= 
            \frac{ 1 + q }{ 2\mspace{1.5mu}q^{2}\.
            \gamma^{2}\.\sigma_{0}^{2}\.\Mcal }
            \pqty{\frac{ ( 1 + q )^{2/5}\Mcal^{2} }
            { q^{1/5} M_{H}^{2} } }^{\mspace{-6mu}1/\gamma}
            \notag
            \displaybreak[1]
            \\
        &\phantom{=\;}\times
            \int_{0}^{1}\dd{a_{1}}\int_{0}^{1}
            \dd{a_{2}}\;
            \Theta
            \big[
                T( a_{1},\mspace{1.5mu}a_{2},\mspace{1.5mu}\chi_{\eff},\mspace{1.5mu}q )
            \big]\,
            T( a_{1},\mspace{1.5mu}a_{2},\mspace{1.5mu}\chi_{\eff},\mspace{1.5mu}q )
            \displaybreak[1]
            \\[4mm]
        &\phantom{=\;}\times
            \frac{1}{a_{1}\mspace{1.5mu}a_{2}}\.
            \prod_{i = 1}^{2}
            P
            \Big[
                a_{i}\,\Big|\, M_{i}( \Mcal,\mspace{1.5mu}q ),\mspace{1.5mu}
                \nu
                \big\{
                    M_{i}( \Mcal,\mspace{1.5mu}q )
                \big\}
            \Big]\,
            P_{\nu}
            \Big[
                \nu
                \big\{
                    M_{i}( \Mcal,\mspace{1.5mu}q )
                \big\}
            \Big]
            \, ,
            \notag
\end{align}
where
\vs{-1mm}
\begin{align}
\begin{split}
    T( a_{1},\mspace{1.5mu}a_{2},\mspace{1.5mu}\chi_{\eff},\mspace{1.5mu}q )
        &= 
            \min\!
            \big[
                a_{1},\mspace{1.5mu}q\.a_{2}
                +
                ( 1 + q )\.\chi_{\eff}
            \big]
            \\[2mm]
        &\phantom{=\;}
            +
            \min\!
            \big[
                a_{1},\mspace{1.5mu}
                q\.a_{2}
                -
                ( 1 + q )\.\chi_{\eff}
            \big]
            \, .
\end{split}
\end{align}
The exponent $\gamma$ is the universal power of the scaling behaviour~\eqref{eq:2-scaling}, and $M_{i}( \Mcal,\mspace{1.5mu}q )$ and $\nu( M )$ are given by
\vs{-1mm}
\begin{subequations}
\begin{align}
    M_{1}( \Mcal,\mspace{1.5mu}q )
        &= 
            q^{-3/5}\.( 1 + q )^{1/5}\.\Mcal
            \, ,
            \displaybreak[1]
            \\[3mm]
    M_{2}( \Mcal,\mspace{1.5mu}q )
        &= 
            q^{\mspace{1mu}2/5}\.( 1 + q )^{1/5}\.\Mcal
            \, ,
\end{align}
\end{subequations}
and
\vs{-2mm}
\begin{align}
    \nu( M )
        = 
            \frac{1}{\sigma_{0}}
            \pqty{
                \frac{M}{M_{H}}
            }^{\mspace{-6mu}1/\gamma}
            +
            \nu_{\uth}
            \, ,
\end{align}
respectively. It is then possible to obtain the two-variable probabilities $p\mspace{1mu}( \chi_{\eff},\mspace{1.5mu}q )$, $p\mspace{1mu}( \Mcal,\mspace{1.5mu}\chi_{\eff} )$ and $p\mspace{1mu}( \Mcal,\mspace{1.5mu}q )$ by integrating Equation~\eqref{eq:p(M-q-chi)}. An example of these probabilities is shown in Figure~\ref{fig: binary PBH spin}.

The first observation is that the probabilities are symmetric under the replacement $\chi_{\eff} \leftrightarrow -\chi_{\eff}$ due to our isotropy assumption. Negative $\chi_{\eff}$ (spin anti-alignment) is understood as an important indicator of the binary environment because in ordinary (non-PBH) astrophysics, the progenitor spins are expected to be almost aligned with their orbital angular momentum if they are isolated. In fact, two candidate events (GW191109\_{0}10717 and GW200225\_{0}60421) suggest negative $\chi_{\eff}$ with significant support~\cite{2021arXiv211103634T}. It is also worth mentioning that $\chi_{\eff}$ has almost no correlation with $q$. Callister {\it et al.}~\cite{2021ApJ...922L...5C} recently suggested that an anti-correlation between the average $\chi_{\eff}$ and $q$. This has been inferred from Gau{\ss}ian analysis, noting noted that this tendency is opposite to standard astrophysical models. Despite the fact that in most scenarios, PBH binaries also do not explain such a $\chi_{\eff}$-$q$ anti-correlation, this characteristics might be important.
\newpage

Above, we neglected the spin evolution through accretion processes. De Luca {\it et al.}~\cite{2020JCAP...04..052D} showed that these can be significant for PBHs with masses exceeding $\Ocal( 10 )\.\Msun$. The reduction of the background fluid pressure due to the QCD phase transition for example (see Section~\ref{sec:Thermal--History--Induced-Mass-Function}) can also enhance the PBH spin~\cite{2017PhRvD..96h3517H}. PBH clustering (see Section~\ref{sec:Clustering-of-Primordial-Black-Holes}) may affect the random-choice assumption. All these effects can be important for distinguishing the origin of binary black holes.
\newpage
\vphantom{.}
\newpage

\section{Quantum Aspects}
\label{sec:Quantum-Aspects-of-Primordial-Black-Holes}
\vs{-3mm}
\lettrine[lines=3, slope=0em, findent=0em, nindent=0.2em, lhang=0.1, loversize=0.1]{T}{} he treatment of primordial black holes has been, and still is, mostly on the classical level. Attempts to go beyond have mainly touched upon semiclassical aspects, prominently resulting in abundance constraints from Hawking radiation (\cf~Reference~\cite{2016PhRvD..94d4029C} for a review). Surprisingly, many of the underlying calculations utilise the semiclassical {\it approximation} even beyond Page time~\cite{1993PhRvL..71.3743P}, after which it should clearly be inapplicable, necessitating a full quantum description. Below, we discuss the consequences of such a treatment.

\subsection{Graviton Condensates}
\label{sec:Graviton-Condensates}
\vs{-1mm}
Before entering the discussion on the consequences of quantum aspects of black holes, it is useful to clarify the terms {\it classical}, {\it semiclassical} and {\it quantum}. First, we note one definite characteristic which all macroscopic bodies share: a large number of constituents. 
An important question to ask then is how the couplings $\alpha_{ij}$ between a pair of constituents $i$ and $j$ behave; particularly: does a universal coupling $\lambda$ exist as in Bose--Einstein condensates, and if yes: which value does it assume?

As first pointed out by Dvali \& Gomez~\cite{Dvali:2011aa}~(see also References~\cite{Dvali:2012rt, 2012PhLB..716..240D, Dvali:2012en, 2012arXiv1212.0765D, 2013PhRvD..88l4041D, 2013arXiv1307.7630D, 2014JCAP...01..023D, 2015NuPhB.893..187D, Dvali:2015cwa, 2016ForPh..64..106D, 2016PhLB..753..173D, 2017arXiv171109079D, 2018arXiv180406154D, Dvali:2018tno, 2018arXiv181002336D, 2020PhRvD.102j3523D, 2022PhRvD.105e6013D, 2022PhRvD.106l5019D}), black holes are indeed macroscopic and {\it inherently quantum} objects, which can be described as {\it Bose--Einstein condensates of $N \gg 1$ gravitons} being {\it at the critical point of a quantum phase transition}, with their coupling
\vs{-2mm}
\begin{align}
\label{eq:lambda-black-hole}
    \lambda
        =
            N \alpha
        =
            N \cdot ( L_{\Prm} / L)^{2}
        =
            N \cdot ( 1/N )
        =
            1
            \, ,
\end{align}
being universal and critical. Here, $L_{\Prm}$ is the Planck length and $L$ is the extent of the black hole, \ie~its Schwarzschild radius. Except for certain aspects, such as their gravitational field beyond some distance, these objects cannot be fully treated (semi)classically{\,---\,}not even approximately. Particularly, they are maximally-entangled systems, which has profound implications on information storage and release (see References~\cite{2013arXiv1307.7630D, 2016ForPh..64..106D, 2022RSPTA.38010071D, 2018arXiv181002336D, 2020PhRvD.102j3523D}), the latter being understood as quantum depletion, which is a $1/N$-effect. As such, due to the large number of gravitons in black holes,\footnote{\setstretch{0.9}For instance, a solar-mass black hole consists of $N \sim ( {\rm km} / L_{\Prm} )^{2} \sim 10^{76}$ gravitons.} it might na{\"i}vely be thought as effectively negligible. However, it is not; it is key to systems which are at the point of a quantum phase transition that their entanglement entropy for the reduced one-particle density matrix becomes maximal and new light modes appear. As we will discuss, this, in principle, provides the possibility to retrieve all information ever captured by a black hole. These and associated phenomena are entirely missed in any classical or semiclassical analysis.

To get a better overview of the increasing levels of quantumness, let us repeat the useful table of Reference~\cite{2012arXiv1212.0765D}:
\vs{1mm}
\begin{tcolorbox}
\begin{itemize}

    \item
        {\bf Ordinary macroscopic objects} (\eg~{\it planets})\\
        $N$ exists, $\lambda$ cannot be defined;

    \item
        {\bf Generic (non-critical) Bose--Einstein condensates}\\
        Both $N$ and $\lambda$ are well-defined, but $\lambda \ne 1$;

    \item
        {\bf Quantum-critical Bose--Einstein condensates} (\eg~{\it black holes})\\
        Both $N$ and $\lambda$ are well-defined, and $\lambda = 1$.

\end{itemize}
\end{tcolorbox}
\vs{1mm}

In particular, this implies that the standard treatment in which black holes are described by a metric $g_{\mu \nu}( x )$, with the effects of quantum gravity (seemingly) accounted for as quantum corrections to $g_{\mu \nu}( x )$, is strictly speaking inconsistent; it can only be used semiclassically. Indeed, the very notion of metric can only be approximate; it needs to be abandoned on the full quantum level since it does not allow to resolve the quantumness of the background itself, \ie~of its constituents. Reference~\cite{2012arXiv1212.0765D} also gives the following clarifying and tabularised overview of the terms {\it classical}, {\it semiclassical} and {\it quantum}:
\vs{1mm}
\begin{tcolorbox}
\begin{itemize}

    \item
        {\bf Classical:} 
        $\hslash = 0$, $1/N = 0$;

    \item
        {\bf Semilassical:} 
        $\hslash \ne 0$, $1/N = 0$;

    \item
        {\bf Quantum:} 
        $\hslash \ne 0$, $1/N \ne 0$.

\end{itemize}
\end{tcolorbox}
\vs{1mm}
\noindent This makes it apparent that the semiclassical approximation reduces quantum effects to $\hslash$-corrections of classical entities, {\it without resolving their constituency}.

Note that at tree-level scattering, the exchange of momentum of a probe with black hole constituents is only suppressed as $1/N$, rather (as semiclassically, and wrongly, thought) as $\erm^{-N}$. These $1/N$-corrections (to planar results) are taking place at {\it each act of emission}, implying that over the half-life time of a black hole, this deviation accumulates to an order-one effect, thereby {\it resolving the information ``paradox"}~\cite{1976PhRvD..13..191H}. This reveals the fundamental mistake in standard semiclassical reasoning: The mentioned $1/N$-corrections provide a sufficient time scale to read out an order-one fraction of the whole information contained within the black hole until reaching the half-evaporation point. This would be impossible with an exponential suppression.\footnote{\setstretch{0.9}One should also note that the Bose--Einstein approach is consistent with global symmetries.}

The above considerations make it clear that black holes are very particular objects. However, they are not unique; other objects exist in quantum field theories, which share the above-mentioned special features. This becomes apparent when considering the entropy $S$ of a general self-sustained object of extent $L$ in a quantum field theory with effective coupling $\alpha$. Unitarity implies the bound~\cite{2021JHEP...03..126D}
\begin{align}
\label{eq:S-Area}
    S
        \leq
            1 / \alpha
            \, .
\end{align} 
Correspondingly, the maximum entropy compatible with unitarity is $S_{\umax} = 1 / \alpha$. The objects saturating this entropy bound are referred to as {\it Saturons}, which, as has been pointed out by Dvali~\cite{2021JHEP...03..126D} (see References~\cite{Dvali:2019jjw, Dvali:2019ulr, Dvali:2020wqi, Dvali:2021rlf, Dvali:2021tez, Dvali:2023xfz} for recent applications), share the universal properties:
\vs{1mm}
\begin{tcolorbox}
\begin{flushleft}
\begin{itemize}

    \item
        Their entropy saturates the bound \eqref{eq:S-Area};

    \item
        If they are unstable, up to $\Ocal( 1 / S )$-corrections, their decay is\\
        thermal and characterised by the temperature $T \sim 1 / L$;

    \item
        If treated semiclassically, they exhibit an information horizon;
    
    \item
        The minimal timescale $t_{\umin}$ required for information retrieval\\
        is bounded from below by $t_{\umin} = L / \alpha = S_{\umax}\.L$.

\end{itemize} 
\end{flushleft}
\end{tcolorbox}
\vs{1mm}
\noindent The last point gives Page's time for a black hole~\cite{1993PhRvL..71.3743P}.\footnote{\setstretch{0.9}The recent Reference~\cite{2023arXiv230109575D} argues that since black holes have maximal information capacity, they might conceivably be used by advanced extra-terrestrial intelligences, with potentially-observable signatures.}

Before we discuss the profound implications the mentioned quantum characteristics have for {\it rotating} PBHs in Section~\ref{sec:Vortices}, the subsequent Subsection is devoted to the impact the fundamentally different information retrieval has on PBH abundance constraints which originate from Hawking evaporation.
\newpage

\subsection{Memory Burden}
\label{sec:Memory-Burden}
\vs{-1mm}
As pointed out by Dvali~\cite{2018arXiv181002336D, 2019JCAP...03..010D} (see also Reference~\cite{2020PhRvD.102j3523D}), black holes admit an enhanced memory capacity which stabilises them. This, in turn, is shown to be maximal at the latest by the time at which half of the energy has been emitted, and the stored information becomes accessible. However, as discussed above, after losing half of its mass, the semiclassical description for black holes is no longer applicable, which Reference~\cite{2020PhRvD.102j3523D} importantly summarises as:
\vs{1mm}
\begin{tcolorbox}
\centering
    {\it An old black hole that lost half of 
    its mass is by no means\\
    equivalent to a young classical black 
    hole of equal mass.} 
\end{tcolorbox}
\vs{1.5mm}

In particular, with increasing time, the black hole evolution deviates increasingly from self-similarity until it becomes entirely different at latest at Page's time. The underlying reason is the phenomenon called {\it memory burden}~\cite{2018arXiv181002336D, 2019JCAP...03..010D, 2020PhRvD.102j3523D}, which describes a certain backreaction of modes within quantum systems. Specifically, on the quantum level, a black hole has a number of modes (so-called {\it memory modes}) in which it can store a large amount of information at practically no energy cost; they are essentially gapless. This gaplessness is {\it exclusively} reached for critical occupation of another mode (the so-called {\it master mode}); any evolution away from criticality, such as through Hawking evaporation, increases the mentioned energy gap, making it more costly for quanta to be emitted from the system, hence {\it slowing down} the rate of emission.

Clearly, the mentioned modification to black hole evaporation dynamics will dramatically impact the part of the PBH constraint landscape which is based on semiclassical Hawking radiation calculations. For instance, the validity of the semiclassical approximation throughout the whole PBH decay process would imply that all PBHs with masses smaller than approximately $M_{*} = 5 \cdot 10^{14}\.\grm$ would have completely evaporated by the present epoch (see Reference~\cite{2016PhRvD..94d4029C}). As discussed in Section~\ref{sec:Evaporation-Constraints}, the non-observation of respective Galactic and extragalactic $\gamma$-rays leads to the formulation of the (seemingly) most stringent constraints on the PBH abundance, but this entirely neglects the memory-burden-induced slow-down of the evaporation, leading to the survival of black holes with mass around $M_{*}$.
\newpage

To date, the evaporation dynamics have not yet been entirely resolved, so it remains unclear how long PBHs will eventually survive, but the calculation of Reference~\cite{2020PhRvD.102j3523D} implies that the lifetime of any black hole is significantly altered. This impacts any abundance constraint of PBHs below a mass $M_{*}$. For instance, this largely alters bounds on the PBH abundance coming from big bang nucleosynthesis (BBN) (see Section II.A of Reference~\cite{2021RPPh...84k6902C}), which originate from violent emission from PBHs with masses between approximately $10^{9}\.\grm$ and $10^{14}\.\grm$.

Of course, primordial black holes of such low mass, which survived until today, are notoriously difficult to detect by gravitational lensing. To date, there appears to be no suitable respective method of observation for these ultracompact objects. However, the fact that these black holes emit energetic quanta motivates the possibility of high-energy cosmic-ray searches.

An illustrative example of the characteristics of the evaporation dynamics of light PBHs has been given in Reference~\cite{2020PhRvD.102j3523D} wherein the exemplary scenario in which small PBHs with a monochromatic mass spectrum peaked at $M \sim 10^{8}\.\grm$, constituting all of dark matter, has been studied. Using the entropy-suppressed decay rate $\tilde{\Gamma} \sim R_{\Srm}^{-1}/S^{2}$, with $R_{\Srm}$ being the Schwarzschild radius and $S$ the entropy, the modified lifetime, $\tilde{\tau}$, of the black holes is correspondingly prolonged as $\tilde{\tau} \gtrsim S^{2}\.\tau$, where $\tau$ is the standard semiclassical result (see Reference~\cite{2021RPPh...84k6902C}). This then leads to $\tilde{\tau} \gtrsim 10^{49}\.\srm$, vastly exceeding the current age of the Universe. Hence, those primordial black holes will be present today.\footnote{\setstretch{0.9}Of course, since such light PBHs as dark matter candidates would be abundant in the Solar system, they have an increased encounter rate with Earth. This might constitute a way to constrain, detect, or even die from primordial black holes~\cite{Loeb-Death-by-PBH-2021}.}

\subsection{Vortices}
\label{sec:Vortices}
\vs{-1mm}
The mentioned correspondence between black holes and generic entropy-saturated systems opens up the possibility of using such saturons as laboratories for understanding well-established black hole properties and for predicting new ones. In Reference~\cite{2022PhRvL.129f1302D} it has been argued that black holes, and other saturons, naturally support vortex structure{\,---\,}an entirely new quantum characteristic of these objects. 

\begin{figure}[t]
	\centering
	\includegraphics[angle=270, width = 0.62 \textwidth]{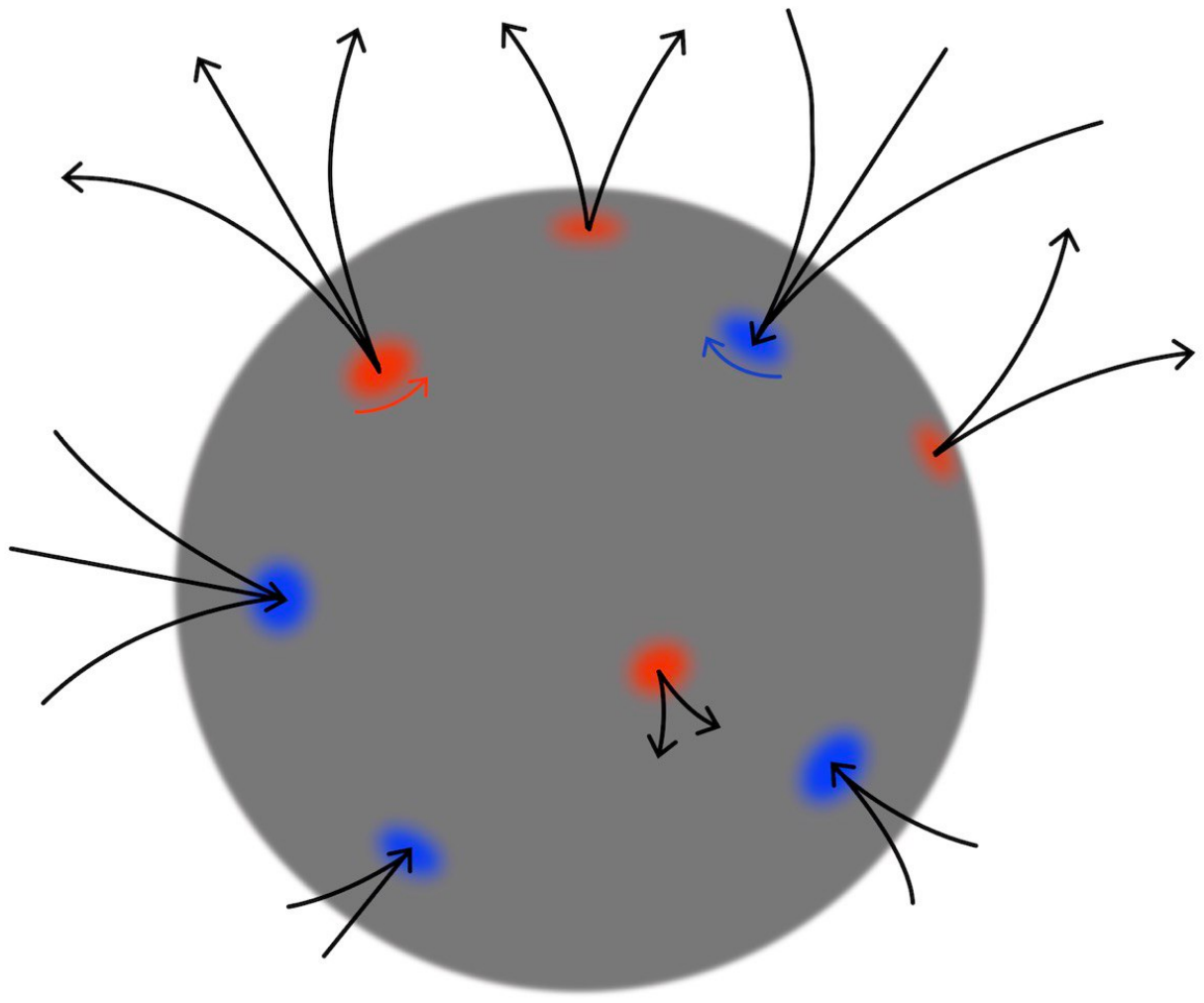}
	\caption{
        Illustration of a black hole with a number of randomly oriented vortex/anti-vortex pairs. Figure from Reference~\cite{2022PhRvL.129f1302D}.}
	\label{fig:vortexes}
\end{figure}

Note that the occurrence of vortices (see Figure~\ref{fig:vortexes} for an illustration) offers a microscopic explanation for the zero temperature of extremal black holes. It is now obvious that such objects cannot evaporate: Evaporation is a process leading to gradual decrease of mass, with a (thermal) spectrum in which emission of quanta of arbitrarily low energy is possible. However, for a black hole with maximal vorticity, such a process is not possible since the winding number of the vortex cannot change continuously due to its topological nature. A key result of Reference~\cite{2022PhRvL.129f1302D} is:
\vs{1mm}
\begin{tcolorbox}
\centering
    {\it Vorticity gives a topological meaning to the stability of extremal black holes.}
\end{tcolorbox}

This in particular implies that light PBHs with large initial spins (such as those produced through the confinement mechanism~\cite{2021PhRvD.104l3507D} (see discussion in Section~\ref{sec:Other-Formation-Scenarios}) will only evaporate until they have reached their vortex threshold, which could lead to remnants evading the evaporation constraints. This characteristic has similar consequences than that originating from memory burden as discussed in the previous Subsection, but both aspects are different in nature and hence need to be distinguished. However, they both contribute to opening up a large, previously strongly-constrained, parameter space of primordial black holes around and below $10^{16}\,\grm$.

Another interesting property of black hole vorticity is that the vortices will trap magnetic field flux{\,---\,}a very general phenomenon, which, in the case of gauge-flux trapping by a global vortex, has been introduced in Reference~\cite{Dvali:1993sg}. Here, vortices trap the magnetic field by interacting with the surrounding neutral plasma of either ordinary or weakly-charged dark matter. This implies that highly-rotating (primordial) black holes can efficiently slow down due to the emission process introduced in Reference~\cite{1977MNRAS.179..433B} with their total emitted power scaling as $P_{\rm BZ} \sim {\rm Flux}^{2}\.\Omega^{2}$. In this way, PBH dark matter might naturally explain the occurrence of primordial magnetic fields (\cf~Reference~\cite{2016RPPh...79g6901S} for a review), which have been argued to be necessary seeds for the observed galactic magnetic fields~\cite{2001SSRv...99..243B}. These then underwent dynamo and/or turbulence enhancement from their initial values (\cf~Reference~\cite{2005PhR...417....1B}). Furthermore, also the weak magnetic fields around $10^{-16}\,$Gau{\ss}, coherent on megaparsec scales, hosted by the intergalactic medium in voids~\cite{2010Sci...328...73N}, might be explained. This has recently been studied in Reference~\cite{PMF-PBH}.

Besides, Reference~\cite{2022PhRvL.129f1302D} points out that the magnetic field lines emerging from vortical PBHs might account for the electromagnetic counterparts to gravitational radiation of merging compact bodies around a solar mass, which are usually thought to be neutron stars~\cite{2017ApJ...848L..20M}. These could conceivably be PBHs, since gravitational-wave observatories are not yet sensitive enough to sufficiently resolve the information on the compactness of these objects, their classification being merely by mass.

Recently, a simulation of actual vortex formation in merger processes has been performed~\cite{Dvali:2023qlk}. It has been found that vorticity can have a large impact on the emitted radiation as well as on the charge and angular momentum of the final black hole. This could potentially leave unique signatures detectable with future gravitational-wave searches, such as suppressed or delayed emission, which in turn could serve as a portal to macroscopic quantum effects in black holes.
\newpage
\vphantom{.}
\newpage

\section{Constraints}
\label{sec:Constraints-on-Primordial-Black-Holes}
\begin{figure}
	\vs{-0.5mm}
	\centering
	\includegraphics[width = 0.95 \textwidth]{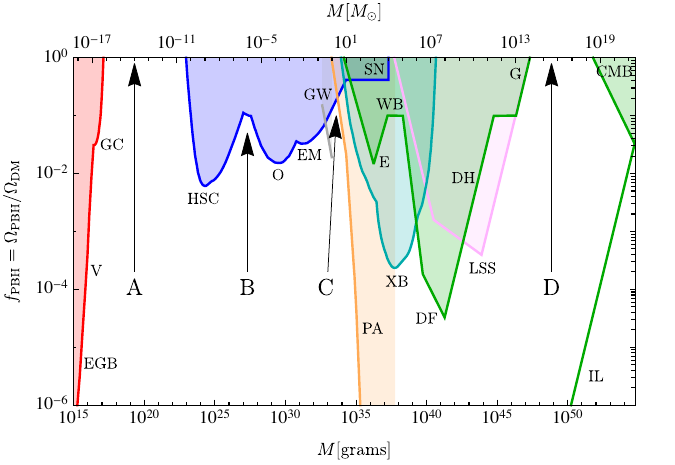}
	\vs{-0.5mm}
	\caption{
        Constraints on the primordial black hole dark matter fraction $f_{\PBH}$ for a {\it monochromatic} mass function. The individual bounds are from evaporation (red), lensing (blue), gravitational waves (GW) (grey), dynamical effects (green), accretion (light blue), CMB distortions (orange) and large-scale structure (purple). The evaporation limits come from the extragalactic $\gamma$-ray background (EGB), the Voyager positron flux (V) and annihilation-line radiation from the Galactic centre (GC). The lensing constraints derive from microlensing of supernov{\ae} (SN) and of stars in M31 by Subaru (HSC), the Magellanic Clouds by the {\it Exp{\'e}rience pour la Recherche d'Objets Sombres} (EROS) and {\it Massive Compact Halo Object} (MACHO) collaborations (EM), and the {\it Galactic bulge by the Optical Gravitational Lensing Experiment} (OGLE) (O). The dynamical bounds are from wide binaries (WB), star clusters in Eridanus II (E), halo dynamical friction (DF), galaxy tidal distortions (G), heating of stars in the Galactic disk (DH) and the cosmic microwave background dipole (CMB). The large-scale structure (LSS) limits are due to the requirement that various cosmological structures do not form earlier than observed. The accretion constraints derive from X-ray binaries (XB) and Planck measurements of cosmic microwave background distortions (PA). Finally, the {\it incredulity limits} (IL) correspond to one PBH per relevant environment (galaxy, cluster, Universe). Note that these are actually {\it lower} bound. The four mass windows (A, B, C, D) indicate regions in which PBHs could have an appreciable density, assuming the validity of the mentioned constraints. Figure from Reference~\cite{2020ARNPS..70..355C} (see also Figure~10 of Reference~\cite{2021RPPh...84k6902C} for a more detailed representation of monochromatic PBH abundance constraints).
        }
	\label{fig:contraints-large}
\end{figure}

\begin{figure}
	\vs{1mm}
	\centering
	\includegraphics[width = 0.9 \textwidth]{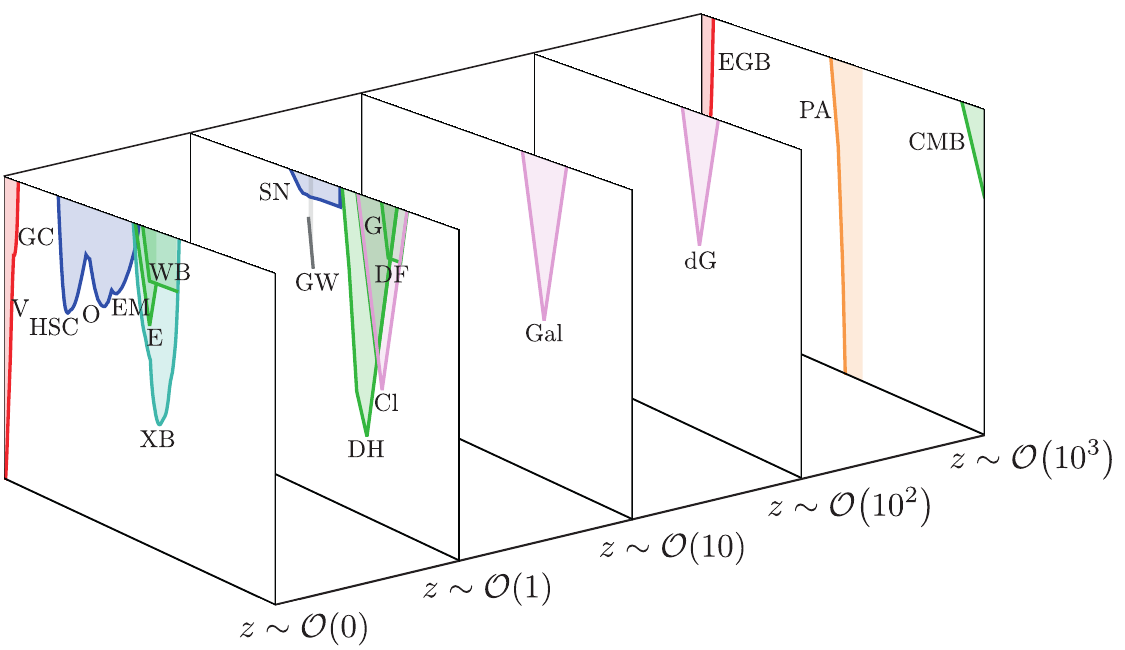}
	\caption{
        Separation of the (monochromatic) primordial black hole constraints of Figure~\ref{fig:contraints-large} into various redshift bins. The individual contributions to the large-scale structure limit are shown separately [clusters (Cl), Milky Way galaxies (Gal) and dwarf galaxies (dG)\hs{-0.3mm}]. See caption of Figure~\ref{fig:contraints-large} for further specifications. Figure (adapted) from Reference~\cite{2020ARNPS..70..355C} (originally inspired by Figure~5 of Reference~\cite{Garcia-Bellido:2018leu}).
        } 
	\label{fig:PBH-constraints-for-different-Redshift}
\end{figure}

\vs{-3mm}
\lettrine[lines=3, slope=0em, findent=0em, nindent=0.2em, lhang=0.1, loversize=0.1]{O}{}\,bservational aspects of primordial black holes involve constraints on their abundance, primarily originating from their (seeming) non-observation. This Section provides a respective overview.\footnote{\setstretch{0.9}For an extensive discussion of PBH constraints, we refer the reader to the specialised review~\cite{2021RPPh...84k6902C} as well as to general PBH reviews, such as References~\cite{Carr:2021bzv, 2020ARNPS..70..355C}. Our discussion partly follows that of the pedagogical Les Houches lecture notes~\cite{Carr:2021bzv}.} Depending on the mass of the PBHs, they manifest themselves through various effects, which have led to the formulation of limits on their abundance. Here, we will mostly focus on those bounds in the mass range $10^{-18}\,\text{--}$ $10^{22}\.\Msun$, which derive from evaporation, gravitational lensing, accretion, and gravitational waves. The limits for monochromatic (single-mass) PBH mass functions are summarised in Figure~\ref{fig:contraints-large}. Figure~\ref{fig:PBH-constraints-for-different-Redshift} details the redshift dependence of the relevant observations.\footnote{\setstretch{0.9}A more detailed form of the constraints can be found in Figure 10 of Reference~\cite{2021RPPh...84k6902C}
.} We should stress that 
\vs{1mm}
\begin{tcolorbox}
\centering
    {\it all constraints have varying degrees of uncertainty and all come with caveats. Some of these limits might be relaxed in the near future, or even disappear entirely.} 
\end{tcolorbox}
\vs{-3mm}

\subsection{Evaporation}
\label{sec:Evaporation-Constraints}
\vs{-1mm}
Amongst the strongest constraints on the PBH abundance are those deduced from non-observation of $\gamma$-rays which originate from PBH evaporation. However, as discussed in the previous Section, most of these constraints assume the validity of semiclassical Hawking radiation for a significant part of the black hole evaporation process. Let us again remark that at latest at Page time~\cite{1993PhRvL..71.3743P} one expects strong deviations from the mentioned semiclassical dynamics (see the discussion in Reference~\cite{2020PhRvD.102j3523D} which indicates that Hawking radiation slows down or might even come entirely to halt). This could cause a significant weakening of the abundance limits, which might even disappear entirely.

If one nevertheless assumes the validity of standard Hawking radiation until complete evaporation, a PBH would evaporate on a timescale $\tau \propto M^{3}$, with $M$ being its initial mass. For masses below $M_{*} \approx 5 \times 10^{14}\.\grm$, this is less than the present age of the Universe~\cite{2016PhRvD..94d4029C}. Observations of the extragalactic $\gamma$-ray background yield very strong constraints on the PBH abundance~\cite{1976ApJ...206....1P}. For $M > 2\.M_{*}$, the instantaneous spectrum for primary (non-jet) photons results in the constraint~\cite{2010PhRvD..81j4019C}
\begin{align}
\label{eq:photon2}
    f( M )
	   < 
            2 \times 10^{-8}
            \big(
				M / M_{*}
			\big)^{\mspace{-2mu}3 + \epsilon}
            \quad
            \text{for $M > M_{*}$}
            \, ,
\end{align}
with $\epsilon$ between $0.1$ and $0.4$. Figure~\ref{fig:contraints-large} shows this constraint for the choice of $\epsilon = 0.2$ (red region). Further evaporation constraints use positron data from Voyager $1$ in order to constrain evaporating PBHs of mass $M < 10^{16}\.\grm$~\cite{2019PhRvL.122d1104B}. Using measurements of the $511\.{\rm keV}$ annihilation-line radiation from the Galactic centre, Laha~\cite{2019PhRvL.123y1101L} and DeRocco \& Graham~\cite{2019PhRvL.123y1102D} constrain $10^{16}\,\text{--}\,10^{17}\.\grm$ PBHs.\footnote{\setstretch{0.9}Anchordoqui {\it et al.}~\cite{2022PhRvD.106h6001A} (see also Reference~\cite{2023PhLB..84037844A}) have recently studied an extra-dimensional scenario, connected to the so-called {\it swampland}~\cite{2005hep.th....9212V}, which also yields a possible explanation for the mentioned $511\.{\rm keV}$ $\gamma$-ray line as well as a relaxation of the PBH evaporation constraints.} Other limits concern $\gamma$-ray and radio observations of the Galactic centre~\cite{2020PhRvD.101l3514L, 2020MNRAS.497.1212C} and the ionising effect of $10^{16}\,\text{--}\,10^{17}\.\grm$ PBHs~\cite{2015JCAP...01..041B}. In Reference~\cite{2020PhRvL.125j1101D}, the effect of PBH spin on these object's evaporation rate has been studied using searches for neutrinos and positrons in the ${\rm MeV}$ energy range, and it has been claimed that spinning PBHs can be probed up to slightly higher masses as compared to non-spinning ones. The authors of Reference~\cite{2022PhRvD.105j3026S} derive constraints and detection prospects of spinning as well as non-spinning PBH dark matter using the global $21$-${\rm cm}$ signal. Detection prospects of asteroidal-mass PBH dark matter with near-future ${\rm MeV}$ telescopes, such as the \emph{All-sky Medium Energy Gamma-ray Observatory} (AMEGO)~\cite{2019BAAS...51g.245M}, have been investigated in Reference~\cite{2021PhRvD.104b3516R}.

\subsection{Lensing}
\label{sec:Lensing-Constraints}
\vs{-1mm}
Observations of Andromeda with the {\it Subaru Hyper Suprime-Camera} (HSC) severely limit the PBH abundance in the mass range $10^{-10} \.\Msun < M < 10^{-6}\.\Msun$. This is shown in Figure~\ref{fig:contraints-large}, which also includes constraints from 
    ({\it i$\mspace{1.5mu}$}) 
        microlensing observations of stars in the Large and Small Magellanic Clouds which probe the fraction of the Galactic halo in PBHs~\cite{1986ApJ...304....1P}, 
    ({\it ii$\mspace{1.5mu}$}) 
        the MACHO project which detected lenses with $M \sim 0.5\.\Msun$ and their halo contribution could be at most $\Ocal( 10 )\mspace{0.5mu}\%$~\cite{2006A&A...454..185H}, 
    ({\it iii$\mspace{1.5mu}$}) 
        the EROS project, which excluded PBHs of mass $6 \times 10^{-8}\.\Msun < M < 15\.\Msun$, as well as 
    ({\it iv}) 
        the OGLE experiment~\cite{2009MNRAS.397.1228W, 2009MNRAS.400.1625C, 2010MNRAS.407..189W, 2011MNRAS.413..493W, 2011MNRAS.416.2949W}, which constrains the PBH abundance in the range $0.1\.\Msun\!<\!M\!<\!20\.\Msun$. Recently, by combining EROS and MACHO data, Reference~\cite{2022A&A...664A.106B} extends the previous limits up to $1000\.\Msun$. Furthermore, Reference~\cite{2009ApJ...706.1451M} suggests a limit~ $f( M ) < 1$ for $10^{-3}\.\Msun < M < 60\.\Msun$, although these surveys may also provide positive primordial black holes evidence. 
The authors of Reference~\cite{2023JCAP...03..043C} studied the gravitational microlensing produced by PBHs which are surrounded by a spike of particle dark matter (see Section~\ref{sec:Primordial-Black-Holes-and-Particle-Dark-Matter}). Using data from OGLE and Subaru/HSC Andromeda observations, improved PBH abundance constraints have been obtained, suggesting that these could be both strengthened and shifted by particle dark matter halos.

\subsection{Dynamical}
\label{sec:Dynamical-Constraints}
\vs{-1mm}
Dynamical constraints have mostly been formulated for heavier black holes~\cite{1999ApJ...516..195C}, whose passage near or through various astronomical objects might lead to their destruction. Let $M_{\crm}$, $R_{\crm}$, $v_{\crm}$ and $t$ be their mass, radius, velocity dispersion and survival time, respectively. Then, PBHs with density $\rho$ and velocity dispersion $v$ yields the constraint~\cite{1999ApJ...516..195C}
\begin{align}
\label{eq:carsaklim}
	f( M ) < 
			\begin{cases}
				M_{\crm}\.v / ( G\.M \rho\,t\.R_{\crm} ) 
					&
                    \text{\!\!\!for $M < M_{\crm}( v / v_{\crm} )$}
                    \, ,
					\\[1.5mm]
				M_{\crm} / ( \rho\,v_{\crm}\.t\.R_{\crm}^{2} )
					&
					\text{\!\!\!for $M_{\crm}(v / v_{\crm} ) < M < M_{\crm}
						( v / v_{\crm} )^{3}$}
                    \, ,
					\\[1.5mm]
				M\.v_{\crm}^{2} / 
				\big(
					\rho\.R_{\crm}^{2}\.v^{3}\.t
				\big)\.
				\exp\!
				\big[
					( M / M_{\crm} )
					( v_{\crm} / V )^{3}
				\big]
					&
					\text{\!\!\!for $M
						 > 
						M_{\crm}( v / v_{\crm} )^{3}$}
                        \, .
			\end{cases}
\end{align}
The above limits correspond to disruption by multiple encounters, one-off encounters and non-impulsive encounters, respectively. As shown by Carr \& Sakellariadou~\cite{1999ApJ...516..195C}, they apply if there is at least one PBH within the relevant environment; this limit is termed {\it incredulity} limit and corresponds to the condition $f( M ) > ( M /\.M_{\Erm} )$, where $M_{\Erm}$ is the mass of the environment. This can be around $10^{12}\.\Msun$ for halos, $10^{14}\.\Msun$ for clusters and $10^{22}\.\Msun$ for the Universe.

The authors of References~\cite{1985ApJ...290...15B, 1987ApJ...312..367W} apply this argument to wide binaries in the Milky Way, since these are particularly vulnerable to disruption by PBHs. Equation~\eqref{eq:carsaklim} gives a constraint $f( M ) < ( M / 500\.\Msun )^{-1}$ for before flattening off at $M \gtrsim 10\.\Msun$ (\cf~Reference~\cite{2014ApJ...790..159M}, and also Reference~\cite{2009MNRAS.396L..11Q} for the original analysis). A related argument for the survival of globular clusters against tidal disruption by passing PBHs yields a limit $f( M ) < ( M / 3 \times 10^{4}\.\Msun )^{-1}$ for $M < 10^{6}\.\Msun$~\cite{1999ApJ...516..195C}. Similarly, using the fact that a star cluster near the centre of the dwarf galaxy Eridanus II has not been disrupted by halo objects, Reference~\cite{2016ApJ...824L..31B} derived an upper limit of $5\.\Msun$. Using Segue 1 as an example, the authors of Reference~\cite{2017PhRvL.119d1102K} exclude the possibility of more than $4\mspace{0.5mu}\%$ of the dark matter being PBHs of around $10\.\Msun$. Figure~\ref{fig:contraints-large} includes this limit.

As shown by Lacey \& Ostriker~\cite{1985ApJ...299..633L}, halo objects will overheat the stars in the Galactic disc unless $f( M ) < ( M / 3 \times 10^{6}\.\Msun )^{-1}$ for $M < 3 \times 10^{9}\.\Msun$, but for $M > 3 \times 10^{9}\.\Msun$, the incredulity limit, $f( M ) < ( M / 10^{12}\.\Msun )$, takes over. A further limit comes from the fact that halo objects will be dragged into the nucleus of the Milky Way by the dynamical friction of various stellar populations, which would lead to excessive nuclear mass unless $f( M )$ is constrained~\cite{1999ApJ...516..195C}, where it bottoms out at $M \sim 10^{7}\.\Msun$ with a value $f \sim 10^{-5}$.

Another class of limits comes from the survival of galaxies in clusters against tidal disruption by giant cluster PBHs, which yields $f( M ) < ( M / 7 \times 10^{9}\.\Msun )^{-1}$ for $M < 10^{11}\.\Msun$~\cite{1999ApJ...516..195C}. This limit flattens off for $10^{11}\.\Msun < M < 10^{13}\.\Msun$ and then rises as $f( M ) < M / 10^{14}\.\Msun$ due to the incredulity limit. This constraint is included in Figure~\ref{fig:contraints-large} with typical values for the mass and the radius of the cluster. As shown in Reference~\cite{1978ComAp...7..161C}, a population of huge intergalactic (IG) PBHs with density parameter $\Omega_{\rm IG}( M )$ results in the limit $\Omega_{\rm IG} < ( M / 5 \times 10^{15}\.\Msun )^{-1/2}$, which gives the limit on the far right of Figure~\ref{fig:contraints-large} and intersects with the cosmological incredulity limit at $M \sim 10^{21}\.\Msun$.

By requiring that various types of structure do not form too early through their `seed' or `Poisson' effect, Carr \& Silk~\cite{2018MNRAS.478.3756C} place limits on the fraction of dark matter in PBHs. For instance, for Milky-Way-type galaxies with a typical mass of $10^{12}\.\Msun$ which must not bind before a redshift $z_{\Brm} \sim 3$, one obtains
\begin{align}
\label{eq:galaxy}
	f( M )
		 < 
			\begin{cases}
				( M / 10^{6}\.\Msun )^{-1}
					&
                    \text{for $10^{6}\.\Msun < M \lesssim 10^{9}\.\Msun$}
                    \, ,
					\\[1.5mm]
				M / 10^{12}\.\Msun
					&
                    \text{for $10^{9}\.\Msun \lesssim M < 10^{12}\.\Msun$}
					\, .
			\end{cases}
\end{align}
\newpage

\noindent Here, the second expression corresponds to having one PBH per galaxy. The above constraint bottoms out at $M \sim 10^{9}\.\Msun$ with a value $f \sim 10^{-3}$. Analogous constraints can be derived for dwarf galaxies and clusters of galaxies. The results are shown in Figure~\ref{fig:contraints-large}. We note that also the Lyman-alpha forest is influenced by the Poisson effect~\cite{2003ApJ...594L..71A, Murgia:2019duy}.
\vs{-4mm}

\subsection{Accretion}
\label{sec:Accretion-Constraints}
\vs{-1mm}
The first study of accretion by primordial black holes dates back to the early 1980s~\cite{1981MNRAS.194..639C}, with numerous subsequent works (see \eg~References~\cite{2007ApJ...665.1277M, 2008ApJ...680..829R, 2007ApJ...662...53R, 2008ApJ...680..829R, 2017PhRvD..95d3534A, 2017PhRvD..96h3524P, 2020PhRvR...2b3204S, 2017PhRvL.118x1101G, 2019JCAP...06..026M, 2017JCAP...10..034I, 2020PhRvD.102d3505D, 2022PhLB..83237265D}) pointing out that the accretion of background gas by PBHs could have a large luminosity which consequently imposes strong constraints on their number density. Particularly, Poulin {\it et al.}~\cite{2017PhRvD..96h3524P, 2020PhRvR...2b3204S} argue for disk instead of spherical accretion, and exclude monochromatic PBH distributions with masses above $2\.\Msun$ as the dominant form of dark matter. This provides the currently strongest accretion constraint, and so we include it in the conservative accretion constraint overview in Figure~\ref{fig:contraints-large}. This figure also includes a constraint coming from PBH interactions with the interstellar medium which would result in a significant X-ray flux, thereby contributing to the observed number density of compact X-ray objects in galaxies. As shown by Inoue \& Kusenko~\cite{2017JCAP...10..034I}, this leads to a constraint on the PBH number density in the mass range from a few to $2 \times 10^{7}\.\Msun$.

\begin{figure}[t!]
	\includegraphics[width = 0.80\linewidth]{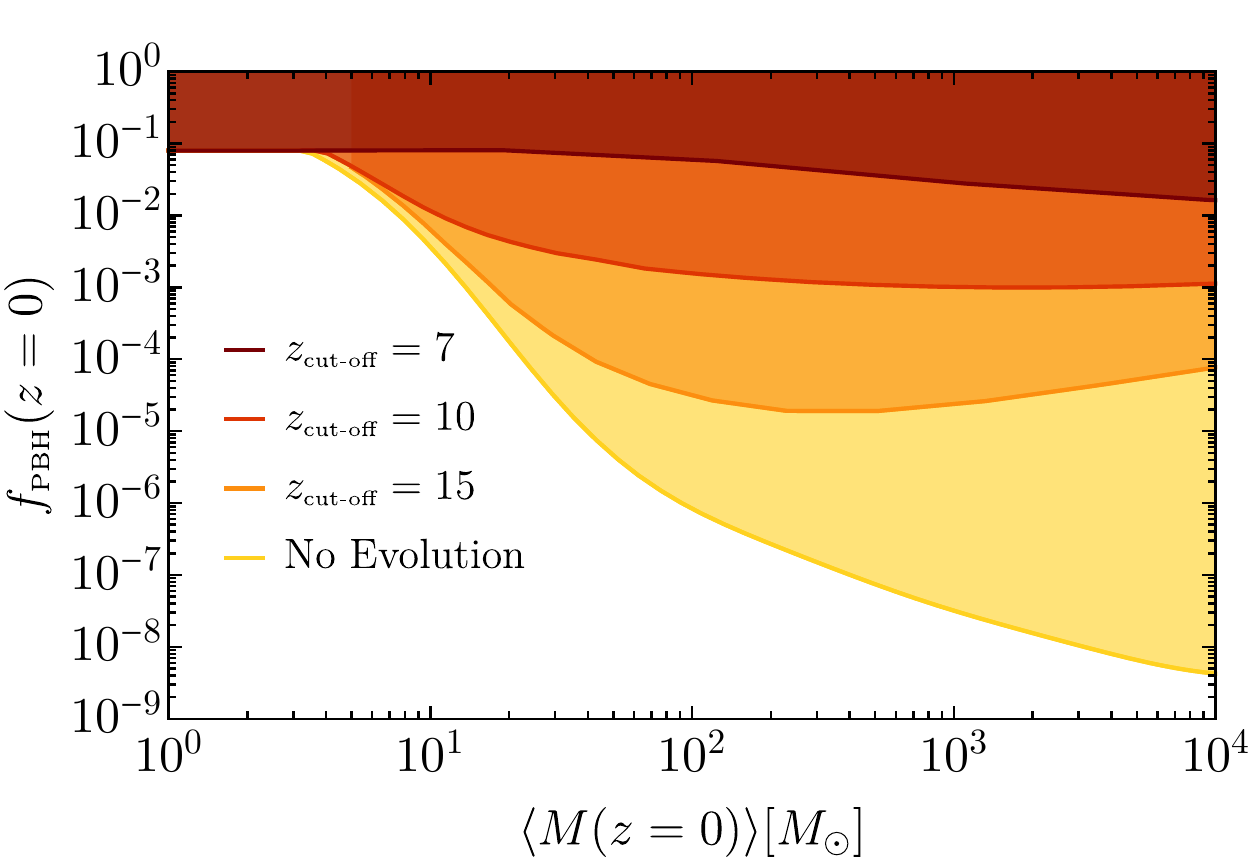}
    \caption{
        Constraints on the current primordial black hole dark matter fraction as function of average mass. Bounds derived using the method of Reference~\cite{2017PhRvD..96b3514C} [\cf~Equation~\eqref{eq:general-constraint}], for different accretion models with redshift cut-offs $z_{\rm cut\text{-}off} = 15$, $10$ and $7$, compared to the case without accretion ('No Evolution'). The mass function is assumed to be extended and to have (at formation) lognormal form with width $\sigma  = 0.5$.$^{\color{midblue}27}$ Figure (adapted) from Reference~\cite{2020PhRvD.102d3505D}.
        \vs{2mm}
        }
	\label{figs:Accretion-redshift-cut-off}
\end{figure}

It must be stressed that accretion constraints are subject to significant levels of uncertainty. Reference~\cite{2020PhRvD.102d3505D} points out that structure formation may lead to an increase in the peculiar velocity of the PBHs when these fall into the potential wells of the merging structure. This increase, together with reionisation and global feedback, then leading to a decrease of the accretion rate~\cite{2008ApJ...680..829R, 2017PhRvD..95d3534A, 2017PhRvD..96l3523A, 2019PhRvD.100h3016H, 2019PhRvD.100h3528I}, which corresponds to an effective accretion cut-off at a certain redshift $z_{\rm cut\text{-}off}$, and consequently to a significant relaxation of the accretion constraints on the PBH abundance. Figure~\ref{figs:Accretion-redshift-cut-off} shows the mentioned effect for the three exemplary values $z_{\rm cut\text{-}off} = 15$, $10$ and $7$ in comparison to the standard case in which accretion is neglected. The first value, which corresponds to the orange curve in Figure~\ref{figs:Accretion-redshift-cut-off}, is argued for in Reference~\cite{2020JCAP...07..022H}.\footnote{\setstretch{0.9}Remarkably, as pointed out by Hasinger~\cite{2020JCAP...07..022H}, many of those accretion constraints are even inconsistent with the {\it observed} abundance of supermassive black holes in galactic centres.} Note that those curves show a sizeable variation; for instance, at current average mass $\langle M( z = 0 ) \rangle \coloneqq  \int \d \ln M\;\psi( M, z = 0 ) = 10^{4}\.\Msun$, the accretion constraints differ by over five orders of magnitude.
\newpage

Furthermore, by using a refined accretion model based on hydrodynamical simulations (\cf~Reference~\cite{2013ApJ...767..163P}), and making conservative assumptions for the emission efficiency, in Reference~\cite{2023PhRvD.107d3537F} it has recently been shown that previous CMB limits~\cite{2017PhRvD..95d3534A, 2017PhRvD..96h3524P} in the mass range between $10$ and $10^{4}~\Msun$ were up to two orders of magnitude too stringent. These findings ensure the viability of the possibility that the black holes detected by the LIGO--Virgo--KAGRA collaboration, and also the seeds for the supermassive black holes in Galactic centres, are primordial in nature. In particular, the thermal-history scenario~\cite{Carr:2019kxo} (see Section~\ref{sec:Thermal--History--Induced-Mass-Function}){\,---\,}even in its simplest formulation without spectral running{\,---\,}does not lead to the violation of any accretion constraint.

\footnotetext{Note that these constraints might change significantly when evaluated for different mass functions such as those naturally shaped by the thermal history of the Universe (see Section~\ref{sec:Thermal--History--Induced-Mass-Function}).}
\newpage

\subsection{Cosmic Microwave Background}
\label{sec:Cosmic-Microwave-Background}
\vs{-1mm}
For light primordial black holes, \ie~less than $10^{9}\.\grm$, Zel'dovich {\it et al.}~\cite{1977PAZh....3..208Z} derive a constraint on their abundance originating from Hawking radiation which contributes to the photon-to-baryon ratio $\eta$. This leads to the constraint on the fraction $\beta( M )$ of collapsed horizon patches:
\begin{align}
    \beta( M )
        <
            10^{-5}\,\eta\,\sqrt{\gamma}\,
            ( g_{*\.\irm} / 106.75 )^{1/4}\,
            ( h / 0.67 )^{2}\,
            ( M / \eta\.\grm )^{-1}
            \, ,
\end{align}
with $h \!\coloneqq \!H_{0} / 100\,{\rm km}\,\srm^{-1}\.{\rm Mpc}^{-1}$, $g_{*\.\irm}$ being the number of relativistic degrees of freedom, and $\gamma$ is the ratio of the horizon mass to the mass of the PBH (see Reference~\cite{2021RPPh...84k6902C} for details). Hence, using the observed value $\eta \approx 10^{-9}$ implies that only PBHs below $10^{4}\.\grm$ could generate the entire cosmic microwave background. Furthermore, the damping of small-scale CMB anisotropies from primordial black holes which evaporate after the time of recombination constrains their abundance in relatively narrow a mass interval around $10^{14}\.\grm$ (see References~\cite{1991ApJ...371..447M, 2007PhRvD..76f1301Z, 2006ARA&A..44..415F, 2010PhRvD..81j4019C, 2017JCAP...03..043P, 2018JCAP...03..018S, 2019arXiv190706485P}).

In addition to the above-mentioned entropy constraints, PBH-induced spectral ($\mu$- and $y$-)distortions of the cosmic microwave background have been subject to intense research (see References~\cite{1977PAZh....3..208Z, 1991MNRAS.248...52B, 1993PhRvD..48..543C} for early work, and References~\cite{2012ApJ...758...76C,  2018PhRvD..97d3525N, 2021JCAP...11..054D, 2020JCAP...02..010A, 2022PhRvD.106d3516Y, 2022PhRvD.105j3535Z} for more recent articles). Concretely, as pointed out by Reference~\cite{2018PhRvD..97d3525N}, if PBHs form at early times directly from inhomogeneities, these will dissipate by Silk damping, leading to $\mu$-distortions of the cosmic microwave background which exclude PBHs unless these form through some mechanism unrelated to the primordial fluctuations or if they are highly non-Gau{\ss}ian. As discussed in Section~\ref{sec:Aspects-of-Inflationary-Quantum-Perturbations} the latter condition might actually be the rule rather than the exception, so the exclusion limits originating from spectral distortions of the cosmic microwave background might actually be rather weak.

\subsection{Gravitational Waves}
\label{sec:Gravitational--Wave-Constraints}
\vs{-1mm}
Like stellar black holes, their primordial pendants can undergo merger processes and in turn emit gravitational radiation as discussed in Section~\ref{sec:Gravitational--Wave-Signatures}. Particularly, and different from stellar black holes, they would generate a stochastic background if constituting a sizeable fraction of the dark matter. The first article on constraints on the PBH abundance from non-observation of those signatures has been written by Carr~\cite{1980A&A....89....6C}. This was followed by increased activity on this topic (see \eg~References~\cite{2009PhRvL.102p1101S, 2010PhRvD..81b3527A, 2011PhRvD..83h3521B, Murgia:2019duy, 2016PhRvL.117f1101S, 2017JCAP...09..037R, 2017PhRvD..96l3523A, Wang:2016ana, 2018JCAP...10..043B, 2017JCAP...09..037R, 2019JCAP...02..018R, 2020PhRvD.101d3015V, 2020PhRvL.124y1101C, 2020JCAP...06..044D, 2020PhRvL.124y1101C, 2021EPJC...81..999H, 2022PhLB..83337332H, 2021Univ....7..398D, 2021JCAP...04..062D, 2022PhRvD.105f2008M, 2023MNRAS.518..149Z, 2022PhRvD.106l3526F}), including various origins of gravitational waves from forming or merging PBHs. Exemplarily, Figure~\ref{fig:contraints-large} includes the constraints obtained by Raidal {\it et al.}~\cite{2017JCAP...09..037R}, who derive strong limits on $f_{\PBH}$ in the mass range $0.5\,\text{--}\,30\.\Msun$ by considering the confirmed binary black hole mergers the first observational run of LIGO/Virgo, and comparing this to the observable merger rate of PBHs. Note that this constraint does not extend up to $f_{\PBH} = 1$. The reason is that for sufficiently high PBH densities, tidal disruption decreases the number of completed merger processes. Spatial (Poisson) clustering of PBHs is the rule rather than the exception (see Section~\ref{sec:Clustering-of-Primordial-Black-Holes}), which implies that concentrations of enhanced PBH density can be expected to occur frequently. Since these have mostly not been taken into account, together with the fact that, due to the complexity of the PBH clustering dynamics, this topic has not been finally addressed with high precision, constraints on the PBH abundance deriving from their non-observation are currently rather uncertain.

\begin{figure}[t!]
\begin{center}
    \includegraphics[width=0.796\textwidth]{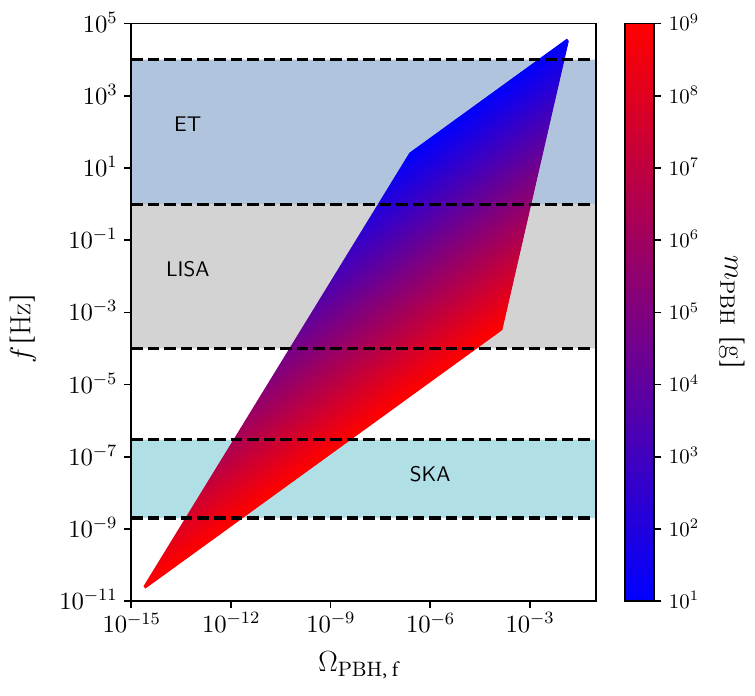}
    \caption{
        Peak frequency of gravitational waves induced by a gas of primordial black holes as a function of their normalised energy density at formation, $\Omega_{\PBH,\mspace{1.5mu}\frm}$ (horizontal axis), and their mass $M$ (colour bar). The coloured region indicates parameters for which the black holes
        ({\it i$\mspace{1.5mu}$}) 
            dominate the Universe for a transient period, \ie~have formed after inflation and then (semiclassically) Hawking-evaporated before big bang nucleosynthesis, and
        ({\it ii$\mspace{1.5mu}$})
            induce gravitational waves which do not lead to a backreaction problem.
        Also included are detection bands of {\it Einstein Telescope} (ET), {\it Laser Interferometer Space Antenna} (LISA) and {\it Square Kilometer Array}~(SKA). Figure from Reference~\cite{2021JCAP...03..053P}.
        }
    \label{fig:GW frequency}
\end{center}
\end{figure}

As regards the ultralight mass range, Reference~\cite{2021JCAP...03..053P} has studied gravitational waves originating from the gravitational potential of a gas of primordial black holes. Utilising Poisson fluctuations in their number density, which underlie small-scale density perturbations and in turn lead to the production of gravitational waves at second order, led the authors to formulate the first constraints on PBHs with masses below $10^{9}\.\grm$ (see Figure~\ref{fig:GW frequency}).

\subsection{Extended Mass Functions}
\label{sec:Constraints-for-Extended-Mass-Functions}
\vs{-1mm}
Most of the constraints on the primordial black hole abundance are derived for monochromatic mass functions, \ie~for the case in which all PBHs have the same (or a very similar) mass. This is also the underlying assumption for the constraints shown in Figure~\ref{fig:contraints-large}. However, {\it this assumption is completely wrong.} None of the many PBH formation scenarios as presented in Section~\ref{sec:Primordial-Black-Hole-Formation-Time} yields a monochromatic power spectrum of primordial density perturbations. Even if they did, the critical nature of the gravitational collapse~\cite{1993PhRvL..70....9C} to PBHs implies that their mass distribution is extended, which broadens {\it any} PBH mass distribution (see \eg~Reference~\cite{2016EPJC...76...93K}), making strict monochromaticity simply impossible. As pointed out in Reference~\cite{2016PhRvD..94f3530G}, even though the very extendedness of the PBH mass function allows to circumvent certain constraints, it might cause violations at other scales.

The first comprehensive reanalysis of constraints for an extended PBH mass function has been performed by K{\"u}hnel \& Freese~\cite{2017PhRvD..95h3508K}, which was followed by the work of Carr {\it et al.}~\cite{2017PhRvD..96b3514C} who utilised the spectral PBH density
\begin{align}
	\psi( M )
		\propto
			M\,\drm n / \drm M
			\, ,
\end{align}
where $n$ is the PBH number density, normalised such that the total PBH dark matter fraction
\vs{-1mm}
\begin{align}
	f_{\PBH}
		\coloneqq
			\frac{ \Omega_{\PBH} }
            { \Omega_{\text{DM}} }
		 = 
			\int_{M_{\umin}}^{M_{\umax}}\drm M\;
            \psi( M )
			\, .
\end{align}
Mean and variance are often useful quantities to characterise the mass function,
\begin{align}
	\log M_{\crm}
		\coloneqq
			\langle \log M \rangle^{}_{\psi}
			\, ,
	\quad
	\sigma^{2}
		\coloneqq
			\langle
				\log^{2} M
			\rangle^{}_{\psi}
			-
			\langle
				\log M
			\rangle_{\psi}^{2}
			\, ,
\end{align}
\newpage

\noindent where
\vs{-1mm}
\begin{align}
	\langle X \rangle^{}_{\psi}
		\coloneqq
			f_{\PBH}^{-1}\,
			\int \drm M\;\psi( M )\,X( M )
			\, .
\end{align}
The characterisation by, and use of, $M_{\crm}$ and $\sigma$ are particularly convenient if $\psi$ is lognormal.\footnote{\setstretch{0.9}In realistic cases, however, two parameters are insufficient to describe a PBH mass function.} For given monochromatic constraints with $f( M ) < f_{\umax}( M )$ one obtains
\vs{-1mm}
\begin{align}
\label{eq:general-constraint}
	\int \drm M\;
	\frac{ \psi( M ) }
	{ f_{\umax}( M ) }
		\leq
			1
			\, .
\end{align}
From $f_{\umax}$ it is possible to apply Equation~\eqref{eq:general-constraint} for any PBH mass function in order to obtain constraints similarly to those for a monochromatic mass function. This can then be plotted in terms of $M_{\crm}$ and $\sigma$, but, of course, has to be reimplemented for each utilised mass function separately.

\subsection{Constraints on Primordial Perturbations}
\label{sec:Constraints-on-Primordial-Perturbations}
\begin{figure}[p]
    \centering
    \includegraphics[width=0.95\hsize]{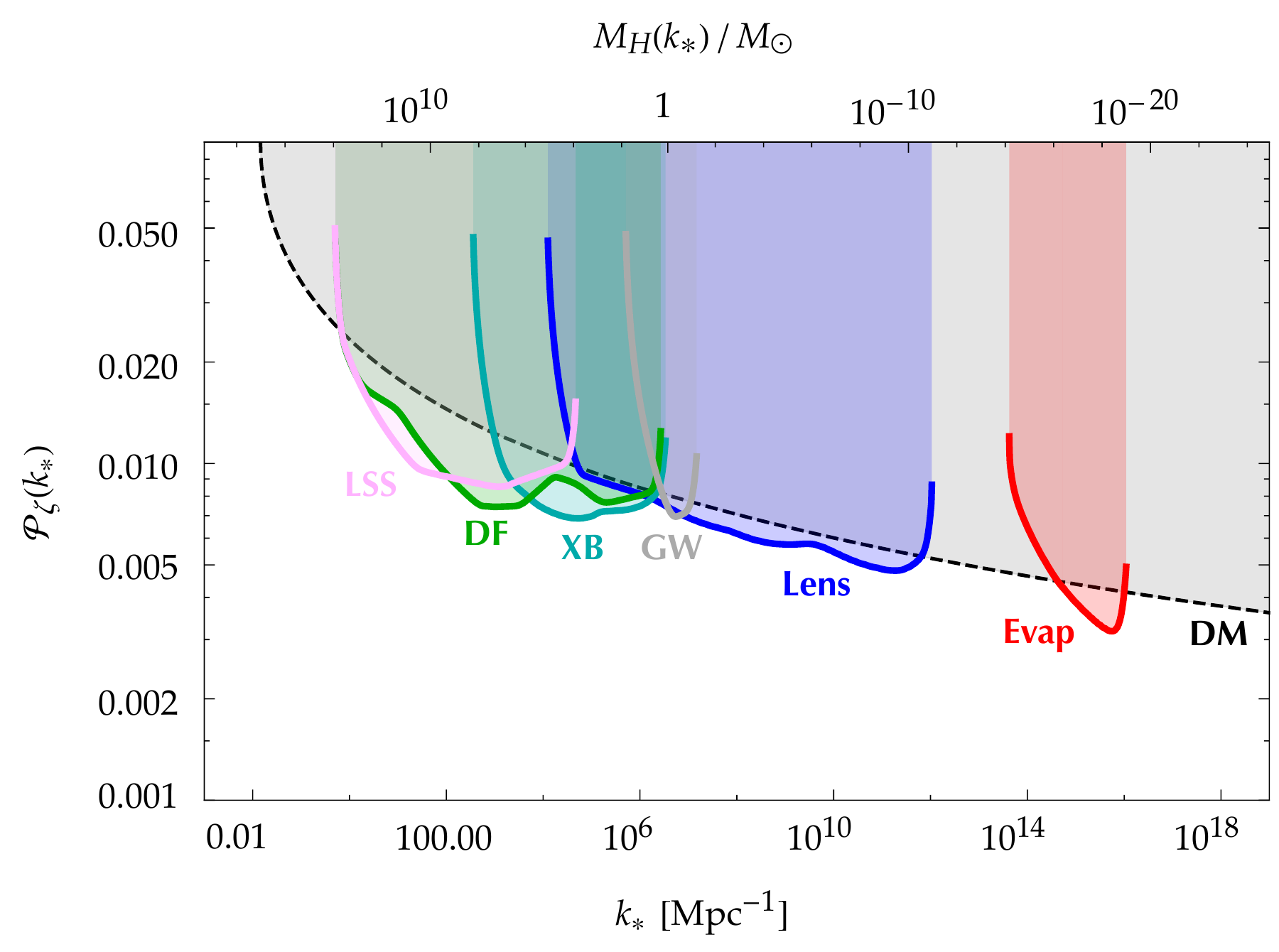}
    \caption{
        Primordial black hole constraints (using same colour coding as in Figure~\ref{fig:contraints-large}) as the upper bound on the amplitude $\Pcal_{\zeta}( k_{*} )$ of the monochromatic primordial perturbation~\eqref{eq: monochromatic-constraint} with respect to the peak scale $k_{*}$. The black dashed line corresponds to $f_{\PBH} = 1$. The horizon mass $M_{H}$~\eqref{eq: horizon mass} associated with $k_{*}$ is also shown as a brief indicator of the PBH mass. This figure corresponds to Figure~19 of Reference~\cite{2021RPPh...84k6902C}, but is updated by the state-of-the-art PBH estimation scheme, \ie~the $q$-parameter criterion (Section~\ref{sec:Case-C:-Non--Gaussian-Contribution-to-zeta}) and the peak-theory procedure (Section~\ref{sec:Peak--Theory-Procedure-with-Curvature-Peaks}).
        }
    \label{fig: Pconst PBH}
\end{figure}

\vs{-1mm}
As primordial perturbations with an amplitude above a certain value inevitably overproduce primordial black holes, observational constraints their abundance can be recast into the constraint on primordial perturbations. For example, the Gau{\ss}ian curvature perturbation with a monochromatic peak power spectrum,
\begin{align}
\label{eq: monochromatic-constraint}
    \Pcal_{\zeta}( k )
        =
            \Pcal_{\zeta}( k_{*} )\.
            \delta\big( \!\ln[ k/k_{*} ] \big)
            \, ,
\end{align}
leads to the PBH mass function shown by the black dashed line in Figure~\ref{fig: fPBHfNL}.\footnote{\setstretch{0.9}The $q$-parameter criterion (Section~\ref{sec:Case-C:-Non--Gaussian-Contribution-to-zeta}) and the peak-theory procedure (Section~\ref{sec:Peak--Theory-Procedure-with-Curvature-Peaks}) have been utilised to estimate this mass function.} With use of the general scheme~\eqref{eq:general-constraint}, the PBH constraints can be translated to constraints on the perturbation amplitude $\Pcal_{\zeta}( k_{*} )$ with respect to the peak scale $k_{*}$ as shown in Figure~\ref{fig: Pconst PBH} using the same colour coding as in Figure~\ref{fig:contraints-large} (see Figure~19 of Reference~\cite{2021RPPh...84k6902C} for an earlier version of a constraint figure for the power spectrum). The horizon mass $M_{H}$~\eqref{eq: horizon mass} corresponding to $k_{*}$ is also shown by the upper horizontal line as a rough indicator of the PBH mass. The black dashed line (DM) corresponds to $f_{\PBH} = 1$; exceeding this line would imply black hole overproduction.

\section{Gravitational-Wave Signatures}
\label{sec:Gravitational--Wave-Signatures}
\vs{-3mm}
\lettrine[lines=3, slope=0em, findent=0em, nindent=0.2em, lhang=0.1, loversize=0.1]{T}{} he first successful direct detection of gravitational waves by the LIGO/Virgo collaboration~\cite{LIGOScientific:2016aoc} has initiated the area of \emph{gravitational-wave astronomy}. Primordial black holes can emit gravitational waves in many ways, being prime observational candidates. This Section is devoted to a respective review.

\subsection{Primordial Black Hole Formation Time}
\label{sec:Primordial-Black-Hole-Formation-Time}
\vs{-1mm}
Already the formation of primordial black holes can be accompanied by gravi-tational-wave emission in several ways. The emitted waves would be in superposition now and could be detected as a stochastic gravitational-wave background. In the case of PBH formation via gravitational collapse of radiation overdensities, the most studied scenario discusses \emph{scalar-induced gravitational waves}~\cite{2009PhRvL.102p1101S, 2010PhRvD..81b3517B, 2010PThPh.123..867S, 2011PhRvD..83h3521B}. While in this case PBH formation is associated with order-unity perturbations, in order for such high peaks to be realised non-negligibly, the typical perturbation amplitudes should also be large enough. If the curvature perturbation is Gau{\ss}ian, the required power spectrum is around $\Pcal_{\zeta} \sim 10^{-2}$, and thus the curvature perturbations typically assume values around $\zeta \sim 0.1$, these being large enough for higher-order perturbative corrections to be relevant. Gravitational waves (tensor perturbations) are decoupled from (scalar) curvature perturbations at linear order in perturbations theory, but can be sourced by second-order effects~\cite{2007PhRvD..75l3518A, 2007PhRvD..76h4019B}. Roughly speaking, the ratio of the induced gravitational-wave energy density to that of the background radiation is of order $\Pcal_{\zeta}^{2}$. This ratio is almost conserved until today as both energy densities decrease as $\propto a^{-4}$, so the current density parameter of induced gravitational waves is estimated as $\Omega_{\GW}\.h^{2} \sim \Pcal_{\zeta}^{2}\,\Omega_{\rrm}\.h^{2} \sim 10^{-9}$, with $\Omega_{\rrm}\.h^{2} \simeq 4.2 \times 10^{-5}$ being the current radiation-density parameter.

The induced gravitational waves have a peak at frequency $\sim f = k / ( 2\mspace{1.5mu}\pi )$ with the wavenumber $k$ of the scalar source perturbation. On the other hand, the horizon mass when the $k$-mode reenters the horizon can be computed as (see \eg~Reference~\cite{2019PhRvD.100b3537T})
\begin{align}
\label{eq: horizon mass}
\begin{split}
    M_{H}( k )
        &\simeq
            \pqty{\frac{g_{*}}{10.75}}^{\mspace{-5mu}-1/6}\mspace{1.5mu}
            \pqty{\frac{k}{4.22 \times 10^{6}\.
            {\rm Mpc}^{-1}}}^{\mspace{-6mu}-2}\,
            \Msun
            \\[3mm]
        &= 
            \pqty{\frac{g_{*}}{10.75}}^{\mspace{-5mu}-1/6}\mspace{1.5mu}
            \pqty{\frac{ f }{ 6.52 \times 10^{9}\.{\rm Hz}}}^{\mspace{-6mu}-2}\,
            \Msun
            \, ,
\end{split}
\end{align}
where $g_{*}$ is the number of effective degrees of freedom for the energy density at horizon reentry, and we assume that it is almost equivalent to that of the entropy density throughout the cosmic history. Through this equation, the PBH mass $\sim M_{H}$ is related to the frequency of induced gravitational waves. In particular, the presently undoubtedly open window in which even PBHs with monochromatic mass functions could constitute the entirety of the dark matter, $\sim\![ 10^{17},\mspace{1.5mu}10^{23} ]\,\grm$, corresponds to $\sim\![ 0.001,\mspace{1.5mu}1 ]\.{\rm Hz}$ which is well covered by the {\it Laser Interferometer Space Antenna} (LISA) (see Section~\ref{sec:Future-Prospects-for-Gravitational--Wave-Searches}).
Note also that stellar-mass PBHs correspond to the nanohertz pulsar-timing-array range. As the NANOGrav collaboration recently reported on a possible common-spectrum signal~\cite{2020ApJ...905L..34A}, primordial black hole explanations are attracting attention (see \eg~References~\cite{2020PhRvL.124y1101C, 2021PhRvL.126e1303V, 2021PhRvL.126d1303D, 2021PhLB..81336040K, 2022SCPMA..6530411D, 2021PhRvL.126m1301I, 2021JCAP...06..022A, 2022JCAP...05..046Y, 2022PhLB..83537542A}).

\begin{figure}
    \centering
    \includegraphics[width = 0.7\hsize]{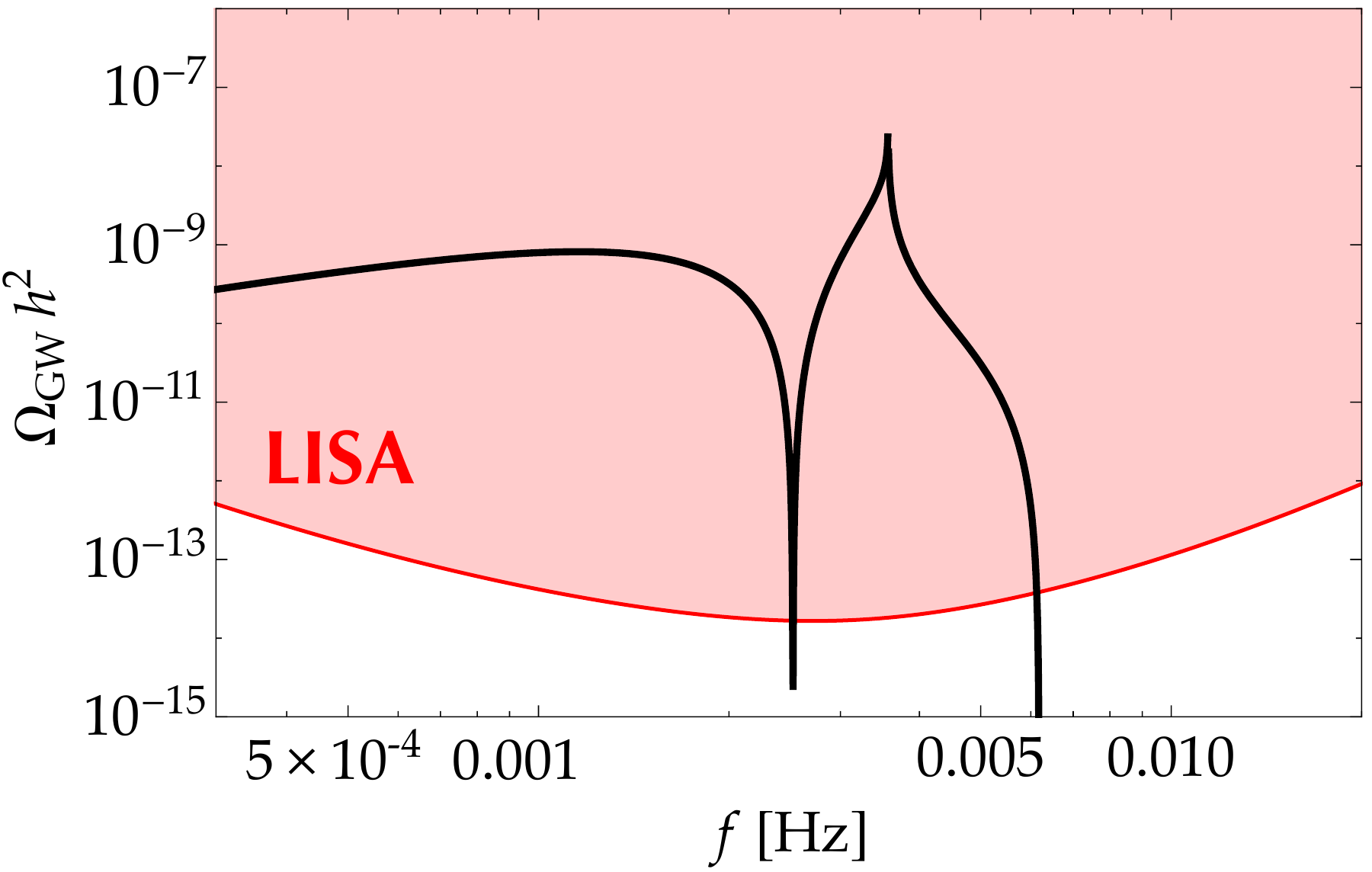}
    \caption{
        Induced gravitational-wave spectrum, $\Omega_{\rm GW}\.h^{2}( f )$, for the monochromatic scalar perturbation~\eqref{eq: monochromatic power} with         $k_{*} = 1.56 \times 10^{12}\.{\rm Mpc}^{-1}$ and $\sigma_{\grm}^{2} = 5.17 \times 10^{-3}$ corresponding to $M_{H} = 10^{22}\.\grm$ and $f_{\PBH} = 1$. The red region indicates the sensitivity of LISA (from Reference~\cite{2020Symm...12.1477S}).
        }
    \label{fig: inducedGW}
\end{figure}

The gravitational-wave amplitude can be calculated as follows (see \eg~References~\cite{2018JCAP...09..012E, 2018PhRvD..97l3532K} for the details). At quadratic order in the scalar perturbation, the linear tensor mode is induced at the quantum-operator level as
\begin{align}
\label{eq: induced GW EoM}
    \partial_{\tau}^{2}
    \hat{h}_{\lambda}( \tau,\mspace{1.5mu}\kbm )
    +
    2\.\Hcal\.
    \partial_{\tau}
    \hat{h}_{\lambda}( \tau,\mspace{1.5mu}\kbm )
    +
    k^{2}\.
    \hat{h}_{\lambda}( \tau,\mspace{1.5mu}\kbm )
        = 
            4\.\hat{S}_{\lambda}( \tau,\mspace{1.5mu}\kbm )
            \, ,
\end{align}
with the source term
\vs{-1mm}
\begin{align}
    \hat{S}_{\lambda}( \tau,\mspace{1.5mu}\kbm )
        = 
            \int\frac{\dd[3]{q}}
            {( 2\mspace{1.5mu}\pi )^{3}}\;
            Q_{\lambda}( \kbm,\mspace{1.5mu}\qbm )\.
            f
            \big(
                \abs{\kbm-\qbm},\mspace{1.5mu}q,\mspace{1.5mu}\tau
            \big)\.
            \hat{\zeta}( \qbm )\.
            \hat{\zeta}( \kbm - \qbm )
            \, .
\end{align}
Here, $\tau$ is the conformal time, $\Hcal = a H = 1 / \tau$ is the conformal Hubble parameter, $\lambda$ represents the two different polarisation patterns '$+$' and '$\times$'. The projection factor $Q_{\lambda}( \kbm,\mspace{1.5mu}\qbm )$ is given by
\vs{-2.5mm}
\begin{align}
    Q_{\lambda}( \kbm,\mspace{1.5mu}\qbm )
        = 
            \frac{q^{2}}
            {\sqrt{2}}\.
            \sin^{2}( \theta )
            \times
            \begin{cases}
                \cos( 2\mspace{1.5mu}\phi )
                    & \text{for $\lambda = +$}
                \, ,
                \\[1mm]
                \sin( 2\mspace{1.5mu}\phi )
                    & \text{for $\lambda = \times$}
                \, ,
    \end{cases}
\end{align}
for the spherical-coordinate expression $\qbm = q\.( \sin\theta\.\cos\phi,\mspace{1.5mu}\sin\theta\.\sin\phi,\mspace{1.5mu}\cos\theta )$ and $\kbm = k\.( 0,\mspace{1.5mu}0,\mspace{1.5mu}1 )$. The function $f( \abs{\kbm-\qbm},\mspace{1.5mu}q,\mspace{1.5mu}\tau )$ includes the linear scalar transfer
\vs{-0.5mm}
\begin{align}
    \Phi( x )
        = 
            -
            \frac{9}{x^{2}}\!
            \left[
                \frac{\sin( x/\sqrt{3}\mspace{1mu} )}
                {x/\sqrt{3}}
                -
                \cos\mspace{-2mu}\big( x/\sqrt{3}\mspace{1mu} \big)
            \right]
            ,
            \\[-9.5mm]
            \notag
\end{align}
as
\vs{-0.5mm}
\begin{align}
    f( p,\mspace{1.5mu}q,\mspace{1.5mu}\tau )
        &= 
            \frac{4}{9}\.
            \Big[
                3\.\Phi( p\.\tau )\.
                \Phi( q\.\tau )
                +
                \Phi^{\prime}( p\.\tau )\.
                \Phi^{\prime}( q\.\tau )
                +
                \Phi( p\.\tau )\.
                \Phi^{\prime}( q\.\tau )
                +
                \Phi^{\prime}( p\.\tau )\.
                \Phi( q\.\tau )
            \Big]
            \, ,
            \\[-3mm]
            \notag
\end{align}
\newpage

\noindent where $\Phi^{\prime}( x ) \equiv \dd{\Phi( x )}/\dd{\ln x}$. The operator equation~\eqref{eq: induced GW EoM} is formally solved as
\begin{align}
    \hat{h}_{\lambda}( \tau,\mspace{1.5mu}\kbm )
        = 
            \frac{4}{a( \tau )}
            \int^{\tau}\dd{\tilde{\tau}}\;
            G_{\kbm}( \tau,\mspace{1.5mu}\tilde{\tau} )\.
            a( \tilde{\tau} )\.
            S_{\lambda}( \tilde{\tau},\mspace{1.5mu}\kbm )
            \, ,
\end{align}
with the Green's function
\begin{align}
    G_{\kbm}( \tau,\mspace{1.5mu}\tilde{\tau} )
        = 
            \sin\!
            \big[
                k\mspace{1.5mu}( \tau - \tilde{\tau} )
            \big] / k
            \, .
\end{align}
Therefore, the tensor two-point function is related to the scalar four-point function:
\begin{align}
    \Big\langle\.
        \hat{h}_{\lambda_{1}}( \tau,\mspace{1.5mu}\kbm_{1} )\.
        \hat{h}_{\lambda_{2}}( \tau,\mspace{1.5mu}\kbm_{2} )
    \Big\rangle
        &= 
            \int\frac{\dd[3]{q_{1}}}{( 2\mspace{1.5mu}\pi )^{3}}
            \int\frac{\dd[3]{q_{2}}}{( 2\mspace{1.5mu}\pi )^{3}}\;
            Q_{\lambda_{1}}( \kbm_{1},\mspace{1.5mu}\qbm_{1} )\.
            Q_{\lambda_{2}}( \kbm_{2},\mspace{1.5mu}\qbm_{2} )
            \notag
            \displaybreak[1]
            \\[2mm]
        &\phantom{\mspace{55mu}}
            \times
            I_{k_{1}}( \abs{\kbm_{1}-\qbm_{1}},\mspace{1.5mu}q_{1},\mspace{1.5mu}\tau )\.
            I_{k_{2}}( \abs{\kbm_{2}-\qbm_{2}},\mspace{1.5mu}q_{2},\mspace{1.5mu}\tau )
            \\[2.5mm]
        &\phantom{\mspace{55mu}}
            \times
            \Big\langle
            \hat{\zeta}( \qbm_{1} )\.
            \hat{\zeta}( \kbm_{1}
            -\qbm_{1} )\.
            \hat{\zeta}( \qbm_{2} )\.
            \hat{\zeta}( \kbm_{2} - \qbm_{2} )
            \Big\rangle
            \, ,
            \notag
\end{align}
with the kernel
\begin{align}
    I_{k}( p,\mspace{1.5mu}q,\mspace{1.5mu}\tau )
        = 
            4\int^{\tau}\dd{\tilde{\tau}}\;
            G_{\kbm}( \tau,\mspace{1.5mu}\tilde{\tau} )\.
            \frac{a( \tilde{\tau} )}
            {a( \tau )}\.
            f( p,\mspace{1.5mu}q,\mspace{1.5mu}\tau )
            \, .
\end{align}
In the subhorizon limit during the radiation-dominated era, the gravitational-wave density parameter is given by
\vs{-2mm}
\begin{align}
\label{eq: OmegaGW in Ph}
    \Omega_{\GW}( \tau,\mspace{1.5mu}k )
        = 
            \frac{1}{24}\.
            \pqty{k\mspace{1.5mu}\tau}^{2}\;
            \overline{\Pcal_{h}( \tau,\mspace{1.5mu}k )}
            \, ,
\end{align}
where the over-line stands for the {\it oscillation average},
\begin{align}
    \overline{X}( t )
        \coloneqq 
            \frac{1}{T}
            \int^{t + T}_{t}\dd{t^{\prime}}\,
            X( t^{\prime} )
            \, ,
\end{align}
with period $T$ of $X$'s oscillation.
Here, we assumed that two polarisation modes $\lambda_{1} = \lambda_{2} = +$ and $\lambda_{1} = \lambda_{2} = \times$ give the same power spectrum $P_{h}(\tau,\mspace{1.5mu}k )$; otherwise ($\lambda_{1} \neq \lambda_{2}$) the two-point function is taken to vanish. In a practical computation, the asymptotic form of the kernel function is useful:
\begin{align}
    k\.\tau\.I_{k}( p,\mspace{1.5mu}q,\mspace{1.5mu}\tau )
        \underset{\tau \to \infty}{\sim}
            \Fcal_{k}( p,\mspace{1.5mu}q )\.
            \Big[
                \Scal_{k}( p,\mspace{1.5mu}q )\sin( k\mspace{1.5mu}\tau )
                +
                \Ccal_{k}( p,\mspace{1.5mu}q )\cos( k\mspace{1.5mu}\tau )
            \Big]
            \, ,
\end{align}
where
\begin{subequations}
\begin{align}
    \Fcal_{k}( p,\mspace{1.5mu}q )
        &= 
            \frac{3\mspace{1.5mu}
            \big(\mspace{0.5mu}
                p^{2}
                +
                q^{2}
                -
                3\mspace{1.5mu}k^{2}
            \big)}
            {p^{3}\mspace{1.5mu}q^{3}}
            \, ,
            \displaybreak[1]
            \\[1.5mm]
    \Scal_{k}( p,\mspace{1.5mu}q )
        &= 
            -\.
            4\.p\.q
            +\!
            \big(\mspace{0.5mu}
                p^{2}
                +
                q^{2}
                -
                3\mspace{1.5mu}k^{2}
            \big)
            \ln
            \abs{
                \frac{3\mspace{1.5mu}k^{2} - ( p + q )^{2}}
                {3\mspace{1.5mu}k^{2} - ( p - q )^{2}}
            }
            \, ,
            \displaybreak[1]
            \\[3.5mm]
        \Ccal_{k}( p,\mspace{1.5mu}q )
            &= 
                -\mspace{1.5mu}
                \pi\mspace{1.5mu}
                \big(\mspace{0.5mu}
                    p^{2} + q^{2}
                    -
                    3\mspace{1.5mu}k^{2}
                \big)\.
                \Theta
                \Big(
                    p + q
                    -
                    \sqrt{3}\.k
                \Big)
                \, .
\end{align}
\end{subequations}
The oscillation average of the cross-correlation hence reads
\begin{align}
    ( k\mspace{1.5mu}\tau )^{2}\;
    \overline{I_{k}( p_{1},\mspace{1.5mu}q_{1},\mspace{1.5mu}\tau )\.
    I_{k}( p_{2},\mspace{1.5mu}q_{2},\mspace{1.5mu}\tau )}
        &= 
            \Big[
                \Scal_{k}( p_{1},\mspace{1.5mu}q_{1} )\.
                \Scal_{k}( p_{2},\mspace{1.5mu}q_{2} )
                +
                \Ccal_{k}( p_{1},\mspace{1.5mu}q_{1} )\.
                \Ccal_{k}( p_{2},\mspace{1.5mu}q_{2} )
            \Big]
            \notag
            \\[2mm]
        &\phantom{=\;}
            \times
            \frac{1}{2}\,
            \Fcal_{k}( p_{1},\mspace{1.5mu}q_{1} )\.
            \Fcal( p_{2},\mspace{1.5mu}q_{2} )
            \, .
\end{align}
It ensures that the tensor density parameter~\eqref{eq: OmegaGW in Ph} converges to a constant in the subhorizon limit. The current tensor-density parameter is roughly estimated by multiplying it by the current radiation density parameter $\Omega_{\rrm}$. An example spectrum using a monochromatic scalar power spectrum is shown in Figure~\ref{fig: inducedGW}.\footnote{\setstretch{0.9}Here, we assume the Bunch--Davies vacuum~\cite{Bunch:1978yq} for the quantum state of the Universe (see Reference~\cite{2022JHEP...03..196F} for gravitational-wave induction in excited states).}

\begin{figure}
    \centering
    \includegraphics[width = 0.74\hsize]{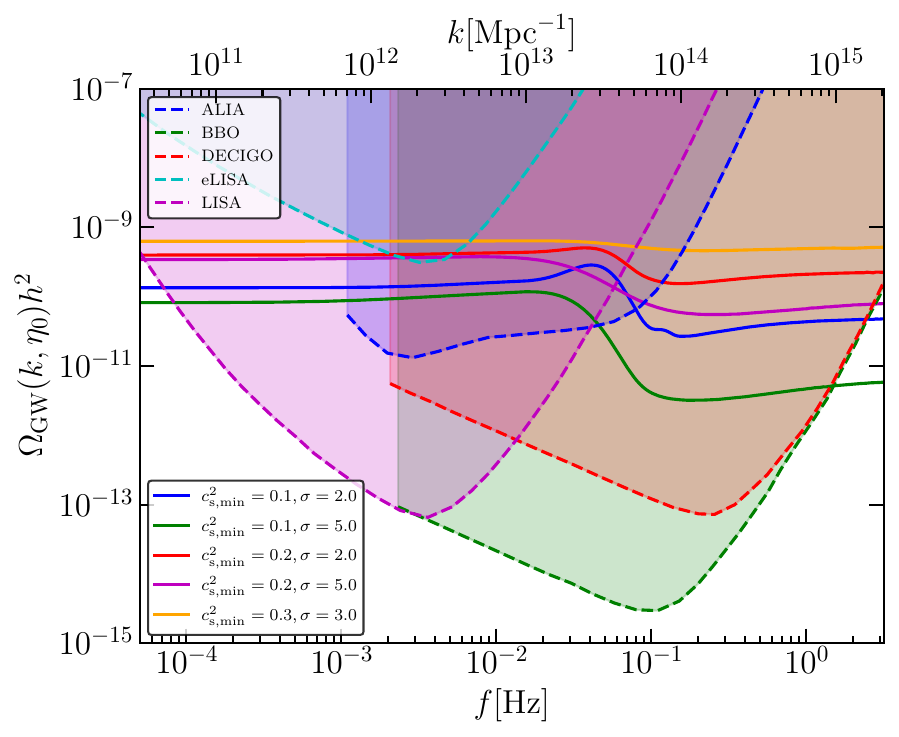}\\[2mm]
    \includegraphics[width = 0.74\hsize]{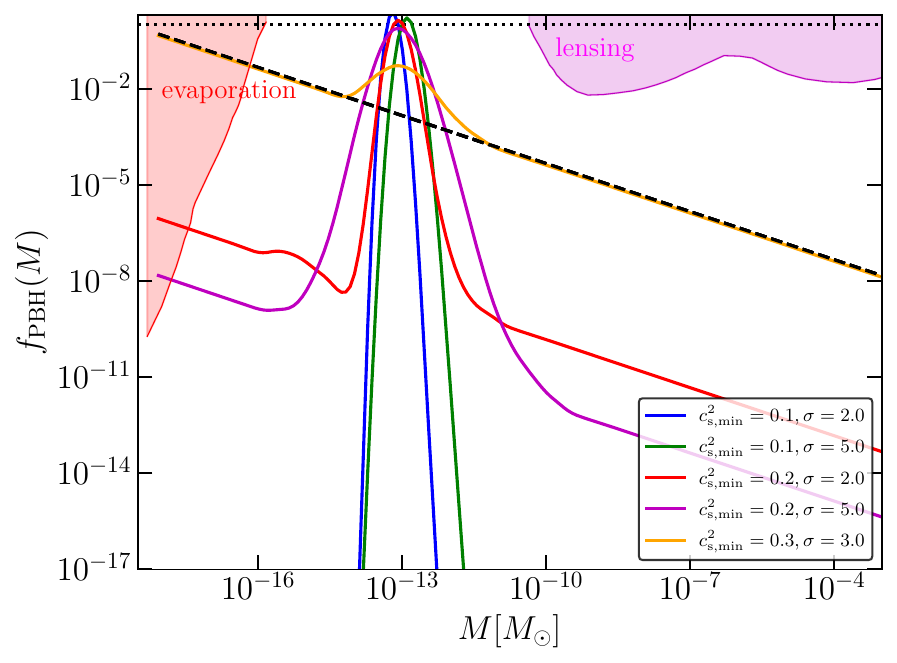}
    \caption{
        Induced gravitational-wave spectrum ({\it top panel}) in the presence of a crossover softening beyond the Standard Model and the corresponding PBH mass function ({\it bottom panel}). The sharp peak in the mass function is caused by the reduction in the threshold due to the softening of $w,\,c_{\srm}^{2}$ (see Reference~\cite{2023arXiv231117760E} for details on the crossover templates for $w,\,c_{\srm}^{2}$). For this particular case, the signal of gravitational waves is detectable with DECIGO/BBO as well as with LISA, with PBHs being compatible with all dark matter. Figures are taken from Reference~\cite{2023arXiv231117760E}.
        }
    \label{fig:inducedGW_thermal}
\end{figure}

Contrary to the abundance of primordial black holes, the primordial non-Gau{\ss}ianity of a scalar perturbation does not directly have a significant effect on the gravitational-wave spectrum. The reason is the following. While the PBH abundance is sensitive to the tail of the scalar probability density, the induced gravitational-wave amplitude is determined by the typical behaviour of the scalar perturbation, and thus the leading-order Gau{\ss}ian contribution dominates. However, the correspondence between the PBH abundance and the gravitational-wave amplitude can be affected. That is, positive/negative non-Gau{\ss}ianity enhances/suppresses the former, requiring a reduction / an increase of the scalar variance $\sigma_{\grm}^{2}$, implying that the corresponding gravitational-wave amplitude becomes smaller/larger. For example, the required amplitude for $f_{\PBH} = 1$ is about $\sigma_{\grm}^{2} \sim 3 \times 10^{-3}$ for $f_{\NL} = 5/2$ while $\sigma_{\grm}^{2} \sim 5 \times 10^{-3}$ in the Gau{\ss}ian case~\cite{2023JCAP...05..044A}. The gravitational-wave amplitude is then reduced by a factor of $~( 3/5 )^{2} \sim 0.4$. Reference~\cite{2019PhRvL.122t1101C} shows that infinitely large $f_{\NL}$ does not infinitely reduce the gravitational-wave amplitude; there is a lower limit because the non-Gau{\ss}ian contribution dominates in this case. This lower limit is still large enough for LISA's sensitivity. Detectable gravitational waves are induced also in the exponential-tail case~\cite{2023JCAP...05..044A}. Therefore, induced gravitational waves are an indirect but somewhat robust test of this class of primordial black hole dark matter scenarios. In addition, it has been recently shown in Reference~\cite{2023arXiv231117760E} that induced gravitational waves (which are affected by the modification of the equation-of-state parameter $w$ and sound speed $c_{\srm}$) can also be used to test the existence of hypothetical crossovers in theories beyond the Standard Model if the curvature fluctuations are sufficiently strong. In particular, it allows us to resolve the existing degeneracy of sharply peaked mass functions caused by peaked power spectra and broad ones in the presence of $w,\,c_{\srm}^{2}$ softening (see Figure~\ref{fig:inducedGW_thermal}).

It is worth noting that the collapse of topological defects such as cosmic loops, domain walls or bubble collisions via a cosmological (phase) transition can also lead to primordial black hole formation (see Section~\ref{sec:Other-Formation-Scenarios}). These events produce stochastic gravitational-wave backgrounds as well, and have been independently studied (see \eg~References~\cite{2020JCAP...04..034A, 2020Symm...12.1477S}).
\newpage

\subsection{Binary Mergers}
\label{sec:Binary-Mergers}
\vs{-1mm}
As the first gravitational waves detected by the LIGO/Virgo collaboration originate from merger events of binary black holes, the important question arises as to whether PBHs could have participated in the formation of those two-body systems~\cite{2016PhRvL.116t1301B, 2017PDU....15..142C, 2016PhRvL.117f1101S}. In fact, PBHs can easily form binaries and exquisitely explain the observed event rate of black hole binary mergers. Below, we show how those could have been formed from primordial black holes.

\subsubsection{Early Binary Formation}
\label{sec:Early-Binary-Formation}
\vs{-1mm}
PBHs formed from collapsed overdensities during the radiation-dominated era are essentially randomly (Poisson-)distributed at rest. They are matter components, and thus their relative density to the background-radiation density grows with time. Let $x$ be the comoving distance between a PBH (PBH$_{1}$) and its closest neighbour (PBH$_{2}$). If the energy density inside the $x$-sphere, $M / \big\{ [ x / ( 1 + z ) ]^{3}\.4\mspace{1.5mu}\pi/3 \big\}$,\footnote{\setstretch{0.9}In this Subsection, we hereafter assume a monochromatic mass function.} becomes comparable to that of the background, $\rho_{\rrm}$, by the time of the matter-radiation equality, these PBHs form a gravitational bound state which decouples from the cosmic background expansion. The second-closest PBH (PBH$_{3}$) in general exerts tidal forces onto the bound PBHs, thereby injecting angular momentum (see References~\cite{2016PhRvL.117f1101S, 2018CQGra..35f3001S} for details).

Assuming an initially random spatial distribution, one can estimate the probability density of the binary major axis $a$ (not to be confused with the scale factor) and eccentricity $e$ as
\begin{align}
\label{eq: probability of a and e}
    \dd{P}
        = 
            \frac{4\mspace{1.5mu}\pi^{2}}
            {3}\.
            n_{\PBH}^{1/2}\.
            ( 1 + z_{\eq} )^{3/2}\.
            f_{\PBH}^{3/2}\,
            a^{1/2}\.e\mspace{1mu}
            \big(
                1 - e^{2}
            \big)^{\mspace{-2mu}-3/2}\.
            \dd{a}\dd{e}
            \, .
\end{align}
Above, $n_{\PBH} = f_{\PBH}\.\Omega_{\DM}\.\rho_{\crm,\mspace{1.5mu}0} / M$ is the comoving PBH number density, with the current critical density $\rho_{\crm,\mspace{1.5mu}0}\mspace{1mu}$, and $z_{\eq} \simeq 2.4 \times 10^{4}\.\Omega_{\mrm}\.h^{2}$ is the redshift at matter-radiation equality. Equation~\eqref{eq: probability of a and e} indicates that the typical PBH binary formed in this way was highly eccentric. Note also that $a$ and $e$ have upper bounds given by
\vs{-2mm}
\begin{subequations}
\begin{align}
    a_{\umax}
        &= 
            \frac{x_{\umax}}{1 + z_{\eq}}
            \, ,
            \label{eq: amax}
            \displaybreak[1]
            \\[3mm]
    e_{\umax}^{2}( a )
        &= 
            1
            -
            \!\pqty{
                \frac{( 1 + z_{\eq} )\.M}
                {\rho_{\crm,\mspace{1.5mu}0}\mspace{4mu}
                \Omega_{\DM}}\,a
            }^{\mspace{-6mu}3/2}\.
            y_{\umax}^{-6}
            \label{eq: emax}
            \, ,
\end{align}
\end{subequations}
where $x_{\umax} = ( f_{\PBH} / n_{\PBH} )^{1/3}$ and $y_{\umax} = \pqty{n_{\PBH}\,4\mspace{1.5mu}\pi / 3}^{-1/3}$. This is due to the conditions that the distance $x$ to PBH$_{2}$ is near enough as $x < x_{\umax}$ in order to escape from the background expansion, and that the distance $y$ to PBH$_{3}$ is near enough so that the expected PBH number in the $y$-sphere will be less than unity: $y < y_{\umax}$. Integrating, the probability yields the $f_{\PBH}$ PBH binary fraction.

\begin{figure}
    \centering
    \includegraphics[width = 0.7\hsize]{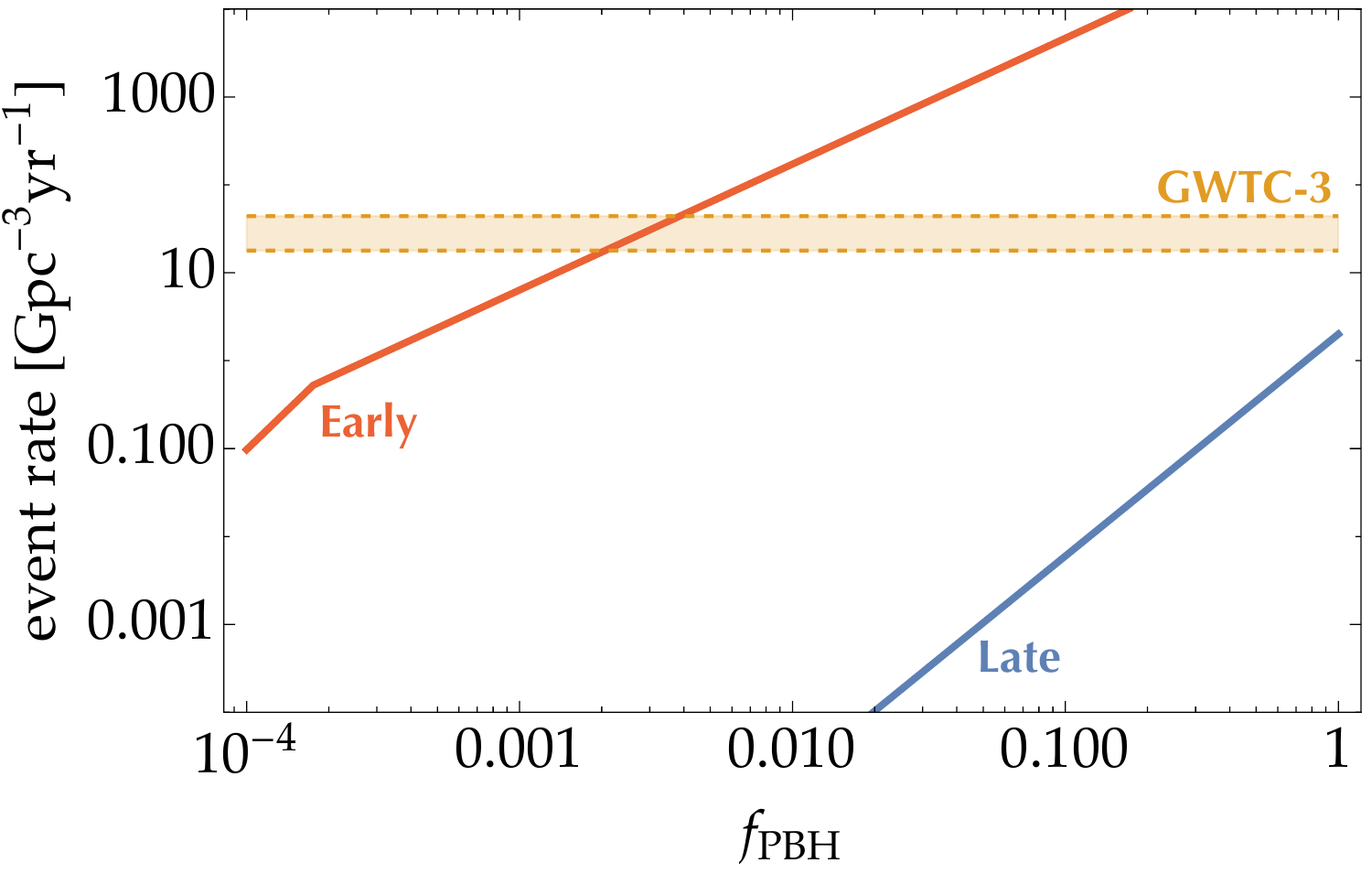}
    \caption{
        Merger event rate $\Rcal$ in terms of the primordial black hole dark matter fraction $f_{\PBH}$ for the early (red) and late (blue) binary-formation scenarios. The orange band shows the estimated merger rate $17.9\.{\rm Gpc}^{-3}\.\.{\rm yr}^{-1} \lesssim \Rcal \lesssim 44\.{\rm Gpc}^{-3}\.{\rm yr}^{-1}$ by the cumulative {\it Gravitational-Wave Transient Catalog} 3 (GWTC-3) of the LIGO--Virgo--KAGRA collaboration~\cite{2021arXiv211103634T}. Figure corresponding to Figure~15 of Reference~\cite{2018CQGra..35f3001S}.
        }
    \label{fig: merger rate}
\end{figure}

A black hole binary system with component masses $m_{1}$ and $m_{2}$ merges due to gravitational-wave emission after the time $t$~\cite{Peters:1964qza},
\vs{-1mm}
\begin{subequations}
\begin{align}
\begin{split}
    t
        &\underset{\phantom{e\approx1}}{ = }
            \frac{15}{304}\.
            \frac{a^{4}}
            {G^{3}\.m_{1}\mspace{1.5mu}m_{2}\.
            ( m_{1} + m_{2} )}
            \bqty{
                \frac{\big( 1 - e^{2} \big)}
                {e^{12/19}}
                \pqty{
                    1
                    +
                    \frac{121}
                    {304}\.
                    e^{2}
                }^{\mspace{-6mu}870/2299}
            }^{\mspace{-2mu}4}
            \\[2mm]
        &\phantom{\underset{\phantom{e\approx1}}{ = }\;}
            \times\int^{e}_{0}
            \dd{\tilde{e}}\;
            \frac{{\tilde{e}}^{\.29/19}}
            {\big( 1 - {\tilde{e}}^{2} \big)^{\mspace{-2mu}3/2}}
            \bigg[
                1
                +
                \frac{121}{304}\.
                {\tilde{e}}^{2}
            \bigg]^{\mspace{-1mu}1181/2299}
\end{split}
            \displaybreak[1]
            \\[3.5mm]
        &\underset{e\.\simeq\.1}{\approx}
            \frac{3}{85}\.
            \frac{1}{G^{3}\.m_{1}\mspace{1.5mu}
            m_{2}\.
            ( m_{1} + m_{2} )}\.
            \big( 1 - e^{2} \big)^{\!7/2}\.a^{4}
            \, .
\end{align}
\end{subequations}
Upon integration of Equation~\eqref{eq: probability of a and e}, the probability density of the merger time for equal masses $m_{1} = m_{2} = M$ is given by
\begin{subequations}
\begin{align}
    \dd{P}
        &= 
            \frac{3}{58}
            \pqty{\frac{t}{T}}^{\mspace{-6mu}3/8}
            \bqty{
                \big(
                    1
                    -
                    e_{\rm upper}^{2}
                \big)^{\mspace{-2mu}-29/16}-1}\,
            \frac{\dd{t}}{t}
            \, ,
            \displaybreak[1]
            \\[2.5mm]
    T
        &\coloneqq
            \frac{3}{170}\.
            \frac{1}{G^{3}M^{3}}
            \bigg[
                \frac{3\.y_{\umax}}
                {4\mspace{1.5mu}\pi
                f_{\PBH}( 1 + z_{\eq} )}
            \bigg]^{4}
            \, .
\end{align}
\end{subequations}
The eccentricity $e_{\rm upper}$ is conditionally defined as
\begin{align}
    e_{\rm upper}
        \coloneqq
            \begin{cases}
                \sqrt{1 - \pqty{t/T}^{6/37}}
                    & \text{for $t < t_{\crm}$}
                \, ,
                \\[3mm]
                \sqrt{1 - \pqty{4\mspace{1.5mu}\pi f_{\PBH} / 3}^{2}\.
                \pqty{t / t_{\crm}}^{2/7}\,} 
                    & \text{for $t \geq t_{\crm}$}
                \, ,
            \end{cases}
\end{align}
where $t_{\crm} = T\.\pqty{ 4\mspace{1.5mu}\pi f_{\PBH} / 3}^{37/3}$. The conditions originate from the entangled integration region given through Equations~(\ref{eq: amax}--\ref{eq: emax}).\vphantom{$_{_{_{_{_{_{_{_{_{_{_{_{_{_{_{_{_{_{_{_{_{_{_{_{_{_{_{_{_{_{_{_{_{_{_{_{_{_{_{_{_{_{_{_{}}}}}}}}}}}}}}}}}}}}}}}}}}}}}}}}}}}}}}}}}}}}$} The binary merger rate $\Rcal$ at time $t$ per unit volume per unit time can be expressed as
\begin{align}
    \Rcal
        = 
            n_{\PBH}\,\dv{P}{t}
        = 
            \frac{3\.n_{\PBH}}{58}\.
            \pqty{\frac{t}{T}}^{\mspace{-6mu}3/8}\,
            \frac{1}{t}\,
            \bqty{
                \big(
                    1
                    -
                    e_{\rm upper}^{2}
                \big)^{\mspace{-2mu}29/16}
                - 1
            }^{-1}
            \, .
\end{align}
Figure~\ref{fig: merger rate} depicts this merger rate for $M = 30\.\Msun$ and $t = 14\.{\rm Gyr}$, compared with the inferred merger rate $17.9\.{\rm Gpc}^{-3}\.{\rm yr}^{-1} \lesssim \Rcal \lesssim 44\.{\rm Gpc}^{-3}\.{\rm yr}^{-1}$ from the cumulative {\it Gravitational-Wave Transient Catalog 3} of the LIGO--Virgo--KAGRA collaboration~\cite{2021arXiv211103634T}. This implies that if primordial black holes at (or near) this mass occupy a subpercent fraction of whole dark matter, $f_{\PBH} \sim 10^{-3}$, the inferred merger rate can be explained in this binary-formation scenario.

\subsubsection{Late Binary Formation}
\label{sec:Late-Mergers}
\vs{-1mm}
In addition to early binary formation, PBHs can form bound states in the current Universe by close encounters in dark halos~\cite{2016PhRvL.116t1301B, 2017PDU....15..142C}. When two PBHs have a near miss, they emit gravitational waves (see Section~\ref{sec:Hyperbolic-Encounters}), and if the associated energy is larger than the kinetic energy of the PBHs, they form a binary. This condition reveals the required smallness of the impact parameter, which, with the relative velocity $v$, can be rewritten in terms of the cross-section as~\cite{2018CQGra..35f3001S}
\begin{align}
    \sigma
        \simeq
            \pqty{
                \frac{85\mspace{1.5mu}\pi}
                {3}
            }^{\mspace{-6mu}2/7}\,
            \frac{\pi\.( 2\.GM )^{2}}
            {v^{18/7}}
            \, .
\end{align}

As an approximation, the binary-formation rate can be assumed to be a measure of the binary-merger rate, since the late binaries discussed in this Section, are expected to merge well within a Hubble time. The binary-formation rate in each halo of mass $M_{\hrm}$ is given by
\begin{align}
    \Rcal_{\hrm}( M_{\hrm} )
        = 
            \int^{R_{\vir}}_{0}\dd{r}\,
            4\mspace{1.5mu}\pi
            \mspace{1.5mu}r^{2}\,
            \frac{1}{2}
            \pqty{
                \frac{\rho_{\PBH}( r )}
                {M}
            }^{\mspace{-6mu}2}\,
            \big\langle
                \sigma\mspace{1.5mu}v_{\PBH}
            \big\rangle
            \. ,
\end{align}
with the virial radius $R_{\vir}$. The density profile $\rho_{\PBH}( r )$ and the velocity distribution are assumed to follow a Navarro--Frenk--White (NFW) profile~\cite{1996ApJ...462..563N} and the Maxwell--Boltzmann distribution, respectively. The total merger rate is calculated as
\vs{-2mm}
\begin{align}
    \Rcal
        = 
            \int_{M_{\umin}}\dd{M_{\hrm}}\;
            \dv{n}{M_{\hrm}}\.
            \Rcal_{\hrm}( M_{\hrm} )
            \, ,
\end{align}
with the minimum halo mass $M_{\umin} \sim 400\.\Msun\,f_{\PBH}^{-1}$ and the halo-mass function $\dd{n} / \dd{M_{\hrm}}$. For $M = 30\.\Msun$, it is roughly given by
\begin{align}
    \Rcal
        \approx
            2\.\alpha\.f_{\PBH}^{53/21}\,
            {\rm Gpc}^{-3}\.{\rm yr}^{-1}
            \, ,
\end{align}
where $\alpha$ is a parameter depending on the halo-mass-function model; the simplest Press--Schechter model gives $\alpha \approx 1$. The corresponding result is shown in Figure~\ref{fig: merger rate}. As the observational estimation of the merger rate has been improved, it was found that PBHs cannot meet the observation solely within the late-binary-formation scenario even if $f_{\PBH} \sim 1$, assuming they are not clustered. However, as discussed in Section~\ref{sec:Clustering-of-Primordial-Black-Holes}, since clustering of PBHs is the rule rather than the exception, the previous conclusion is certainly based on an incorrect assumption. The eventual merger rate estimation hence needs further investigation.

It should be also recalled that regardless of their primordial pendants, there will always be a population of astrophysical black holes. The late binary-formation mechanism can be applied to both of them in principle, and thus the formation of mixed binary systems is also possible. Their merger rate is however calculated to be small compared to that of pure binaries~\cite{2022ChPhC..46e5103C}. Similarly, binary formation of systems with one neutron star and one PBH is also possible. Reference~\cite{2022ApJ...931....2S} has estimated the corresponding merger rate to be less than the observationally-inferred one. Therefore, one is led to conclude that black holes in the observed black hole\./\.neutron star binaries have astrophysical origin with high probability. Black hole\./\.white dwarf binaries would also be interesting targets. Their merger frequencies are laid at a sweet spot of $\Ocal( 10^{-2}\,\text{--}\,10^{-1} )\,\si{Hz}$ of space-based gravitational wave detectors such as DECIGO~\cite{2001PhRvL..87v1103S, 2021PTEP.2021eA105K} or BBO~\cite{2005PhRvD..72h3005C} if the paired black holes are subsolar, and hence their origin is ensured to be primordial. Reference~\cite{2024arXiv240100044Y} shows that the expected number of PBH\./\.white dwarf merger detections in three years by DECIGO is as small as $\sim 10^{-6}$, but there could be some enhancement mechanism without contradicting the current constraints on subsolar PBH binaries.

\subsubsection{Long Duty-Cycle Inspirals}
\label{sec:Long-Duty--Cycle-Inspirals}

\begin{figure}[t]
    \centering
    \includegraphics[width = 0.93\textwidth]{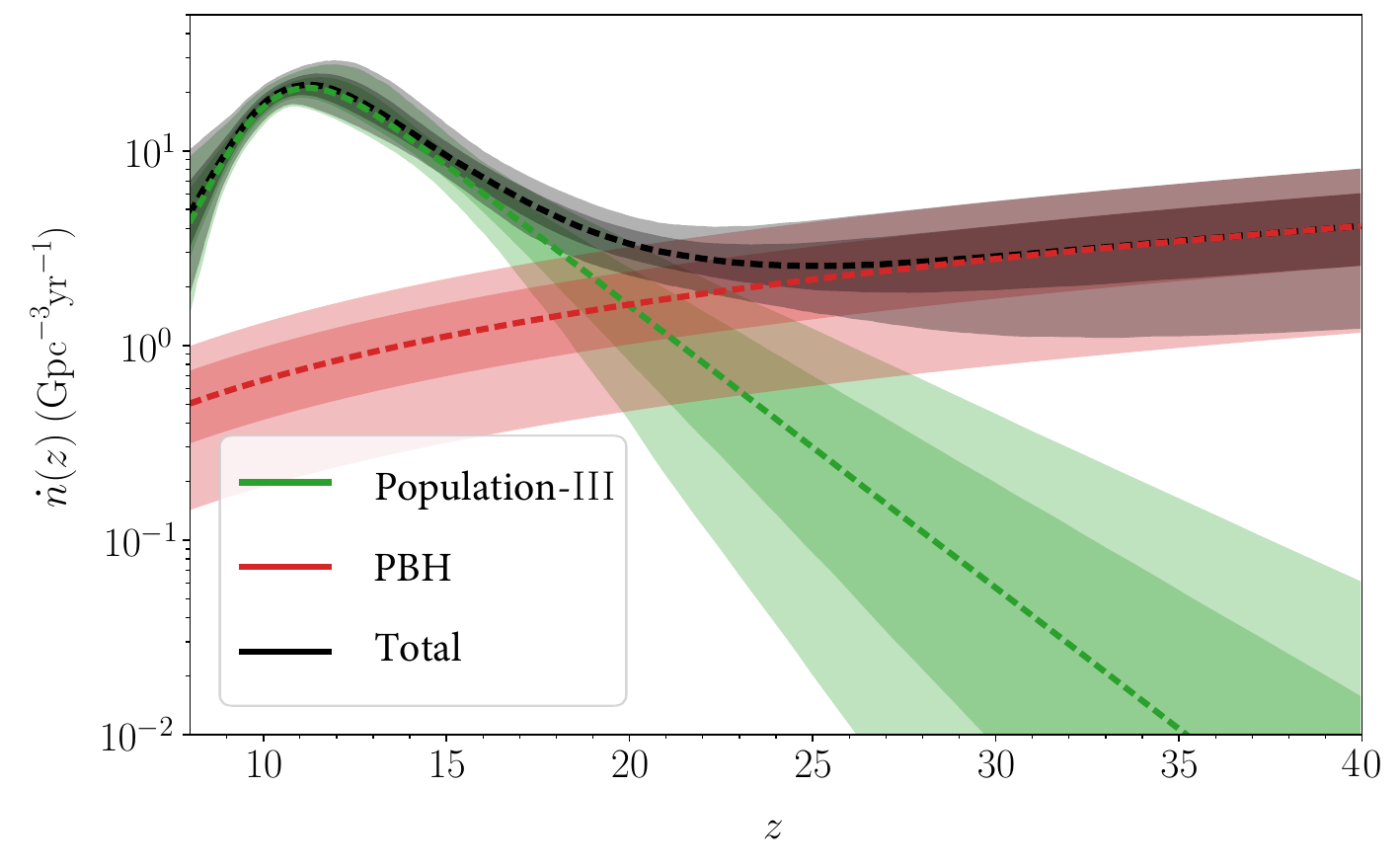}
    \caption{
        Rate densities of the total mergers (black), as well as the individual contributions from primordial black holes (green), and Population-III stars (red). A PBH dark matter fraction of $f_{\PBH} = 10^{-4}$ is assumed. Darker (light) coloured bands indicate $68\mspace{0.5mu}\%$ ($95\mspace{0.5mu}\%$) confidence intervals. Median curves of inferred rate densities are shown by dashed lines. Figure (adapted) from Reference~\cite{2022ApJ...933L..41N}.
        }
    \label{fig:RateDensities}
\end{figure}

Astrophysical and primordial black holes have several distinct properties as sources of gravitational waves. Strong and near-enough mergers can be identified as independent events for which the merger rate can be measured. The redshift dependence of this rate is different for astrophysical and primordial black holes because the latter exist as binaries already in the early Universe, while astrophysical black holes do not form strictly before the star-formation epoch (see Figure~\ref{fig:RateDensities}). Weak or far mergers cannot be identified as independent events, but their superposition (particularly of gravitational waves in the final inspiral phase) can be detected as a stochastic gravitational-wave background. Their primary characteristic is the associated power spectrum. Here, the frequency-dependence of the amplitude for astrophysical and primordial black holes is hard to discriminate, unless the latter are subsolar. Another characteristic, called \emph{duty cycle}, has attracted attention, since it allows us to distinguish the origin of binary black holes~\cite{2006NewAR..50..461C, 2008CQGra..25r4018R, 2012PhRvD..85j4024W, 2020MNRAS.491.4690M, 2023MNRAS.tmp..108B, 2023PhRvD.107d4032Y}.

The frequencies of the gravitational waves emitted in the inspiral phase are time-evolving. The duration $\dd{\tau}$ with signal frequency in the range $[ f,\mspace{1.5mu}f + \dd{f} ]$ is given by
\vs{-2mm}
\begin{align}
    \dd{\tau}
        = 
            \frac{ 5 }{ 96\.\pi^{8/3} }\.
            \big[
                G\mspace{1mu}
                \Mcal( z )
            \big]^{-5/3}\.
            f^{-11/3}\,
            \dd{f}
            \, ,
\end{align}
where the redshift dependence has been recast into the chirp mass $\Mcal( z )$.$\vphantom{_{_{_{_{_{_{_{_{_{_{_{_{_{_{_{_{_{_{_{_{_{_{_{_{_{_{_{_{_{_{_{_{_{_{_{_{_{_{_{_{_{_{_{_{_{_{_{_{_{_{_{_{_{_{_{_{_{_{_{_{_{_{1}}}}}}}}}}}}}}}}}}}}}}}}}}}}}}}}}}}}}}}}}}}}}}}}}}}}}}}}}}}}}}}$ The duty cycle for this frequency bin is then defined by~\cite{2006NewAR..50..461C}
\begin{align}
    \dv{D}{f}
        \equiv
            \int\dd{z}\;
            \dv{\Rcal}{z}\dv{\tau}{f}
            \, ,
\end{align}
with the merger rate $\Rcal$. That is, the duty cycle is the ratio of the signal duration to the typical merger-event period. If $\dd{D}/\dd{f} \gtrsim 1$, the signal duration is longer than the typical period and thus the gravitational-wave signal is observed as \emph{continuous}. Otherwise, the signal is seen as pulse-like and is called ``\emph{popcorn}" signal. Braglia {\it et al.}~\cite{2023MNRAS.tmp..108B} showed that astrophysical black hole binaries correspond to a popcorn background because they appear only in the low-redshift Universe, while primordial ones have high-redshift contributions and can generate continuous signals, depending on the PBH mass function (such as that induced by the thermal history of the Universe; see Section~\ref{sec:Thermal--History--Induced-Mass-Function}).
\newpage

\subsubsection{Spin Enhancement}
\label{sec:Spin-Enhancement}
\vs{-1mm}
There are several situations in which primordial black holes are produced with large initial spins. We already saw one in Section~\ref{sec:Other-Formation-Scenarios}, where PBHs form from confinement of quark/anti-quark pairs. Another one occurs when black hole formation happens during an epoch of matter domination (\cf~Reference~\cite{2017PhRvD..96h3517H}). Here, a crucial difference compared to radiation domination is the absence of any pressure-gradient force. Adopting the theory of angular momentum in structure formation~\cite{1969ApJ...155..393P, 1996MNRAS.282..436C}, it has been found~\cite{2017PhRvD..96h3517H} that most of these PBHs were already rapidly rotating at the time of their formation and would to a large extent continue to do so until now.

As pointed out in Reference~\cite{2020EPJC...80..243K}, this leads to an enhancement of the detectability of the stochastic gravitational-wave background from their mergers. These results are based on the observation of Reference~\cite{2013PhRvD..88f4014H} that when those black holes merge, depending on how their spins are oriented towards each other, the amount of energy emitted in gravitational waves will be either larger or smaller as compared to the non-spinning case, being maximal for aligned and minimal for anti-aligned spins. Concretely, the authors of Reference~\cite{2013PhRvD..88f4014H} have conducted a numerical study of black hole mergers and found that the fraction of energy radiated away through gravitational waves, $E_{\rm GW} \equiv 1 - M_{f} / M_{i}$, can be approximated by $E_{\rm GW}( \chi_{i} ) \approx 0.00258 - 0.07730 / ( 1.6939 - \chi_{i} )$, where $\chi_{i} \equiv \norm{\vec{\chi}_{i}}$, $M_{f}$ is the mass of the final black hole after the merging process has completed, and $M_{i}$ is the sum of the initial (Christodoulou) masses (see Reference~\cite{1970PhRvL..25.1596C}).

The above case holds when both initial spins $\vec{\chi}_{i}$ are equal to and (anti-)aligned with each other. Hence, despite the fact that the black holes are assumed to have maximal spin, depending on their relative orientation, the radiated energy varies. Averaging over the spin orientations of an ensemble of randomly-oriented spinning black holes will in turn lead to a surplus of radiation as compared to the non-spinning case.

A $50\mspace{0.5mu}\%$ increase was found in Reference~\cite{2020EPJC...80..243K}, using the scenario outlined in Reference~\cite{2017PhRvD..96f3507C}. This consists of two scalar fields which together determine the spectrum of perturbations: a light scalar ``spectator'' field, being energetically subdominant during the period of inflation, then giving the dominant component at small scales, and the inflaton field, providing the dominant component to the curvature spectrum at large scales. For a range of conceivable parameter choices, the PBHs produced by the spectator-field model under consideration are expected to have close to maximal spin~\cite{2017PhRvD..96f3507C}.

\begin{figure}
    \vs{-1.5mm}
	\includegraphics[width = 0.8\textwidth]{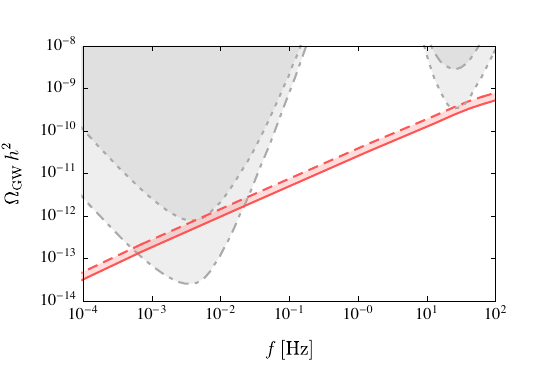}
	\caption{
        Exemplary stochastic gravitational-wave spectrum for the spectator-field model of Reference~\cite{2017PhRvD..96f3507C}. The red dashed curve shows the result including spin; the red solid curves depict the case in which spin has been neglected. The two dotted and dot-dashed grey curves at lower frequencies represent the expected sensitivities for LISA for the worst and best experimental designs, respectively~\cite{2016JCAP...12..026B}. On the higher-frequency side, the limits from the O2 (dotted) and the O5 run (dot-dashed; forecast) of Advanced LIGO (\cf~Figure~3 of Reference~\cite{2017PDU....15..142C}) are included. Figure from Reference~\cite{2020EPJC...80..243K}.
		}
	\label{fig:Constraints}
\end{figure}

Figure~\ref{fig:Constraints} displays the stochastic gravitational-wave background amplitude $\Omega_{\rm GW}( f )$, which is amplified accordingly. This hence provides an explicit example of a scenario in which a large number of highly-spinning black holes are produced with their stochastic gravitational-wave background signal being significantly enhanced. It must be stressed that these results are expected to hold generically for highly-spinning PBHs, such as those formed within an era of matter domination.

\subsubsection{Gravitational-Wave Imprint of Dark Matter Halos}
\label{sec:Gravitational--Wave-Imprint-of-Dark-Matter-Halos}
\vs{-1mm}
The presence of particle dark matter halos around black holes (see discussion in Section~\ref{sec:Primordial-Black-Holes-and-Particle-Dark-Matter}) alters the merger rate as well as the emission of gravitational waves when those ``dressed" black holes merge~\cite{2014PhRvD..89j4059B, 2015JPhCS.610a2044B, 2013ApJ...774...48M, 2013PhRvL.110v1101E, 2015PhRvD..91d4045E, 2018PhRvD..98b3536K, 2019arXiv190710610B, 2019CQGra..36n3001B, 2020PhRvD.102h3006K, 2022PhRvD.105d3009C}. Kavanagh {\it et al.}~\cite{2018PhRvD..98b3536K} found that the merger rate of such PBHs only differs slightly from those of their ``naked" pendants. Similarly, the effect on the inspiral time is also relatively small, although not unobservable, as Reference~\cite{2020PhRvD.102h3006K} points out, thereby correcting previous overestimations. For instance, for a mass ratio of the two black holes around $10^{-3}$, a five-year inspiral in vacuum would be reduced by a few days (as opposed to earlier estimates of around one year). Using a Bayesian analysis, Coogan {\it et al.}~\cite{2022PhRvD.105d3009C} show that LISA should be able to 
    ({\it i}$\mspace{1.5mu}$) 
        distinguish ``dressed" black hole binaries from ``naked" ones, and 
    ({\it ii}$\mspace{1.5mu}$) 
        to characterise the dark matter environments around astrophysical and primordial black holes for a wide range of  parameters.

\subsection{Hyperbolic Encounters}
\label{sec:Hyperbolic-Encounters}
\vs{-1mm}
Much more frequent than mergers of two black holes (which generically involve multi-body processes, \cf~Reference~\cite{1998PhRvD..58f3003I}) are their gravitational scatterings. These {\it hyperbolic encounters} emit gravitational Bremsstrahlung which might be detectable with future gravitational-wave observatories as individual events~\cite{2017PDU....18..123G, 2018PDU....21...61G, 2022PDU....3500932M} or as a stochastic background~\cite{2022PDU....3601009G}, besides possibly increasing the black hole spin~\cite{2021PDU....3400882J}. Actually, several of the LIGO/Virgo candidates might be due to hyperbolic PBH encounters instead of binary black hole merger events~\cite{2022PDU....3500932M}.

\begin{figure}[t]
    \centering
    \includegraphics[width = 0.75\textwidth]{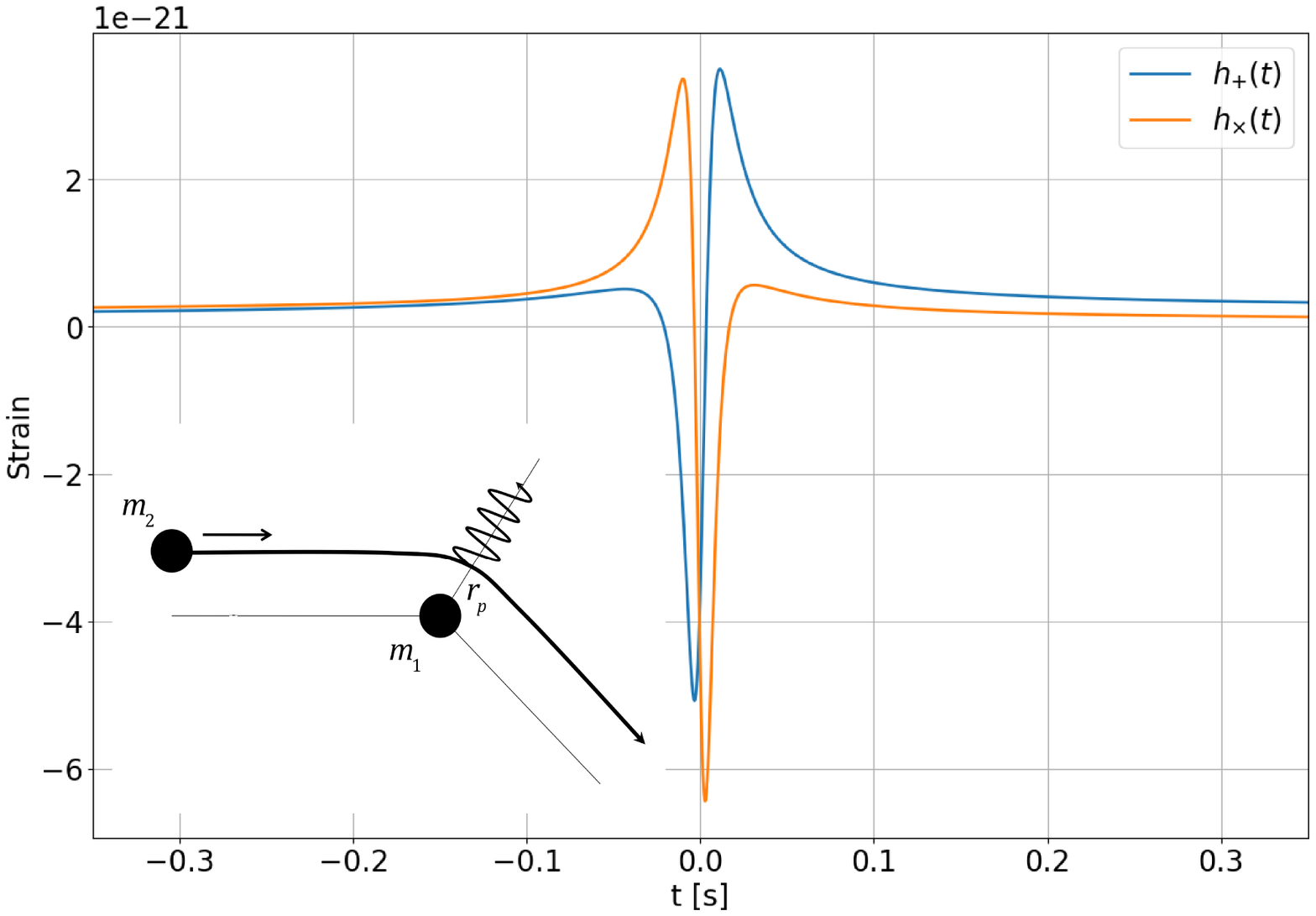}
    \caption{
        {\it Outer panel}\hs{0.2mm}: 
            Time behaviour of the gravitational-wave strain due to emission from a hyperbolic encounter of two black holes with masses $m_{1} = 10\.\Msun$ and $m_{2} = 20\.\Msun$ at a distance $r_{\prm} = 20\.{\rm Mpc}$ and with an orbital inclination of $\Theta = 45^{\circ}$. Figure from Reference~\cite{2022PDU....3500932M}. 
        {\it Inner panel}\hs{0.2mm}: 
            Illustration of how the scattering process induces the emission of gravitational waves. Figure (simplified) from Reference~\cite{2018PDU....21...61G}.
        }
    \label{fig:hyperbolic}
\end{figure}

In detail, as shown by the authors of Reference~\cite{2018PDU....21...61G}, hyperbolic encounters of PBHs with relative velocities of $\Ocal( 0.1 )\.c\mspace{1mu}$, which happen at relative distances of around $10^{-4}\.{\rm AU}$ and at redshift between $z = 0$ and $z = 0.5$, could produce gravitational-wave bursts being well detectable with LISA. Since the associated waveforms significantly differ from that of merging black holes, it will be possible to clearly distinguish these two classes of events.

Moreover, even if the two black holes which hyperbolically encounter each other are initially spinless, non-zero angular momentum can actually be induced onto both holes~\cite{2021PDU....3400882J}. If these are of unequal mass, the heavier one is most impacted. Besides on mass, the amount of induced spin depends on the relative distance and velocity of the black holes, and can in principle be large, \ie~leading to effective spin parameters up to $\chi \sim 0.8$. However, since most of the hyperbolic encounters occur at impact parameters many times the Schwarzschild radii as well as at low relative velocities, the induced spin will be at most moderate for the majority of the black holes, implying that the distribution of $\chi$ will peak at significantly lower values. In Reference~\cite{2021PDU....3400882J} it is argued that this might explain the observed spin distribution of the mergers found by LIGO/Virgo.

\begin{figure}[t]
    \includegraphics[width = 0.78\linewidth]{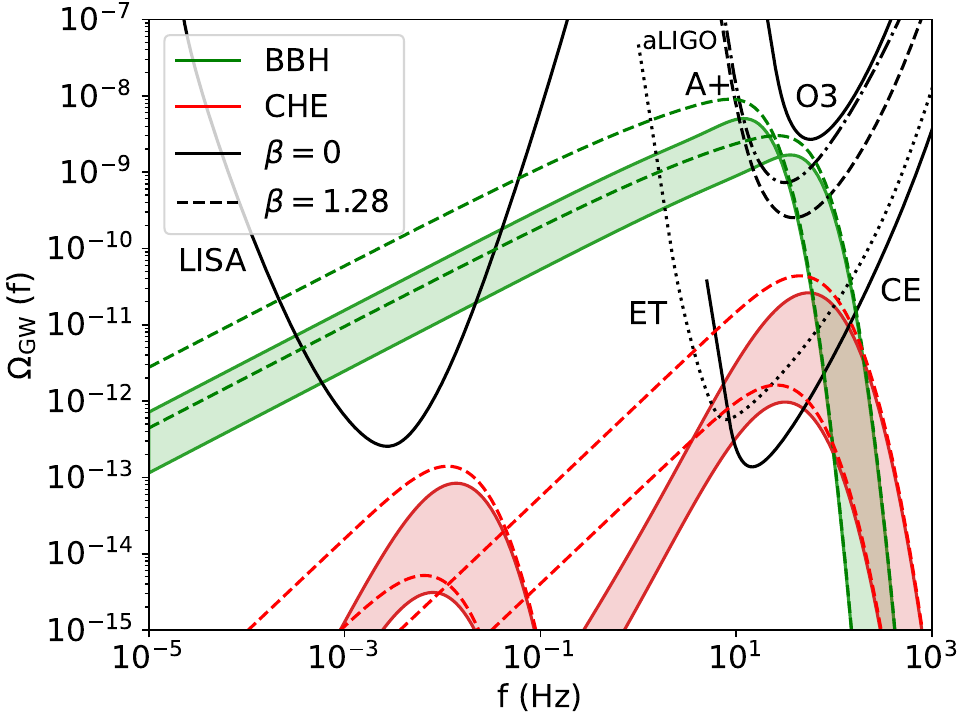}
    \caption{
        Stochastic gravitational-wave backgrounds from {binary black holes} (BBHs) and {close hyperbolic encounters} (CHE), for equal masses in the range $100\,\text{--}\,300\.\Msun$. The solid coloured lines indicate the case of constant merger rate $\tau$, which scales as $\tau \propto ( 1 + z )^{\beta}$, \ie~for $\beta = 0$, while the dashed coloured lines assume $\beta = 1.28$ (\cf~References~\cite{2019JCAP...02..018R, 2020MNRAS.491.4690M}). Also shown are the power-law integrated sensitivity curves of various gravitational-wave detectors for a signal-to-noise ratio of ten and an observation time of one year. Figure from Reference~\cite{2022PDU....3601009G}.
        }
    \label{fig:SGWB}
\end{figure}

If primordial black holes constitute a significant fraction of the dark matter, the superposition of gravitational waves from their hyperbolic encounters might become relevant. The authors of Reference~\cite{2022PDU....3601009G} studied such stochastic gravitational-wave backgrounds and find that these might be well detectable with gravitational-wave interferometers such as the Event Horizon Telescope or the Einstein Telescope. As for the individual events, the detectability of hyperbolic encounters strongly depends on the clustering characteristics of the black hole population. Exemplary, Figure~\ref{fig:SGWB} shows a double-comparison of stochastic gravitational-wave backgrounds from 
    ({\it i}$ \mspace{1.5mu}$)
        binary black holes versus close hyperbolic PBH encounters, and 
    ({\it ii}$ \mspace{1.5mu}$)
        astrophysical versus primordial black holes. As regards the latter, it can be observed that their event rate evolves very differently with time, particularly regarding the slope of its low-frequency tail (\cf~Reference~\cite{2011PhRvL.106x1101A}). This could help to disentangle both contributions and to derive their relative abundance, in particular together with supplementary information regarding spectral shape~\cite{2020MNRAS.491.4690M, 2018JCAP...11..038K, 2019JCAP...11..017C} or anisotropy~\cite{2020MNRAS.493L...1C, 2020PhRvD.102d3502C, 2021PhRvD.104b2005A}.

\subsection{Non-Stellar Black Hole Merger}
\label{sec:Non--Stellar-Black-Hole-Merger}
\vs{-1mm}
The LIGO/Virgo observations have revealed black hole mergers whose progenitor masses strongly challenge standard astrophysical explanations. The reason for this is threefold. Firstly, already in the first three observational runs, around ten events have component masses in the range from $61\.\Msun$ to $107\.\Msun$, \ie~within the so-called {\it upper mass gap}. Unless (comparatively unlikely) multi- or hierarchical merger processes are assumed, those progenitor masses strongly point towards a primordial origin for the following reason: Above a certain mass, the temperature in the core of stars triggers electron-positron pair production which leads to a reduction of the pressure and to core collapse. In turn, the stars explode as {\it remnantless} supernov{\ae}. As a result, stars are not expected to directly form black holes between $\sim 60\.\Msun$ and $150\.\Msun$, being the established {\it pair-instability mass gap}.

Secondly, four merger events have at least one progenitor within the so-called {\it lower mass gap} between $2.5\.\Msun$ and $5\.\Msun$, wherein neither neutron stars nor black holes from stellar collapse are expected. This is supported by microlensing observations of OGLE/Gaia towards the Galactic centre~\cite{2020A&A...636A..20W}. The lower mass gap is still under debate~(\cf~Reference~\cite{2019ApJ...887...53F}) but if confirmed, either stellar models need to be substantially revised or the observed black hole mergers have a primordial component. Interestingly, as mentioned above, the cosmic QCD transition induces a pronounced peak which overlaps the lower mass gap (see Figure~\ref{fig:fPBH}).

Thirdly, there are several merger events with very small mass ratios. This is spearheaded by the event GW190814, which is exceptional not only because its secondary component lies within the lower mass gap, but also because of its relatively low mass ratio $q \coloneqq m_{2} / m_{1} \approx 0.1$. Of course, there is nothing strictly excluding the existence of such asymmetric binaries for astrophysical black hole populations, but it appears unlikely that their merger rate is comparable to that of binaries with similar masses. Indeed, the LIGO/Virgo collaboration even writes in the abstract of their article~\cite{2020ApJ...896L..44A} that an asymmetric binary like GW190814 {\it challenges all current (astrophysical) models of the formation and mass distribution of compact-object binaries}. Interestingly, such binaries frequently occur in thermal-history-induced mass functions (see Section~\ref{sec:Thermal--History--Induced-Mass-Function} and Reference~\cite{Carr:2019kxo}).

\begin{table*}
\begin{center}
    {\bf Advanced LIGO and Advanced Virgo}\\[3.5mm]
    ---\,{\bf Observing Run O2}\,---
\end{center}
\begin{ruledtabular}
    \begin{tabular}{c c c c c c}
		FAR [yr$^{-1}$]
			& $m_{1}$ $[\Msun]$
				& $m_{2}$ $[\Msun]$
					& $\mathrm{SNR}_{\rm total}$
						& $\mathrm{SNR}_{\rm Hanford}$
							& $\mathrm{SNR}_{\rm Livingston}$ \\
		\hline\\[-2.5mm]
		0.2
			& 3.1
				& 0.9
					& 8.5
						& 8.5
							& ---\\[0.5mm]
		0.2
			& 2.1
				& 0.3
					& 8.2
					 	& ---
							& 8.2\\[0.5mm]
		\bf 0.4
			& \bf 4.8
				& \bf 0.8
					& \bf 8.7
						& \bf 6.3
							& \bf 6.0\\[0.5mm]
		1.2
			& 2.3
				& 0.7
					& 8.5
						& 6.3
							& 5.7\\[1mm]
	\end{tabular}
\end{ruledtabular}
\vs{1.25mm}
\begin{center}
    ---\,{\bf Observing Run O3b}\,---
\end{center}
\begin{ruledtabular}
    \begin{tabular}{c c c c c c c}
		FAR [yr$^{-1}$]
			& $m_{1}$ $[\Msun]$
				& $m_{2}$ $[\Msun]$
					& $\mathrm{SNR}_{\rm total}$
						& $\mathrm{SNR}_{\rm Hanford}$
							& $\mathrm{SNR}_{\rm Livingston}$
                                & $\mathrm{SNR}_{\rm Virgo}$ \\
		\hline\\[-2.5mm]
		0.2
			& 0.8
				& 0.2
					& 8.9
                        & 6.3
                            & 6.3
                                & ---\\[0.5mm]
		1.4
			& 0.4
				& 0.2
					& 10.3
                        & 6.6
                            & 5.3
                                & 5.8\\[0.5mm]
		1.6
			& 1.5
				& 0.4
					& 9.1
                        & 6.8
                            & 6.1
                                & ---\\[0.5mm]
	\end{tabular}
\end{ruledtabular}
    \caption{
        Subsolar candidate events with a false-alarm rate (FAR) $< 2\.{\rm yr}^{-1}$. Reported are FAR [${\rm yr}^{-1}$], masses $m_{1}$ and $m_{2}$ $[\Msun]$ and various detector SNRs. Tables (adapted) from Reference~\cite{2021arXiv210511449P} ({\it O2 run}) and~Reference~\cite{2022arXiv221201477T} ({\it O3b run}), respectively. The most significant candidate event is highlighted (see Reference~\cite{2023PDU....4201285M} and main text).
        }
        \vs{-2mm}
    \label{tab:candidate}
\end{table*}

Whilst the observations mentioned above might have an astrophysical origin (even though this arguably appears rather unlikely), detection of black holes {\it below solar mass} would certainly be regarded as a decisive evidence for primordial black holes. So far, none of these has been observed, but there are already seven strong candidates as reanalyses of the data from the second observing run (O2) of Advanced LIGO~\cite{2021arXiv210511449P} as well as from the second part of Advanced LIGO's and Advanced Virgo's third observing run (O3b)~\cite{2022arXiv221201477T} have shown. These have (network) {signal-to-noise ratios} (SNRs) above eight and {false-alarm rates} (FARs) below two per year (see Table~\ref{tab:candidate}), being the thresholds usually considered by the LIGO/Virgo collaboration for claiming merger detection. Note that while some of the candidates did not trigger all detectors, which might make it difficult to eliminate a possible noise origin, several candidates are indeed seen in all detectors.

Based on the analysis of Reference~\cite{2023PhRvD.107b3027M}, Morr{\'a}s {\it et al.}~\cite{2023PDU....4201285M} have recently reinvestigated the most significant candidate event reported by Phukon {\it et al.}~\cite{2021arXiv210511449P}, 
corresponding to the third entry in the upper part of Table~\ref{tab:candidate}. Here, the authors performed an improved analysis, estimating the compact-binary-coalescence parameters with the state-of-the-art waveform families $\texttt{IMRPhenomPv2}$~\cite{2019PhRvD.100b4059K} and $\texttt{IMRPhenomXPHM}$~\cite{2021PhRvD.103j4056P}. Assuming that the signal comes from a real gravitational-wave event, the trigger is consistently identified in both LIGO detectors with mass of the lightest progenitor, $m_{2} = 0.8\.\Msun$ ($90\mspace{0.5mu}\%$ credible interval), being below one and $1.2$ solar masses at $83.8\mspace{0.5mu}\%$ and $92.7\mspace{0.5mu}\%$~confidence level, respectively.

Despite the fact that the observational data and the search of Reference~\cite{2021arXiv210511449P} do not yield enough significance to claim a firm gravitational-wave observation, the analysis of Morr{\'a}s {\it et al.}~\cite{2023PDU....4201285M} shows that the signal, assuming it originated from a gravitational-wave event, is consistent with the participation of a subsolar-mass{\,---\,}therefore primordial{\,---\,}black hole.\footnote{\setstretch{0.9}Any of the other astrophysical attempts to explain this event appear strongly disfavoured. For instance, invoking a neutron star would only be possible for an inconceivable deviation from the standard matter equation of state.} In turn, data from the (planned) forth and fifth observing runs of Advanced LIGO and Advanced Virgo, with improved sensitivity, will provide excellent conditions for discoveries of additional subsolar-mass candidate events, and appear likely to strongly increase the statistical significance for the existence of subsolar-mass primordial black holes.

\begin{figure}[t!]
    \centering
    \includegraphics[width=0.7\textwidth]{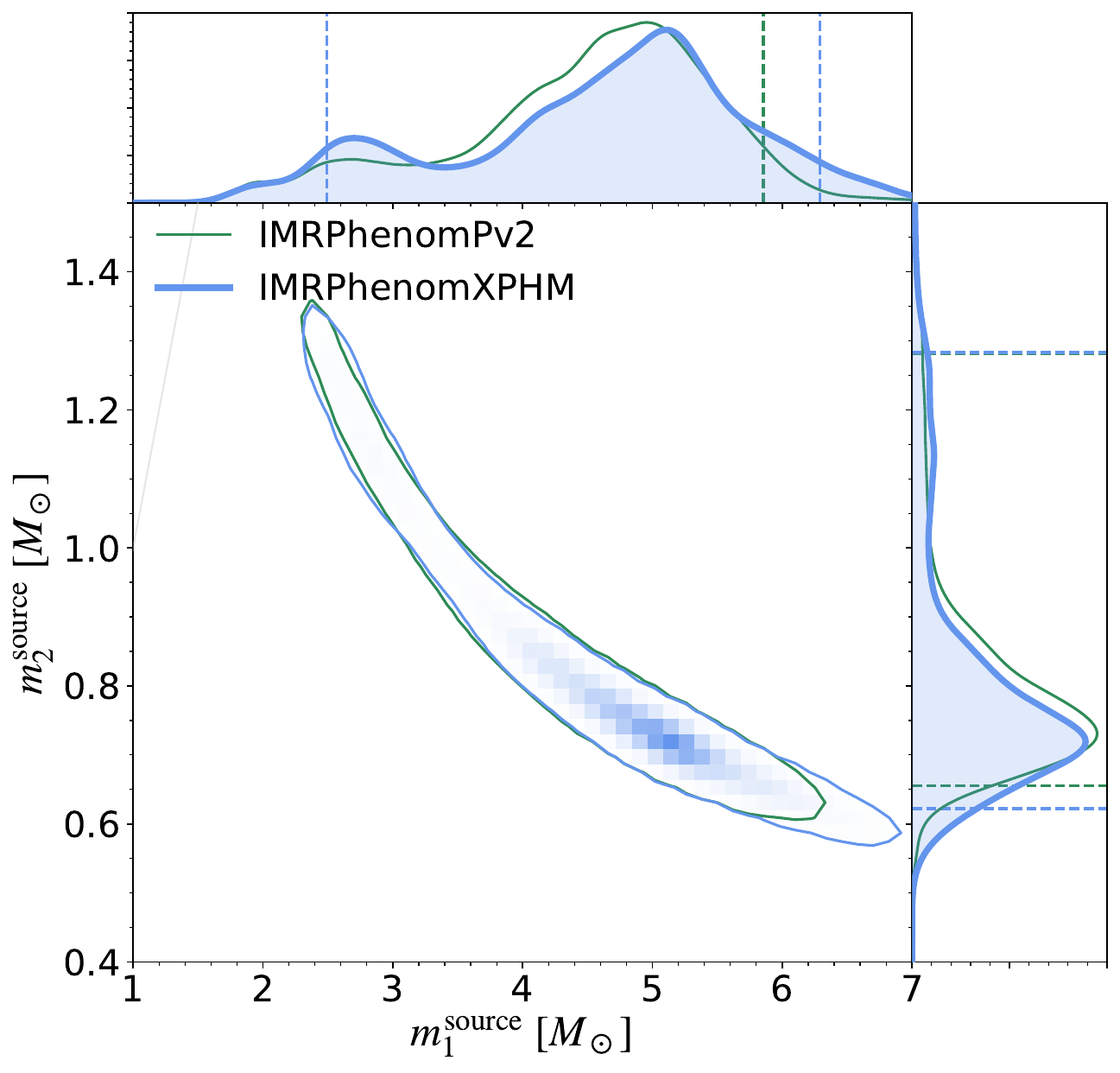}
    \caption{
        Posterior distributions for the primary and secondary mass in the source frame for the candidate event corresponding to the third entry in the upper part of Table~\ref{tab:candidate} reported by Phukon {\it et al.}~\cite{2021arXiv210511449P}. Solid contours in the joint distribution indicate the $90\mspace{0.5mu}\%$ credible regions; in the marginalised distributions, these are shown by the dashed vertical and horizontal lines. Figure from Reference~\cite{2023PDU....4201285M}
    }
    \label{fig:lalinference_m1m2corner}
\end{figure}

\subsection{Transmuted Black Holes}
\label{sec:Transmuted-Black-Holes}
\begin{figure*}[tb]
    \includegraphics[width=0.82\linewidth]{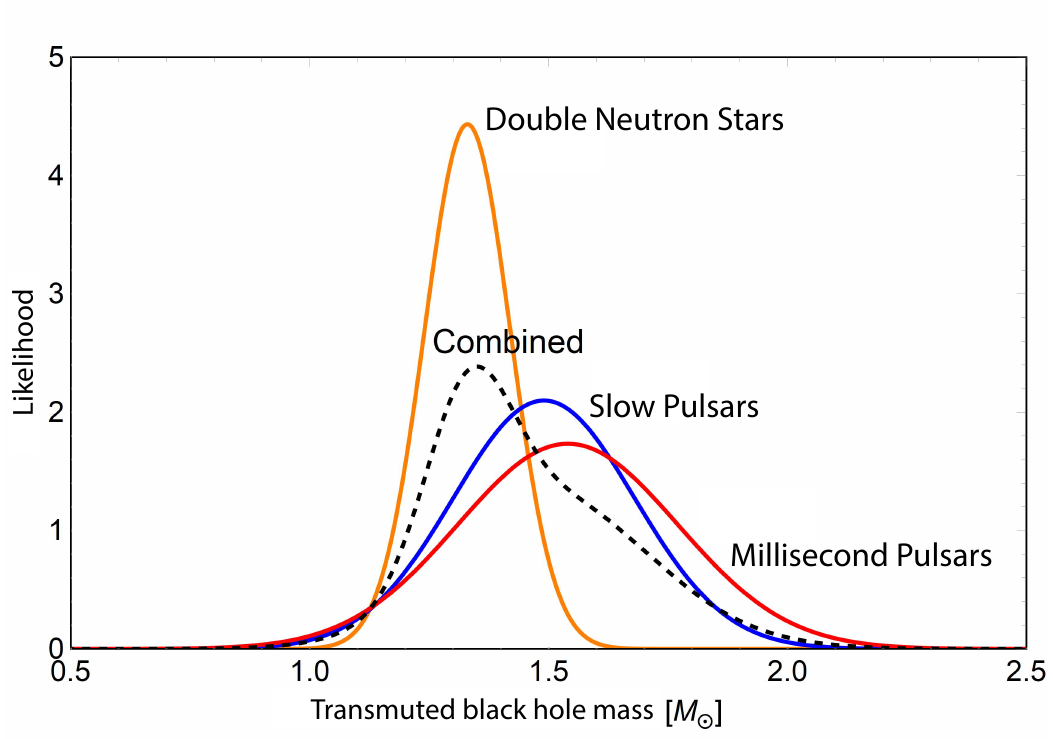}
    \caption{
        Expected mass distribution of transmuted solar-mass black holes, assuming that these track their neutron star progenitors, for the subpopulations originating from slow pulsars (blue), recycled fast-rotating millisecond pulsars (red) and double neutron stars (orange). Also included is the combined distribution (black dashed). Input parameters for the Gau{\ss}ian distributions of the neutron star populations have been taken from Reference~\cite{2013ApJ...778...66K, 2016ARA&A..54..401O}. Figure (adapted) from Reference~\cite{2021PhRvL.126g1101T}.
        }
\label{fig:nsmass}
\end{figure*}

\vs{-1mm}
Takhistov {\it et al.}~\cite{2021PhRvL.126g1101T} have suggested one potential detection strategy of primordial black holes using neutron stars. The idea is that small-mass PBHs can be captured by neutron stars and subsequently consume them, thereby converting them into $1\,\text{--}\,2$ solar-mass black holes (see also References~\cite{2013PhRvD..87l3524C, 2017PhRvL.119f1101F, 2018PhLB..782...77T} for earlier works on PBH-induced implosions of neutron stars). The resulting mass distribution (depicted in Figure~\ref{fig:nsmass}, broken down into its subcomponents) has been shown to be very different from those resulting in standard astrophysical scenarios. Due to its characteristic shape, the authors of Reference~\cite{2021PhRvL.126g1101T} concluded that the observation of a population of $1\,\text{--}\,2\.\Msun$ black holes would be a strong indication for their primordial nature.

In connection, as pointed out by Flores {\it et al.}~\cite{Flores:2023lll}, this might explain the recently discovered {\it G} objects in the Galactic centre~\cite{s41586-019-1883-y}. These may be clouds of gas bound by the gravitational field of stellar-mass black holes. Specifically, the authors argue that light PBHs of mass in the range $10^{-16}\,\text{--}\,10^{-10}\.\Msun$ could have transmuted neutron stars into a population of $1\,\text{--}\,2\.\Msun$ black holes which in turn develop gaseous atmospheres. This could explain all currently observed {\it G} object properties such as their avoidance of tidal disruption from the supermassive black hole at the Galactic centre, while generating the observed emission.

Besides neutron stars, ordinary stars could be strongly affected by PBHs as well. In particular, some of them might be born harbouring such a black hole, which could have observable consequences, as first suggested by Hawking~\cite{Hawking:1971ei}. For a range of stellar masses and metallicities, Bellinger {\it et al.}~\cite{2023ApJ...959..113B} have investigated the evolution of stars hosting a PBH in their core. While the lightest holes leave no evolutionary imprint, the heavier ones can partially consume the star, with strong consequences for its spectrum. For instance, a solar-type star could develop a convective core, which is not possible via ordinary stellar evolution. Furthermore, a giant star would exhibit the characteristic feature of pure-pressure-mode pulsations. These effects might even be asteroseismologically detectable~\cite{2023ApJ...959..113B}.
\newpage

\subsection{Future Prospects}
\label{sec:Future-Prospects-for-Gravitational--Wave-Searches}

As has recently been pointed out in Reference~\cite{2022ApJ...933L..41N}, observations of binary black hole mergers at high redshift offer a promising way to discriminate their origin (see also References~\cite{2022JCAP...08..006M, 2022ApJ...931L..12N, 2023PhRvD.107b4041N}). While a population of PBH mergers would rise monotonically with redshift, the number of mergers of black holes from collapsed Population-III stars~\cite{2000MNRAS.317..385S, 2002ApJ...571...30S, 2003Natur.422..869S, 2014MNRAS.442.2963K, 2016MNRAS.456.1093K, 2016MNRAS.460L..74H, 2017MNRAS.471.4702B} rapidly decreases. As shown in Figure~\ref{fig:RateDensities}, around a redshift of $z = 20$, the difference between the two populations might be clearly visible. In particular, the very first stars formed at redshifts $z \lesssim 50$~\cite{2006ApJ...642..382B, 2007MNRAS.382.1050T, 2009ApJ...694..879T, 2011A&A...533A..32D, 2017PhRvL.119v1104K, 2020MNRAS.494.2027M}, leading to the first mergers of black holes from these Population-III stars around $z \sim 40$ (peaking at $z \sim 10$)~\cite{2014MNRAS.442.2963K, 2016MNRAS.456.1093K, 2016MNRAS.460L..74H, 2017MNRAS.471.4702B, 2017MNRAS.468.5020I, 2020ApJ...903L..40L, 2020MNRAS.495.2475L, 2020MNRAS.498.3946K, 2021ApJ...910...30T, 2022ApJ...926...83T}{\,---\,}in stark contrast to primordial black hole mergers.

Future observations of high-redshift mergers will be possible with the next-generation ground-based gravitational-wave detectors~\cite{2021arXiv211106990K} {\it Cosmic Explorer} (CE) \cite{2017CQGra..34d4001A, 2019BAAS...51g..35R, CEHS} and {\it Einstein Telescope} (ET)~\cite{Punturo:2010zz, 2020JCAP...03..050M}. These are designed to observe binary black holes with total mass of $\Ocal( 10\,\text{--}\,100 )\.\Msun$ up to redshifts $z \sim 100$~\cite{2019CQGra..36v5002H}. This will allow to discriminate a potential dark matter abundance of PBHs from Population-III stars in the most sensitive mass range of those observatories~\cite{2021arXiv210909882E}. In this regard, Reference~\cite{2020JCAP...08..039C} investigates how the Einstein Telescope and Cosmic Explorer can be used to distinguish primordial from astrophysical black holes, particularly in the subsolar mass range.

Besides Earth-bound observational facilities, spaceborn instruments like LISA \cite{2020GReGr..52...81B} will provide precision gravitational-wave observations at frequencies several orders of magnitude lower, in particular, when used together with the {\it Taiji Program in Space} (Taiji)~\cite{Hu:2017mde} as a network~\cite{2022MNRAS.512.6217Y}. This has high potential to discover gravitational waves from merging primordial black holes~\cite{deAraujo:2006zz, Aguiar:2006kk, 2016JCAP...12..026B, 2017JPhCS.840a2030R, 2017arXiv170510361K, 2018JCAP...11..034B, 2019PhRvD..99b3001G, 2019PhRvD..99j3521B, 2019PhRvL.122u1301B, 2019JCAP...07..024B, 2020EPJC...80..627K, 2020GReGr..52...81B, 2022PhRvL.128k1104B, 2022JCAP...07..020F}, in particular their stochastic gravitational-wave backgrounds.

Furthermore, LISA will be able to probe the innermost region of the Milky Way dark matter halo if a population of light primordial black holes of mass $M$ spikes near the central black hole~\cite{2020EPJC...80..627K}. Due to their large mass difference ($M_{{\rm Sgr\.A}^{\!*}} / M \gg 1$), these compact bodies would collectively contribute to stochastic gravitational waves since the merger time for these extreme mass-ratio inspirals easily exceeds the present Hubble time by a large multiple. Observability of the mentioned signal depends on if and how the halo's innermost region possibly exceeds that of an NFW profile~\cite{1996ApJ...462..563N}, which is given by $\rho_{\rm NFW}( r ) = \rho_{\srm}\.r_{\srm} / r\.( 1 + r / r_{\srm} )^{2}$, with 
$r_{\srm}\!=\!24.42\.{\rm kpc}$ and density $\rho_{\srm} = 0.184\,{\rm GeV}\.{\rm cm}^{-3}$ (see Reference~\cite{2011JCAP...03..051C}).
\newpage

Gondolo \& Silk~\cite{1999PhRvL..83.1719G} investigated the enhancement of the dark matter halo profile near the Galactic centre, as compared to an NFW profile, due to adiabatic accretion of dark matter by the central black hole (see Reference~\cite{2013PhRvD..88f3522S} for relativistic corrections). This led the authors to estimate the dark matter spike density as $\rho_{\rm sp}( r ) \approx ( 1 - \epsilon )\.\rho_{R}\.( 1 - 2\.R_{\Srm} / r )^{3}\,( \mspace{-1mu}R_{\rm sp} / r )^{\tilde{\gamma}_{\rm sp}}$. Here, $\epsilon = 0.15$, $2\.R_{\Srm} < r < R_{\rm sp}$, $R_{\Srm} = 2\.G\mspace{1mu}M_{{\rm Sgr\.A}^{\!*}} / c^{2} \simeq 3\.( M_{{\rm Sgr\.A}^{\!*}} / \Msun )\.{\rm km}$ is the Schwarzschild radius of Sgr\.${\Arm}^{\!*}$, and $R_{\rm sp} \coloneqq \alpha_{\tilde{\gamma}}\,r_{0}\.[ M_{{\rm Sgr\.A}^{\!*}} / ( \rho_{0}\.r_{0}^{3} ) ]^{1 / ( 3 - \tilde{\gamma} )}$, with the normalisation $\alpha_{\tilde{\gamma}}$ being numerically derived for each power-law index $\tilde{\gamma}$. Above, $\rho_{R} \!\coloneqq \!\rho_{0}\.( \mspace{-1mu}R_{\rm{sp}} / r_{0} )^{- \tilde{\gamma}}$, where $\tilde{\gamma}_{\rm sp} \coloneqq ( 9 - 2\.\tilde{\gamma} ) / (4 - \tilde{\gamma} )$ (see References~\cite{1999PhRvL..83.1719G, 2019PhRvD..99d3533N}).

\begin{figure}[t]
    \vs{-7mm}
	\includegraphics[width = 0.8\textwidth]{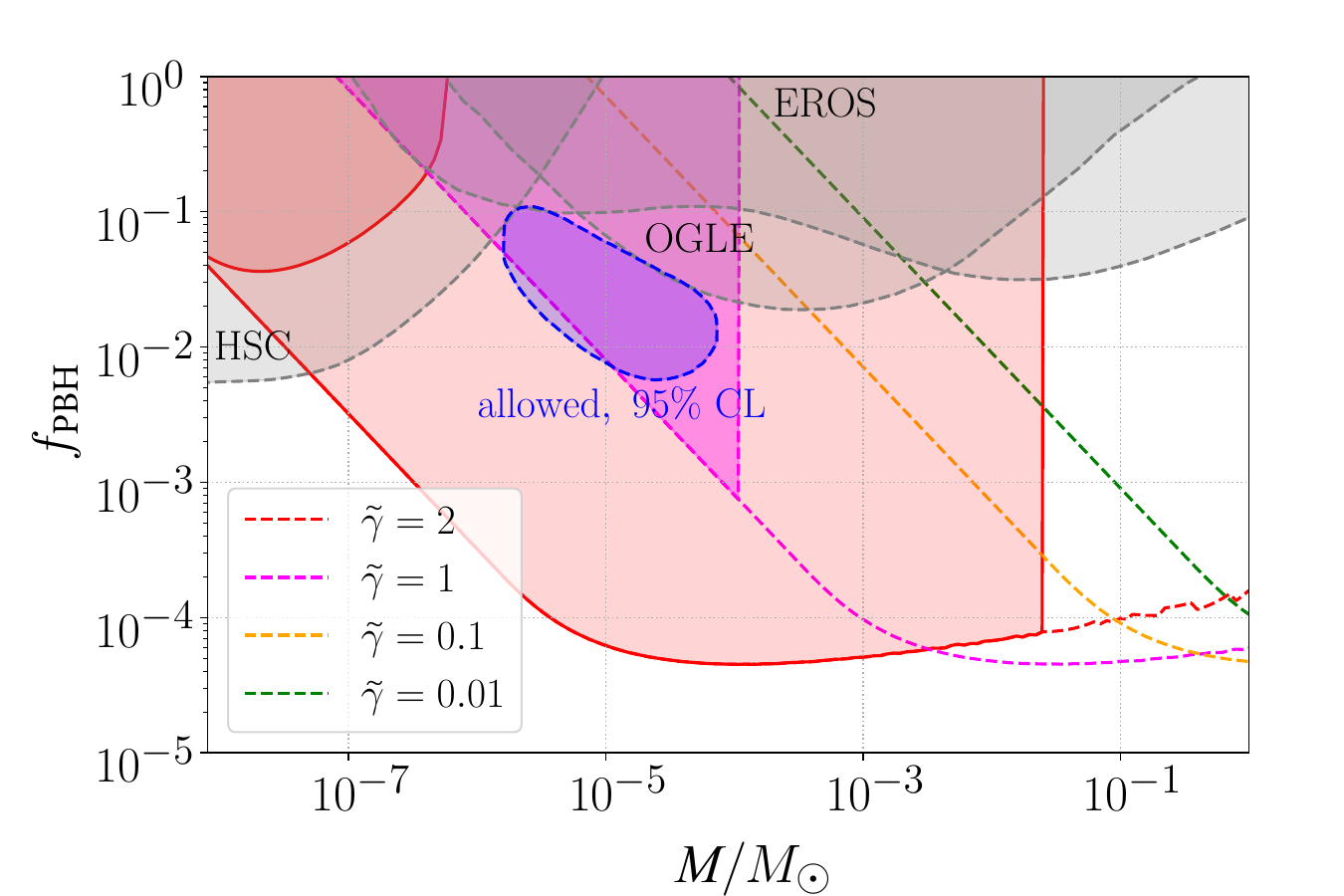}
	\caption{
        Minimum value of the (monochromatic) primordial black hole dark matter fraction $f_{\PBH}$, which LISA will be able to detect, as a function of mass $M$ (red lines) for various dark matter spike parameters $\tilde{\gamma}$ (see main text for details). The upwards-bending branch of the $\tilde{\gamma} = 2$ (red, solid) curve utilises $\tau_{\umin} = H_{0}^{-1}$; the straight lines are for $\tau_{\umin} = 5\.{\rm years}$. The solid lines and filled regions indicate the regions of parameter space in which the signal is well-described as a gravitational-wave background. Individual sources may be resolvable in areas indicated by dotted lines. Microlensing constraints from HSC/Subaru~\cite{2019NatAs...3..524N}, EROS~\cite{2007A&A...469..387T} and OGLE~\cite{2009MNRAS.397.1228W, 2017Natur.548..183M} are depicted in grey. The blue region reflects the positive detection of ultrashort-timescale events attributable to planetary-mass objects between $10^{-6}$ and $10^{-4}\.\Msun$~\cite{2019PhRvD..99h3503N}, which plausible could be PBHs contributing about $\Ocal( 1 )\mspace{0.5mu}\%$ of the dark matter~\cite{2019A&A...624A.120V}. Figure from Reference~\cite{2020EPJC...80..627K}.
        \vs{3mm}
	    }
	\label{fig:constraints}
\end{figure}

Figure~\ref{fig:constraints} shows the minimum value of the primordial black hole dark matter fraction $f_{\PBH}$ which LISA will be able to detect, assuming a signal-to-noise threshold of ten. Also shown are microlensing constraints in the same mass range, as well as recently-reported positive detection of ultrashort-timescale events corresponding to planetary-mass objects between $10^{-6}$ and $10^{-4}\.\Msun$~\cite{2019PhRvD..99h3503N}, which are most plausibly attributed to PBHs. As shown in Reference~\cite{Carr:2019kxo}, these can naturally be explained by thermal-history-induced enhancements (here at the electroweak scale) of the PBH mass function. It can be observed that LISA may be an excellent tool to detect subsolar PBHs as well as to potentially determine the innermost shape of the dark matter halo.

Recently, Miller {\it et al.}~\cite{2021PDU....3200836M} have further elaborated on the possibility of using gravitational waves as probes of planetary-mass primordial black holes. Considering a range of possible signals from inspiraling PBHs (\ie~emission of continuous gravitational waves, quasi-monochromatic signals, or transient continuous waves) detection forecasts of PBHs within the Milky Way have been derived, showing that current detectors should already be able to detect subsolar PBHs with $f_{\PBH} \approx 1$ in the (chirp) mass range $4 \cdot 10^{-5}\,\text{--}\,10^{-3}\.\Msun$; for the Einstein Telescope, it should be possible to observe PBH mergers with chirp masses between $10^{-4}\.\Msun$ and $10^{-3}\.\Msun\mspace{1mu}$, having $f_{\PBH} \approx 10^{-2}$ in that mass window.

\begin{sidewaysfigure}[t]
	\centering
	\includegraphics[width = 0.95\textwidth]{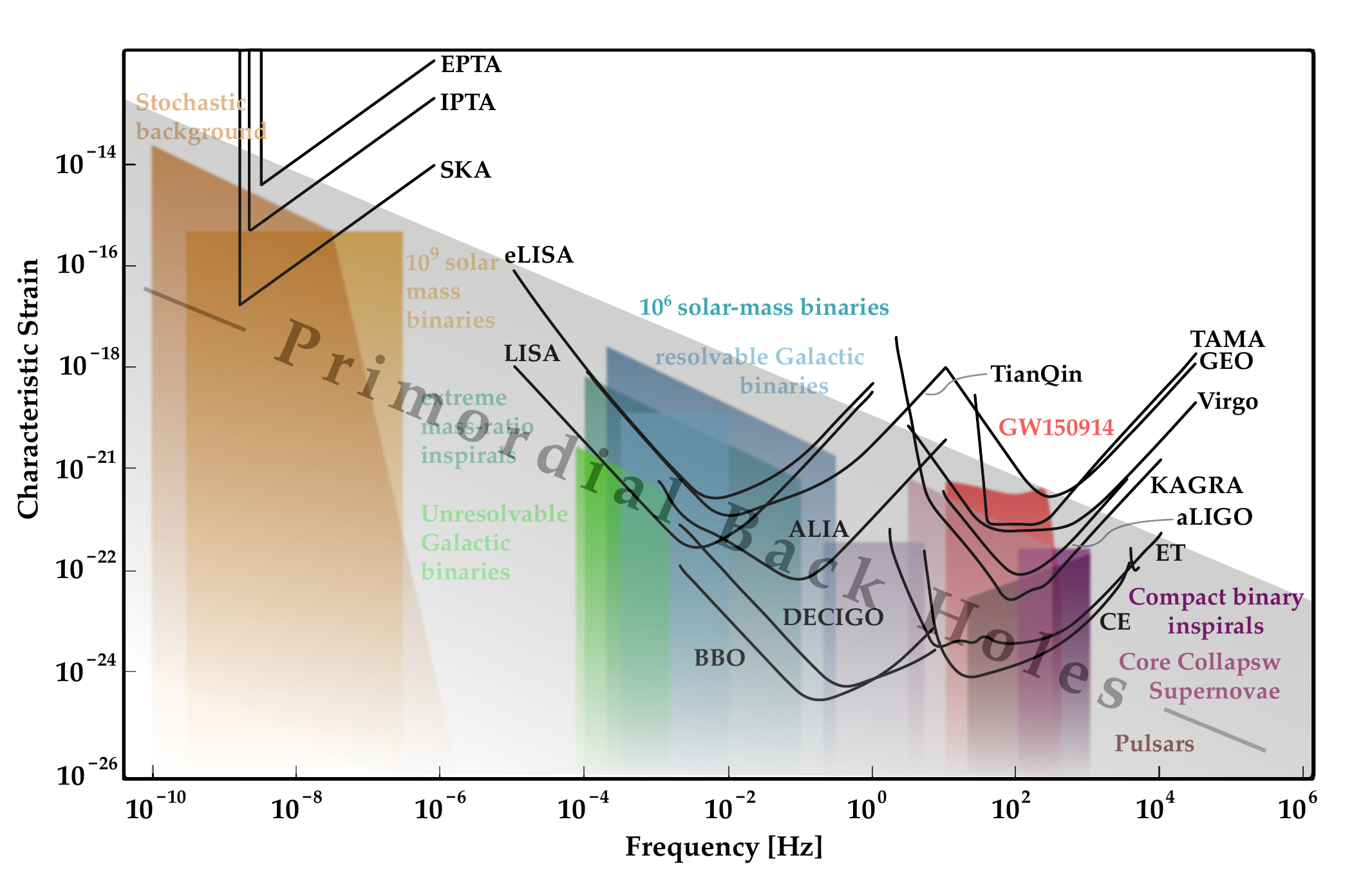}
	\vs{-2mm}
	\caption{
	    Characteristic strain versus frequency (in Hz) for a variety of gravitational-wave detectors (black lines). Included are several astrophysical sources and their characteristic signal regions (coloured bands). Primordial black holes could be responsible for signals across the entire frequency range. Figure originally generated using the online application of the website \href{http://gwplotter.com}{http://gwplotter.com}, then further modified by the authors (\cf~Reference~\cite{2015CQGra..32a5014M} for a similar, but non-PBH figure).
		}
	\label{fig:GW-Sensitivity-Curves-Overview}
\end{sidewaysfigure}

\begin{figure}[t]
    \centering
    \vs{-1mm}
    \includegraphics[width = 0.85\textwidth]{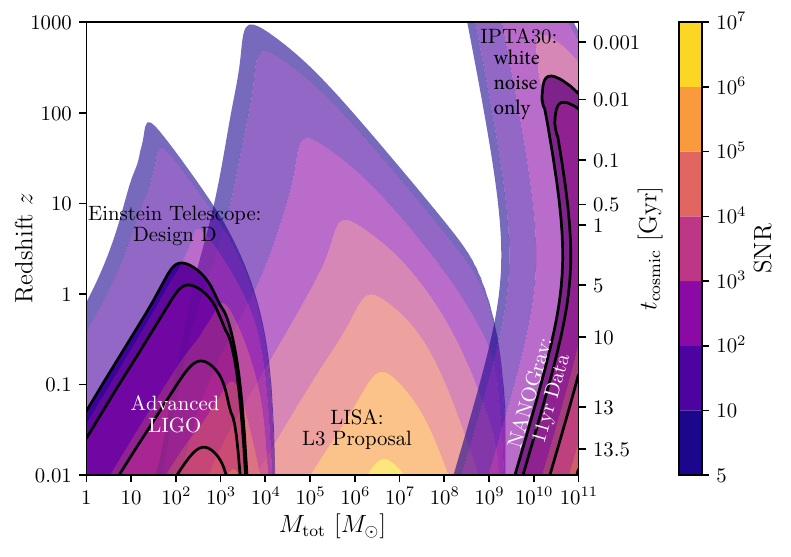}
    \caption{
        Signal-to-Noise Ratios (SNRs) as function of redshift and total mass of the source black hole binaries for various gravitational-wave detectors: Einstein Telescope (design model D), advanced LIGO, LISA (L3 proposal), IPTA30 (white noise only) and NANOGrav ($11\.{\rm yr}$). Figure (adapted) from Reference~\cite{2021CQGra..38e5009K}.
        }
    \label{fig:full-waterfall-plots-lb}
\end{figure}

Figure~\ref{fig:GW-Sensitivity-Curves-Overview} provides an overview of the sensitivity curves of a larger number of ongoing and planned gravitational-wave observatories. It also includes several astrophysical sources as coloured bands. Remarkably, the manifold possible emissions from primordial black holes span over the entire observational range (grey band). Complementary, Figure~\ref{fig:full-waterfall-plots-lb} summarises PBH observation prospects at high redshifts for a selection of detectors.
\newpage
{\color{white}.}
\newpage

\section{Other Signatures}
\label{sec:Other-Signatures}
\vs{-3mm}
\lettrine[lines=3, slope=0em, findent=0em, nindent=0.2em, lhang=0.1, loversize=0.1]{B}{} \hs{-0.5mm}esides gravitational waves, there are multiple ways through which primordial black holes could manifest themselves. These depend on their mass ranges, and might for instance be due to cosmic microwave background distortions induced by accretion onto the PBHs~\cite{1981MNRAS.194..639C, 2007ApJ...665.1277M, 2008ApJ...680..829R, 2007ApJ...662...53R, 2017PhRvD..95d3534A, 2020PhRvR...2b3204S, 2017PhRvL.118x1101G, 2019JCAP...06..026M, 2017JCAP...10..034I, 2021ApJ...908L..23L, 2020PhRvD.102d3505D}, various dynamical effects~\cite{1985ApJ...299..633L, 2020MNRAS.492.5218B, 2017ApJ...839L..13S, 2017PhRvL.119f1101F, 2018PDU....22..137C, 2018ApJ...868...17A, 2020MNRAS.492.5247S}, X-ray/infrared/radio backgrounds~\cite{2013ApJ...769...68C, 2005Natur.438...45K, 2018RvMP...90b5006K, 2016ApJ...823L..25K, 2020JCAP...07..022H, 2021MNRAS.508.5709Y, 2021PhRvD.104f3528Y, 2022MNRAS.517.2454A, 2022MNRAS.510.4992M}, gravitational lensing~\cite{1973ApJ...185..397P, 1979Natur.282..561C, 1981ApJ...243..140G, 1986A&A...166...36K, 1986ApJ...304....1P, 1996ApJ...461...84A, 1993Natur.366..242H, 1998A&A...340L..23H, 2000ApJ...541..270A, 2011MNRAS.415.2744H, 2015A&A...575A.107H, 2018MNRAS.479.2889C, 2018ApJ...862..123M, 2019PhRvD..99h3503N, 2020A&A...636A..20W, 2020A&A...633A.107H, 2020A&A...643A..10H, 2022MNRAS.512.5706H}, bursts from disruptive events such as neutron stars~\cite{2013PhRvD..87l3524C}, or white dwarfs~\cite{2013PhRvD..87b3507C, 2022arXiv221100013S} (see Reference~\cite{Carr:2023tpt} for an extended discussion on positive evidence for PBHs). Most of these have been used to constrain the PBH abundance, but many originally reported on unexpected observations which have been attributed to compact objects (well including primordial black holes).

\subsection{Gravitational Lensing}
\label{sec:Gravitational-Lensing}
\vs{-1mm}
Amongst all observational categories for the detection of primordial black holes, gravitational lensing may be regarded as the currently most decisive. The first discussion that compact bodies might be detected by observing microlensing of distant sources dates back to the early 1970s~\cite{1973ApJ...185..397P} and was taken on in the subsequent decades (\cf~References~\cite{1979Natur.282..561C, 1981ApJ...243..140G, 1986ApJ...304....1P}) with a first outstanding success by the MACHO collaboration~\cite{1996ApJ...461...84A, 2000ApJ...541..270A}. Even though their results are usually and mistakenly interpreted against the possibility of compact-object dark matter, they might actually be regarded as the first positive detection of solar-mass PBHs~(\cf~Reference~\cite{2015A&A...575A.107H}). A broad mass function~\cite{2017PhRvD..96d3020G} as well as clustering~\cite{2018MNRAS.479.2889C} weaken the concerns against a PBH explanation for the MACHO results, leading to the conclusion that their detected events are not only consistent with up to $40\mspace{0.5mu}\%$ PBH dark matter but that this possibility is likely to be realised.

Importantly, contrary to common claims in the literature and the manifold (even most recent) use of supposedly implied primordial black hole constraints from the mentioned results of the MACHO collaboration, Hawkins has shown \cite{2015A&A...575A.107H} that those limits are unreliable. The analysis leading to those constraints relies on the assumption of a heavy Milky Way halo with flat rotation curve, which has been demonstrated to be incorrect~\cite{2005MNRAS.364..433B, 2008ApJ...684.1143X, 2013PASJ...65..118S}. Later observations clearly show a steadily declining rotation curve which implies a drastically reduced optical depth to microlensing{\,---\,}reinvigorated the hypothesis of a primordial black hole dark matter halo~\cite{2015A&A...575A.107H}.
\newpage

Besides the radial form of the Milky Way rotation curve, estimates of the detection efficiency are crucial for deriving reliable limits of the maximally-allowed fraction of compact dark matter bodies. As demonstrated by Hawkins~\cite{2015A&A...575A.107H}, the utilised efficiencies show very little consistency, due to the crowded nature of the Magellanic Cloud star fields.

Further early support for the primordial black hole dark matter hypothesis came from microlensing of quasar light curves~\cite{1993Natur.366..242H, 2011MNRAS.415.2744H}. Recently, it has been found that there are several systems which show a significant misalignment of the microlensed quasar images and the stellar population of the lensing galaxy, such that the probability of stellar lensing is very low (even down to $10^{-4}$), leading to conclude that the only plausible origin of those lenses is PBHs~\cite{2020A&A...633A.107H, 2020A&A...643A..10H, 2022MNRAS.512.5706H}.

Regarding the observation of PBHs in the Galactic bulge, the OGLE collaboration has detected a sample of dark lenses which overlaps the lower mass gap (\cf~Reference~\cite{2020A&A...636A..20W}, and also the respective discussion in the previous Section). Data from the same survey have been reanalysed by Niikura {\it et al.}~\cite{2019PhRvD..99h3503N} and unambiguously six Earth-mass microlenses could be identified. Their only astrophysical explanation would be free-floating planets, but in order for this to work, they would need to assume $\Ocal( 1 )\mspace{0.5mu}\%$ of the dark matter, being inconceivably large. Hence the authors conclude that the most plausible explanation for these bodies is a population of Earth-mass primordial black holes.

For sufficiently large optical depths, caustic features emerge in the amplification pattern of microlensing events, having characteristic shapes and structures (see Reference~\cite{1986A&A...166...36K} for illustrative simulations). For sufficiently large samples, the identification of caustic-crossing events in quasar light curves has been shown to be an unambiguous signature of PBH dark matter~\cite{1998A&A...340L..23H}. Here, the typical time for a compact body to cross a caustic is around ten to twenty years. Using a sample of a thousand quasar observation made between the years 1975 and 2002, it was possible to find evidences for caustic crossings in individual light curves~\cite{1998A&A...340L..23H} as well as statistically~\cite{2018ApJ...862..123M}, these necessarily implying a cosmological distribution of non-stellar lenses. Due to the required large optical depth, the lenses must make up a large part of the dark matter, with the most conceivable candidates being primordial black holes~\cite{2020A&A...633A.107H}.

Besides caustic crossings, the observations of microlensing events in the light curves of multiply-lensed quasar systems provide one of the most convincing evidences for compact-body dark matter. The first detection of such a gravitational lens dates back to the late 1970s~\cite{1979Natur.279..381W}, and many more follow-up analyses have been made (\cf~References~\cite{2009ApJ...706.1451M, 2012ApJ...744..111P}), revealing that the observed microlensing events are consistent with a halo population of compact objects of around $10\mspace{0.5mu}\%$. A particularly clear example is constituted by the observed microlensing of the quasar images J1004+4112 in the Sloan Digital Sky Survey. As summarised in Reference~\cite{2020A&A...633A.107H}, the probability that these events are due to stars is less than $10^{-4}$, strongly pointing towards a primordial origin.

Pixel lensing~\cite{1995ApJ...455...44G} (see Reference~\cite{1996ApJ...470..201G} for an introduction, and also References~\cite{1996AJ....112.2872T, 1997ASPC..117..281G, 1997ApJ...487...55W, 1998ApJ...497...62G, 1999ApJ...521..602A, 2001MNRAS.323...13K, 2001ApJ...553L.137A, 2009MNRAS.399..219I, 2010GReGr..42.2101C} for further reading), which is gravitational lensing of unresolved stars, has been intensively used to detect compact objects, such as primordial black holes. Correspondingly, the collaboration {\it Pixel-lensing Observations with the Isaac Newton Telescope-Andromeda Galaxy Amplified Pixels Experiment} (POINT-AGAPE) has detected six microlensing events in the Andromeda Galaxy~\cite{2005A&A...443..911C}. These vastly exceed the number expected in standard astrophysical scenarios (without compact dark matter), which has led to argue that about $20\mspace{0.5mu}\%$ of the Galactic halo mass in the direction of Andromeda consists of PBHs within the mass range $0.5$ -- $1\.\Msun$. Recently, by carrying out an observation of Andromeda with the Subaru Hyper Suprime-Camera (HSC), Niikura {\it et al.}~\cite{2019NatAs...3..524N} have identified a gravitational-lensing event caused by a compact object within the mass range $10^{-11}\,\text{--}\,10^{-5}\.\Msun$. This, as well as all mentioned events reported by the POINT-AGAPE collaboration, can naturally be explained by the thermal history of the Universe~\cite{Carr:2019kxo} (see Section~\ref{sec:Thermal--History--Induced-Mass-Function}).

\subsection{X-ray, Infrared and Radio Backgrounds}
\label{sec:X--Ray,-Infrared-and-Radio-Backgrounds}
\vs{-1mm}
A cosmological population of primordial black holes can generate various radiation backgrounds due to accretion of matter onto them. The kind of radiation depends on the size of the holes and the details of the accretion processes, and can yield sizeable imprints, for instance on the cosmic X-ray and infrared backgrounds (for intermediate-mass black holes) as well as on the radio background (concerning the supermassive range). These imprints constitute an important diagnostic tool for primordial black hole dark matter.

The spatial coherence of the X-ray and infrared source-subtracted cosmological backgrounds have been studied in detail in References~\cite{2013ApJ...769...68C, 2005Natur.438...45K, 2018RvMP...90b5006K, 2016ApJ...823L..25K}, and it has been strongly argued for the existence of solar-mass primordial black holes. Poisson fluctuations in their number density could have easily caused an overabundance of high-redshift halos in which stars form and emit in the infrared; X-ray emission could be due to accretion onto the PBHs.
\newpage

Hasinger~\cite{2020JCAP...07..022H} has estimated the contribution of baryon accretion onto the overall PBH population to the cosmic X-ray and infrared backgrounds, and ana-lysed their crosscorrelation using deep {\it Chandra} and {\it Spitzer} survey data~\cite{2013ApJ...769...68C}. Assuming Bondi capture and advection-dominated disk accretion, he finds that a population of $10^{-8}\,\text{--}\,10^{10}\.\Msun$ PBHs to be consistent with the residual X-ray fluctuation signal, peaking at redshifts $z \approx 17\,\text{--}\,30$. Furthermore, he argues that PBHs could have an important bearing on a number of phenomena such as 
    ({\it i$\mspace{1.5mu}$})
        amplifying primordial magnetic fields, 
    ({\it ii$\mspace{1.5mu}$})
        modifying the reionisation history of the Universe (while being consistent with recent {\it Planck} measurements~\cite{2020A&A...641A...6P}),\;\;
    ({\it iii$\mspace{1.5mu}$})
        impacting on X-ray heating, thereby providing a contribution to the entropy floor observed in groups of galaxies~\cite{1999Natur.397..135P}, as well as 
    ({\it iv})
        certain $21$-${\rm cm}$ absorption-line features~\cite{2018Natur.555...67B} which could be connected to radio emission from PBHs.

\begin{figure}[t]
	\centering
	\includegraphics[width=0.95\textwidth]{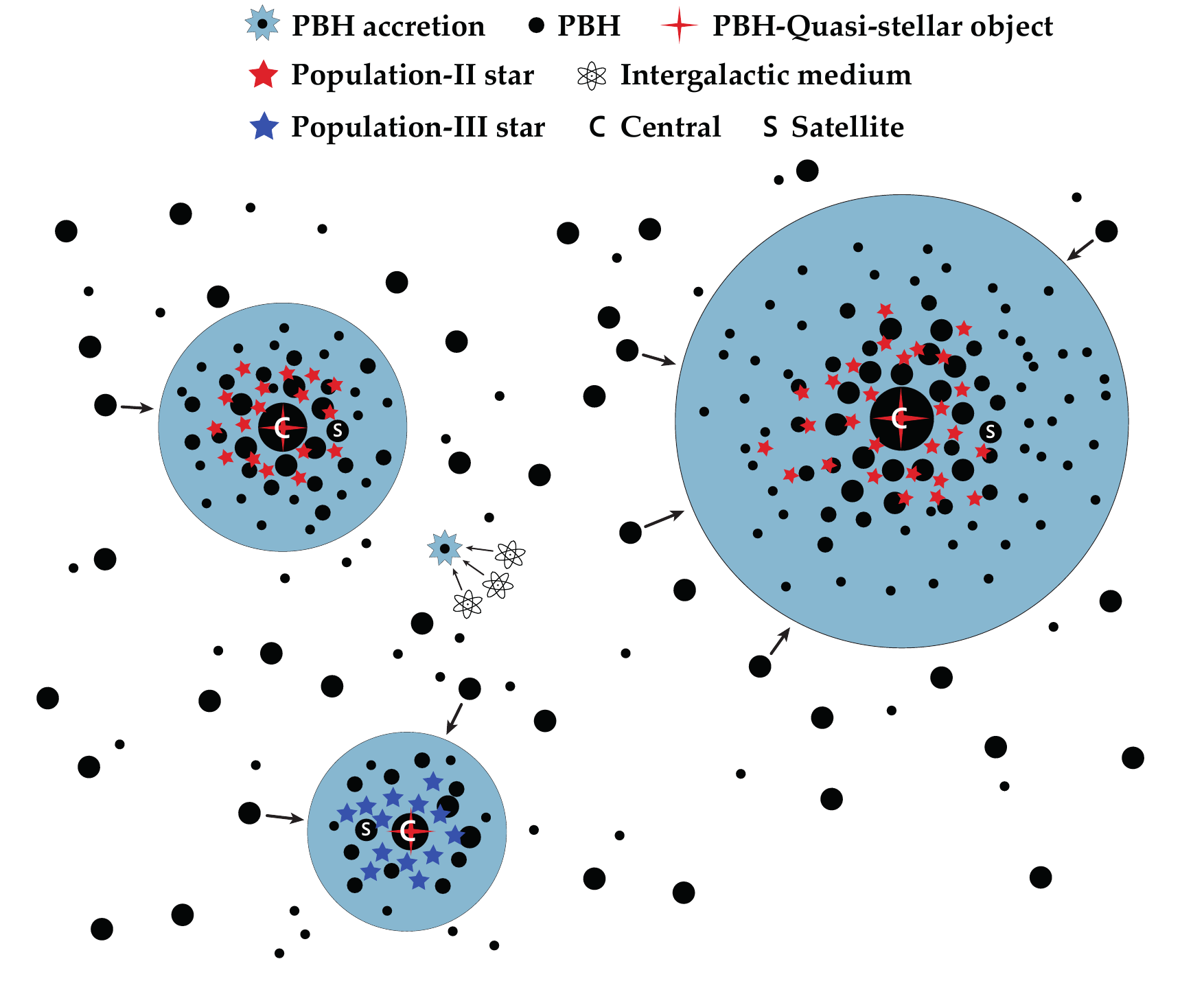}
    \vs{-4mm}
	\caption{
		Sketch of primordial black hole clustering around redshifts $z = 10\,\text{--}\,15$. PBHs (black dots) initially capture baryons during accretion processes and in turn contribute to the cosmic X-ray background as well as (moderately) to the cosmic infrared background. Later on, lighter PBHs cluster around more massive ones, thereby initiating star formation: first of Population-III stars in lowest-mass halos, and then of Population-II stars in higher-mass halos. The respective most massive, central supermassive PBH continues accretion and merging with other lighter PBHs, thereby counting as the central source of infrared and X-ray emission. Lighter PBHs as well as stars fill the halo as satellites. Figure (adapted) from Reference~\cite{2022ApJ...926..205C}.
        \vs{3mm}
        }
	\label{fig:figZ2}
\end{figure}

In a recent related study, Cappelutti {\it et al.}~\cite{2022ApJ...926..205C} have explored high-redshift properties of PBH dark matter with an extended mass spectrum induced by the thermal history of the Universe (see Reference~\cite{Carr:2019kxo} and Section~\ref{sec:Thermal--History--Induced-Mass-Function}). Their main results are summarised in Figure~\ref{fig:figZ2}. Further findings are
    ({\it i$\mspace{1.5mu}$})
        a secondary peak of star formation at $z \sim 15\,\text{--}\,20$ (beyond the well-established observed peak at $z \sim 3$), being driven by mini halos, which are likely to host the first episode of Population-III star formation, 
    ({\it ii$\mspace{1.5mu}$})
        a significant enhancement of the X-ray background fluctuations, and in turn of the unresolved cosmic X-ray and infrared-background cross-power spectrum with only a minor effect on the cosmic infrared-background fluctuations alone, 
    ({\it iii$\mspace{1.5mu}$})
        that while the required
        measured signal~\cite{2016MNRAS.455..282H, 2018RvMP...90b5006K}
        cannot be fully account for by non-PBH cosmologies,
        beyond-solar-mass PBHs could well achieve it, and
    ({\it iv})
        that the X-ray spectral-energy distribution of the cosmic X-ray and infrared-background cross-correlation signal also contains information about their production mechanism (\cf~Reference~\cite{2018ApJ...864..141L}).

\subsection{Supernova Ignition}
\label{sec:Supernova-Ignition}
\vs{-1mm}
Compact dark matter, such as primordial black holes, can trigger explosions of white dwarfs, leading to potentially-observable signatures (see References~\cite{2006PhRvL..96u1302F, 2015PhRvL.115n1301B, 2019PhRvD.100d3020A, 2021ApJ...914..138C, 2021ApJ...914..138C, 2018PhRvD..98k5027G, 2022PhRvD.105b3012A, 2022arXiv221100013S}). Interestingly, a recently-observed supernova population, so-called {\it Calcium-rich gap transients}~\cite{2010Natur.465..322P, 2017ApJ...836...60L}, does not trace the stellar density but is rather located much off-centre from the host galaxies. Furthermore, they appear to originate from white dwarfs with masses at around $\sim 0.6\.\Msun$~\cite{2015A&A...573A..57M}, \ie~well below the Chandrasekhar limit, and predominantly to occur in old systems.
\newpage

\begin{figure}[t!]
    \centering
    \includegraphics[width=0.98\columnwidth]{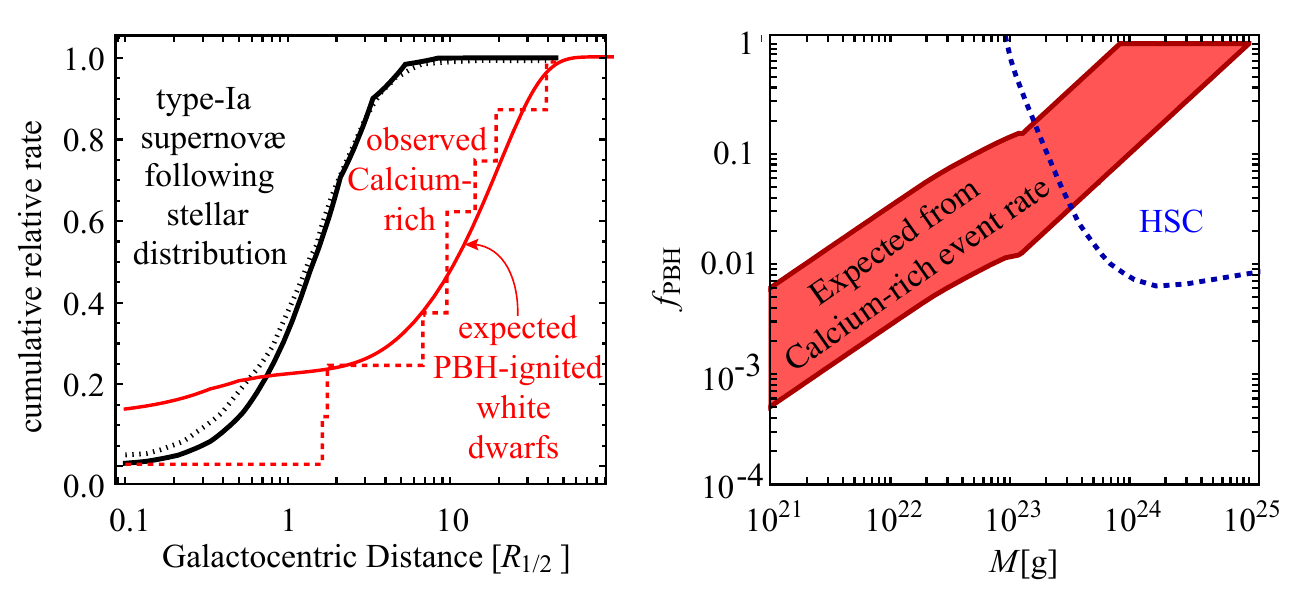}
    \caption{
        {\it Left panel:}
            Radial distributions of standard type-Ia supernov{\ae} (black dashed) and Calcium-rich gap transients (red dashed) within their host galaxies (as determined by Reference~\cite{2017ApJ...836...60L}) in comparison with the spatial distribution of galactic stars (black solid), and the expected spatial distribution of PBH/white dwarf interactions (red solid).
        {\it Right panel:}
            Required PBH dark matter fraction as a function of mass to produce a triggered supernova rate compatible with Calcium-rich gap transient observations. Also shown (blue dashed) is the $95\mspace{0.5mu}\%$ confidence-level exclusion region from stellar microlensing in M31 by the Subaru Hyper Suprime-Camera (HSC)~\cite{2019NatAs...3..524N, 2020PhRvD.101f3005S}. Figures (adapted) from Reference~\cite{2022arXiv221100013S}.
            \vs{1mm}
        }
    \label{fig:datacompare}
\end{figure}

The authors of Reference~\cite{2022arXiv221100013S} argue that these transient events could have well been ignited by asteroidal-mass PBHs, and that the associated event rate to lie within reach of current or near-future microlensing surveys~\cite{2019NatAs...3..524N, 2020PhRvD.101f3005S}. The left panel of Figure~\ref{fig:datacompare} shows the cumulative type-Ia supernova event rate normalised to the half-light radius, $R_{1/2}$, of the host galaxies, as determined by Reference~\cite{2017ApJ...836...60L}. Whilst the observed locations of standard type-Ia supernov{\ae} closely follow the stellar distribution, this is not the case for Calcium-rich gap transients. Interestingly, the authors of Reference~\cite{2022arXiv221100013S} find that the galactocentric distance profile of these events well follows the expected distribution of PBH/white dwarf interactions. This, together with the unique chemical properties and atypical progenitors of Calcium-rich gap transients, gives strong support to the hypothesis that these events originate from interactions of primordial black holes with white dwarfs.

The modelling of Reference~\cite{2022arXiv221100013S} suggests that PBHs of mass between approximately $10^{21}\.\grm$ and $10^{24}\.\grm$ with $f_{\PBH}$ within $[10^{-3},\.0.1]$ are the most plausible triggers for these events (see the right panel of Figure~\ref{fig:datacompare}).\footnote{\setstretch{0.9}As pointed out by Goobar~\cite{2023-Goobar-Private-Communication}, the PBH mass function, originating from a top-hat primordial power spectrum tuned to fit the stochastic gravitational-wave background attributed to recent NANOGrav observations as presented in Reference~\cite{2021PhRvL.126d1303D}, overlaps with both the primordial black hole mass range and number density that could explain the Calcium-rich gap transients discussed in Reference~\cite{2022arXiv221100013S}.} Future observations will further explore this possibility. Interestingly, the {\it Zwicky Transient Facility}, which is the largest ongoing systematic survey, classified eight new events in its first 16 months of operation~\cite{2020ApJ...905...58D}. The {\it Large Synoptic Survey Telescope} (LSST) is expected to achieve an increase in search volume by more than two orders of magnitude. Furthermore, observations with the {\it James Webb Space Telescope} (JWST) should be able to resolve dwarf spheroidal galaxies for the current sample of Calcium-rich gap transients~\cite{2020ApJ...905...58D}, which they seem to closely follow.

\subsection{Fast Radio Bursts}
\label{sec:Fast-Radio-Bursts}
\vs{-1mm}
Fast radio bursts are bright radio transient events at GHz frequencies and with millisecond pulse width. To date, their nature remains unknown~\cite{2021Univ....7...76N}. Interestingly, all of those events are extragalactic and most of them are non-repeating. There have been multiple explanations, including cataclysmic events involving merging neutron stars and stellar black holes, neutron star seismic activity, black hole accretion and active galactic nuclei (see References~\cite{2018PrPNP.103....1K, 2019A&ARv..27....4P, 2021SCPMA..6449501X} for recent reviews). Perhaps the most plausible explanation for fast radio bursts is by PBH collisions with neutron stars, as has for instance been explored by Fuller {\it et al.}~\cite{2017PhRvL.119f1101F} and Abramowicz {\it et al.}~\cite{2018ApJ...868...17A}, who argue that this could apply if PBHs of mass around $10^{23}\.\grm$ constitute around $1\mspace{0.5mu}\%$ of the dark matter. 

It is interesting that the PBH explanation for fast radio bursts may also resolve the so-called missing-pulsar problem: Despite the firm standard astrophysical prediction of a large pulsar population in the Galactic centre, none have been detected within the innermost $\sim 20\.{\rm pc}$, and there is a lack of old pulsars at much larger distances~\cite{2018ASPC..517..793B}. If the Galactic halo comprises PBHs, these sink into the centres of the pulsars due to dynamical friction and in turn consume them. Since the (PBH) dark matter concentration has a maximum around the Galactic centre, the missing-pulsar problem may be entirely resolved.

Besides, this mechanism could also be responsible for the production of certain $r$-process elements, as suggested in Reference~\cite{2017PhRvL.119f1101F}. Strikingly, the thermal-history model (see Section~\ref{sec:Thermal--History--Induced-Mass-Function}) exactly provides the required conditions.
\vs{3mm}

\subsection{Primordial Black Holes and Particle Dark Matter}
\label{sec:Primordial-Black-Holes-and-Particle-Dark-Matter}
\vs{-1mm}
Dark matter might consist of more than one component. This could be {\it microscopic} (in the form of particles, such as WIMPs) and {\it macroscopic} (such as PBHs) at the same time. In this case, there are nontrivial interactions which could even significantly enhance detection prospects~\cite{2016AstL...42..347E, 2018JCAP...07..003B, 2018AstL...44..491B, 2019PhRvD.100b3506A, 2020IJMPA..3540046E, 2021MNRAS.506.3648C, 2021MNRAS.501.2029C, 2021PhRvD.103f3025H, 2021JCAP...03..057C, 2022JCAP...03..045K, 2021PhRvD.103l3532T, 2022arXiv220709481U, 2022arXiv220907541C}. But even if sizeable overdensities are generated, which are large, but somewhat below the PBH formation threshold, so-called ultracompact mini-halos (UCMHs) could be formed~\cite{2009ApJ...707..979R, 2009PhRvL.103u1301S, 2010ApJ...720L..67L, 2010PhRvD..82h3527J, 2012PhRvD..85l5027B}, with distinct features such as enhanced particle annihilation rate. Below, we will focus on the former case, namely that the dark matter consists of PBHs and WIMPs.

In this case, the latter will be accreted by the former already during the radia-tion-dominated era, as first shown by Eroshenko~\cite{2016AstL...42..347E}. Here, a low-velocity subset of the WIMPs will accumulate around the PBHs as density spikes shortly after kinetic decoupling of the WIMPs from the background plasma. The associated annihilation will then give rise to bright $\gamma$-ray signals which can be compared with respective observations, such as with Fermi. In doing so, Eroshenko was able to derive stringent constraints on $\Omega_{\PBH}$ for PBH masses $M > 10^{-8}\.\Msun$ which are several orders of magnitude stronger than previous ones if one assumes a WIMP mass of $m_{\chi} \sim \Ocal( 100 )\.\GeV$ and the standard value of $\langle \sigma v \rangle^{}_{\Frm} = 3 \times 10^{-26}\,{\rm cm}\,\srm^{-1}$ for the velocity-averaged annihilation cross-section. In turn, Boucenna {\it et al.}~\cite{2018JCAP...07..003B} have investigated this scenario for a larger range of values for $\langle \sigma v \rangle$ and $m_{\chi}$ and reach similar conclusions. Several authors have consecutively refined and extended the mentioned analyses~\cite{2019PhRvD.100b3506A, 2021MNRAS.506.3648C, 2022arXiv220709481U}. As we will see below, standard WIMPs and PBHs are mutually exclusive for a large part of the parameter space, leading the authors of Reference~\cite{2021MNRAS.506.3648C} to call their paper ``{\it Black holes and WIMPs: all or nothing or something else}".

One mechanism behind the growth of the density spike is secondary infall~\cite{1985ApJS...58...39B} around heavier PBHs, which yields the constraint $f_{\PBH} \lesssim \Ocal( 10^{-9} )$ for $\langle \sigma v \rangle = 3 \times 10^{-26}\.{\rm cm}^{3} / \srm$ and $m_{\chi} = 100\,\GeV$. This result was obtained by Adamek {\it et al.}~\cite{2019PhRvD.100b3506A} for solar-mass PBHs. The argument has been extended to the entire PBH mass range from $10^{-18}\.\Msun$ to $10^{15}\.\Msun$ and for a large range of WIMP masses by Carr {\it et al.}~\cite{2021MNRAS.506.3648C}, including so-called {\it stupendously large black holes} (SLABs)~\cite{2021MNRAS.501.2029C}.

\begin{figure}[t]
    \vs{-3mm}
	\includegraphics[width = 0.62\linewidth]{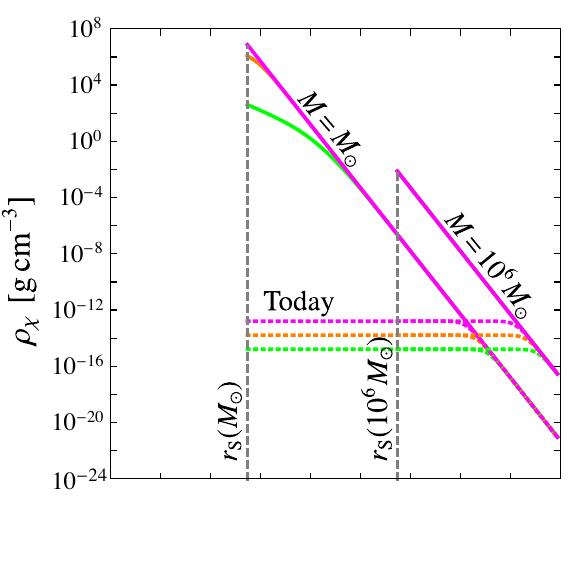}\\[-6.5mm]
	\hs{1.8mm}\includegraphics[width = 0.62\linewidth]{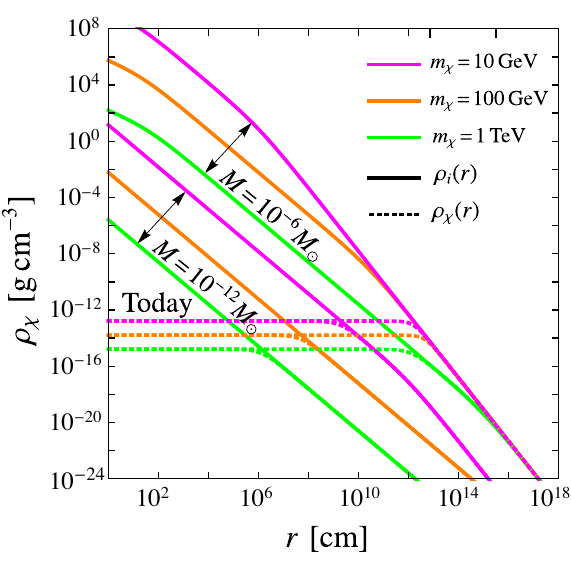}
    \vs{-3mm}
    \caption{
        Density profile of WIMPs bound to a PBH of mass $M = 10^{-12}\.\Msun$ or $M = 10^{-6}\.\Msun$ ({\it upper panel}) and $M = 1\.\Msun$ or $M = 10^{6}\.\Msun$ ({\it lower panel}) for $f_{\chi} \simeq 1$. The utilised WIMP masses are $m_{\chi} = 10\.{\rm GeV}$ (magenta), $m_{\chi} = 100\.{\rm GeV}$ (orange) and $m_{\chi} = 1\.{\rm TeV}$ (green). The density profiles before WIMP annihilations, $\rho_{i}( r )$, are indicated by solid lines. The density profiles after annihilations, $\rho_{\chi}( r )$, are shown by dotted lines, carrying the label ``Today''. Figures (adapted) from Reference~\cite{2021MNRAS.506.3648C}.
		}
	\label{fig:rhochi}
\end{figure}

In order to derive precision constraints, the WIMP halo profile needs to be accurately calculated, and the dynamical evolution of the halo needs to be taken into account. In particular, WIMP annihilations change its profile significantly from its initial form. Figure~\ref{fig:rhochi} demonstrates this, showing the presence of three initial scaling regimes,
\vs{-2mm}
\begin{align}
\label{eq:densityprofile2}
	\rho_{\rm \chi,\mspace{1.5mu}spike}( r )
		\propto
			\begin{cases}
				f_{\chi}\.\rho_{\rm KD}\,
				r^{-3/4} &\text{(innermost)}
                \, ,
				\\[2mm]
				f_{\chi}\.\rho_{\eq}\,
				M^{3/2}\,r^{-3/2}
                    &\text{(intermediate)}
                \, ,
				\\[2mm]
				f_{\chi}\.\rho_{\eq}\.
				M^{3/4}\,r^{-9/4}
                    &\text{(outermost)}
				\, ,_{_{_{_{_{_{_{_{_{}}}}}}}}}
				\\[-1mm]
            \end{cases}
\end{align}
as well as the later emergence of a flat core due to annihilation. Above, $f_{\chi}$ is the WIMP dark matter fraction, $\rho_{\rm KD}$ and $\rho_{\eq}$ are the cosmological densities at kinetic decoupling and at matter-radiation equality, respectively. The derivation of this result and further details can be found in Reference~\cite{2021MNRAS.501.2029C}. Boudaud {\it et al.}~\cite{2021JCAP...08..053B} have numerically confirmed the three scaling regimes of Equation~\eqref{eq:densityprofile2}.

\begin{figure}
    \vs{-2mm}
	\includegraphics[width = 0.6\linewidth]{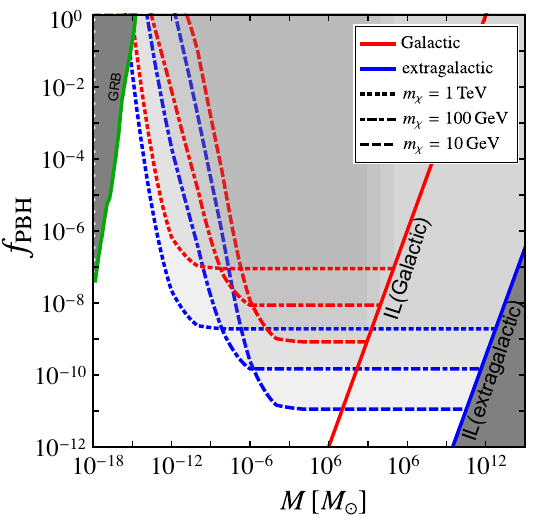}\\[3mm]
	\includegraphics[width = 0.75\linewidth]{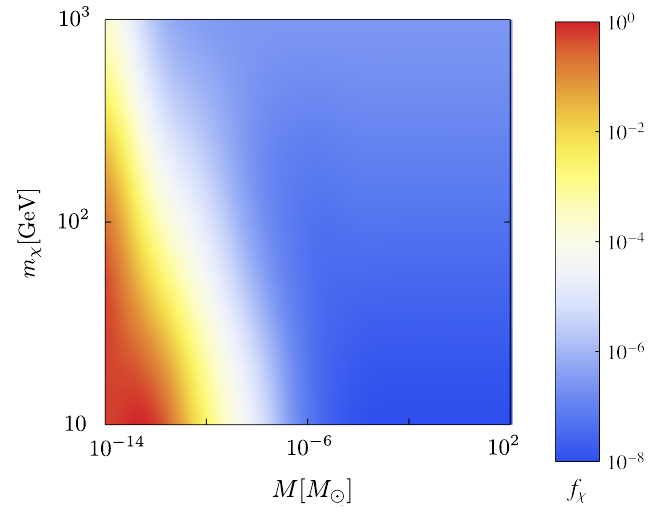}
    \vs{-2mm}
	\caption{
        {\it Upper panel:}
	       Constraints on primordial black hole dark matter fraction $f_{\PBH}$ {\it (monochromatic)} as a function of PBH mass from Galactic (red) or extragalactic (blue) $\gamma$-ray background. Shown are the results for $m_{\chi} = 10\.{\rm GeV}$ (dashed lines), $m_{\chi} = 100\.{\rm GeV}$ (dot-dashed lines) and $m_{\chi} = 1\.{\rm TeV}$ (dotted lines), setting $\sv = 3 \times 10^{-26}\.{\rm cm}^{3}/\srm$. Also included are the Galactic (red solid line) and the extragalactic incredulity limits (blue solid line).
        {\it Lower panel:}
            Density plot of the WIMP dark matter fraction $f_{\chi} = 1 - f_{\chi}$ (colour bar) as a function of PBH mass $M$ (horizontal axis) and WIMP mass $m_{\chi}$ (vertical axis). 
        Figures (adapted) from Reference~\cite{2021MNRAS.506.3648C}.
		}
	\label{fig:fPBH-all}
\end{figure}

As shown in Reference~\cite{2021MNRAS.501.2029C}, the strongest constraints come from extragalactic observations. Here, the differential flux of the $\gamma$-rays is produced by {\it collective} annihilations of WIMPs around PBHs over a large range of redshifts~\cite{2002PhRvD..66l3502U},
\begin{align}
\label{eq:flux}
	\frac{ \drm\Phi_{\gamma} }{ \drm E\.\drm\Omega }
	\bigg|_{\rm eg}
		 \! = 
		 	\int\limits_{0}^{\,\infty}\drm z\;
			\frac{ \erm^{-\tau^{}_{\Erm}(z,\mspace{1.5mu}E)} }
			{ 8\mspace{1mu}\pi H( z ) }
			\frac{ \drm N_{\gamma} }{ \drm E }
			\int\!\drm M\;
			\Gamma( z )\.
			\frac{ \drm n_{\PBH}( M ) }
			{ \drm M }
			\, ,
\end{align}
where $H( z )$ is the Hubble rate at redshift $z$, ``eg'' indicates extragalactic and $n_{\PBH}$ is the PBH number density. Also, $\Gamma( z ) = \Gamma_{0}\,[ h( z ) ]^{2/3}$, where $\Gamma_{0} = \Upsilon\.f_{\chi}^{1.7}\,\mspace{1.5mu}M / \Msun$ is the WIMP annihilation rate around each PBH, and $\tau^{}_{\Erm}$ is the optical depth at redshift $z$ resulting from
	({\it i$\mspace{1.5mu}$})
        photon-photon scattering,
	({\it ii$\mspace{1.5mu}$})
        photon-matter pair production, and
	({\it iii$\mspace{1.5mu}$})
        photon-photon pair production~\cite{2010NuPhB.840..284C, 2009PhRvD..80d3526S}.
The numerical expressions for both the optical depth and the energy spectrum $\drm N_{\gamma} / \drm E$ can be found in Reference~\cite{2011JCAP...03..051C}. Integrating over energy and angles yields the flux
\begin{align}
\label{eq:flux-normalised}
	\Phi_{\gamma,\mspace{1.5mu}{\rm eg}}
		 = 
			\frac{ f_{\PBH}\,
			\rho_{\text{DM}} }{ 2\mspace{1.5mu}H_{0}\.\Msun }\,
			\Upsilon\.f_{\chi}^{1.7}
			\tilde{N}_{\gamma}( m_{\chi} )
			\, ,
\end{align}
where $\tilde{N}_{\gamma}$ is the number of produced photons:
\begin{align}
\label{eq:tildeNgamma}
	\tilde{N}_{\gamma}( m_{\chi} )
		\coloneqq
			\int_{z_{\star}}^{\infty}\!\drm z\;
			\int_{E_{\rm th}}^{m_{\chi}}\!\drm E\;
			\frac{ \drm N_{\gamma} }{ \drm E }
			\frac{ \erm^{-\tau^{}_{\Erm}( z,\mspace{1.5mu}E )} }
			{ [ h( z ) ]^{1/3} }
			\, ,
\end{align}
where the lower limit in the redshift integral corresponds to the epoch of galaxy formation, assumed to be $z_{\star} \sim 10$.\footnote{\setstretch{0.9}The analysis becomes more complicated at times earlier than $z_{\star}$.} This flux can be compared with the Fermi sensitivity $\Phi_{\rm res}$, yielding
\begin{align}
\label{eq:egbound}
	f_{\PBH}
		&\lesssim
			\frac{ 2\mspace{1.5mu}M\.H_{0}\,\Phi_{\rm res} }
			{ \rho_{\text{DM}}\,
			\Gamma_{0}\,\tilde{N}_{\gamma}( m_{\chi} ) }
            \notag
            \displaybreak[1]
			\\[2mm]
		&\approx
			\begin{cases}
				2 \times 10^{-9}\,
				( m_{\chi} /\mspace{1mu}{\rm TeV} )^{1.1}
					& \hbox{($M \gtrsim M_{*}$)}
					\, ,
					\\[3mm]
				\!1.1\times 10^{-12}
                \left(
					m_{\chi} /\mspace{1mu}{\rm TeV}
				\right)^{-5.0}
				\big[
					M / ( 10^{-10}\.\Msun )
				\big]^{\!-2}
					& \hbox{($M \lesssim M_{*}$)}
					\, ,_{_{_{_{_{_{_{_{_{}}}}}}}}}
				\end{cases}
\end{align}
with $M_{*} \approx 2 \times 10^{-12}\.\Msun\,( m_{\chi} /\mspace{1mu}{\rm TeV} )^{-3.0}$. The full constraint (from extragalactic observations) is shown by the blue curves in the upper panel of Figure~\ref{fig:fPBH-all} for a WIMP mass of $10\.{\rm GeV}$ (dashed line), $100\.{\rm GeV}$ (dot-dashed line) and $1\.{\rm TeV}$ (dotted line), using the fit for $\tilde{N}_{\gamma}( m_{\chi} )$ as obtained by the authors of Reference~\cite{2021MNRAS.506.3648C}.

It has been possible to extend the above analysis to the case in which WIMPs do not provide most of the dark matter~\cite{2021MNRAS.506.3648C}. The lower panel of Figure~\ref{fig:fPBH-all} shows the results on the allowed WIMP and PBH dark matter fractions, with the values of the former being indicated by the coloured scale as a function of $M$ (horizontal axis) and $m_{\chi}$ (vertical axis). It is important to realise that even a small value of $f_{\PBH}$ can imply a strong upper limit on $f_{\chi}$. Interestingly, there is a large part of the $M$--$m_{\chi}$ parameter space in which the fractions of PBHs and WIMPs are both much less then unity. This motivates the existence of a third dark matter candidate. Note that there are of course several particles which are not produced through the mechanisms discussed above or which avoid annihilation; these include axion-like particles~\cite{Abbott:1982af, Dine:1982ah, Preskill:1982cy},\footnote{\setstretch{0.9}As shown by Dvali~\cite{1995hep.ph....5253D}, within an inflationary scenario, the upper bound on the axion scale is essentially removed (see Reference~\cite{2023PhRvD.107i5009K} for a recent application).} sterile neutrinos~\cite{1994PhRvL..72...17D, 1999PhRvL..82.2832S}, ultralight or ``fuzzy'' dark matter~\cite{2000PhRvL..85.1158H, 2014NatPh..10..496S}.

\begin{figure}
	\centering
	\includegraphics[width = 0.85\textwidth]{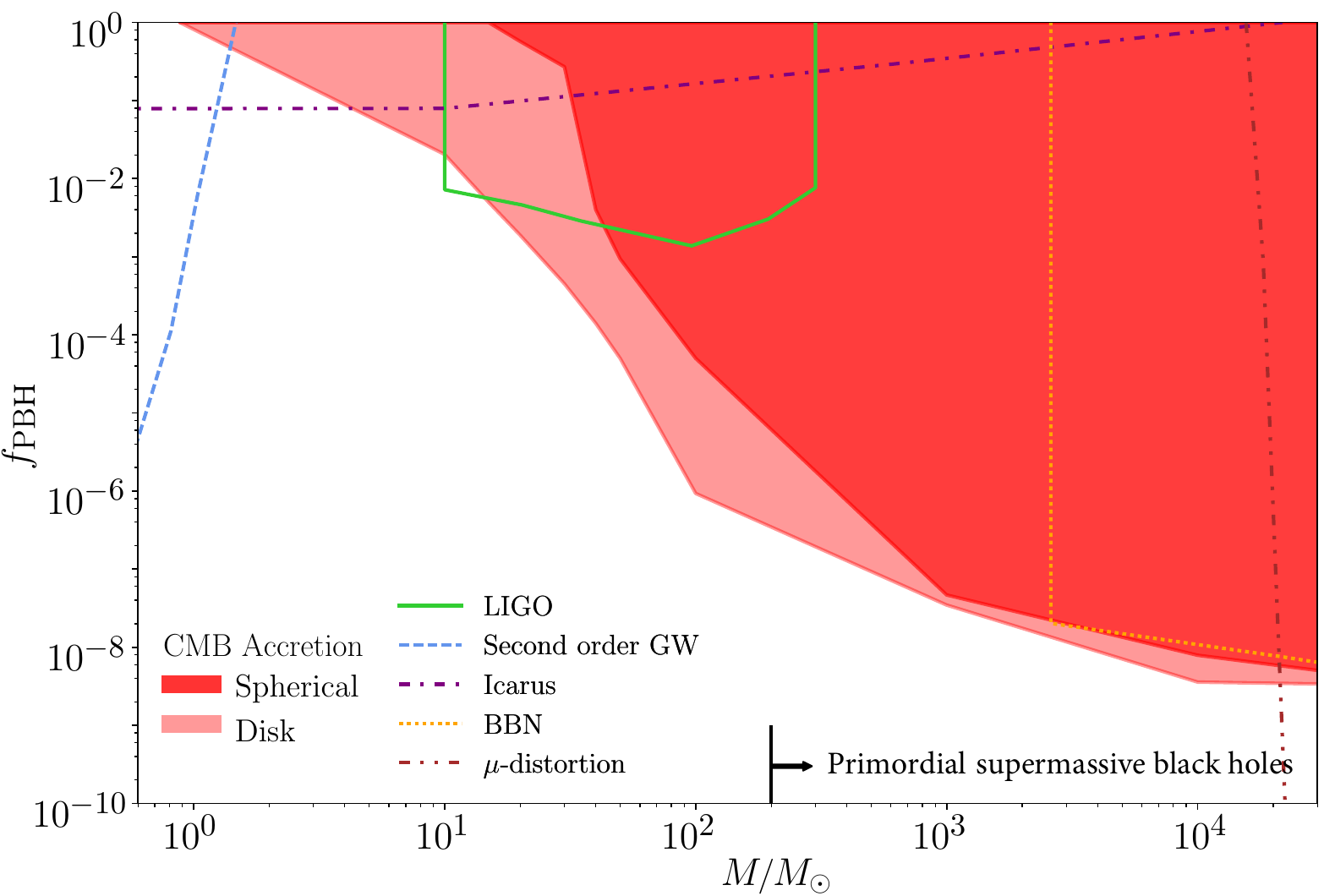}
    \caption{
        Constraints on the (monochromatic) primordial black hole dark matter fraction $f_{\PBH}$ assuming disk accretion (light shade) and spherical accretion (dark shade). Also shown are various bounds in the same mass regions, from 
            ({\it i}$\mspace{1.5mu}$)
                second-order gravitation waves~\cite{2021RPPh...84k6902C}, 
            ({\it ii}$\mspace{1.5mu}$)
                Icarus~\cite{2018PhRvD..97b3518O}, 
            ({\it iii}$\mspace{1.5mu}$) 
                LIGO~\cite{2017PhRvD..96l3523A}, 
            ({\it iv}) 
                big bang nucleosynthesis (BBN)~\cite{2016PhRvD..94d3527I} and 
            ({\it v}) 
                spectral CMB distortions~\cite{2014PhRvD..90h3514K}.
        Figure (adapted) from Reference~\cite{2020PhRvR...2b3204S}.
        }
    \label{fig:Serpico-summary}
\end{figure}

Recently, Serpico {\it et al.}~\cite{2020PhRvR...2b3204S} performed a general analysis of combined dark matter scenarios including PBHs and generic particle dark matter. The increased gravitational potential of the combined system fosters baryonic accretion whose luminosity can be strongly constrained by the cosmic microwave background. This reference then proves that these constraints dominate over other constraints available in the literature at masses $M \gtrsim 20\,\text{--}\,50\.\Msun$, and reach a level of $f_{\PBH} < 3 \times 10^{-9}$ around $M\sim 10^{4}\.\Msun$. Of course, these bounds depend on the accretion dynamics (see Figure~\ref{fig:Serpico-summary}). Despite being relatively stringent, these bounds still allow PBH seeds for the supermassive black holes in galactic centres.

\subsection{Future Prospects}
\label{sec:Future-Prospects-for-Non--Gravitational--Wave-PBH-Searches}
\vs{-1mm}
There are numerous ongoing and planned observations which will test the hypothesis of PBH dark matter besides the gravitational-wave searches discussed in Section~\ref{sec:Future-Prospects-for-Non--Gravitational--Wave-PBH-Searches}. Concretely, this concerns the wide-field surveys {\it Euclid}, {\it Nancy Grace Roman Wide Field Infrared Survey Telescope} (WFIRST-Roman), {\it extended Roentgen Survey with an Imaging Telescope Array}~(eROSITA) and {\it Advanced Telescope for High Energy Astrophysics}~(ATHENA), but also the {\it James Webb Space Telescope}~(JWST) as well as the {\it Square Kilometer Array}~(SKA), which promise to probe the largely unknown reionisation history which is strongly affected by PBH dark matter.

\begin{figure}
    \centering
    \vs{-1mm}
    \includegraphics[width = 0.68\textwidth]{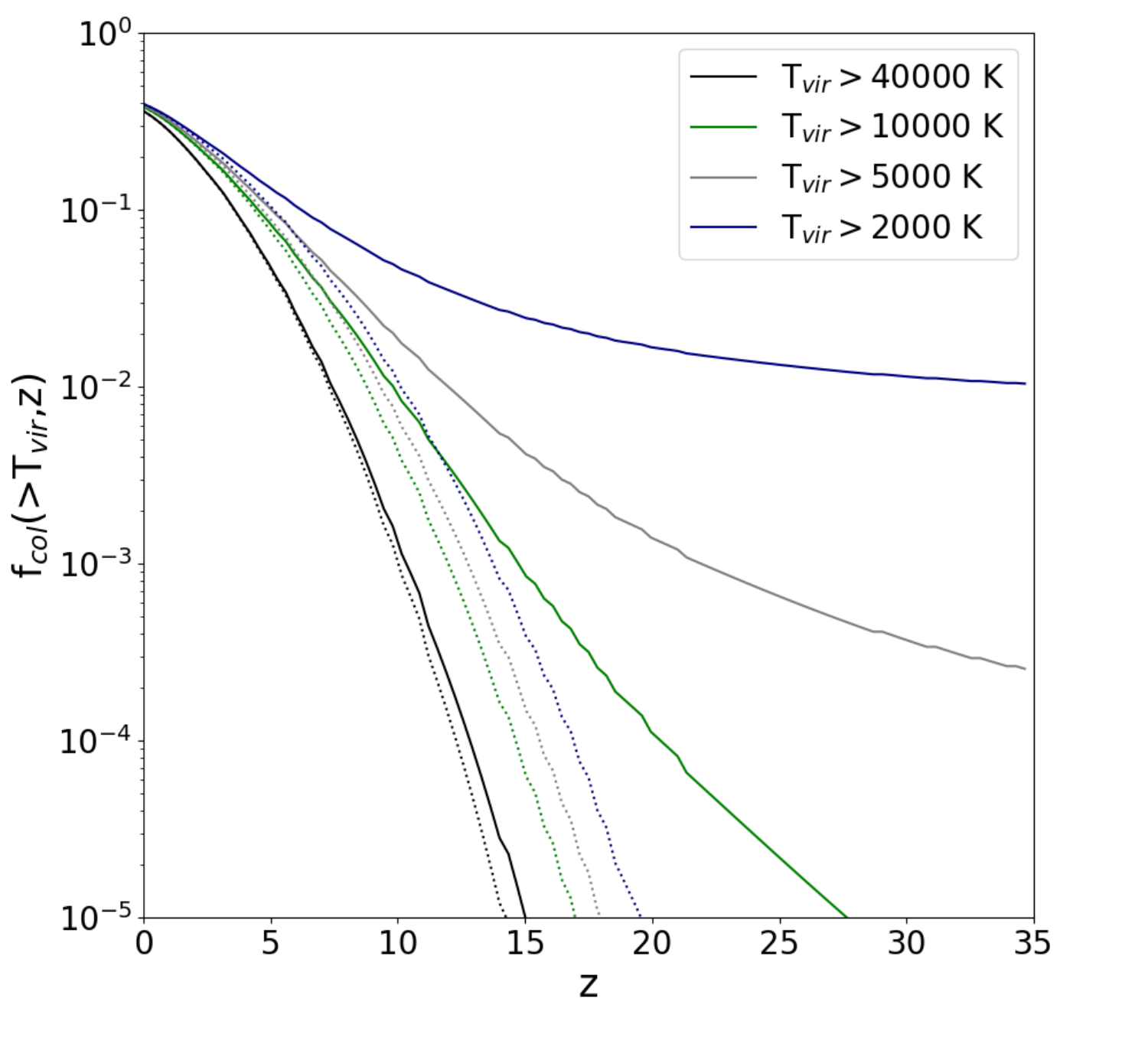}\\
    \includegraphics[width = 0.68\textwidth]{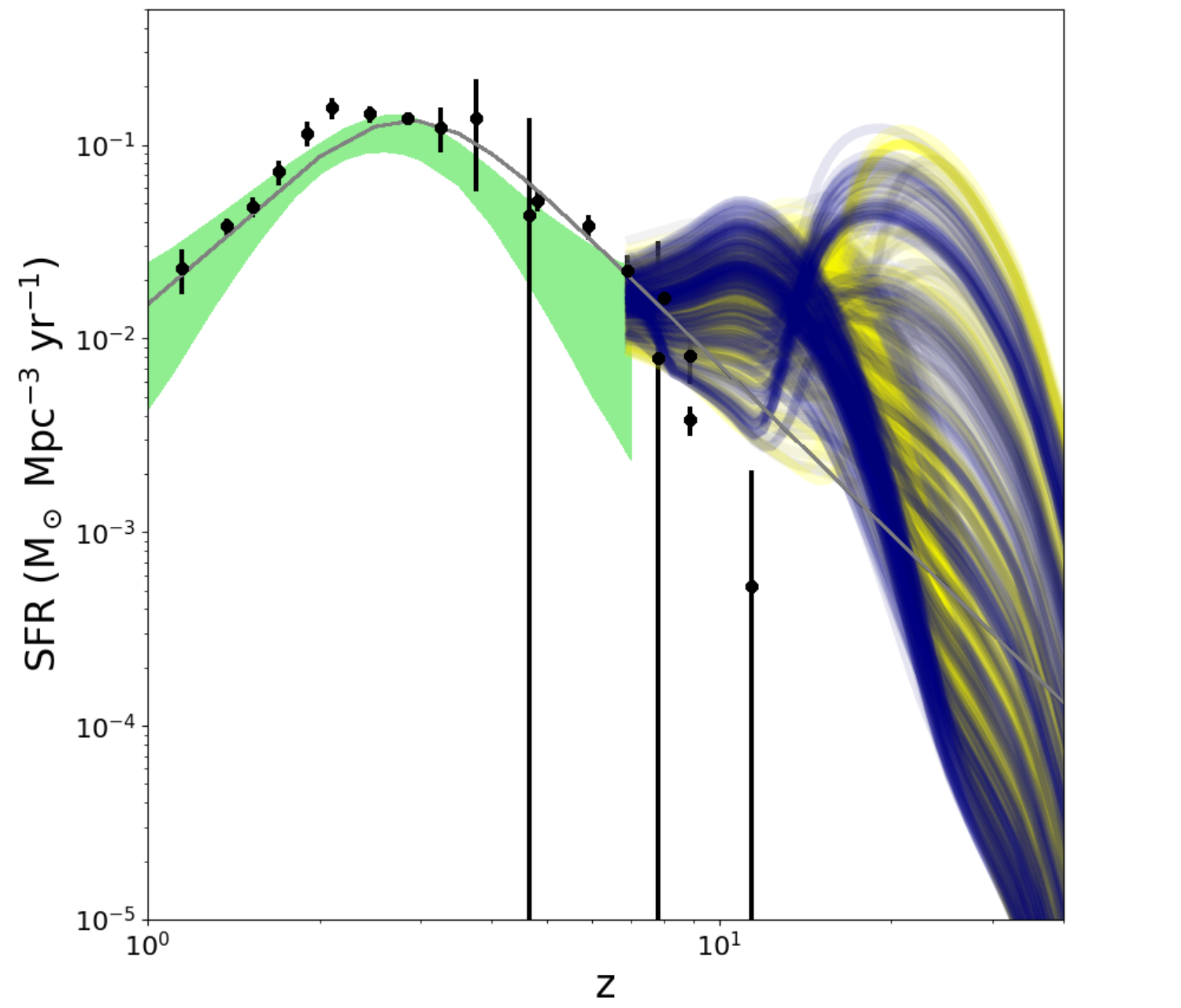}
    \caption{
        {\it Upper panel}\hs{0.2mm}: 
            Fraction of collapsed halos, $f_{\rm col}$, as a function of redshift $z$ for various virial temperatures $T_{\rm vir}$. The solid lines indicate the behaviour for PBH dark matter, while the dashed lines represent ordinary (cold) particle dark matter. 
        {\it Lower panel}\hs{0.2mm}: 
            Star-formation rate (SFR) as a function of $z$. The green band indicates local measurements from extragalactic background light~\cite{2018Sci...362.1031F} and high-redshift surveys~\cite{2020ApJ...902..112B}. Also shown are data points from Reference~\cite{2014ARA&A..52..415M} as well as the respective best fit as a grey continuous line.
        Figures from Reference~\cite{2022ApJ...926..205C}.
        }
    \label{fig:Halo-and-SFRD}
\end{figure}

As pointed out by the authors of Reference~\cite{2022ApJ...926..205C}, deep JWST data will soon allow to explore star formation and early growth of active galactic nuclei up to redshifts around $z = 15$. This will provide an excellent test for PBH dark matter, in particular arising in models in which the thermal history of the Universe shapes the PBH mass function, as first discussed in Reference~\cite{Carr:2019kxo}. Figure~\ref{fig:Halo-and-SFRD} shows how this scenario strongly differs from the standard particle dark matter scenario in two important ways{\,---\,}in the redshift dependence of the fraction of collapsed halos, $f_{\rm col}$ ({\it upper panel}), as well as of the star-formation rate ({\it lower panel}) (see Reference~\cite{2022ApJ...926..205C}).

\begin{figure}[t]
    \centering
    \includegraphics[width=0.72\columnwidth]{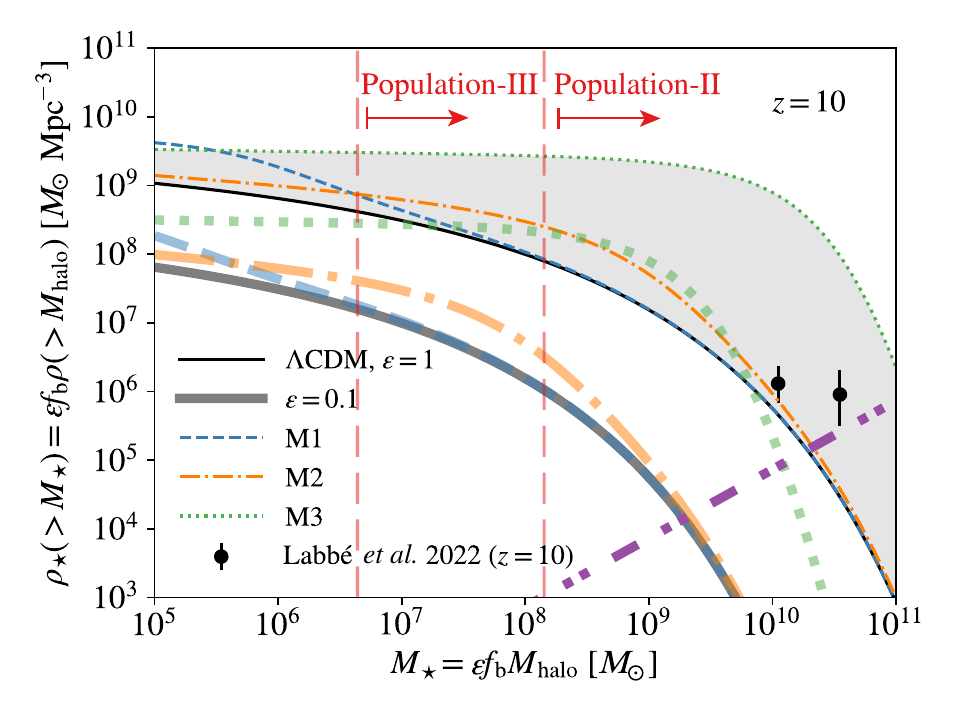}\\[-8.5mm]
    \includegraphics[width=0.72\columnwidth]{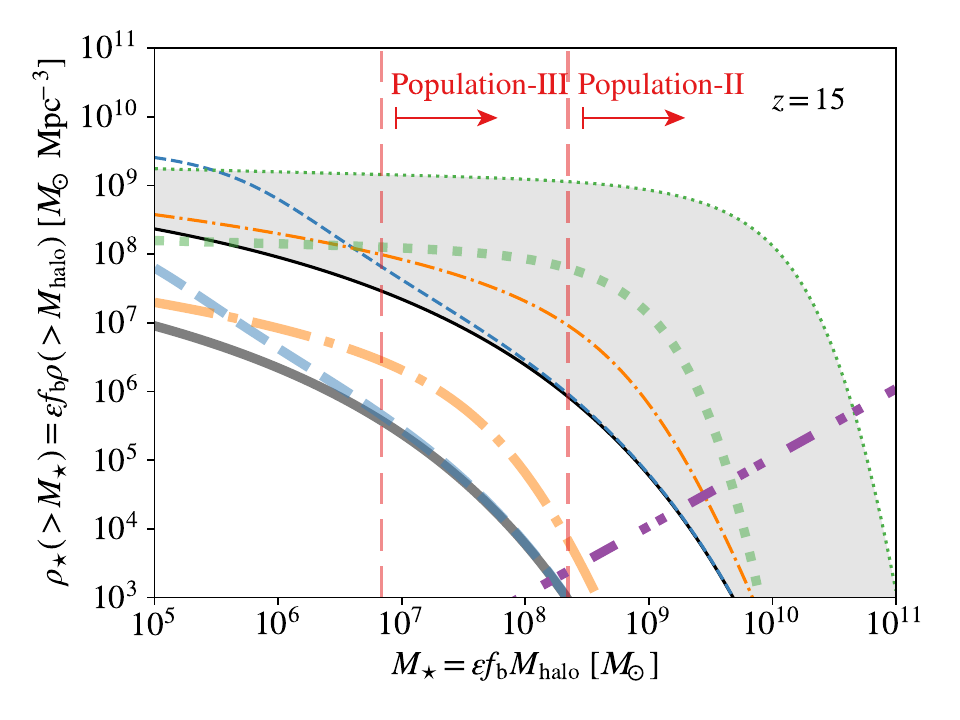}
    \vs{-2mm}
    \caption{
        Cumulative (comoving) stellar mass density, $\rho_{\star}(\.>\!M_{\star} )$, in galaxies with mass exceeding $M_{\star}$ at redshift $z = 10$ ({\it upper panel}) and $z = 15$ ({\it lower panel}). The results for the standard particle dark matter scenario ($\Lambda$CDM) are shown by solid lines, while the PBH models M1, M2 and M3 are depicted by dashed, dashed-dotted and dotted curves, respectively, for $\epsilon = 1$ (thin) and 0.1 (thick). The shaded band denotes the region which can only be populated in PBH cosmologies. The results of Labb{\'e} {\it et al.}~\cite{2023Natur.616..266L}, inferred from JWST observations at redshift $z \sim 10$, are shown by black data points with vertical error bars. The lower mass limits for Population-III and -II galaxies, which are detectable via JWST, are denoted by long-dashed vertical lines. Also plotted (dashed-dotted-dotted line) is the number-count limit of $M_{\star} / V_{\rm com}$, given the comoving volume $V_{\rm com}$ for the CEERS survey. Above, $f_{\brm}$ is the cosmic baryonic mass fraction. Figures (adapted) from Reference~\cite{2022ApJ...937L..30L}.}
    
    \label{fig:smd}
\end{figure}

Recently, Liu \& Bromm~\cite{2022ApJ...937L..30L} investigated the effect of primordial black holes on early massive galaxy formation in view of high-redshift observations with JWST. These have revealed unusually massive galaxy candidates at $z \gtrsim 10$, with inferred stellar masses $\gtrsim 10^{9}\.\Msun$~(see \eg~References~\cite{2023MNRAS.519.1201A, 2022ApJ...940L..55F, 2023ApJS..265....5H, 2022arXiv220802794N, 2023ApJ...942L...9Y, 2023MNRAS.518.6011D}), including similarly (over)massive galaxy candidates detected at $z \simeq 10$ with masses up to $\sim 10^{11}\.\Msun$~\cite{2023Natur.616..266L}. These are rather challenging to reconcile with the expectation from standard particle dark matter scenarios, because the required star-formation efficiency would need to be too high~\cite{2023NatAs...7..731B, 2022ApJ...938L..10I, 2023MNRAS.518.2511L}. Using an analysis based on linear perturbation theory and the Press--Schechter formalism (see Section~\ref{sec:Press---Schechter-Formalism}), Liu \& Bromm demonstrated that the mentioned observed galaxy candidates can be explained with a conceivable lower star-formation efficiency, if structure formation is fostered by PBHs of mass $\gtrsim 10^{9}\.\Msun$ and abundance $f_{\PBH} \sim 10^{-6}\,\text{--}\,10^{-3}$.\footnote{\setstretch{0.9}In a related study, the authors of Reference~\cite{2023PhRvD.107d3502H} find that the tension of standard particle dark matter cosmologies with the recent high-redshift observation is resolved if the dark matter is composed of $4 \times 10^{6}\.(0.005 / \fPBH)\.\Msun$ primordial black hole clusters.}

Figure~\ref{fig:smd} shows their results for the cumulative stellar mass density in galaxies with mass exceeding $M_{\star}$, for the standard particle dark matter ($\Lambda{\rm CDM}$) model (solid) compared to three different (monochromatic) PBH scenarios, assuming different masses and abundances: M1 ($f_{\PBH} = 3 \times 10^{-4}$ and $M = 3 \times 10^{5}\.\Msun$; dashed), M2 ($f_{\PBH} = 10^{-5}$ and $M = 10^{9}\.\Msun$; dashed-dotted) and M3 ($f_{\PBH} = 10^{-4}$ and $M = 10^{10}\.\Msun$; dotted), for $\epsilon = 1$ (thin) and 0.1 (thick) at $z = 10$ ({\it upper panel}) and $z = 15$ ({\it lower panel}). As can clearly be observed, the stellar mass density is significantly increased by PBHs, with the effect being larger at higher redshift, being strongest for halos with masses $M_{\rm halo} \sim 1\,\text{--}\,10\.M$. Considering the area ($\sim 40\ \rm arcmin^{2}$) of the {\it Cosmic Evolution Early Release Science} (CEERS) survey, corresponding to the number-count limits in Figure~\ref{fig:smd} (dashed-dotted-dotted lines) the authors of Reference~\cite{2022ApJ...937L..30L} conclude that the detected galaxies above $10^{8}\.\Msun$ at $z \gtrsim 15$ already require overly-high star-formation efficiencies ($\epsilon \gtrsim 0.1$, for Population-II star formation) in standard particle dark matter cosmologies. On the contrary, PBH-accelerated structure formation only requires an easily achievable (moderate) value lower (for instance $\epsilon \sim 0.01$ in the M3-scenario). 

\begin{figure}
    \centering
    \includegraphics[width = 0.9\textwidth]{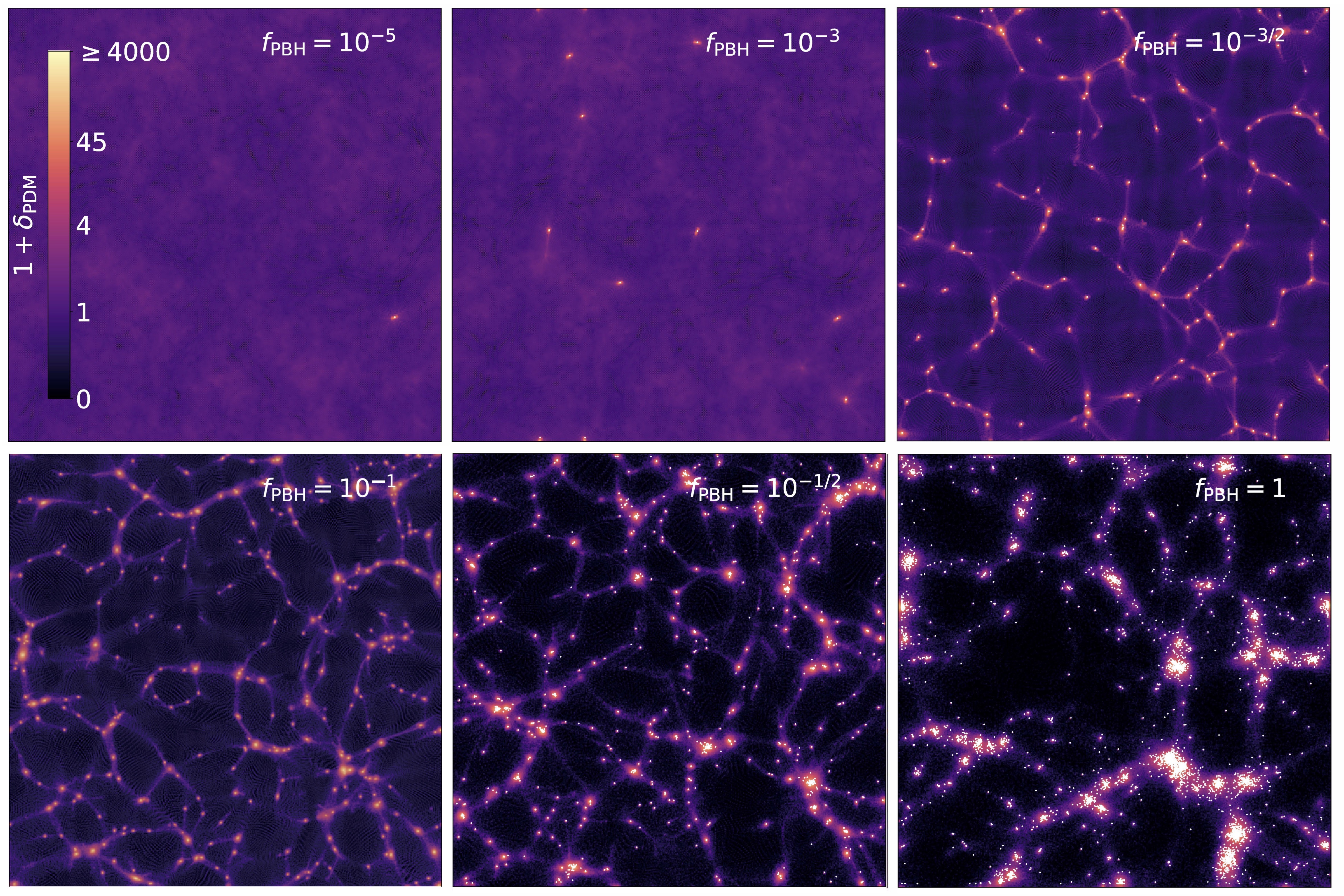}
    \caption{
        Simulation of the matter-density field at redshift $z = 99$ for various primordial black hole dark matter fractions $f_{\PBH}$ in a two-component scenario in which the dark matter complement consists of weakly-interacting massive particles. Its respective density field is represented by the colour map; white points indicate PBH locations. The linear box size is approximately $2\.{\rm kpc}/h$. Figure (adapted) from Reference~\cite{2019PhRvD.100h3528I}.
        }
    \label{fig:PBH-Structure-Formation}
\end{figure}

Using $N$-body simulations starting well within the radiation-dominated era down to redshifts around $100$ (see Figure~\ref{fig:PBH-Structure-Formation}), Inman \& Ali-Ha{\"i}moud~\cite{2019PhRvD.100h3528I} have studied the formation of the first structures in scenarios in which the dark matter consists of two components, PBHs and particles (\cf~Section~\ref{sec:Primordial-Black-Holes-and-Particle-Dark-Matter}).\footnote{\setstretch{0.9}The effects of a sizeable primordial black hole dark matter fraction on the formation of structure in the early Universe has been intensively discussed, for instance in References~\cite{Ivanov:1995uyu, 2019PhRvD.100h3528I, 2021PhRvD.104f3522T, 2021PhRvD.103d3530K, 2021PhRvD.104f3522T, 2022MNRAS.514.2376L, 2023PhRvD.107l3513I}.} Their results depend sensitively on the PBH dark matter fraction. If the latter is less than approximately $10^{-4}$, the PBHs are found to be generally isolated from one another and acquire particle dark matter halos. For larger values of $f_{\PBH}$, clustering of PBHs amongst themselves becomes relevant, where the halo masses are found to be nearly linearly proportional to the PBH number density with their halo-mass function being described by Poisson statistics.

\newpage

Key findings of Reference~\cite{2019PhRvD.100h3528I} are that isolated halos which contain only a single PBH form steep power-law dark matter distributions, regardless of the value of $f_{\PBH}$. On the contrary, when the halos contain many virialised PBHs, such steep profiles do not occur. Furthermore, while in the absence of PBHs, the first stars are expected to typically form when the halos reach a mass of $10^{6}\.\Msun$~\cite{2012Natur.487...70V}, the presence of PBHs encompasses the "{\it possibility that the first lights in the Universe turn on earlier}". Kashlinsky~\cite{2016ApJ...823L..25K} has analytically studied such PBH-induced modifications to early star formation and found that these could explain the cosmic infrared-background fluctuations which cannot be understood via observed galaxies.

\begin{figure}
    \centering
    \vs{-2mm}
    \includegraphics[width = 0.9\textwidth]{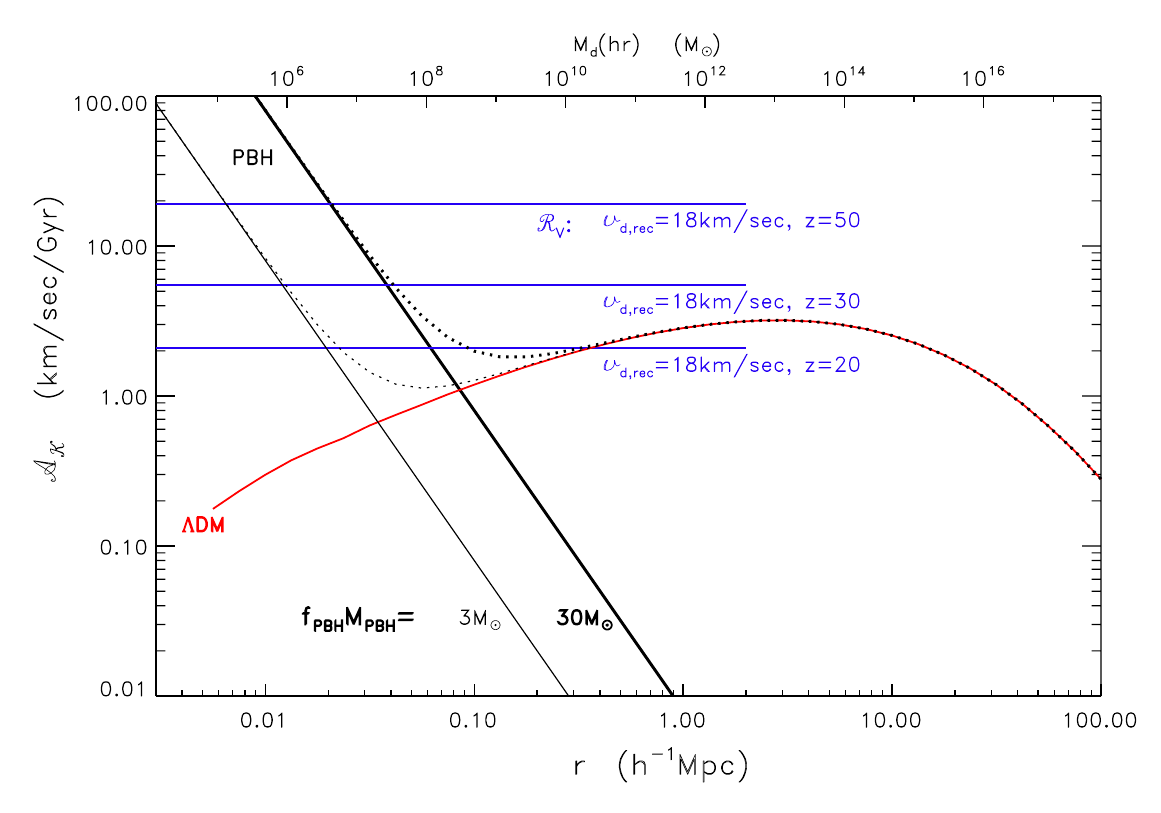}
    \vs{-2mm}
    \caption{
        Advection rate $\Acal_{\Kcal}$ as a function of scale for the individual particle (red solid) and primordial black hole (black solid) dark matter components; dotted lines depict the total rates. The (reduction) rate $\Rcal_{V}$, at which the cosmic expansion equalises the baryon dark matter velocities at recombination, $v_{\drm,\.{\rm rec}}$, is shown in blue colour for given values of $v_{\drm,\.{\rm rec}}$ at three different redshifts. Figure from Reference~\cite{2021PhRvL.126a1101K}.
        \vs{2mm}
        }
    \label{fig:Advection}    
\end{figure}

In a recent study~\cite{2021PhRvL.126a1101K} of cosmological advection processes at redshifts $z \simeq 100$, arising at second order from PBHs dark matter, Kashlinsky has shown that these processes foster early formation of compact objects. This, in turn, makes it easier to explain the existence of supermassive black holes observed in quasars at redshifts $z > 7$. Figure~\ref{fig:Advection} shows the obtained net advection rate $\Acal_{\Kcal}$ as a function of scale. Opposed to the standard cosmological particle dark matter model, in PBH dark matter scenarios, $\Acal_{\Kcal}$ possesses a minimum for dark matter mass scales $\lesssim 10^{9}\.\Msun$ and subsequently rises to a maximum around $10^{12}\.\Msun$. This is likely to have important consequences for early galaxy formation. The results of Reference~\cite{2021PhRvL.126a1101K} have recently been confirmed and further investigated by Atrio-Barandela~\cite{2022ApJ...939...69A}.

Summarised, recent high-redshift observations already appear to strongly hint at primordial black hole cosmologies. As pointed out in Reference~\cite{2022ApJ...937L..30L}, the mentioned results by Liu \& Bromm are also supported by observations of (proto-) galaxy clusters which show an excess of strong-lensing sources~\cite{2020Sci...369.1347M, 2022A&A...668A.188M} and star formation~\cite{2023ApJ...950..191R}, both of which are challenging to explain in particle dark matter cosmologies. This hence further substantiates the paradigm shift from {\it micro-} to {\it macroscopic} dark matter, which has already been numerously heralded. That road will be further paved by JWST, Euclid and the Square Kilometre Array, as well as the Einstein Telescope and the Laser Interferometer Space Antenna.
\newpage

\section{Conclusions}
\label{sec:Conclusions}
\vs{-3mm}
\lettrine[lines=3, slope=0em, findent=0em, nindent=0.2em, lhang=0.1, loversize=0.1]{P}{} rimordial black holes may be regarded as the most natural and most plausible of all dark matter candidates. They emerge organically{\,---\,}{\it a priori} without the need for additional degrees of freedom beyond standard inflationary cosmology. Remarkably, the thermal history of the Universe naturally imprints pronounced peaks in their mass function at scales at which there already is a plethora of strong observational hints for their existence. These many strands of observations might already be sufficient to claim firm evidence for detection. The coming years will shed enough light onto the dark matter to decisively reveal its potential macroscopic nature. This will then open the new area of using primordial black holes as probes for the conditions present during a fraction of a second after the Big Bang, which is unavailable by any other observation or experiment, thereby allowing us to study the physics close to the very birth of our Universe.
\newpage

\section{Acknowledgements}
\vs{-3mm}
\lettrine[lines=3, slope=0em, findent=0em, nindent=0.2em, lhang=0.1, loversize=0.1]{I}{} t is a pleasure to thank Adam Brown, S{\'e}bastien Clesse, Gia Dvali, Juan Garc{\'i}a-Bellido, {Ariel Goobar, G{\"u}nther Hasinger, David Kaiser, Alexander Kashlinsky, }Alexander Kusenko, Ranjan Laha, Andrew Miller, Marc Moniez, Shi Pi, James Rich, Gerasimos Rigopoulos, Teruaki Suyama, Vincent Vennin, Shuichiro Yokoyama and Michael Zantedeschi for helpful comments and insightful remarks. We are most indebted to Bernard Carr and Gia Dvali for showing us the way. A.\,E.~acknowledges support from the Belgian Francqui Foundation and the JSPS Postdoctoral Fellowships for Research in Japan (Graduate School of Sciences, Nagoya University). Y.\,T.~is supported by JSPS KAKENHI Grant No.~JP21K13918.
\newpage

\section{Acronyms}

\begin{tabular}{l@{\hskip 4mm}l}

    AH
        & {\bf a}pparent {\bf h}orizon
    \\[0.8mm]
    ALIA
        & {\bf A}dvanced {\bf L}aser {\bf I}nterferometer {\bf A}ntenna
    \\[0.8mm]
    aLIGO
        & {\bf a}dvanced {\bf L}aser {\bf I}nterferometer {\bf G}ravitational-Wave {\bf O}bservatory
    \\[0.8mm]
    AMEGO
        & {\bf A}ll-sky {\bf M}edium {\bf E}nergy {\bf G}amma-ray {\bf O}bservatory
    \\[0.8mm]
    AR
        & {\bf a}coustic {\bf r}eheating
    \\[0.8mm]
    ATHENA
        & {\bf A}dvanced {\bf T}elescope for {\bf H}igh {\bf En}ergy {\bf A}strophysics
    \\[0.8mm]
    BBH
        & {\bf b}inary {\bf b}lack {\bf h}ole
    \\[0.8mm]
    BBN
        & {\bf b}ig {\bf b}ang {\bf n}ucleosynthesis
    \\[0.8mm]
    BBO
        & {\bf B}ig {\bf B}ang {\bf O}bserver
    \\[0.8mm]
    CE
        & {\bf C}osmic {\bf E}xplorer
    \\[0.8mm]
    CEERS
        & {\bf C}osmic {\bf E}volution {\bf E}arly {\bf R}elease {\bf S}cience
    \\[0.8mm]
    CL
        & {\bf c}onfidence {\bf l}evel
    \\[0.8mm]
    CHE
        & {\bf c}lose {\bf h}yperbolic {\bf e}ncounters
    \\[0.8mm]
    CMB
        & {\bf c}osmic {\bf m}icrowave {\bf b}ackground
    \\[0.8mm]
    DECIGO
        & {\bf DEC}i-Hertz {\bf I}nterferometer {\bf G}ravitational Wave {\bf O}bservatory
    \\[0.8mm]
    DF
        & {\bf d}ynamical {\bf f}riction
    \\[0.8mm]
    DH
        & {\bf d}isk {\bf h}eating
    \\[0.8mm]
    DM
        & {\bf d}ark {\bf m}atter
    \\[0.8mm]
    dS
        & {\bf d}e {\bf S}itter
    \\[0.8mm]
    EGB
        & {\bf e}xtragalactic {\bf g}amma-ray {\bf b}ackground
    \\[0.8mm]
    eLISA
        & {\bf e}volved {\bf L}aser {\bf I}nterferometer {\bf S}pace {\bf A}ntenna
    \\[0.8mm]
    EM
        & {\bf E}ROS and {\bf M}ACHO collaborations
    \\[0.8mm]
    EPTA
        & {\bf E}uropean {\bf P}ulsar {\bf T}iming {\bf A}rray
    \\[0.8mm]
    EROS
        & {\bf E}xp{\'e}rience pour la {\bf R}echerche d'{\bf O}bjets {\bf S}ombres
    \\[0.8mm]
    eROSITA
        & {\bf e}xtended {\bf Ro}entgen {\bf S}urvey with an {\bf I}maging {\bf T}elescope\\
        & {\bf A}rray
    \\[0.8mm]
    ET
        & {\bf E}instein-{\bf T}elescope
    \\[0.8mm]
    FAR
        & {\bf f}alse-{\bf a}larm {\bf r}ate
    \\[0.8mm]
    FLRW
        & {\bf F}riedmann--{\bf L}ema{\^i}tre--{\bf R}obertson--{\bf W}alker
    \\[0.8mm]
    FPT
        & {\bf f}irst {\bf p}assage {\bf t}ime 
    \\[0.8mm]
    GC
        & {\bf G}alactic {\bf c}entre
    \\[0.8mm]
    GW
        & {\bf g}raviational {\bf w}ave
    \\[0.8mm]

\end{tabular}
\newpage

\begin{tabular}{l@{\hskip 4mm}l}

    GWTC
        & {\bf G}raviational-{\bf W}ave {\bf T}ransient {\bf C}atalog
    \\[0.8mm]
    HSC
        & {\bf H}yper {\bf S}uprime-{\bf C}amera
    \\[0.8mm]
    HYK
        & {\bf H}arada-{\bf Y}oo-{\bf K}ohri
    \\[0.8mm]
    IG
        & {\bf i}nter{\bf g}alactic
    \\[0.8mm]
    IL
        & {\bf i}incredulity {\bf l}imit
    \\[0.8mm]
    IPTA
        & {\bf I}nternational {\bf P}ulsar {\bf T}iming {\bf A}rray
    \\[0.8mm]
    IR
        & {\bf i}nfra{\bf r}ed
    \\[0.8mm]
    JWST
        & {\bf J}ames {\bf W}ebb {\bf S}pace {\bf T}elescope
    \\[0.8mm]
    KAGRA
        & {\bf Ka}mioka {\bf Gra}vitational Wave Detector\\[0.5mm]
    LIGO
        & {\bf L}aser {\bf I}nterferometer {\bf G}ravitational-Wave {\bf O}bservatory
    \\[0.8mm]
    LISA
        & {\bf L}aser {\bf I}nterferometer {\bf S}pace {\bf A}ntenna
    \\[0.8mm]
    LVK
        & {\bf L}IGO--{\bf V}irgo--{\bf K}AGRA
    \\[0.8mm]
    LSS
        & {\bf l}arge-{\bf s}cale {\bf s}tructure
    \\[0.8mm]
    LSST
        & {\bf L}arge {\bf S}ynoptic {\bf S}urvey {\bf T}elescope
    \\[0.8mm]
    MACHO
        & {\bf ma}ssive {\bf c}ompact {\bf h}alo {\bf o}bject
    \\[0.8mm]
    NANOGrav
        & {\bf N}orth {\bf A}merican {\bf N}anohertz {\bf O}bservatory for {\bf Grav}itational\\
        & Waves
    \\[0.8mm]
    NFW
        & {\bf N}avarro-{\bf F}renk-{\bf W}hite
    \\[0.8mm]
    OGLE
        & {\bf O}ptical {\bf G}ravitational {\bf L}ensing {\bf E}xperiment
    \\[0.8mm]
    PBH
        & {\bf p}rimordial {\bf b}lack {\bf h}ole
    \\[0.8mm]
    POINT-AGAPE
        & {\bf P}ixel-lensing {\bf O}bservations with the {\bf I}saac {\bf N}ewton\\
        &{\bf T}elescope-{\bf A}ndromeda {\bf G}alaxy {\bf A}mplified {\bf P}ixels\\
        & {\bf E}xperiment
    \\[0.8mm]
    QCD
        & {\bf q}uantum {\bf c}hromo {\bf d}ynamics
    \\[0.8mm]
    SFR
        & {\bf s}tar-{\bf f}ormation {\bf r}ate
    \\[0.8mm]
    SLABs
        & {\bf S}tupendously {\bf LA}rge {\bf B}lack holes
    \\[0.8mm]
    SKA
        & {\bf S}quare {\bf K}ilometer {\bf A}rray
    \\[0.8mm]
    SN
        & {\bf s}uper{\bf n}ova
    \\[0.8mm]
    SNR
        & {\bf s}ignal-to-{\bf n}oise {\bf r}atio
    \\[0.8mm]
    UCMHs
        & {\bf u}ltra-{\bf c}ompact {\bf m}ini-{\bf h}alos
    \\[0.8mm]
    UV
        & {\bf u}ltra{\bf v}iolet
    \\[0.8mm]
    WB
        & {\bf w}ide {\bf b}inaries
    \\[0.8mm]
    WFIRST
        & {\bf W}ide {\bf F}ield {\bf I}nfra{\bf r}ed {\bf S}urvey {\bf T}elescope
    \\[0.8mm]
    WIMP
        & {\bf w}eakly-{\bf i}nteracting {\bf m}assive {\bf p}article
    \\[0.8mm]
    XB
        & {\bf X}-ray {\bf b}inaries
\end{tabular}
\newpage

\interlinepenalty=10000 
\section{Bibliography}
\setlength{\bibsep}{6pt}
\bibliography{PBH-References}

\end{document}

%% file: PBH-header.tex
\usepackage{siunitx}
\usepackage{physics}
\usepackage{braket}

\usepackage{amsmath}	
\usepackage{amsthm}		
\usepackage{amssymb}	

\usepackage{mathrsfs, mathtools}

\usepackage{letltxmacro} 

\usepackage{color}
\usepackage{bm}

\usepackage{tgpagella}

\usepackage{graphicx}

\usepackage{fancyhdr}

\usepackage{lastpage}

\usepackage{titlesec}

\usepackage{rotating}
\usepackage{setspace}

\usepackage[printonlyused]{acronym}
\usepackage{aas_macros}

\usepackage{datetime}

\usepackage{lettrine}

\usepackage{anyfontsize}

\usepackage[colorlinks=true
,urlcolor=DARKBLUE
,anchorcolor=DARKBLUE
,citecolor=DARKBLUE
,filecolor=DARKBLUE
,linkcolor=DARKBLUE
,menucolor=DARKBLUE
]{hyperref}

\titleformat{\section}
       {\centering\normalfont\fontsize{14}{17}\bfseries}{\thesection}{1em}{}

\numberwithin{equation}{section}

\makeatletter 
    \renewcommand{\thefigure}{{\bf\@arabic\c@figure}}
\makeatother

\usepackage{tcolorbox}

\definecolor{grey}{rgb}{0.4,0.4,0.4}
\definecolor{dullmagenta}{rgb}{0.4,0,0.4}
\definecolor{darkblue}{rgb}{0,0,0.4}
\definecolor{midblue}{rgb}{0,0,0.7}
\definecolor{midred}{rgb}{0.5,0,0}
\definecolor{orange}{rgb}{1,0.5,0}
\definecolor{lightbrown}{rgb}{0.75,0.5,0.25}
\definecolor{tan}{cmyk}{0.14,0.42,0.56,0}
\definecolor{djunglegreen}{cmyk}{0.99,0,0.52,0}
\definecolor{lightgreen}{rgb}{0,1,0}
\definecolor{olivegreen}{cmyk}{0.64,0,0.95,0.40}
\definecolor{midgreen}{rgb}{0.0,0.675,0.0}
\definecolor{darkgreen}{rgb}{0,0.5,0}
\definecolor{pink}{rgb}{1,0.078,0.57}

\definecolor{MONZA}{HTML}{CF000F}
\definecolor{DARKBLUE}{HTML}{00008b}
\definecolor{DARKMAGENTA}{HTML}{8b008b}

\newcommand{\vs}{\vspace}
\newcommand{\hs}{\hspace}
\renewcommand{\.}{\hspace{0.5mm}}

\makeatletter
\let\oldr@@t\r@@t
\def\r@@t#1#2{%
\setbox0=\hbox{$\oldr@@t#1{#2\,}$}\dimen0=\ht0
\advance\dimen0-0.2\ht0
\setbox2=\hbox{\vrule height\ht0 depth -\dimen0}%
{\box0\lower0.4pt\box2}}
\LetLtxMacro{\oldsqrt}{\sqrt}
\renewcommand*{\sqrt}[2][\ ]{\oldsqrt[#1]{#2}}
\makeatother

\newcommand{\la}{\ensuremath{\leftarrow}}

\newcommand{\Arm}{\ensuremath{\mathrm{A}}}
\newcommand{\Brm}{\ensuremath{\mathrm{B}}}

\newcommand{\Erm}{\ensuremath{\mathrm{E}}}
\newcommand{\Frm}{\ensuremath{\mathrm{F}}}
\newcommand{\Grm}{\ensuremath{\mathrm{G}}}

\newcommand{\Lrm}{\ensuremath{\mathrm{L}}}

\newcommand{\Nrm}{\ensuremath{\mathrm{N}}}

\newcommand{\Prm}{\ensuremath{\mathrm{P}}}

\newcommand{\Srm}{\ensuremath{\mathrm{S}}}

\newcommand{\Wrm}{\ensuremath{\mathrm{W}}}

\newcommand{\brm}{\ensuremath{\mathrm{b}}}
\newcommand{\crm}{\ensuremath{\mathrm{c}}}
\newcommand{\drm}{\ensuremath{\mathrm{d}}}
\newcommand{\erm}{\ensuremath{\mathrm{e}}}
\newcommand{\frm}{\ensuremath{\mathrm{f}}}
\newcommand{\grm}{\ensuremath{\mathrm{g}}}
\newcommand{\hrm}{\ensuremath{\mathrm{h}}}
\newcommand{\irm}{\ensuremath{\mathrm{i}}}

\newcommand{\mrm}{\ensuremath{\mathrm{m}}}

\newcommand{\prm}{\ensuremath{\mathrm{p}}}

\newcommand{\rrm}{\ensuremath{\mathrm{r}}}
\newcommand{\srm}{\ensuremath{\mathrm{s}}}
\newcommand{\trm}{\ensuremath{\mathrm{t}}}

\newcommand{\wrm}{\ensuremath{\mathrm{w}}}

\newcommand{\Acal}{\ensuremath{\mathcal{A}}}

\newcommand{\Ccal}{\ensuremath{\mathcal{C}}}

\newcommand{\Fcal}{\ensuremath{\mathcal{F}}}

\newcommand{\Hcal}{\ensuremath{\mathcal{H}}}

\newcommand{\Kcal}{\ensuremath{\mathcal{K}}}
\newcommand{\Lcal}{\ensuremath{\mathcal{L}}}
\newcommand{\Mcal}{\ensuremath{\mathcal{M}}}
\newcommand{\Ncal}{\ensuremath{\mathcal{N}}}
\newcommand{\Ocal}{\ensuremath{\mathcal{O}}}
\newcommand{\Pcal}{\ensuremath{\mathcal{P}}}

\newcommand{\Rcal}{\ensuremath{\mathcal{R}}}
\newcommand{\Scal}{\ensuremath{\mathcal{S}}}

\newcommand{\Rbb}{\ensuremath{\mathbb{R}}}

\newcommand{\kbm}{\ensuremath{\bm{k}}}

\newcommand{\qbm}{\ensuremath{\bm{q}}}

\newcommand{\wbm}{\ensuremath{\bm{w}}}
\newcommand{\xbm}{\ensuremath{\bm{x}}}
\newcommand{\ybm}{\ensuremath{\bm{y}}}

\renewcommand{\d}{\ensuremath{\mathrm{d}}}

\newcommand{\sign}{\ensuremath{\mathrm{sign}}}

\newcommand{\GeV}{\ensuremath{\mathrm{GeV}}}

\newcommand{\eg}{e.g.}
\newcommand{\ie}{i.e.}

\newcommand{\cf}{cf.}
\newcommand{\be}{\begin{equation}}
\newcommand{\ee}{\end{equation}}
\newcommand{\ba}{\begin{eqnarray}}
\newcommand{\ea}{\end{eqnarray}}

\smallskipamount
\settimeformat{ampmtime}

\def\ga{\mathrel{\raise.3ex\hbox{$>$\kern-.75em\lower1ex\hbox{$\sim$}}}}
\def\la{\mathrel{\raise.3ex\hbox{$<$\kern-.75em\lower1ex\hbox{$\sim$}}}}

\def\Msun{M_\odot}

\def\fPBH{f_{\rm PBH}}

\newcommand{\sv}{\langle\sigma v \rangle}

\newcommand{\com}{\Ccal}

\newcommand{\Mpl}{M_\mathrm{Pl}}
\newcommand{\dS}{\mathrm{dS}}
\newcommand{\ns}{n_\mathrm{s}}
\newcommand{\cs}{c_\mathrm{s}}
\newcommand{\CMB}{\mathrm{CMB}}
\newcommand{\NL}{\mathrm{NL}}
\newcommand{\IR}{\mathrm{IR}}
\newcommand{\UV}{\mathrm{UV}}
\newcommand{\FP}{\mathrm{FP}}
\newcommand{\FPT}{\mathrm{FPT}}
\newcommand{\bw}{\mathrm{bw}}
\newcommand{\lin}{\mathrm{lin}}
\newcommand{\uth}{\mathrm{th}}
\newcommand{\eff}{\mathrm{eff}}
\newcommand{\PBH}{\mathrm{PBH}}
\newcommand{\GW}{\mathrm{GW}}
\newcommand{\eq}{\mathrm{eq}}
\newcommand{\umax}{\mathrm{max}}
\newcommand{\umin}{\mathrm{min}}
\newcommand{\DM}{\mathrm{DM}}
\newcommand{\vir}{\mathrm{vir}}

\DeclareMathOperator{\sinc}{sinc}
\DeclareMathOperator{\erfc}{erfc}

\makeatletter
\def\l@subsubsection#1#2{}
\makeatother

\headwidth=\textwidth

